\documentclass[12pt]{article}

\usepackage{amsmath,bbm,latexsym,epsfig}

\usepackage{amsmath,bbm,latexsym}
\usepackage{epsfig}
\usepackage{epstopdf}

\newcommand{\be}{\begin{equation}}
\newcommand{\ee}{\end{equation}}
\newcommand{\ba}{\begin{eqnarray}}
\newcommand{\ea}{\end{eqnarray}}
\newcommand{\bs}{\begin{subequations}}
\newcommand{\es}{\end{subequations}}
\newcommand{\no}{\nonumber\\}
\newcommand{\ra}{\rightarrow}
\newcommand{\etal}{$\!${\it et al.}\ }
\newcommand{\etall}{$\!${\it et al.}}
\newcommand{\ie}{{\it i.e.}\ }
\newcommand{\viz}{{\it viz.}\ }

\newcommand{\cf}{{\it cf.}\ }

\newcommand{\imag}{\mbox{Im}}
\newcommand{\real}{\mbox{Re}}
\newcommand{\tr}{\mbox{tr}}
\newcommand{\grts}{\raise.3ex\hbox{$>$\kern-.75em\lower1ex\hbox{$\sim$}}}
\newcommand{\lets}{\raise.3ex\hbox{$<$\kern-.75em\lower1ex\hbox{$\sim$}}}
\newcommand{\ls}{{\cal L}}
\newcommand{\hc}{\mbox{H.c.}}
\newcommand{\sfrac}[2]{{\textstyle \frac{#1}{#2}}}

\newcommand{\sud}{\mbox{SU(2)}}
\newcommand{\sot}{\mbox{SO(3)}}
\newcommand{\sod}{\mbox{SO(2)}}
\newcommand{\od}{\mbox{O(2)}}

\newcommand*{\tvec}[1]{\ensuremath{\boldsymbol{\mathrm{#1}}}}
\newcommand*{\trans}{\mathrm{T}}

\begin{document}

\title{
\LARGE Theory and phenomenology \\ of two-Higgs-doublet models}

\author{
\large
G.~C.~Branco,$^{(1)}$\thanks{E-mail: gbranco@ist.utl.pt } \
P.~M.~Ferreira,$^{(2,3)}$\thanks{E-mail: ferreira@cii.fc.ul.pt} \
L.~Lavoura,$^{(1)}$\thanks{E-mail: balio@cftp.ist.utl.pt} \
\\
\large M.~N.~Rebelo,$^{(1)}$\thanks{E-mail: rebelo@ist.utl.pt} \
Marc Sher,$^{(4)}$\thanks{E-mail: mtsher@wm.edu} \
and Jo\~{a}o P.~Silva$\, ^{(2,5)}$\thanks{E-mail: jpsilva@cftp.ist.utl.pt}
\\*[5mm]
\small $^{(1)}$ Departamento de F\'\i sica and Centro de F\'\i sica Te\'orica de Part\'\i culas,\\
\small Instituto Superior T\'ecnico, Technical University of Lisbon \\
\small 1049-001 Lisboa, Portugal
\\*[2mm]
\small $^{(2)}$ Instituto Superior de Engenharia de Lisboa \\
\small 1959-007 Lisboa, Portugal
\\*[2mm]
\small $^{(3)}$ Centro de F\'\i sica Te\'orica e Computacional,
University of Lisbon  \\
\small 1649-003 Lisboa, Portugal
\\*[2mm]
\small $^{(4)}$ High Energy Theory Group, College of William and Mary\\
\small Williamsburg, Virginia 23187, U.S.A.
\\*[2mm]
\small $^{(5)}$ Centro de F\'\i sica Te\'orica de Part\'\i culas,\\
\small Instituto Superior T\'ecnico, Technical University of Lisbon \\
\small 1049-001 Lisboa, Portugal
}

\date{\today}

\maketitle

\begin{abstract}
We discuss theoretical and phenomenological aspects of two-Higgs-doublet
extensions of the Standard Model.  In general, these extensions have scalar
mediated flavour changing neutral currents which are strongly constrained
by experiment.   Various strategies are discussed to control these flavour changing
scalar currents and their phenomenological consequences are analysed.
In particular, scenarios with natural flavour conservation are investigated,
including the so-called type I and type II models as well as lepton-specific
and inert
models.  Type III models are then discussed, where scalar flavour changing
neutral currents are present at tree level, but are suppressed by either
specific ansatze for the Yukawa couplings or by the introduction of family
symmetries leading to a natural suppression mechanism.
We also consider the phenomenology of charged scalars in these models.
Next we turn to the role of symmetries in the
scalar sector. We discuss the six symmetry-constrained
scalar potentials and their extension into the fermion sector.
The vacuum structure of the scalar potential is analysed, including a study
of the vacuum stability conditions on the potential and  the
renormalization-group improvement
of these conditions is also presented.
The stability of the tree level minimum of the scalar potential in connection
with  electric charge conservation and its behaviour under CP is analysed.
The question of CP violation is addressed in detail,  including the cases
of explicit CP violation and spontaneous CP violation. We present a detailed
study of weak basis invariants which are odd under CP. These invariants allow
for the possibility of studying the CP properties of any two-Higgs-doublet
model in an arbitrary Higgs basis. A careful study of spontaneous CP violation
is presented, including an analysis of the conditions which have to be
satisfied in order for a vacuum to violate CP. We present minimal models of CP
violation where the vacuum phase is sufficient to generate a complex CKM
matrix, which is at present a requirement for any  realistic model
of spontaneous CP violation.
\end{abstract}

\tableofcontents

\newpage

\section{Introduction}
\label{sec:int}

The gauge boson and fermion sectors
of the Standard Model of the electroweak interactions
have been extremely well probed phenomenologically;
yet,
its scalar sector has not yet been directly explored.
In the Standard Model (SM)
the simplest possible scalar structure---just one SU(2) doublet---is
assumed~\cite{11883,50073,12291,12292,51165};
on the contrary,
the fermion structure,
with more than one family and with family mixing,
is not simple at all.

One critical piece of evidence
about the scalar structure is the parameter~$\rho$.
In the SU(2)$\times$U(1) gauge theory, if there are
$n$ scalar multiplets $\phi_i$,
with weak isospin $I_i$,
weak hypercharge $Y_i$,
and vacuum expectation value (vev) of the neutral components $v_i$,
then the parameter~$\rho$ is,
at tree level~\cite{Langacker:1980js},
\be
\rho = \frac{{\displaystyle \sum_{i=1}^n} \left[
I_i \left( I_i+1 \right) - \frac{1}{4}\, Y_i^2 \right] v_i}
{{\displaystyle \sum_{i=1}^n}\, \frac{1}{2}\, Y_i^2 v_i}.
\label{jduei}
\ee
Experimentally~\cite{Nakamura:2010zzi} $\rho$ is very close to one.
According to eq.~(\ref{jduei}),
both SU(2) singlets with $Y = 0$ and SU(2) doublets with $Y = \pm 1$
give $\rho=1$,
since they both have $I \left( I+1 \right) = \frac{3}{4}\, Y^2$.
Other scalars with vevs in much larger SU(2) multiplets,
scalars with small or null vevs,
and models with triplets
and a custodial SU(2) global symmetry~\cite{Chanowitz:1985ug},
are compatible with $\rho = 1$;
but such scalar sectors tend to be large and complex---the simplest extension
of the SM consists in simply adding scalar doublets and singlets.

In this review we focus on one of the simplest possible extensions
of the SM---the two-Higgs-doublet model (2HDM)~\cite{Lee:1973iz}.
There are many motivations for 2HDMs.
The best known motivation is supersymmetry~\cite{Haber:1984rc}.
In supersymmetric theories
the scalars belong to chiral multiplets
and their complex conjugates belong to multiplets of the opposite chirality;
since multiplets of different chiralities
cannot couple together in the Lagrangian,
a single Higgs doublet is unable to give mass simultaneously
to the charge $2/3$ and charge $-1/3$ quarks.
Moreover,
since scalars sit in chiral multiplets together with chiral spin-1/2 fields,
the cancellation of anomalies also requires that
an additional doublet be added.
Thus,
the Minimal Supersymmetric Standard Model (MSSM)
contains two Higgs doublets.

Another motivation for 2HDMs comes from axion models~\cite{Kim:1986ax}.
Peccei and Quinn~\cite{Peccei:1977hh} noted that
a possible CP-violating term in the QCD Lagrangian,
which is phenomenologically known to be very small,
can be rotated away if the Lagrangian contains a global U(1) symmetry.
However,
imposing this symmetry is only possible if there are two Higgs doublets.
While the simplest versions of the Peccei--Quinn model
(in which all the New Physics was at the TeV scale)
are experimentally ruled out,
there are variations with singlets at a higher scale that are acceptable,
and the effective low-energy theory for those models
still requires two Higgs doublets \cite{Kim:1986ax}.

Still another motivation for 2HDMs is the fact that
the SM is unable~\cite{Trodden:1998qg} to generate
a baryon asymmetry of the Universe of sufficient size.
Two-Higgs-doublet models can do so,
due to the flexibility of their scalar mass spectrum \cite{Trodden:1998qg}
and the existence of additional sources of CP violation.
There have been many works on baryogenesis in the
2HDM~\cite{Turok:1990zg,Joyce:1994zt,Funakubo:1993jg,Davies:1994id,
Cline:1995dg,Cline:1996mga,Laine:2000rm,Fromme:2006cm,arXiv:1106.0790,arXiv:1107.3559}.
Exciting new possibilities for explicit or spontaneous CP violation
constitute one of the attractive features of 2HDMs.

With the Large Hadron Collider (LHC) starting to produce data,
time seems appropriate for a review of 2HDMs.
The Higgs sector of the Standard Model is very predictive, with the Higgs
mass being the only free parameter, and it will be tested at the LHC over
the entire theoretically preferred mass ranges within the next few months.
In contrast, due to the larger number of free parameters in the 2HDM, it
will take much longer to probe the entire parameter space of the various models.
Should the Higgs not be seen at the LHC in the next few months, the 2HDM will
be one of the simplest alternatives. With charged Higgs bosons, pseudoscalars
and different decay modes and branching ratios, the experimental challenges will
be quite different than in the Standard Model. While it may not be possible
to completely probe the entire parameter space of the various 2HDMs at the LHC,
most of the parameter space can be probed, and this is further incentive for a
review of the various forms of the 2HDM and their experimental signatures.

We shall explicitly exclude supersymmetric models from this review.
The Higgs sector of supersymmetric models is extremely well-studied
and Djouadi~\cite{Djouadi:2005gj}
has written a very comprehensive review of it.
We shall also not include models with scalar SU(2) singlets
in addition to the two doublets,
since those models usually include many additional parameters.

In general,
the vacuum structure of 2HDMs is very rich.
The most general scalar potential contains 14 parameters
and can have CP-conserving,
CP-violating,
and charge-violating minima.
In writing that potential one must be careful in defining the various bases
and in distinguishing parameters which can be rotated away
from those which have physical implications.
However,
most phenomenological studies of 2HDMs make several simplifying assumptions.
It is usually assumed that CP is conserved in the Higgs sector
(only then can one distinguish between scalars and pseudoscalars),
that CP is not spontaneously broken,
and that discrete symmetries eliminate from the potential
all quartic terms odd in either of the doublets; however,
usually one considers all possible real quadratic coefficients,
including a term which softly breaks these symmetries.
We shall also make those assumptions in the early chapters of this report
but will subsequently discuss relaxing them.
Under those assumptions,
the most general scalar potential for two doublets $\Phi_1$ and $\Phi_2$
with hypercharge $+1$ is
\ba
V &=&
m^2_{11}\, \Phi_1^\dagger \Phi_1
+ m^2_{22}\, \Phi_2^\dagger \Phi_2 -
 m^2_{12}\, \left(\Phi_1^\dagger \Phi_2 + \Phi_2^\dagger \Phi_1\right)
+ \frac{\lambda_1}{2} \left( \Phi_1^\dagger \Phi_1 \right)^2
+ \frac{\lambda_2}{2} \left( \Phi_2^\dagger \Phi_2 \right)^2
\no & &
+ \lambda_3\, \Phi_1^\dagger \Phi_1\, \Phi_2^\dagger \Phi_2
+ \lambda_4\, \Phi_1^\dagger \Phi_2\, \Phi_2^\dagger \Phi_1
+ \frac{\lambda_5}{2} \left[
\left( \Phi_1^\dagger\Phi_2 \right)^2
+ \left( \Phi_2^\dagger\Phi_1 \right)^2 \right],
\ea
where all the parameters are real. For a region of
parameter space,
the minimization of this potential gives
\be
\langle \Phi_1 \rangle_0
= \left( \begin{array}{c} 0 \\ \displaystyle{\frac{v_1}{\sqrt{2}}} \end{array} \right),
\quad
\langle \Phi_2 \rangle_0
= \left( \begin{array}{c} 0 \\ \displaystyle{\frac{v_2}{\sqrt{2}}} \end{array} \right).
\ee
With two complex scalar SU(2) doublets there are eight fields:
\be
\Phi_a = \left( \begin{array}{c} \phi_a^+ \\
\left(v_a +  \rho_a + i \eta_a \right) \left/ \sqrt{2} \right.
\end{array} \right),
\quad a= 1, 2.
\ee
Three of those get `eaten' to give mass to the $W^\pm$ and $Z^0$ gauge bosons;
the remaining five are physical scalar (`Higgs') fields.
There is a charged scalar,
two neutral scalars,
and one pseudoscalar.
With the above minimum,
the mass terms for the charged scalars are given by
\be
{\mathcal L}_{\phi^\pm\, {\rm mass}} =
\left[ m^2_{12} - (\lambda_4 + \lambda_5)v_1 v_2 \right]
\left( \begin{array}{cc} \phi_1^-, & \phi_2^- \end{array} \right)
\left( \begin{array}{cc} {\displaystyle \frac{v_2}{v_1}} & - 1 \\
- 1 &  {\displaystyle \frac{v_1}{v_2}} \end{array} \right)
\left( \begin{array}{c} \phi_1^+ \\ \phi_2^+ \end{array} \right).
\ee

There is a zero eigenvalue
corresponding to the charged Goldstone boson $G^\pm$
which gets eaten by the $W^\pm$.
The mass-squared of the `charged Higgs' is \\
$m^2_+ = [ m^2_{12} / (v_1 v_2)
- \lambda_4 - \lambda_5 ]
\left( v^2_1+v^2_2 \right)$.
The mass terms for the pseudoscalars are given by
\be
{\mathcal L}_{\eta\, {\rm mass}} =
\frac{m^2_A}{v_1^2 + v_2^2}
\left( \begin{array}{cc} \eta_1, & \eta_2 \end{array} \right)
\left( \begin{array}{cc} v^2_2 & - v_1 v_2 \\
- v_1 v_2 & v^2_1 \end{array} \right)
\left( \begin{array}{c} \eta_1 \\ \eta_2 \end{array} \right).
\ee
This gives a pseudoscalar Goldstone mode
together with the mass-squared of the physical pseudoscalar,
$m^2_A = \left[ m^2_{12} / (v_1 v_2) - 2 \lambda_5 \right]
\left( v_1^2 + v_2^2 \right)$.
Note that,
when $m_{12}^2 = 0$ and $\lambda_5 = 0$,
the pseudoscalar becomes massless.
This is due to the existence,
in that limit,
of an additional global U(1) symmetry which is spontaneously broken.
Finally,
the mass terms for the scalars are given by
\be
{\mathcal L}_{\rho\, {\rm mass}} =
- \left( \begin{array}{cc} \rho_1, & \rho_2 \end{array} \right)
\left( \begin{array}{cc} m^2_{12}
\displaystyle{\frac{v_2}{v_1}} + \lambda_1 v^2_1
& -m^2_{12} + \lambda_{345} v_1 v_2 \\
-m^2_{12} +  \lambda_{345} v_1 v_2 &
m^2_{12} \displaystyle{\frac{v_1}{v_2}}
+ \lambda_2 v^2_2 \end{array} \right)
\left( \begin{array}{c} \rho_1 \\ \rho_2 \end{array} \right),
\ee
with $\lambda_{345} = \lambda_3 + \lambda_4 + \lambda_5 $.
The mass-squared matrix of the scalars can be diagonalized
and the angle $\alpha$ is defined to be the rotation angle
that performs that diagonalization.

Perhaps the single most important parameter
in studies of 2HDMs is
\be
\tan{\beta} \equiv \frac{v_2}{v_1}.
\ee
The angle $\beta$ is the rotation angle
which diagonalizes the mass-squared matrices
of the charged scalars and of the pseudoscalars.
If one redefines the doublets as
$H_1 = \cos{\beta}\, \Phi_1 + \sin{\beta}\, \Phi_2$
and $H_2 = - \sin{\beta}\, \Phi_1 + \cos{\beta}\, \Phi_2$,
one finds that the lower component of $H_1$
has a (real and positive) vev
$v \left/ \sqrt{2} \right.$,
where $v \equiv \left( v_1^2 + v_2^2 \right)^{1/2}$,
while $H_2$ has null vev.
The two parameters $\alpha$ and $\beta$
determine the interactions of the various Higgs fields
with the vector bosons and
(given the fermion masses)
with the fermions;
they are thus crucial in discussing phenomenology.
Still,
one should keep in mind the assumptions that were made in defining them.

In this review,
we shall begin by discussing the phenomenology
of the above restricted version of the 2HDM.
A feature of general 2HDMs
is the existence of tree-level flavour-changing neutral currents (FCNC).
One can avoid these potentially dangerous interactions
by imposing discrete symmetries in several possible ways.
In chapter~\ref{sec:nfc},
the phenomenological analyses of 2HDMs
without tree-level FCNCs are presented,
including decays and production of neutral scalars and pseudoscalars,
bounds from LEP and the Tevatron,
and expectations for the LHC.

If one does not impose discrete symmetries,
then there are tree-level FCNCs in the 2HDM.
A prototype model of this kind,
the so-called type~III 2HDM,
is discussed in chapter~\ref{sec:fcnc},
along with other such models,
such as the Branco--Grimus--Lavoura (BGL) model
and models of Minimal Flavour Violation.

All 2HDMs have charged scalar bosons (`charged Higgses').
An analysis of charged-Higgs production and decay,
for all the 2HDMs of chapter~\ref{sec:nfc},
is presented in chapter~\ref{sec:charg};
this is followed by a discussion of charged-Higgs phenomenology
in models with tree-level FCNC.

In chapter~5 we relax the rather strict assumptions
made in this Introduction and in chapters~\ref{sec:nfc} to~\ref{sec:charg}.
An analysis of the full scalar potential,
including the vacuum structure with or without CP violation and
the possible symmetries one can impose on the potential,
is presented.

As noted earlier,
the 2HDM offers possibilities for new sources of CP violation.
This is analysed in some detail in chapter~\ref{sec:cpv}.
In particular we discuss the weak-basis invariant conditions for the
2HDM Lagrangian to be CP invariant and we present a minimal realistic
extension of the Standard Model, with spontaneous CP violation.

In chapter~\ref{sec:conc} we briefly summarize some of our conclusions.
Some isolated topics that the reader
may want to consult separately---constraints on the parameters
of the potential from unitarity,
renormalization-group running of the parameters of 2HDMs,
and contributions to the oblique parameters
from the scalar sector of 2HDMs---are left to appendices.

\newpage

\section{Models with natural flavour conservation}
\label{sec:nfc}

The most serious potential problem facing
all 2HDMs\footnote{All multi-Higgs-doublet models
in general face this potential problem.}
is the possibility of tree level flavour-changing neutral currents(FCNC).
For example,
the Yukawa couplings of the $Q=-1/3$ quarks will,
in general,
be
\begin{equation}
{\cal L}_Y = y^1_{ij}\bar{\psi}_i\psi_j\Phi_1
+ y^2_{ij}\bar{\psi}_i\psi_j \Phi_2,
\end{equation}
where $i,j$ are generation indices.
The mass matrix is then
\begin{equation}
M_{ij} = y^1_{ij}\frac{v_1}{\sqrt{2}} + y^2_{ij}\frac{v_2}{\sqrt{2}}.
\end{equation}
In the Standard Model,
diagonalizing the mass matrix automatically diagonalizes
the Yukawa interactions,
therefore there are no tree-level FCNC.
In 2HDMs,
however,
in general $y^1$ and $y^2$  will not be simultaneously diagonalizable,
and thus the Yukawa couplings will not be flavour diagonal.
Neutral Higgs scalars $\phi$ will mediate FCNC of the form,
for example,
$\overline{d}s\phi$.

These FCNC can cause severe phenomenological difficulties.
The $\overline{d}s\phi$ interaction,
for example,
will lead to $K$--$\overline{K}$ mixing at tree level.
If the coupling is as large as the b-quark Yukawa coupling,
the mass of the exchanged scalar would have to exceed
10 TeV~\cite{McWilliams:1980kj,Shanker:1981mj}.
Nonetheless,
under reasonable assumptions,
models with these FCNC may still be viable.
They will be discussed in the next chapter.
In this chapter,
however,
we will assume that tree level FCNC are completely absent,
due to a discrete or continuous symmetry.

It is easy to see that
if all fermions with the same quantum numbers
(which are thus capable of mixing)
couple to the same Higgs multiplet,
then FCNC will be absent.
This was formalized by the Paschos-Glashow--Weinberg
theorem~\cite{Glashow:1976nt,Paschos:1976ay}
which states that a necessary and sufficient condition
for the absence of FCNC at tree level
is that all fermions of a given charge and helicity
transform according to the same irreducible representation of $SU(2)$,
correspond to the same eigenvalue of $T_3$
and that a basis exists in which they receive their contributions
in the mass matrix from a single source.
In the Standard Model with left-handed doublets and right-handed singlets,
this theorem implies that all right-handed quarks of a given charge
must couple to a single Higgs multiplet.
In the 2HDM,
this can only be ensured by the introduction
of discrete or continuous symmetries.

Looking at the quark sector of the 2HDM,
there are only two possibilities.
In the type~I 2HDM,
all quarks couple to just one of the Higgs doublets
(conventionally chosen to be $\Phi_2$).
In the type~II 2HDM,
the $Q=2/3$ right-handed (RH) quarks couple to one Higgs doublet
(conventionally chosen to be $\Phi_2$)
and the $Q=-1/3$ RH quarks couple to the other ($\Phi_1$).
The type~I 2HDM can be enforced with a simple
$\Phi_1\rightarrow -\Phi_1$ discrete symmetry,
whereas the type~II 2HDM is enforced with a
$\Phi_1\rightarrow -\Phi_1, d_R^i \rightarrow -d_R^i$ discrete symmetry.
Note that the original Peccei--Quinn models as well as supersymmetric models
give the same Yukawa couplings as in a type~II 2HDM,
but do it by using continuous symmetries.

We will in this chapter consider that there is no CP violation
in the vacuum expectation values (vevs) of the scalar doublets $\Phi_{1,2}$.
This means that $v_{1,2}$ will be assumed to be both real and
(without loss of generality) non-negative.
Thus
\begin{equation}
\Phi_j = \left( \begin{array}{c} \phi_j^+ \\
\left( v_j + \rho_j + i \eta_j \right) \left/ \sqrt{2} \right.
\end{array} \right),
\end{equation}
with $v_1 = v \cos{\beta}$ and $v_2 = v \sin{\beta}$.
Then,
the neutral Goldstone boson is
$G^0 = \eta_1 \cos{\beta} + \eta_2 \sin{\beta}$.
The linear combination of the $\eta_j$ orthogonal to $G^0$
is the physical pseudoscalar
\begin{equation}
A = \eta_1 \sin{\beta} - \eta_2 \cos{\beta}.
\end{equation}
The physical scalars are a lighter $h$ and a heavier $H$,
which are orthogonal combinations of $\rho_1$ and $\rho_2$:
\begin{eqnarray}
h &=& \rho_1 \sin{\alpha} - \rho_2 \cos{\alpha},
\\
H &=& - \rho_1 \cos{\alpha} - \rho_2 \sin{\alpha}.
\end{eqnarray}
Notice that the Standard-Model Higgs boson would be
\begin{eqnarray}
H^\mathrm{SM} &=& \rho_1 \cos{\beta} + \rho_2 \sin{\beta}
\nonumber \\ &=& h \sin{\left( \alpha - \beta \right)}
- H \cos{\left( \alpha - \beta \right)}.
\end{eqnarray}

As shown by Carena and Haber~\cite{Carena:2002es},
one can,
without loss of generality,
assume that $\beta$ is in the first quadrant,
\textit{i.e.}\ that both $v_1$ and $v_2$ are non-negative real;
also,
one can add $\pi$ to $\alpha$,
\textit{i.e.}\ invert the sign of both the $h$ and $H$ fields,
without affecting any physics.
In the tree-level MSSM,
$\alpha$ is in the fourth quadrant,
but this is not the case in the general 2HDM,
therefore we will choose $\alpha$
to be either in the first or the fourth quadrant.
We will choose our independent variables to be $\tan\beta$ and $\alpha$,
which are single valued over the allowed range.

It is conventionally assumed,
in discussions of type~I and type~II 2HDMs,
that the right-handed leptons satisfy the same discrete symmetry
as the $d_R^i$ and thus
the leptons couple to the same Higgs boson as the $Q=-1/3$ quarks.
However,
the Glashow--Weinberg theorem does not require this,
and there are two other possibilities.
In the ``lepton-specific'' model,
the RH quarks all couple to $\Phi_2$ and the RH leptons couple to $\Phi_1$.
In the ``flipped'' model,
one has the $Q=2/3$ RH quarks coupling to $\Phi_2$ and
the $Q=-1/3$ RH quarks coupling to $\Phi_1$,
as in the type~II 2HDM,
but now the RH leptons couple to $\Phi_2$.
The phenomenology of these models is,
as we will see,
quite different.
In one of the earliest papers~\cite{Barger:1989fj}, the names ``Model III" and
``Model IV" were used for the flipped and lepton-specific models, respectively.
The term ``Model III", however, has become associated with the 2HDM with tree-level
FCNCs (the subject of Chapter III). In other early
papers~\cite{Grossman:1994jb,Akeroyd:1994ga,Akeroyd:1996di}, the terms ``Model I" and
``Model II" were used for the lepton-specific and flipped models respectively, and in even
earlier works~\cite{Barnett:1983mm,Barnett:1984zy}, the terms IIA and IIB were used.
More recently~\cite{Aoki:2009ha}, the terms type X and type Y were used for the
lepton-specific and flipped models.
The four models which lead to natural flavour conservation
are presented in Table \ref{tab:3_models}.
It is straightforward to find a $Z_2$ symmetry
which will ensure that only these interactions exist.
\begin{table}
\begin{center}
\begin{tabular}{|c|p{0.8in}|p{0.8in}|p{0.8in}|}  \hline
Model & $u_R^i$     & $d_R^i$& $e_R^i$     \\
\hline
Type I     & $\Phi_2$   &$ \Phi_2$ & $\Phi_2$      \\
\hline
Type II     & $\Phi_2$       &$\Phi_1$           & $\Phi_1$    \\
\hline
Lepton-specific     &$\Phi_2$    & $\Phi_2 $         & $\Phi_1$    \\
\hline
Flipped     & $\Phi_2$     & $\Phi_1$          & $\Phi_2$     \\
\hline
\end{tabular}
\end{center}
\caption{Models which lead to natural flavour conservation.
The superscript $i$ is a generation index.
By convention,
the $u^i_R$ always couple to $\Phi_2$.}
\label{tab:3_models}
\end{table}

In a somewhat related work,
Pich and Tuz\'on~\cite{Pich:2009sp,Tuzon:2010vt}
simply assumed that the Yukawa coupling matrices of $\Phi_1$ and $\Phi_2$
in flavour space are proportional.
This then eliminates all
tree-level
FCNC,
and gives three arbitrary proportionality constants.
Note that this assumption is {\it ad hoc} and,
in general,
is not radiatively stable~\cite{Ferreira:2010xe} -
one would obtain FCNC couplings being generated radiatively,
as was analysed recently in Ref.~\cite{arXiv:1111.5760}.
However, Ser\^odio has recently proposed
a UV completion of the Pich--Tuz\'on
model~\cite{Serodio:2011hg}.
And Varzielas~\cite{arXiv:1104.2601} has studied how family
symmetries in multi-Higgs doublet models may give a justification for
the alignment hypothesis.
Each of the four models
(as well as the Inert Doublet model discussed later)
then arises as a specific choice of the proportionality constant
(and only these choices allow for a symmetry~\cite{Ferreira:2010xe}).
Another recent, very general, formulation in which the various models
are special cases is shown in Ref.~\cite{DiazCruz:2010yq}.
One should keep in mind that even if a 2HDM without FCNC is correct,
it will take some time to determine all of the couplings to determine which
2HDM it is, and the Pich-Tuzon parametrization might be a valuable guide
for phenomenologists.
In addition, the Pich-Tuzon parametrization might arise in other models;
for example, the three doublet model of Cree and Logan~\cite{arXiv:1106.4039}
reproduces the
Pich-Tuzon model in its charged Higgs Yukawa couplings.
%
Of particular interest is the fact that
if the proportionality constants are complex,
one has CP violating effects.
It has been noted~\cite{Tuzon:2010vt,Braeuninger:2010td}
that loop corrections induce
flavour changing currents of the Minimal Flavour Violation form,
and  bounds on the charged-Higgs mass were discussed.
A similar approach was recently used
by Mahmoudi and Stal~\cite{Mahmoudi:2009zx},
who studied the constraints on
the charged-Higgs mass
from meson decays and FCNC transitions,
using a more general model-independent approach,
getting results in the four models as special cases.

The Yukawa couplings can now be determined.
In the Standard Model,
the coupling of the fermion $f$
to the Higgs boson is
$m_f/v$.
Following the notation of Aoki {\it et al.}~\cite{Aoki:2009ha},
we define the parameters $\xi^f_h, \xi^f_H, \xi^f_A$
through the Yukawa Lagrangian
\begin{eqnarray}
{\mathcal L}_\text{Yukawa}^\text{2HDM} &=&
- \sum_{f=u,d,\ell} \frac{m_f}{v} \left(
\xi_h^f {\overline f}f h
+ \xi_H^f{\overline f}f H
- i \xi_A^f {\overline f}\gamma_5f A
\right)
\nonumber \\ & &
- \left\{\frac{\sqrt2V_{ud}}{v}\,
\overline{u} \left( m_u \xi_A^u \text{P}_L
+ m_d \xi_A^d \text{P}_R \right)
d H^+
+\frac{\sqrt2m_\ell\xi_A^\ell}{v}\,
\overline{\nu_L^{}}\ell_R^{}H^+
+\text{H.c.}\right\} \label{Eq:Yukawa}
\end{eqnarray}
where $P_{L/R}$ are projection operators for left-/right-handed fermions,
and the factors $\xi$ are presented in Table \ref{tab:3_couplings}.
\begin{table}
\begin{center}
\begin{tabular}{|c|p{1.2in}|p{1.2in}|p{1.2in}|p{1.2 in}|}  \hline
{} & Type I  & Type II & Lepton-specific & Flipped     \\
\hline
$\xi^u_h $    & $\cos\alpha/\sin\beta$   & $\cos\alpha/\sin\beta$ & $\cos\alpha/\sin\beta$ & $\cos\alpha/\sin\beta$      \\
\hline
$\xi^d_h $    & $\cos\alpha/\sin\beta$        & $-\sin\alpha/\cos\beta$            & $\cos\alpha/\sin\beta$  & $-\sin\alpha/\cos\beta$   \\
\hline
$\xi^\ell_h $ & $\cos\alpha/\sin\beta$     & $-\sin\alpha/\cos\beta$          & $-\sin\alpha/\cos\beta$   & $\cos\alpha/\sin\beta$   \\
\hline\
$\xi^u_H $    & $\sin\alpha/\sin\beta$     & $\sin\alpha/\sin\beta$   & $\sin\alpha/\sin\beta$ & $\sin\alpha/\sin\beta$   \\
\hline
$\xi^d_H $    & $\sin\alpha/\sin\beta$     & $\cos\alpha/\cos\beta$   & $\sin\alpha/\sin\beta$ & $\cos\alpha/\cos\beta$   \\
\hline
$\xi^\ell_H $    & $\sin\alpha/\sin\beta$     & $\cos\alpha/\cos\beta$   & $\cos\alpha/\cos\beta$ & $\sin\alpha/\sin\beta$   \\
\hline
$\xi^u_A $    & $\cot\beta$     & $\cot\beta$   & $\cot\beta$ & $\cot\beta$   \\
\hline
$\xi^d_A $    &
$- \cot\beta$     &
$\tan\beta$   & $-\cot\beta$ & $\tan\beta$   \\ \hline
$\xi^\ell_A $    &
$- \cot\beta$     &
$\tan\beta$   & $\tan\beta$ & $-\cot\beta$   \\ \hline
\end{tabular}
\end{center}
\caption{Yukawa couplings of $u,d,\ell$ to the neutral Higgs bosons $h,H,A$
in the four different models.
The couplings to the charged Higgs bosons follow Eq. \ref{Eq:Yukawa}.}
\label{tab:3_couplings}
\end{table}

In all models,
the coupling of the neutral Higgs bosons to the $W$ and $Z$ are the same:
the coupling of the light Higgs,
$h$,
to either $WW$ or $ZZ$ is the same as the Standard-Model coupling
times $\sin(\beta-\alpha)$ and the coupling of the heavier Higgs,
$H$,
is the same as the Standard-Model coupling
times $\cos(\alpha-\beta)$.
The coupling of the pseudoscalar,
$A$,
to vector bosons vanishes.

In this section,
we will summarize some of the work done on these four models,
and will follow with a more detailed discussion in the following sections.

There are relatively few studies which directly compare all four models.
One of the earliest papers to mention all four models was by Barger,
Hewett and Phillips~\cite{Barger:1989fj},
who studied the charged-Higgs phenomenology
but assumed fairly light top quarks.
The famous Higgs Hunter's Guide~\cite{Gunion:1989we}
mentions all four,
but concentrates only on the type~I and type~II 2HDMs.
Grossman~\cite{Grossman:1994jb} also discusses all four models,
but focuses on models with more than two doublets, and concentrates
on the on the charged Higgs sector.
Akeroyd has several papers in which all four models are discussed.
In an early paper with Stirling~\cite{Akeroyd:1994ga},
the phenomenology of the charged Higgs boson at LEP2
was analysed in each model,
and this was followed~\cite{Akeroyd:1996di} by
a study of the neutral sector at LEP2.
In addition,
he looked~\cite{Akeroyd:1998sv} at LHC phenomenology in all four models,
focusing in particular on the Higgs branching ratios to $\gamma\gamma$
and $\tau\tau$.
More recently,
Barger,
Logan and Shaughnessy~\cite{Barger:2009me}
performed a comprehensive analysis of the couplings in all models
with natural flavour conservation,
including doublets and singlets;
the four models appear as special cases.

There are two recent papers comparing Higgs decays in all four models.
Aoki {\it et al.}~\cite{Aoki:2009ha} study
the decays of the Higgs bosons in each model,
summarize current phenomenological constraints and look
at methods of distinguishing the models at colliders,
although they focus on the type~II
and lepton-specific models
and assume that the heavy Higgs bosons are not too heavy
(typically with masses below 200 GeV).
Arhrib {\it et al.}~\cite{Arhrib:2009hc}
study the decays of the light Higgs in each model,
although the main point of their work concerns
double-Higgs production at the LHC.

Recently,
a new computer code was written by
Eriksson {\it et al.}~\cite{Eriksson:2009ws}.
The code allows one to input any of the different $Z_2$ symmetries,
or even more general couplings,
and calculates all two-body and some three-body Higgs boson decays,
and the oblique
parameters $S$, $T$ and $U$
and other collider constraints.

The least studied model is the flipped model
(the word was coined in Ref.~\cite{Barger:2009me});
even works that discuss all four models generally focus less
on this structure than the others.
The only paper dedicated entirely to the flipped model
was the very recent article of Logan and MacLennan~\cite{Logan:2010ag}.
They studied the charged-Higgs phenomenology in that model,
including branching ratios and indirect constraints
and analyse prospects at the LHC.

The lepton-specific model was first discussed in two papers
by Barnett {\it et al.}~\cite{Barnett:1983mm,Barnett:1984zy}
in the context of extremely light Higgs scalars.
The model was recently analysed extensively by Su and Thomas~\cite{Su:2009fz}.
They studied theoretical and experimental constraints on the model
and showed that there can be substantial enhancement of the couplings
between the charged leptons and the neutral Higgs scalar.
Logan and MacLennan~\cite{Logan:2009uf} considered the constraints on
the charged-Higgs mass,
with bounds arising from lepton flavour universality and direct searches,
and discuss prospects at the LHC.
Goh, Hall and Kumar~\cite{Goh:2009wg}
discussed whether a lepton-specific model
could explain the leptonic cosmic-ray signals seen by PAMELA and ATIC,
and studied the implications for the LHC.
A very recent analysis of this possibility, including analyses of
astrophysical results and direct dark matter detection, can be found
in the work of Boucenna and Profumo~\cite{arXiv:1106.3368}.
In another analysis,
Cao {\it et al.}~\cite{Cao:2009as} assumed that
the $3\sigma$ discrepancy~\cite{Jegerlehner:2009ry}
between theory and experiment in the $g-2$ of the muon
is primarily due to the lepton-specific model---this requirement
substantially reduces the available parameter space,
forcing the model to have a very light pseudoscalar
and very large values of $\tan\beta$,
and they analysed this parameter space.
Finally,
Aoki {\it et al.}~\cite{Aoki:2008av}
looked at neutrino masses and dark matter in the lepton-specific model,
but did add singlets.

The type~I 2HDM~\cite{Haber:1978jt} is the second most
studied.\footnote{Recent developments
in string phenomenology~\cite{Ambroso:2008kb}
suggest that a type~I 2HDM is generic among the vacua of the heterotic string,
providing new motivation for study of this model.}
In the quark sector,
it is identical to the lepton-specific model,
thus many results from studies of the type~I 2HDM
apply to the lepton-specific as well.
A special limit of the type~I 2HDM is $\alpha=\pi/2$,
in which case the fermions all completely decouple from the lightest Higgs;
this limit is referred to as the fermiophobic limit.
Note that even in this limit,
the coupling does reappear at the one-loop level,
but it will in any event be very small.   Later in this chapter, it will be shown that in the inert doublet models, this limit can be obtained exactly.
The earliest discussions
of the fermiophobic limit in the context of imminent Tevatron data
were those of
Stange {\it et al.}~\cite{Stange:1993ya},
of Diaz and Weiler~\cite{Diaz:1994pk}
and of Barger {\it et al.}~\cite{hep-ph/9211234},
who looked at Higgs production and decay through photon loops.
Shortly thereafter,
Akeroyd~\cite{Akeroyd:1995hg}
studied the other phenomenological implications
during the early Tevatron runs.
Br\"ucher and Santos~\cite{Brucher:1999tx}
mentioned all four models,
but then focused on the fermiophobic  limit of the type~I 2HDM,
studying the decays of the various Higgs bosons
and the constraints on the model from LEP 2.
The same authors together with Barroso~\cite{Brucher:2000qc,Barroso:1999bf}
used the possibility of a fermiophobic Higgs
to look at $h\rightarrow \gamma\gamma$
in two different versions of the type~I 2HDM,
showing that this decay could distinguish between them.

Moving away from the fermiophobic limit,
there are many papers looking at the type~I 2HDM.
A very early discussion of the fermiophobic,
gauge-phobic and fermiophilic limits was given by Pois,
Weiler and Yuan~\cite{Pois:1993ay},
who studied top production below the $t\bar{t}$ threshold
in Higgs decays.
The Higgs Hunter's Guide~\cite{Gunion:1989we}
contains some analysis
(although they focus on the type~II 2HDM and the MSSM),
but this was before the top mass was known to be very heavy.
In a series of papers,
Akeroyd and
collaborators~\cite{Akeroyd:1998dt,Akeroyd:1998uw,Akeroyd:2003xi,Akeroyd:1998ui,Akeroyd:2003bt}
considered Higgs decays into lighter Higgs bosons,
charged-Higgs decays into a $W$ and a pseudoscalar,
double-Higgs production,
Higgs decays to $\gamma\gamma, \tau^+\tau^-$ at the LHC,
and the possibility of a very light Higgs,
respectively,
all in the context of the type~I 2HDM.
There have also been studies
of the contribution to the anomalous magnetic moment of the muon
in the type~I 2HDM~\cite{Dedes:2001nx,Krawczyk:1996sm}.

The type~II 2HDM is by far the most studied,
since it is the structure
present
in supersymmetric models.
A voluminous Physics Reports review article in 2008
by Djouadi~\cite{Djouadi:2005gj}
analyses the Higgs bosons of the MSSM in great detail.
We will not discuss the MSSM Higgs structure
and phenomenology in this work,
and refer the reader to Djouadi's review article.
Here,
we will only focus on the differences between the general type~II 2HDM
and the MSSM.

The most crucial difference is that the general type~II 2HDM
does not have a strict upper bound on the mass of the lightest Higgs boson,
which is an important characteristic of the MSSM.
In addition,
the scalar self-couplings are now arbitrary.
Another important difference is that the mixing parameter $\alpha$,
which in the MSSM is given in terms of $\tan\beta$
and the scalar and pseudoscalar masses,
is now arbitrary.
Finally,
in the MSSM the charged-scalar and pseudoscalar masses
are so close that the decay of the charged Higgs into a pseudoscalar
and a real W is kinematically forbidden,
while it is generally allowed in the type~II 2HDM
(although see Ref.~\cite{Akeroyd:2001in,Litsey:2009rp}
for possible exceptions).

The Higgs Hunter's Guide has numerous phenomenological bounds
on the type~II 2HDM,
but they have become quite outdated.
There have been several more recent
works~\cite{Ciuchini:1997xe,Borzumati:1998tg,Misiak:2006zs}
on $B\rightarrow X_s\gamma$;
the work by Misiak~\cite{Misiak:2006zs}
would also apply to the lepton-specific model.
A recent summary of bounds on the charged-Higgs mass
in the type~II 2HDM is by Krawczyk and Sokolowska~\cite{Krawczyk:2007ne},
and detailed analyses of charged-Higgs production
at hadron colliders can be found
in Refs.~\cite{Berger:2003sm,Alwall:2004xw,Plehn:2002vy}.
Asakawa, Brein and Kanemura~\cite{Asakawa:2005nx}
compare  associated $W$ plus charged Higgs production in the type~II 2HDM
with that in the MSSM.
As noted above,
the decay of the charged Higgs into a pseudoscalar and a $W$ is forbidden
in the MSSM,
but not in the more general type~II 2HDM,
and a detailed study of that decay
is found in Ref.~\cite{Akeroyd:1998uw},
which also discusses the type~I 2HDM.
Krawczyk and Temes~\cite{Krawczyk:2004na}
studied constraints from leptonic tau decays.
Finally,
a very recent study
by Kaffas, Osland and Ogreid~\cite{WahabElKaffas:2007xd}
looked at all these (and other) processes
in a very comprehensive analysis of the type~II 2HDM parameter space.
There are many additional papers on the type~II 2HDM;
their results will be discussed in subsequent sections.

In addition,
there is another model which has natural flavour conservation,
in which the quarks and charged leptons all couple to $\Phi_2$,
but the right-handed neutrino couples to $\Phi_1$.
In this model,
there are Dirac neutrino masses,
and the vacuum expectation value of $\Phi_1$ must be O(eV).
The only way to have such a small vev is through an approximate symmetry.
The model originally used a $Z_2$ symmetry
which was softly~\cite{Ma:2000cc}
or spontaneously~\cite{Gabriel:2006ns} broken,
but this allows for right-handed neutrino masses.
Extending the symmetry to a $U(1)$ and breaking it softly
(to avoid a Goldstone boson)
gives the model of Davidson and Logan~\cite{Davidson:2009ha}.
An interesting question concerns the effects of quantum corrections
to the small vacuum expectation value and that is discussed in
Ref.~\cite{arXiv:1107.1026}.
The phenomenology of this model
(which is basically the type~I 2HDM with a right-handed neutrino added)
is quite interesting,
especially in the charged-Higgs sector,
and will be discussed subsequently.

In this Chapter,
we will first discuss the decays of the neutral Higgs bosons
in the various models,
followed by an analysis of the production of the neutral Higgs.
Then the constraints due to both collider bounds
as well as lower energy processes
(such as $B$ decays,
$b$ production,
the anomalous magnetic moment of the muon,
etc.)
will be presented.
Finally,
the ``inert'' doublet model will be discussed.

\newpage

\subsection{Higgs decays}

\begin{figure}[ht]
\centerline{\epsfysize=15cm \epsfbox{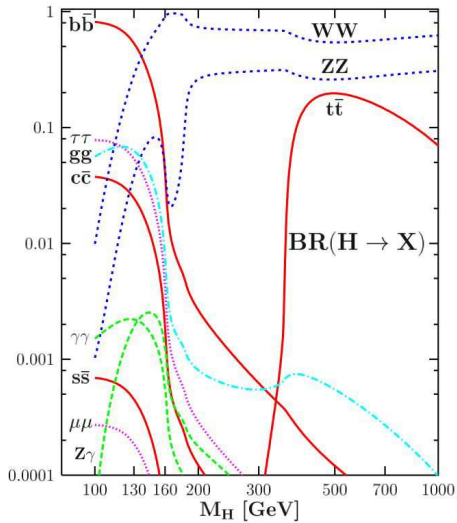} }
\vskip -7cm
\caption{The branching ratios for the decay
of the SM Higgs boson as a function of its mass.}
\label{fig:3SMbrs}
\end{figure}

The key methods of distinguishing
the various 2HDMs
from each other and from the Standard Model
involve the branching ratios in Higgs decays.
In this section,
we will focus on the decays of the neutral scalars ($h,H,A$) of the 2HDM.
In the Standard Model,
the Tevatron is sensitive to the $\bar{b}b, WW, ZZ$ Higgs decays,
whereas at the LHC the decays $\gamma\gamma, WW, ZZ$
are more important.
The branching ratios and total width of the Standard Model Higgs
are very well studied;
the most comprehensive analysis
is found in the recent review of Djouadi~\cite{Djouadi:2005gi}.
The results are in Fig.~\ref{fig:3SMbrs}.
One can see that the $WW$ and $ZZ$ decays
are dominant for Higgs masses above 160
GeV; they would provide
the best signature at either the Tevatron or the LHC.
Below $160$ GeV,
the $b\bar{b}$ branching ratio is more important at the Tevatron,
but that decay mode is swamped by large backgrounds at the LHC,
which must rely on the $\gamma\gamma$ mode
in the low-mass range.

We now must look at branching ratios in 2HDMs.
As shown in the last section,
the branching ratios will not depend exclusively on the masses,
but also on $\alpha$ and $\beta$.
This makes plots of the various branching ratios necessarily incomplete.
Ignoring,
for now,
the possibility of heavy Higgs bosons decaying into lighter ones,
one has branching ratios depending on $m_H$,
$\alpha$ and $\beta$ for each Higgs ($h,H,A$)
and for each model.
This makes a comprehensive analysis difficult.

How have other authors dealt with this previously?
In the comprehensive review of all four models
by Aoki {\it et al.}~\cite{Aoki:2009ha},
they plotted the branching ratios of $H$ and $A$
for each of the four models as a function of $\tan{\beta}$,
setting the mass to 150 GeV
and $\cos(\alpha-\beta)=0$.
In this limit,
the decays of the $H$ and $A$ to $WW$ and $ZZ$ vanish.
While certainly an interesting and important limit,
this does correspond to a single point in the
two-dimensional mass--$\cos(\alpha-\beta)$ parameter space.
Akeroyd~\cite{Akeroyd:1998sv}
plotted the branching ratio of the light Higgs
as a function of $\tan\beta$,
but also chose only a few points
in the parameter space.
In their analysis of the lepton-specific model,
Su and Thomas chose a specific benchmark point
for $\alpha$ and $\beta$
and plotted the branching ratios as a function of the mass.
These are just some examples.

In order to cut down on the complexity of the graphs,
we will not display decay modes
which can never be seen at the LHC and Tevatron
(at least within the next decade),
although they will be included in the computation of the branching ratios.
Long before these modes could be measured,
many of the parameters of the 2HDM will be known,
and more comprehensive analyses will be done.
Thus,
we will only consider decays into $t\bar{t}$,
$b\bar{b}$,
$WW$,
$\tau\tau$ and $\gamma\gamma$
(and possibly into other Higgs bosons).
In fact,
\begin{itemize}
\item In the type~I and type~II 2HDMs
the ratio of the $\tau\tau$ to $b\bar{b}$ modes
is fixed at $m^2_\tau / 3 m^2_b$.
Using the fact that one must use
the running $b$-quark mass at the 100 GeV scale
(which is approximately 3.0 GeV)
and including radiative corrections,
this ratio is fixed at approximately $10\%$,
and thus we will not explicitly include the $\tau\tau$ mode in the figures.
\item The decay width into $ZZ$
is related to the one into $WW$
by only the weak mixing angle and phase-space factors,
and thus will have the same ratio as in the Standard Model.
As a result,
the branching ratio into $ZZ$ will not be directly displayed.
\item The decay into gluons
(while important in production)
can't be directly measured due to very large backgrounds.
\item Other decay modes, such as $c\bar{c}$, $Z\gamma$ and $\mu\mu$,
have either too small branching ratios or too large backgrounds.
\end{itemize}
Again,
only neutral Higgs fields will be discussed in this Chapter.

In all four of the 2HDMs,
a lower bound on $\tan\beta$ of roughly $0.3$
can be obtained from the requirement that
the top-quark Yukawa coupling be perturbative.
Since large Yukawa couplings have positive beta functions,
if they start out large then they will exceed the perturbative limit
(as well as unitarity)
at a relatively low scale
(see section~\ref{2_sec:ufb} and appendix~\ref{2_sec:uni}).
It is hard to see how a $\tan\beta$ near or below $0.3$
can be accommodated.
At the other extreme,
in the type~II 2HDM,
the bottom-quark Yukawa coupling
will be nonperturbative if $\tan\beta$ exceeds roughly 100.
In the context of the MSSM,
Barger {\it et al.}~\cite{Barger:1992ac}
and Carena {\it et al.}~\cite{Carena:1994bv}
showed that perturbative unification
could be achieved for $\tan\beta < 60$,
but the MSSM has beta functions very different from those of the 2HDM.
Kanemura and collaborators~\cite{Kanemura:1999tg,Kanemura:1999xf}
looked at bounds on $\tan\beta$ from perturbative unification
in the non-supersymmetric 2HDM,
and Akeroyd, Arhrib and Naimi~\cite{Akeroyd:2000wc,Arhrib:2000is}
looked for
violations of tree-level unitarity
in a large number of processes
and concluded that values of $\tan\beta$ greater than $30$ are disfavored,
although there are some regions of parameter space in which they are allowed.
More recently,
Arhrib {\it et al.}~\cite{Arhrib:2009hc}
argued that perturbative and unitarity constraints,
in all four models,
require that $\tan\beta < 6$ for all but a very small region
of parameter space,
and this is  confirmed
in the work of Kaffas, Osland and Ogreid~\cite{WahabElKaffas:2007xd}.
Thus,
we will focus on values of $\tan\beta$ between 1 and 6,
but will mention the effects of larger $\tan\beta$ in a few instances.

One
region of interest is the decoupling region.
This is the region of parameter space in which the $H$, $A$ and charged Higgs
all are much heavier than the $h$,
and it is thus possible to integrate out the heavy fields.
The resulting effective theory is then
like the Standard-Model Higgs sector,
with corrections to the various couplings due to the heavy sector.
This was discussed in detail by Gunion and Haber~\cite{Gunion:2002zf}.
Later,
Mantry {\it et al.}~\cite{Mantry:2007ar} and Randall~\cite{Randall:2007as}
examined Higgs decays in the decoupling region
and showed how one could obtain a measurable sensitivity
to the high scale even if the heavy scalars are inaccessible at the LHC,
pointing out the importance of accurate measurements
of the branching fractions.

\subsubsection{Higgs decays in the type~I 2HDM}

The type~I 2HDM
has the simplest discrete symmetry and its
couplings can be easily described.
The coupling of the light neutral Higgs,
$h$,
to fermions is the same as in the Standard Model
but multiplied by $\cos\alpha/\sin\beta$
while its couplings to $WW$ and $ZZ$ are multiplied by
$\sin(\alpha-\beta)$.
For the heavy neutral Higgs,
$H$,
these factors are $\sin\alpha/\sin\beta$
and $\cos(\alpha-\beta)$,
respectively.
Thus one can determine the widths of the various decays
by simple multiplication,
with the exception of the $\gamma\gamma$ decay,
in which the contribution of the $W$ loop and that of the fermion loops
are multiplied by their respective factors.

There are a few interesting limits.
If $\sin(\alpha-\beta)$
($\cos(\alpha-\beta)$)
vanishes,
then the $h$
($H$)
field is
gauge-phobic, \textit{i.e.}\ it does not
couple to $WW$ and $ZZ$,
radically changing the phenomenology of Higgs decays.
If $\cos\alpha$
($\sin\alpha$) vanishes,
then the $h$
($H$)
is fermiophobic.
This is a particularly interesting limit,
since then the $\gamma\gamma$ decay can become dominant
well below the $WW$ threshold.
Although there is no symmetry that can enforce this limit
(and a nonzero $\sin(\alpha-\beta)$ would be generated
at one-loop if it is set equal to zero at tree level),
the phenomenological implications are so dramatic that
study of the limit is warranted~\footnote{Notice, though,
that this limit corresponds to the inert model, in which only
one of the doublets gains a vev.}.

\begin{figure}[ht]
\centerline{\epsfysize=10cm
\epsfbox{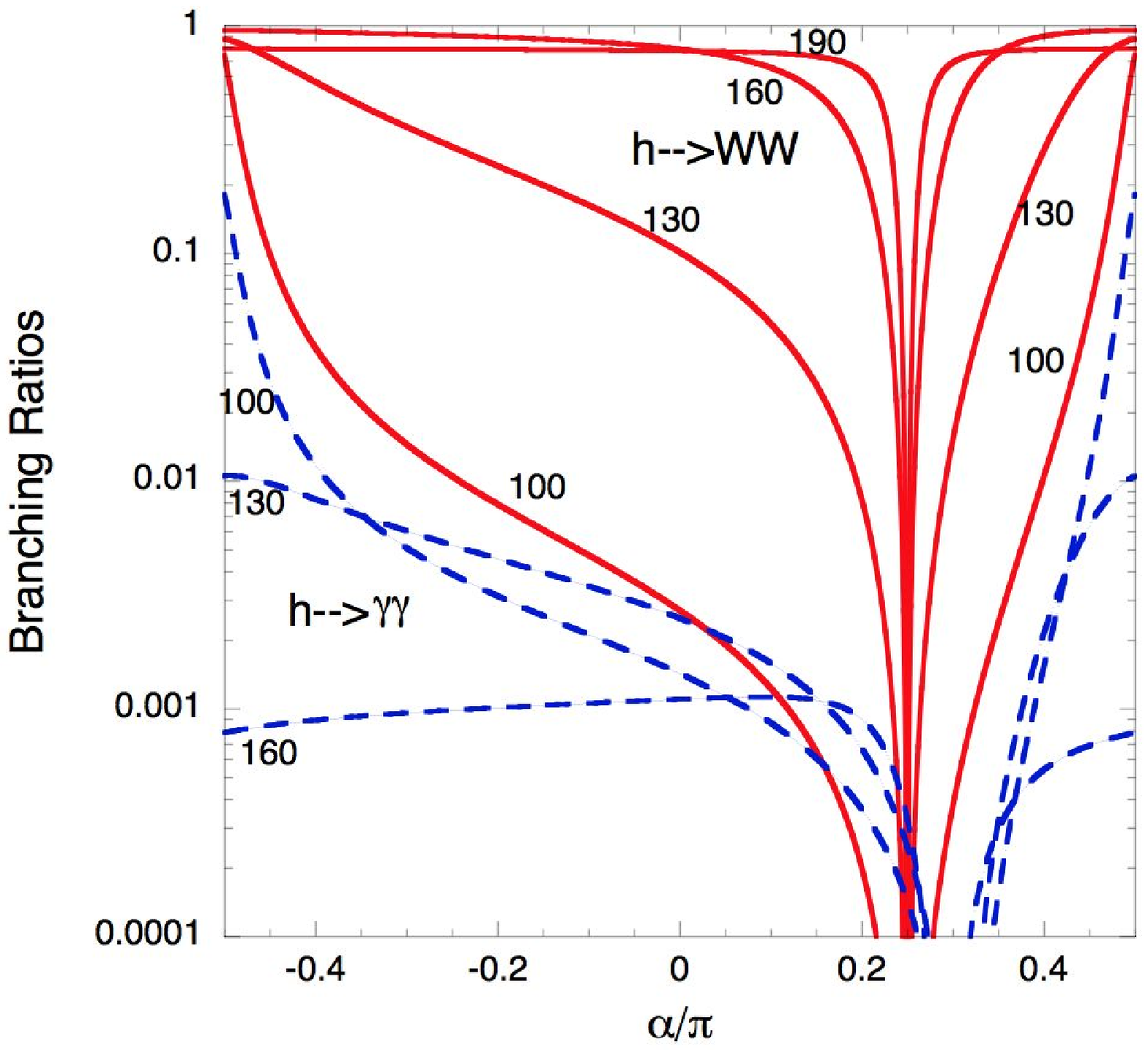}\epsfysize=10cm \epsfbox{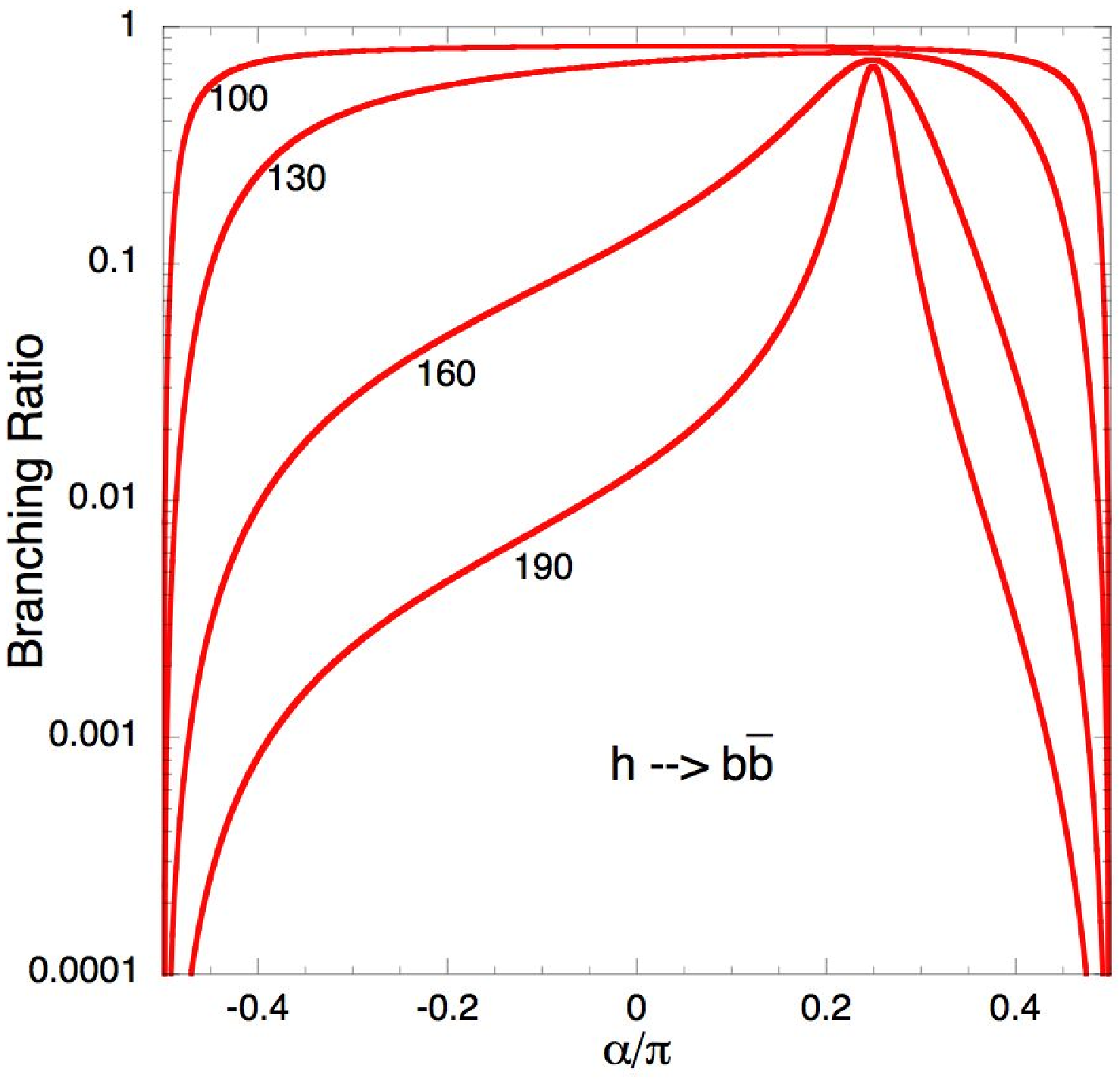} }
\vskip -2cm
\caption{The type-I 2HDM
light-Higgs branching ratios into $W$ pairs,
diphotons and $b\bar{b}$
are plotted as a function of $\alpha$
for $\tan\beta=1$
and for various values of the Higgs mass (in GeV).
In the left figure, the solid lines correspond to $h\rightarrow WW$
and the dashed lines to $h\rightarrow \gamma\gamma$.}
\label{fig:3hww1}
\end{figure}

For $\tan\beta=1$,
the branching ratios of the light Higgs
have been plotted in Fig.~\ref{fig:3hww1}
for various values of the Higgs mass.
One can clearly see that $\alpha=\pm\pi/2$ is the fermiophobic limit,
where the branching ratio to fermions vanishes,
and that $\alpha=\beta$ is the gauge-phobic limit,
where the branching
ratios to $WW$ and to $ZZ$ vanish.
Note that
the branching ratio to $\gamma\gamma$
at the gauge-phobic point
does not
quite vanish since there is a small contribution
from top-quark loops.

For the light Higgs $h$
we have considered masses ranging from 100 GeV to 190 GeV.
An $h$ heavier than 190 GeV may have problems with
electroweak precision results~\cite{Amsler:2008zzb},
although in the 2HDM, such a heavy Higgs could be made compatible
with these
results, as discussed in Ref.~\cite{arXiv:1107.0975}.
An $h$ lighter than 100 GeV
might be allowed by the LEP data~\cite{Schael:2006cr}
if its coupling to $ZZ$ were suppressed
by a sufficiently small value of $\sin(\alpha-\beta)$;
for instance,
a 70 GeV $h$
would be allowed~\cite{hep-ph/9506291} if $\sin^2(\alpha-\beta)$ were no larger than 3\%.
For a recent discussion,
see the work of Gupta and Wells~\cite{Gupta:2009wn}.
Since phenomenological consistency is obtained only
for a narrow range of $\alpha$ around the gauge-phobic point,
we will not allow for $m_h < 100$ GeV in the plots,
but any interesting physics of a lighter $h$ will be discussed in the text.

In the Standard Model,
for a Higgs mass of 100 GeV,
the  decay to $WW$ is heavily suppressed since at least one of the $W$s must be far off-shell,
and the $b\bar{b}$ mode is dominant.
Thus,
as seen in Fig.~\ref{fig:3hww1},
one must be very close to the fermiophobic limit
in order for the $b\bar{b}$ branching ratio to be small.
As the Higgs becomes heavier,
the $WW$ decay mode is less suppressed
and eventually dominates; for larger
Higgs masses,
the decay into $b\bar{b}$ is very small
except very close to the gauge-phobic point.
Note that the $b\bar{b}$ branching ratio never reaches unity,
even at the gauge-phobic point,
due to the contribution of the gluonic decay
(which depends on the coupling of the Higgs to the top quark).
Comparing with the Standard Model decays,
one can see that the relative branching ratios can differ substantially.

For larger values of $\tan\beta$,
the results look very similar,
with slightly different slopes.
The main difference is that the gauge-phobic point
is at $\alpha=\beta$
and thus moves to the right.
In the  case of $\tan\beta=30$,
the gauge-phobic and fermiophobic points are only two degrees apart,
leading to very steep slopes in the curves
(since the gauge-phobic and fermiophobic points cannot coincide,
the Higgs will certainly decay,
but in this case its width will be quite narrow).

The diphoton decay,
$h\ra\gamma\gamma$,
has contributions from both $W$ loops
and $t$ loops,
therefore it does not vanish
in either the gauge-phobic or fermiophobic limit
(the contribution of $W$ loops in the Standard Model
is substantially larger than
that of top-quark loops,
so the gauge-phobic limit does cause
a suppression).
An analysis of this mode in the type-I 2HDM
was carried out
by Posch~\cite{Posch:2010hx},
who found that an enhancement over the Standard Model branching ratio
of as much as $70\%$ was possible.

The fermiophobic limit is of special interest.
Although the mass limit
on the Standard-Model Higgs boson is 114.4 GeV~\cite{Barate:2003sz},
this assumes that the coupling
of the Higgs boson to $ZZ$ is not suppressed,
and the bound can be much weaker if one is fairly close
to the gauge-phobic limit.
Many studies have been done
concerning the possibility of a Higgs much lighter than 100 GeV
in the fermiophobic limit.
Note that the decay into $WW$ drops off dramatically
as the Higgs mass is below the $W$ mass,
therefore
the fermiophobic limit
leads to a dominant decay into $\gamma\gamma$ for these masses.
CDF~\cite{Aaltonen:2009ga} and D0~\cite{:2008it}
have published bounds on a fermiophobic Higgs
(by looking for the diphoton mode)
which are slightly over 100 GeV,
but they still assume
that the coupling to $ZZ$ is not suppressed,
which may not be the case
(note that if $\alpha=\pm\pi/2$,
then $\sin(\alpha-\beta)=\pm\cos\beta$
which can be quite small in the large $\tan\beta$ limit).

In a comprehensive analysis,
Akeroyd and Diaz~\cite{Akeroyd:2003bt} noted that
if the coupling to gauge bosons is suppressed,
then the coupling of the light Higgs to a charged Higgs and a $W$
is not suppressed
(one scales as $\sin(\alpha-\beta)$ and the other as $\cos(\alpha-\beta)$)
thus one can pair-produce the charged Higgs
and from there a pair of light Higgs bosons.
D0~\cite{Landsberg:2007mc} has looked at this and found a lower bound
on the light-Higgs mass,
which is approximately 80 GeV
for a charged-Higgs mass below 100 GeV
and approximately 50 GeV
for a charged-Higgs mass below 150 GeV.
This is based on less than a single fb${}^{-1}$,
and will be improved substantially in the near future.
Subsequently,
Akeroyd,
Diaz and Pacheco~\cite{Akeroyd:2003xi}
looked at this process at the LHC,
which would be particularly relevant
if the charged Higgs boson were heavier.
Finally,
it was shown~\cite{Akeroyd:2007yh,Brucher:1999tx,Brucher:2000qc}
that charged-Higgs loops
can substantially alter the diphoton branching ratio
in 2HDMs,
although this depends on unknown
scalar self-couplings, \textit{viz.}\ on the coupling $h H^+ H^-$.
Since one can see from Fig.~\ref{fig:3hww1}
that the dominance of the diphoton decay mode
(especially for a very light Higgs)
will only occur if $\alpha$ is extremely close to $\pm\pi/2$,
the region of parameter space is very small,
but the signature is sufficiently dramatic that searches should continue.
Note that the fermiophobic limit is only relevant
for the type~I 2HDM.

\begin{figure}[ht]
\vskip -0.5cm
\centerline{\epsfysize=15cm \epsfbox{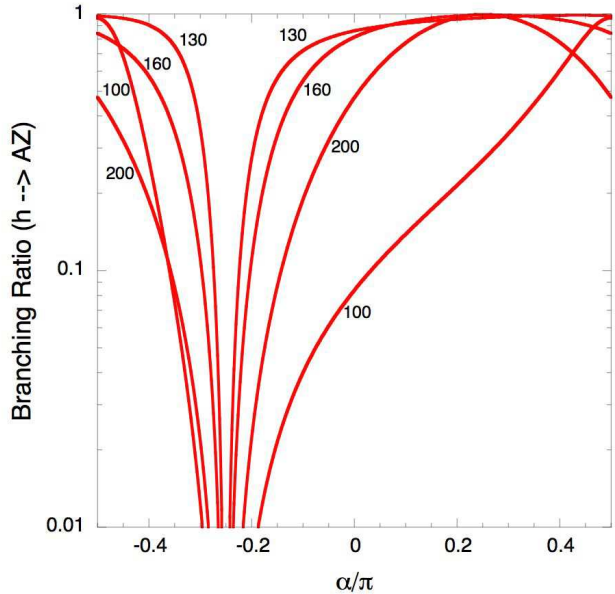} }
\vskip -3cm
\caption{Branching ratios of the light Higgs
boson $h$ into a $Z$ and a pseudoscalar,
for various values of the mass (in GeV) of $h$.
The value of $\tan\beta$ is chosen to be 1
and the mass of the pseudoscalar is chosen to be 5 GeV.}
\label{3fig:hza}
\end{figure}

Throughout the above we have neglected the possibility that
the light Higgs can decay into other Higgs.
In fact,
in both the type~I and type~II 2HDMs
there is a decay that could be important for a range of parameter space,
and yet it has not,
to our knowledge,
been explored substantially.
The range of parameter space will occur if the pseudoscalar mass $M_A$
is less then $m_h - m_Z$.
Specifically,
one can look for $h\rightarrow ZA$.
This could occur if the pseudoscalar
is very light (a few GeV)
and the Higgs is near the current bound
(this would fit electroweak precision tests better),
or else if
the pseudoscalar mass is
comparable to the gauge-boson masses
and the lightest Higgs scalar is considerably heavier.
The $h \rightarrow ZA$ decay has generally not been mentioned,
and we know of no experimental searches for this mode
(although it has been mentioned by DELPHI~\cite{Abdallah:2004wy}).
The primary reason for this lack of attention is that
the rate is proportional to $\cos^2(\alpha-\beta)$,
which is very small in the MSSM.
But in the general 2HDM
there is no reason that $\cos(\alpha-\beta)$ is small.
The width is given by~\footnote{Throughout this work,
$\pi$ refers to the mathematical constant~\cite{ahmes},
not the pion field.}
\begin{equation}
\Gamma (h\rightarrow ZA) =
\frac{g^2 m_h^3
\cos^2(\alpha-\beta)}
{64 \pi m_W^2}\  \lambda^{3/2}
\end{equation}
where $\lambda =  (1 - (m_Z^2 - m_A^2)/m_h^2)^2
- 4 m_Z^2 m_A^2/m_h^4 $.
One can compare this with the more well-known decay
of the light Higgs to $WW$,
which is (for decays into real gauge bosons)
\begin{equation}
\Gamma(h\rightarrow WW) = \frac{g^2 m_h^3\sin^2(\alpha-\beta) }{64 \pi M^2_W } \lambda^\prime
\end{equation}
where $\lambda^\prime=\sqrt{1-4x}(1-4x+12x^2)$, with $x=M^2_W/m_h^2$.
For $\cos^2(\alpha-\beta) > 1/2$,
$h \to Z A$ will actually {\it dominate} Higgs decays.
The branching ratio is plotted
in Fig.~\ref{3fig:hza}, for a range of $h$ masses,
choosing $\tan\beta=1$ and
$m_A = 5$ GeV.
One can see the dip at $\cos(\alpha-\beta)=0$;
this is the region expected in supersymmetric models.
At $\alpha=\beta$,
one has the gauge-phobic limit,
and thus the $h\rightarrow ZA$ decay dominates
(except at a Higgs mass of 100 GeV,
where the $b\bar{b}$ decay is still substantial).
For larger $\tan\beta$,
as before,
the maxima and minima simply shift to the right.
For larger $m_A$,
the only change will be in $\lambda$,
but as long as the light Higgs is reasonably heavier than $m_Z+m_A$,
there will be a range of parameters where this decay is large.

One sees that there is a substantial range of parameters
for which this relatively unstudied decay mode could be substantial,
and even dominate.
The only discussion of it that we are aware of is the analysis,
in the type~I 2HDM,
by  Akeroyd~\cite{Akeroyd:1998dt},
who looked at the decay
of a neutral scalar into a \emph{virtual} $Z$ and a pseudoscalar.
There has been no analysis of detector capabilities.
Even if the $Z$ is real,
the decay  may be challenging~\cite{conway}.
At a hadron collider,
the $h$ longitudinal boost will not be known,
so the $Z$ will not be
monoenergetic;
moreover, the decays of the pseudoscalar may not be easy to observe
(especially if it decays into
$b$ quarks).
But if this is the dominant decay,
further analysis is needed.
As we will see below,
the decay involving the heavier of the neutral scalars,
$H\rightarrow ZA$ has been discussed in detail,
since this can also be substantial in supersymmetric models.

What about
Higgs decays into pairs of scalars?
By definition,
the $H$ is heavier,
but the pseudoscalar could be lighter,
leading to $h\rightarrow AA$.
There has been substantial discussion of the possibility of a very light $A$,
especially in a series of papers by Dermisek and
Gunion~\cite{Dermisek:2005ar,Dermisek:2005gg,Dermisek:2006wr,Dermisek:2006py,Chang:2008cw}.
One motivation is that precision electroweak fits
prefer the light Higgs to be
lighter
than the 114 GeV bound,
and this could be allowed if the
$h\rightarrow AA \rightarrow 4\tau$ or 4 jet signature exists.
LEP has published~\cite{Abbiendi:2002qp} a bound
of 82 GeV for the light Higgs,
independent of decay modes
(assuming one is not near the gauge-phobic limit),
but there is still a substantial allowed window.
If the $A$ is heavier than twice the $b$ quark mass,
then $b\bar{b}$ decays will dominate.
So bounds generally can be found for $A$ masses between 4 and 10 GeV
and Higgs masses between 80 and 115 GeV.
The bounds are generally expressed in terms of $\xi$, where
\begin{equation}
\xi^2 = \left(
\frac{g_{hVV}}{g^\mathrm{SM}_{hVV}}
\right)^2
\mathrm{BR} \left( h\rightarrow AA \right)
\left[ \mathrm{BR} \left( A\rightarrow\tau^+\tau^- \right) \right]^2,
\end{equation}
and in the bounds the range $\left[ 0.1, 1 \right]$
is typically~\cite{Schael:2010aw} considered for $\xi^2$.
In the type~I 2HDM,
however,
$\xi^2$ will generally be smaller.
The reason is that the $A$ can,
in the type~I 2HDM,
decay into $c\bar{c}$.
Since the running $c$-quark mass at the 10 GeV scale
isn't that much smaller than the $\tau$ mass,
the rates of $c\bar{c}$ and $\tau^+\tau^-$ will be similar.
Thus $\mathrm{BR} \left( A\rightarrow\tau^+\tau^- \right)$ and,
consequently,
$\xi$ will be smaller.
This will not be the case in the
type~II 2HDM.
We will not have much to say about this mode here,
since it will depend on
the completely arbitrary quartic couplings of the scalar potential
(unlike the MSSM or NMSSM,
where
those couplings
are specified).
The experimental possibility of measuring these quartic couplings
in the context of the four models discussed in this section
was considered by Arhrib {\it et al.}~\cite{Arhrib:2009hc}.
The possibility that this decay mode
could suppress the other ones should be kept in mind.

We now look at the other neutral Higgs bosons.
The heavier neutral scalar,
$H$,
has a coupling to fermions proportional to $\sin\alpha/\sin\beta$
and a coupling to vector bosons proportional to $\cos(\alpha-\beta)$.
These are \emph{identical}
to the couplings of the light Higgs,
$h$,
if one
shifts $\alpha$ to $\alpha - \pi/2$.
As a result,
for the decay modes of $H$ one may still use Fig.~\ref{fig:3hww1}
after shifting the graph to the left by $\pi/2$
(recall that a shift by $\pi$ in $\alpha$ does not affect any physics).
So the fermiophobic point will be
$\alpha=0$ and, for $\tan\beta=1$,
the gauge-phobic point will be at $\alpha=-\pi/4$.
As $\tan\beta$ increases,
just as in the case for $h$,
the gauge-phobic point moves towards the fermiophobic point.
It should be emphasised that
if there is a detection of a single scalar field with a mass of,
say,
190 GeV,
then it will be impossible to determine if it is an $h$ or an $H$
by looking at the decays,
since they are identical after a redefinition of $\alpha$.

\begin{figure}[ht]
\centerline{\epsfysize=15cm \epsfbox{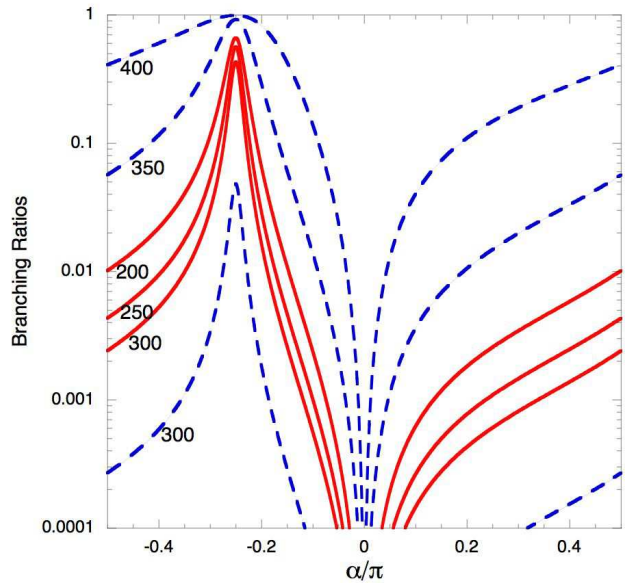} }
\vskip -3cm
\caption{Branching ratios of the heavy Higgs boson into
$b \bar b$ (solid lines) and $t \bar t$ (dashed lines),
for various values of the
heavy-Higgs mass. We have chosen $\tan{\beta} = 1$.}
\label{fig:3heavy}
\end{figure}

Unlike the light Higgs field,
however,
the $H$ can be substantially heavier.
As can be seen from Fig.~\ref{fig:3hww1},
the branching ratio into $WW$
will be fairly constant for heavier masses,
with a very sharp dip at the gauge-phobic point
(which is now at $\alpha=-\pi/4$ for $\tan{\beta} = 1$).
But,
the branching ratio into fermions
will change as the
top-quark threshold
is reached.
In Fig.~\ref{fig:3heavy}
we have plotted the branching ratio of the $H$ into
$b \bar b$ and $t \bar t$.
Note that the curve for $H$ decays into $b\bar{b}$
at a Higgs mass at 200 GeV
is
identical to that for $h$ decays
in Fig.~\ref{fig:3hww1},
shifted by $\pi/2$.
As the Higgs mass increases,
the branching ratio to $b\bar{b}$ decreases
(except at the gauge phobic point)
until the
top-quark threshold
is reached.
As the threshold is reached,
top quarks dominate
(with a branching ratio of nearly $100\%$ at the gauge-phobic point).

Our previous discussion about Higgs bosons
decaying into other Higgs bosons still applies.
In this case,
one can also have $H\rightarrow hh$
if $h$ is light enough.
Since this depends on unknown scalar self-couplings,
we will not have much to say about this mode.
The same is true for the $H\rightarrow AA$ mode
if the pseudoscalar is very light.

It was noted above that the decay of the light Higgs into a $Z$
and a light pseudoscalar has not been studied,
primarily because it
is
small in the MSSM.
However,
the decay of the heavy Higgs into
$ZA$
has been discussed in detail by Dermisek and
Gunion~\cite{Dermisek:2008uu} and by Kao {\it et al.}~\cite{Kao:2003jw},
who note that it can,
for a reasonable range of parameters,
dominate the decays  of the $H$.

An interesting scenario that has, to our knowledge, not been studied
is that the $h$ and $H$ could be very close in mass (a few GeV mass difference
would be within the LHC experimental resolution). While fine-tuned, this
possibility leads to the question of whether this could be distinguished
from the Standard Model Higgs.    If the decays into $WW$ or $ZZ$ are
considered, then the $h$ ($H$) will have a $\cos^2(\alpha-\beta)$
($\sin^2(\alpha-\beta)$) factor in the decay rate, leading to a Standard Model
decay rate if both are added.  However, the production rate will depend on
the top-quark coupling (since gluon fusion dominates), as will the $\gamma\gamma$
decay rate, leading to a more complicated and model dependent picture.
This possibility deserves further analysis.

Finally,
one can look at the decays of the pseudoscalar.
Here,
all fermion couplings are multiplied by $\cot\beta$,
and there are no couplings to a pair of vector bosons.
Thus,
the branching ratios will be independent of $\alpha$ and $\beta$,
and are just given by fermion mass ratios and phase space.
It should be remembered that all fermions are accessible (in the
type~I 2HDM)
and thus the branching ratio into $\tau^+\tau^-$
will always be similar
(within a factor of two) to the one into $c\bar{c}$.
If the $A$ is heavier than 10 GeV,
then decays into $b\bar{b}$ will dominate
(until the
top-quark threshold
is reached).
Since it is difficult to detect $b\bar{b}$ at a hadron collider,
one can look at other decays,
such as $\tau^+\tau^-$,
$W^\pm H^\mp$,
$Zh$,
$ZH$,
$gg$,
$\gamma\gamma$,
$Z\gamma$.
A very comprehensive analysis of all
these decays
in the
type~I 2HDM
can be found in the article by Kominis~\cite{Kominis:1994fa}.

If the
mass of the pseudoscalar
is above $m_h+M_Z$,
but below 350 GeV
(when the decays into $t \bar t$ will start to dominate),
a possible decay mode is $A \rightarrow hZ$.
Given that there are no decays to a pair of vector bosons,
and the decay into bottom quarks is suppressed by $m_b^2$,
this decay could be dominant over the
region~\cite{Kominis:1994fa, Baer:1992uu,Abdullin:1996as}.
The decay width is given in the Higgs Hunter's Guide~\cite{Gunion:1989we}.
It does depend on
$\cos(\alpha-\beta)^2$ and one would expect the decay
to dominate if it is kinematically accessible
and if that quantity is not small.
A recent discussion about detection at the LHC is in Ref.~\cite{Gupta:2009wn}.
The other decays mentioned above
are loop effects and will generally be small,
but may be experimentally easier to detect.

Kominis~\cite{Kominis:1994fa} has studied the diphoton decay channel
and the $Zh$ channel in the
type~I and type~II 2HDMs
and compared it with the results of the MSSM.
He found that in the lower end of the mass range,
due to the difficulty in seeing $b\bar{b}$ pairs at a hadron collider,
the diphoton mode would be
the most promising one.
The process $A\rightarrow hZ$ provides a very clear signature
in the intermediate mass range
in which the $h$ is between 40 and 160 GeV.
These two modes are thus complementary.
This work was done in 1994,
and an updated analysis would be
welcome.

Recently, Bar-Shalom, Nandi and Soni~\cite{arXiv:1105.6095} studied
2HDMs in which dynamical electroweak symmetry breaking is triggered
by condensation of 4th generation fermions. Their models have one ``heavy"
doublet with a large vev and a ``light" doublet with a much smaller vev.
The heavy doublet couples to the 4th generation fermions and, in one of the
models to the 3rd generation of fermions, in another to the top quark only,
and in another to only the 4th generation fermions. The phenomenology is
similar to the type I 2HDM. They study the phenomenology of the models,
including precision electroweak measurements, rare decays and collider
implications.

\subsubsection{Higgs decays in the type~II 2HDM}

The type~II 2HDM
is the most studied one,
since the couplings of the MSSM are a subset of the couplings
of the type~II 2HDM.
The coupling of the light neutral Higgs,
$h$,
to fermions depends on the fermion charge.
The coupling of the $Q=2/3$ quarks is the same as in the type~I 2HDM,
\textit{i.e.}\ it is the Standard-Model coupling
times $\cos\alpha/\sin\beta$.
On the other hand,
the coupling of the $Q=-1/3$ quarks
and of the leptons is the Standard-Model coupling
times $-\sin\alpha/\cos\beta$.
In the large $\tan\beta$ scenario,
this means that the couplings of the $Q=-1/3$ quarks
and of the leptons are much larger than in the type~I 2HDM.
In fact,
one can see that the ratio of the bottom quark Yukawa coupling
to that of the top quark
is approximately $\tan\alpha\tan\beta$ times the same ratio for the type~I 2HDM,
and this can drastically affect the phenomenology.
The couplings to gauge bosons are the same as in the type~I 2HDM.
Therefore, one
still has the gauge-phobic limit for $h$
if $\sin(\alpha-\beta)=0$,
but there is no fully fermiophobic limit.
If $\alpha=\pm\pi/2$ ($\alpha=0$),
then the $h$ won't couple to the $Q=2/3$ quarks
($Q=-1/3$ quarks and leptons),
but no choice of $\alpha$ will eliminate all couplings to fermions.

\begin{figure}
\vskip -3cm
\centerline{\epsfysize=12cm
\epsfbox{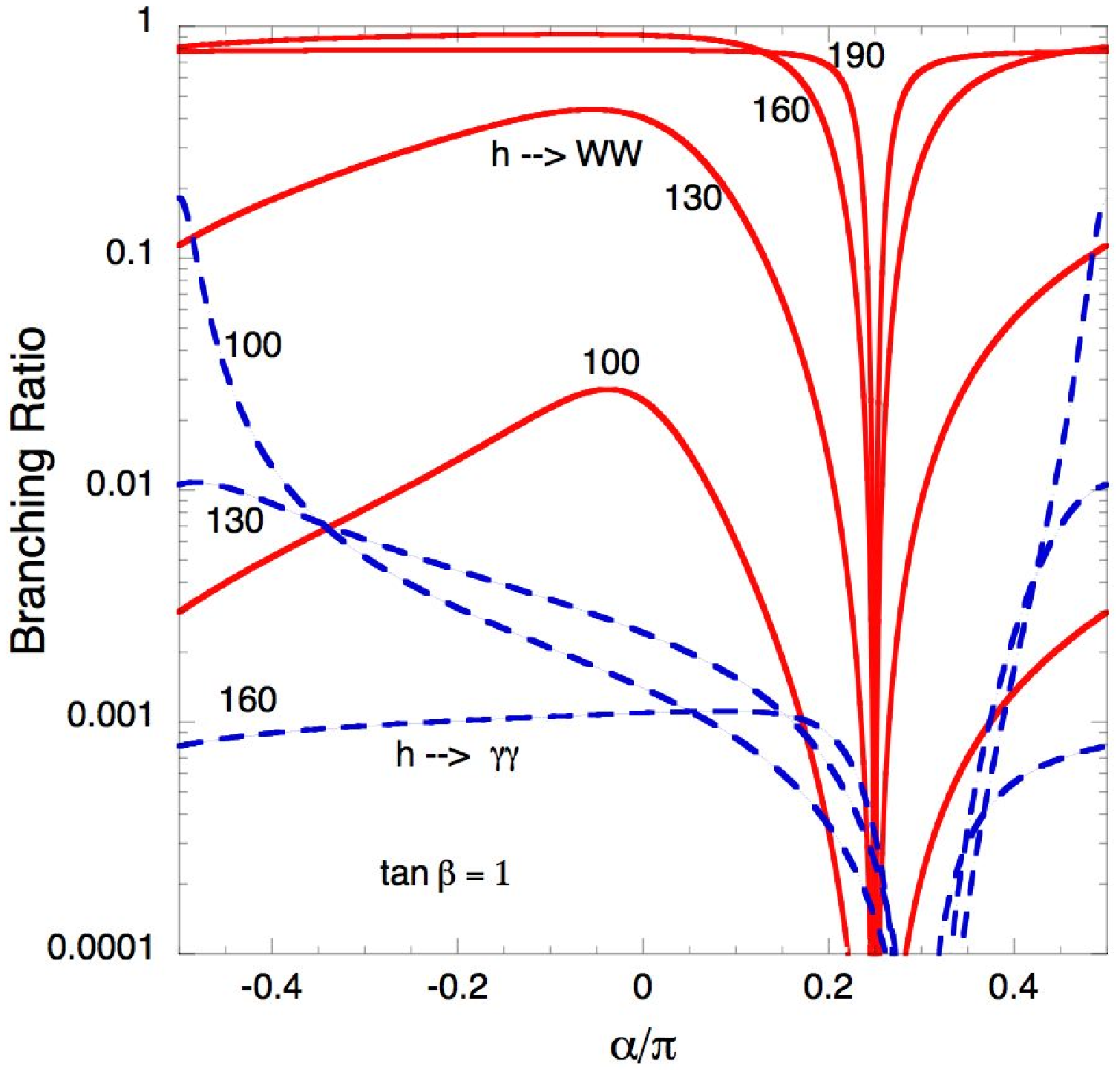}\epsfysize=12cm \epsfbox{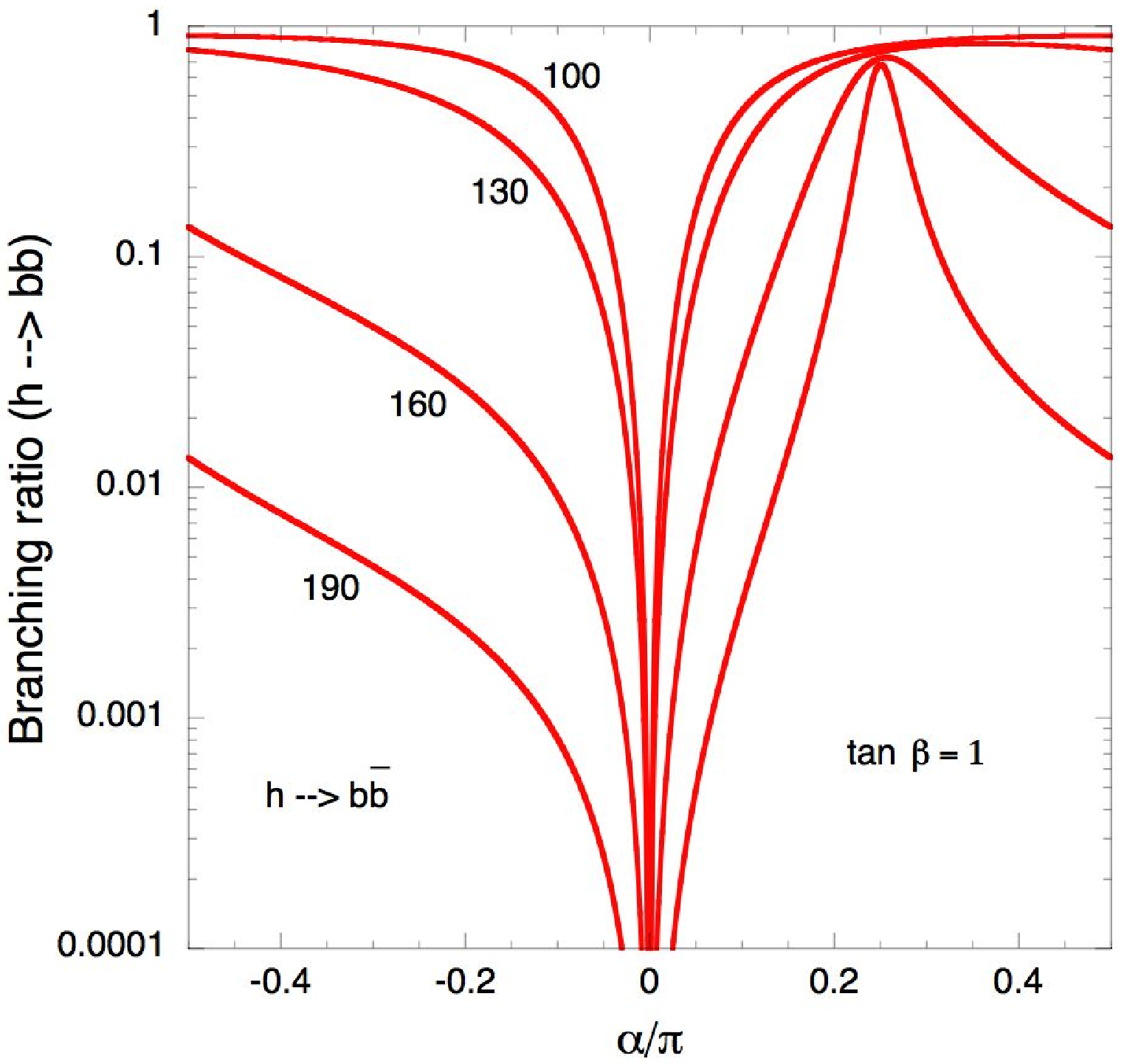} }
\vskip -2cm
\centerline{\epsfysize=12cm
\epsfbox{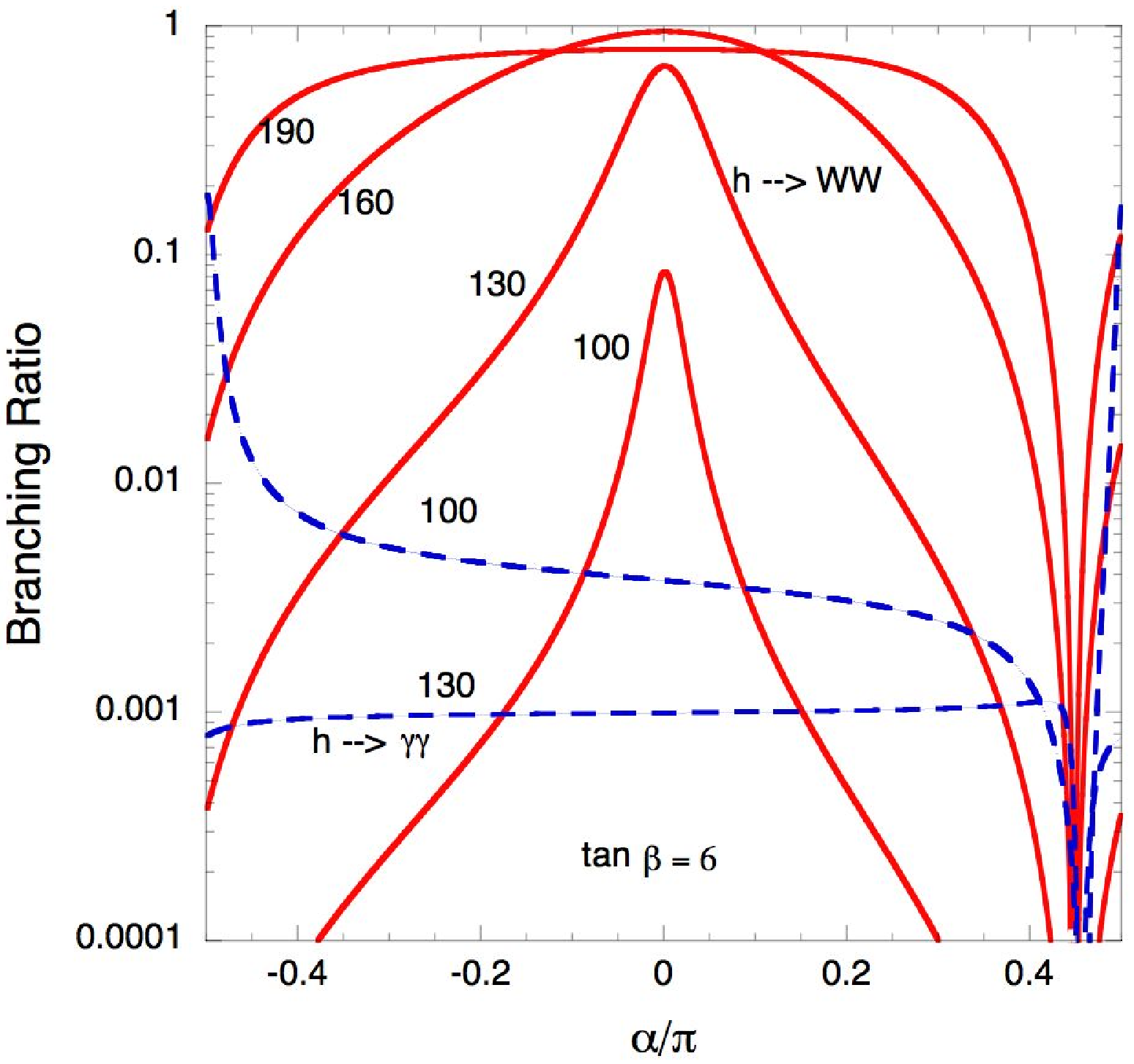}\epsfysize=12cm \epsfbox{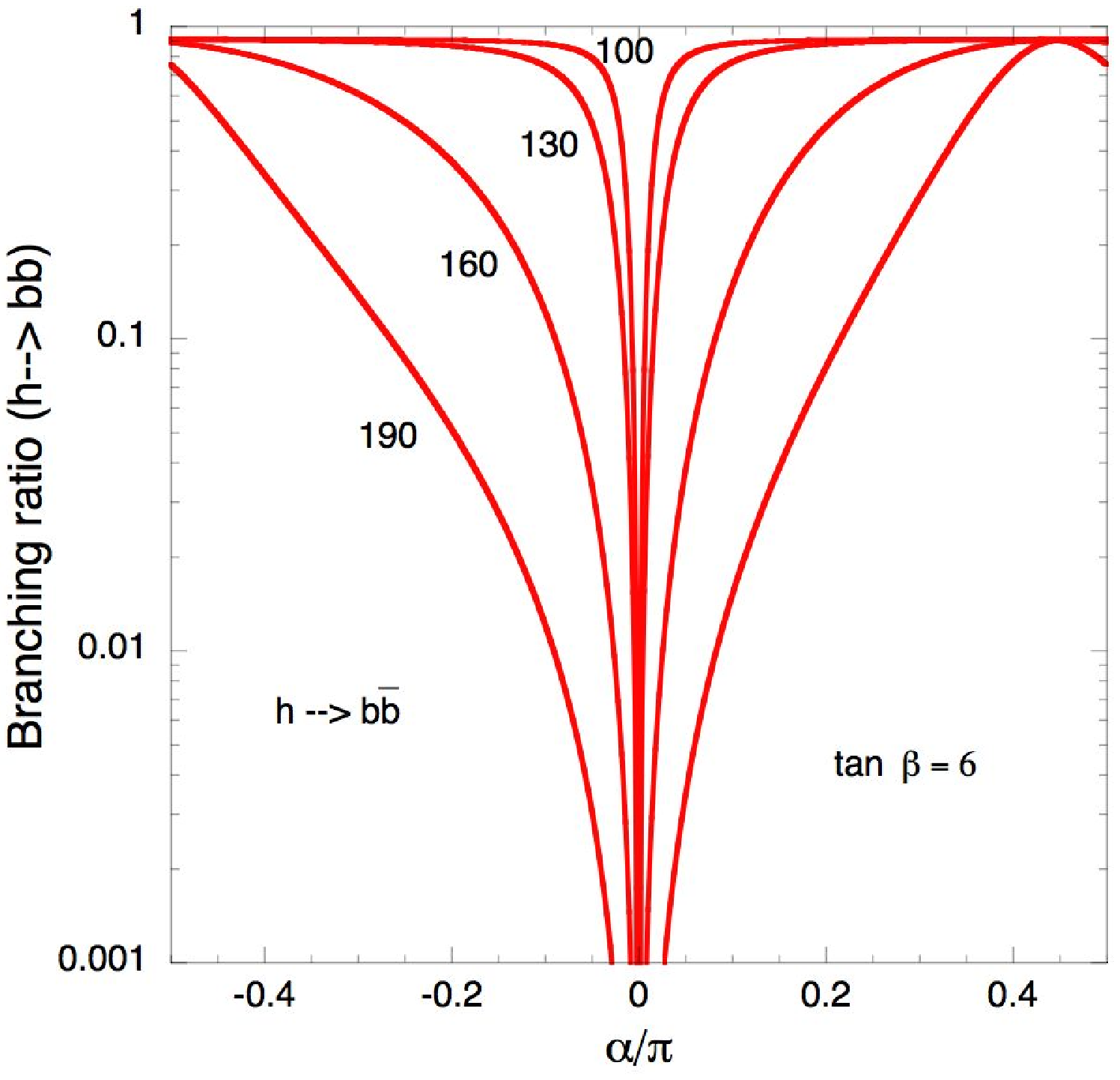} }
\vskip -2cm
\caption{The type~II 2HDM
light-Higgs branching ratios
into $W$ pairs,
diphotons and $b\bar{b}$ are plotted as a function of $\alpha$
for $\tan\beta=1$ and $\tan\beta=6$
and for
various values of the Higgs mass (in GeV).
In the left-hand figures, the solid lines
correspond to $h\rightarrow WW$
and the dashed lines to $h\rightarrow \gamma\gamma$.  The branching ratio into $Z$ pairs
has the same ratio to
the one into $W$ pairs
as in the Standard Model.}
\label{fig:3hww2}
\end{figure}

The branching ratios of the light neutral Higgs
are plotted in Fig.~\ref{fig:3hww2}
for $\tan\beta=1$ and $\tan\beta=6$.
Note that,
as expected,
the curves for $\tan\beta=1$ are very similar to those
in Fig.~\ref{fig:3hww1} if one shifts $\alpha$ by $\pi/2$
and flips the sign.
The $b$-phobic value of $\alpha$ is zero in the type~II 2HDM.
At the $b$-phobic point,
say for a Higgs mass of 100 GeV,
neither the $b\bar{b}$ nor the $WW, ZZ$ modes are substantial.
In this case,
the $c\bar{c}$ and gluon-gluon decays dominate.
Since neither can be seen at the LHC,
we have not included them here,
although a very similar figure with these modes included
appears in Arhrib {\it et al.}~\cite{Arhrib:2009hc}
for $m_h = 110$ GeV and $\tan\beta=1$.

Unlike the type~I 2HDM,
however,
there is in this case a strong dependence on $\tan\beta$.
In Fig.~\ref{fig:3hww2}
we also show the decays for $\tan\beta=6$.
One sees that,
as expected,
the gauge-phobic point $\alpha=\beta$ moves to the right,
and the $b$-phobic point remains at $\alpha=0$.
For a relatively light $h$,
with a mass of 100 GeV,
the enhanced coupling of the $b$ quark when $\tan{\beta} = 6$
causes the phase-space suppression of the $WW$ decay mode
to be quite dramatic,
with virtually no $WW$ pairs except at the $b$-phobic point.
Even at that point,
the $WW$ branching ratio is only around $10\%$,
since the gluon-gluon decay goes through
a top-quark loop
which is not suppressed at the $b$-phobic point,
and thus becomes dominant.
Still,
slightly away from the $b$-phobic point,
the $b\bar{b}$ mode already dominates again although, for a light $h$,
the diphoton mode is not negligible.
It is interesting to note that
this model contains a sizable region
of parameter space
in which a light Higgs with a mass of 160 GeV
will still predominantly decay into
$b$ quarks rather than $W$ pairs.

For $\tan\beta=30$,
the gauge-phobic point
is
extremely close to $\alpha=\pi/2$,
and the trend becomes even more dramatic,
with the $m_h = 190$ GeV curve
becoming close to the $m_h = 130$ GeV curve of the $\tan\beta = 6$ plot.
At this point,
even a relatively heavy $h$
will have negligible couplings to $WW$ or $ZZ$
for most values of $\alpha$.

It should be kept in mind that the branching ratio into $\tau^+\tau^-$
is approximately $10\%$ of
the one
into $b\bar{b}$.
The backgrounds
for the $b\bar{b}$ mode are huge at the LHC,
rendering that decay mode very difficult to
observe,
while the backgrounds
for $\tau^+\tau^-$ are much
less severe and therefore,
even if the $b\bar{b}$ mode cannot be seen at the LHC,
$h$ could still be detected through its decay to $\tau^+ \tau^-$.

As noted earlier,
the recent review article by Djouadi~\cite{Djouadi:2005gj}
studies the Higgs phenomenology of the MSSM in great detail,
and that is a special case of the type~II 2HDM.
Much of the analysis in that article will be relevant here.
In the general case,
there are some differences.
In the MSSM,
the light Higgs cannot be much heavier than 130 GeV,
as it can here.
In addition,
$\alpha$ is determined in terms of other Higgs masses and $\beta$
and is thus much more restricted.
Nonetheless,
the basic features shown in Fig.~\ref{fig:3hww2} are,
for $h$ masses at 130 GeV and below,
also present in the MSSM.
For large $\tan\beta$,
the $b\bar{b}$ decay mode dominates for most of parameter space.
However,
the MSSM (at tree level) does not allow $\alpha=0$,
and thus the $b$-phobic region does not occur.
The major differences,
then,
are
that the general type~II 2HDM allows
for a heavier $h$
and that it also allows for the possibility that a fairly light $h$
could still
predominantely
decay into $W$ pairs.

In the discussion of the type~I 2HDM,
we also discussed the decays of the light Higgs
into a $Z$ plus a pseudoscalar
and of a light Higgs into two pseudoscalars.
Nothing much changes in the type~II 2HDM.
One still has a region of parameter space in which $h\ra ZA$
can be substantial,
and $h\ra AA$ can be significant if $A$ is light.
The only difference is that if the $A$ has a mass between 4 and 10 GeV,
then its branching ratio into $\tau^+\tau^-$
will be larger for large $\tan\beta$,
facilitating detection.

\begin{figure}
\vskip -3cm
\centerline{\epsfysize=12cm \epsfbox{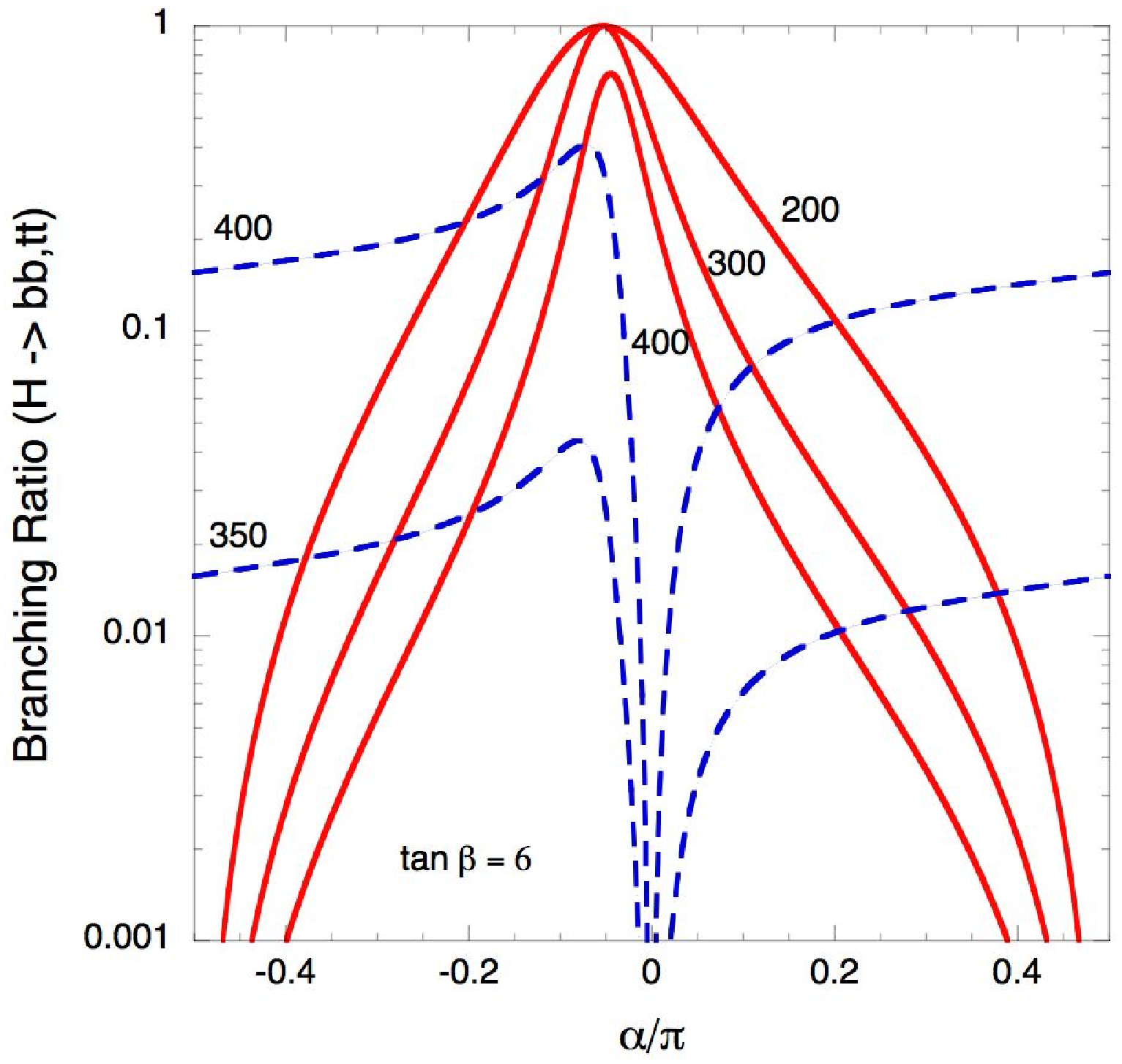} \epsfysize=12cm \epsfbox{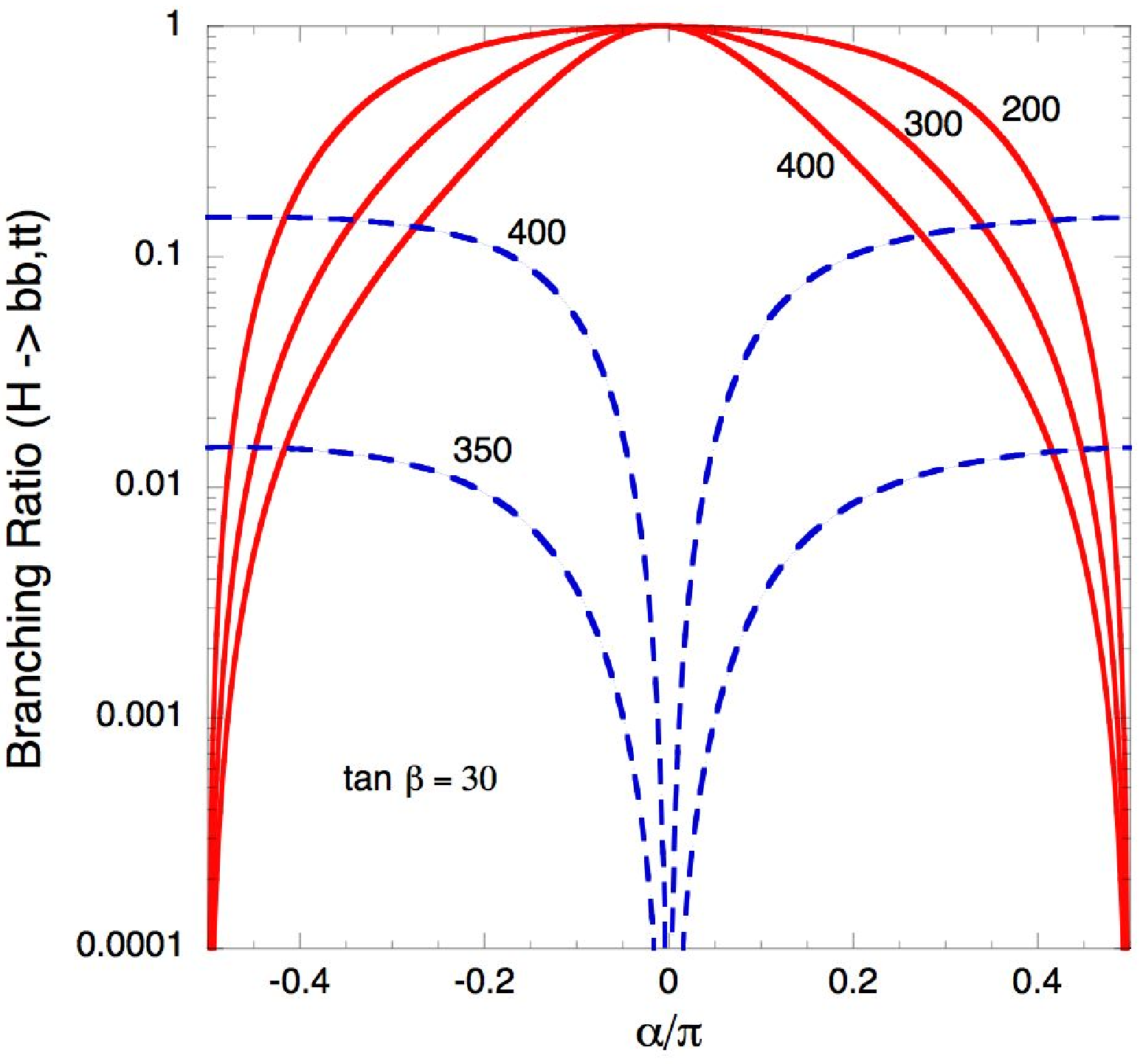} }
\vskip -2cm
\caption{Branching ratios of the heavy Higgs boson
into fermions for $\tan\beta=6,30$.
The solid (dashed) lines are the branching ratios
into $b\bar{b}$ ($t\bar{t}$)
for various values of the heavy-scalar mass.}
\label{fig:3heavy2}
\end{figure}

Turning to the other neutral Higgs bosons,
one can see that the couplings
of the heavier scalar $H$
are identical to those of the $h$
if one replaces $\alpha$ by $\alpha-\pi/2$.
Thus the decay modes of the $H$,
if it is lighter than 200 GeV,
will be identical to those in Fig.~\ref{fig:3hww2}
after $\alpha$ is shifted to the left by $\pi/2$.
For heavier $H$ fields,
however,
there is a substantial difference between the type~I and type~II 2HDMs.
This is because in the type~II 2HDM,
the ratio of the branching fraction into bottom quarks compared to
the one
into top quarks
varies as $\tan^2\beta\tan^2\alpha$ compared with the type~I 2HDM.
In addition,
the $b$-phobic and $t$-phobic points do not coincide.
For $\tan\beta=1$,
the results for the $t\bar{t}$ mode
are identical to those in Fig.~\ref{fig:3heavy},
and the results for the $b\bar{b}$ are the same
but with $\alpha$ shifted by $\pi/2$
(so the $b$-phobic points are at $\alpha=\pm \pi/2$).
In Fig.~\ref{fig:3heavy2},
we plot the branching ratios of the heavy Higgs into fermions
for $\tan\beta=6$ and $\tan\beta=30$.
Again,
one sees that the $t$-phobic and $b$-phobic points are now different.
There is here
a striking difference between the type~I and type~II 2HDMs.

The pseudoscalar does not decay into gauge bosons,
and thus it will decay into the heaviest fermions accessible,
as in the type~I 2HDM.
However,
now the decay to $Q=-1/3$ quarks ($Q=2/3$ quarks)
will be multiplied by $\tan^2\beta$ ($\cot^2\beta$),
and for large $\tan\beta$ this will strongly suppress
the coupling to top quarks.
In fact,
for $\tan\beta$ greater than about 6--7,
the decay into bottom quarks will always exceed that into top quarks.

Thus,
for $\tan\beta$ near $1$,
the results for pseudoscalar decay are identical
to those of the type~I 2HDM.
For a pseudoscalar mass in between 4 and 10 GeV,
the main decay mode is $\tau^+\tau^-$,
with a comparable branching ratio into $c\bar{c}$.
Above 10 GeV,
the principal decay is into $b\bar{b}$
(with $\tau^+\tau^-$ being approximately $10\%$
of $b\bar{b}$).
As discussed in detail by Kominis~\cite{Kominis:1994fa},
it might be easier,
in this mass range,
to look for the one-loop diphoton mode.
Once the mass exceeds 200 GeV,
$A\ra hZ$ becomes possible,
and then above 350 GeV,
decays into top quarks dominate.

For $\tan\beta$ much larger,
say larger than $10$,
the decay of a 4--10 GeV pseudoscalar
is almost entirely into $\tau^+\tau^-$,
due to the absence of charm decays.
Above 10 GeV,
$b\bar{b}$ dominates,
and this domination continues for all
masses; since the $b \bar b$ mode is difficult to observe at the LHC,
it may be necessary
to look at the diphoton mode to
find the pseudoscalar.
Kominis~\cite{Kominis:1994fa} discusses these possibilities in detail.

It is straightforward to see
(as discussed above)
how the results of this section can be expanded
to larger values of $\tan\beta$---the results do not change substantially,
except for the ratio of $h$ decays into bottom quarks relative to top quarks.
For a much more recent discussion with many references,
which look at both
the CP-conserving and CP-violating models,
see the papers of
Kaffas, Osland and Ogreid~\cite{ElKaffas:2007rq,WahabElKaffas:2007xd}.

There is a class of models in which the electroweak symmetry is broken by the
condensation of a strongly coupled fermion sector. Although this sector could
come from a fourth generation, it need not. As noted originally by Luty~\cite{EFI-89-66},
if this strongly interacting sector respects isospin invariance, then the resulting
low energy theory is a two-Higgs doublet model. Using an RG-improved Nambu-Jona-Lasinio
model, Burdman and Haluch~\cite{arXiv:1109.3914} studied the effective low-energy
scalar sector. They found that, not surprisingly, the scalars are all fairly heavy
(in the 600-800 GeV region), but also found that the pseudo-scalar is light, with a
mass ranging from 10 to 120 GeV. They discuss the phenomenology of the model, including
precision electroweak fits, and find it similar to a type II 2HDM with an
unusual mass spectrum and with $\tan\beta\sim 1$.

\subsubsection{Higgs decays in the lepton-specific 2HDM}

The couplings of the quarks to the Higgs bosons
in the lepton-specific (LS) 2HDM
are identical to those in the type~I 2HDM,
but
the couplings of the leptons are quite different.
In the previous models,
the branching ratio into $\tau^+\tau^-$ was roughly $10\%$
of the branching ratio into $b\bar{b}$ for all values of the parameters.
This is not the case
in the LS~2HDM.
There are two major differences between the LS 2HDM and the type~I 2HDM.
Firstly,
the branching ratio of the $h$ into $\tau^+\tau^-$ can be much larger and,
compared with the
branching ratio into
$b\bar{b}$,
grows as $\tan^2\beta\cot^2\alpha$.
Secondly,
the $b$-phobic value of $\alpha$ is $\pm\pi/2$,
whereas in the LS~2HDM,
the $\tau$-phobic value of $\alpha$ is $0$.
This can dramatically affect
the phenomenology---the $\tau^+\tau^-$ branching ratio
can exceed that of $b\bar{b}$
even for values of $\tan\beta$ near unity.

As in the type~I 2HDM,
the $h$ field ($H$ field) will be gauge-phobic
if $\sin(\alpha-\beta)$ ($\cos(\alpha-\beta)$) vanishes.

\begin{figure}
\vskip -4cm
\centerline{\epsfysize=12cm \epsfbox{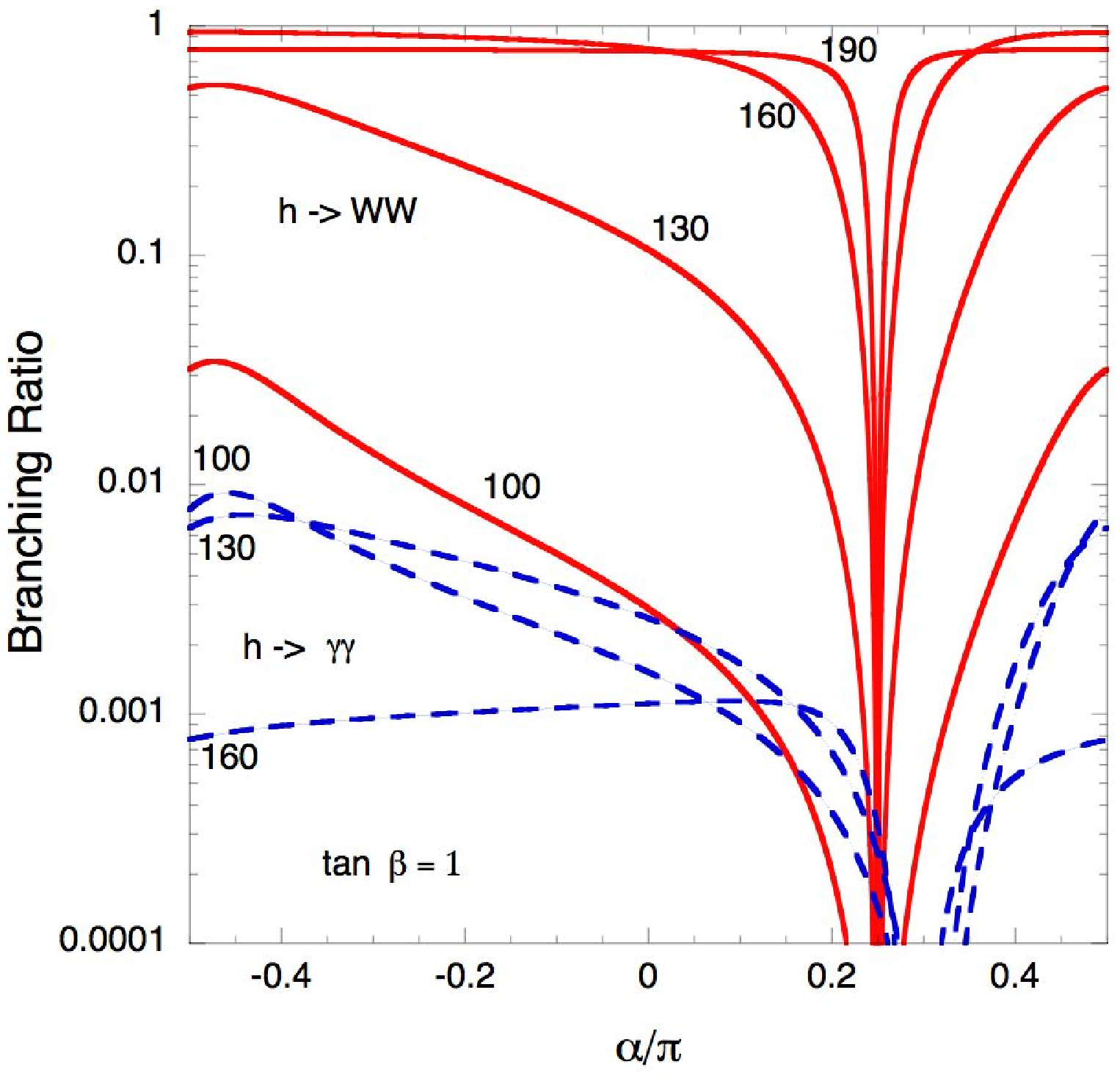}
\epsfysize=12cm \epsfbox{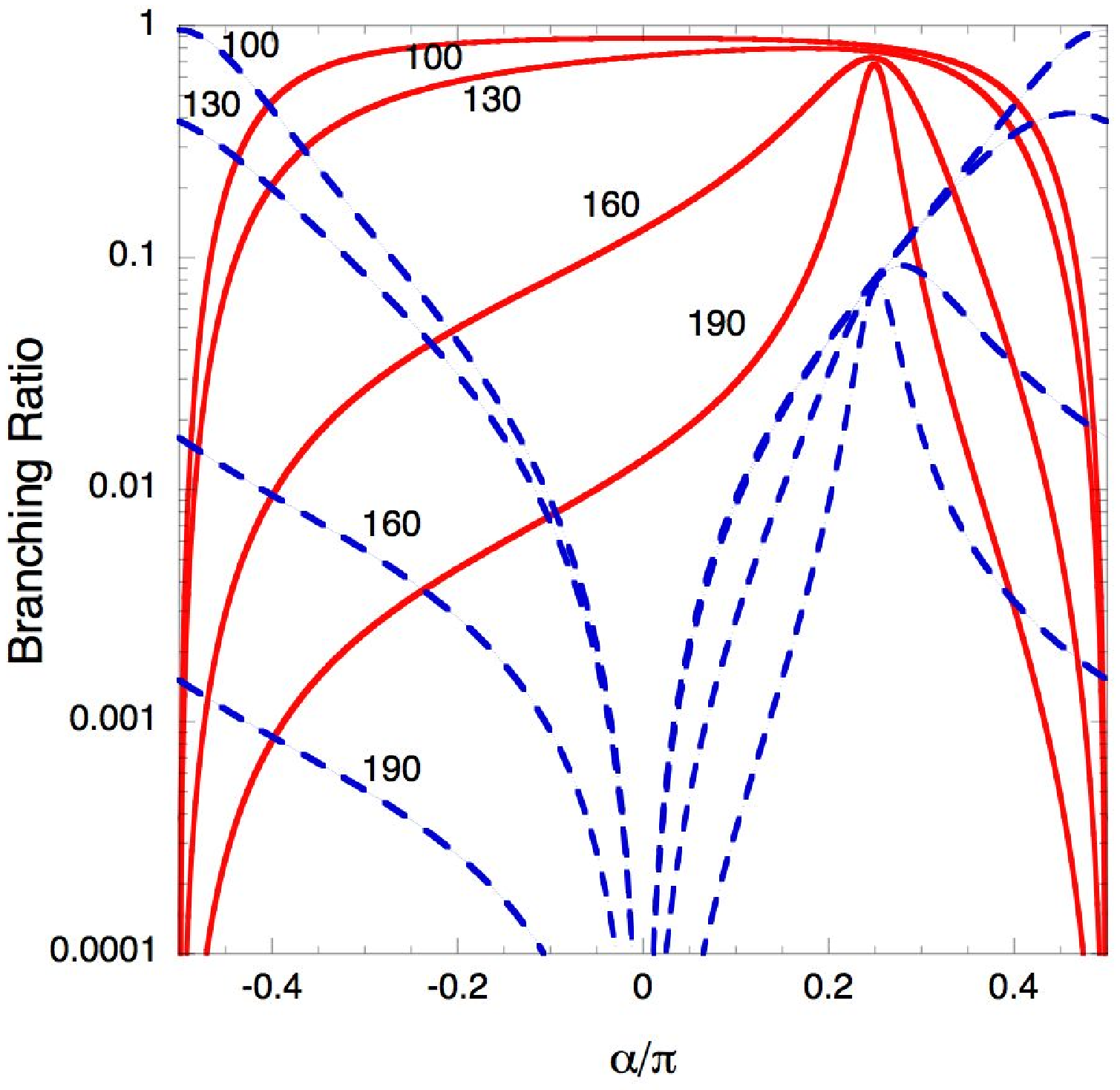} }
\vskip -2cm
\caption{Branching ratios of the light Higgs boson
in the lepton-specific 2HDM. We have taken $\tan{\beta} = 1$.
In the left figure,
the solid (dashed) lines are the branching ratios into $WW$
($\gamma\gamma$); in the right figure,
the solid (dashed) lines are the branching ratios into $b\bar{b}$
($\tau^+\tau^-$)
for various values of the
Higgs-boson mass.}
\label{fig:33tb1}
\end{figure}

In Fig.~\ref{fig:33tb1},
we have plotted the branching ratios of the light Higgs,
$h$,
into $WW$,
$\gamma\gamma$,
$b\bar{b}$ and $\tau^+\tau^-$ for various $h$ masses,
assuming $\tan\beta=1$.
With such a low $\tan\beta$,
one might expect that there would be relatively little contribution
from the $\tau^+\tau^-$ mode.
However,
since the $\tau$-phobic and $b$-phobic points are different,
one can see that near the $b$-phobic point,
the $\tau^+\tau^-$ mode dominates the decays of the $h$
for $m_h = 100$ GeV.
This is understandable,
since the $WW$ mode is phase-space suppressed
and one is near the $b$-phobic point.
Thus,
near $\alpha=\pm\pi/2$,
the $\tau^+\tau^-$ mode
supersedes the other branching ratios.
Note that away from this point,
the curves look virtually identical to those in the type~I 2HDM,
Fig.~\ref{fig:3hww1},
as expected.
Su and Thomas~\cite{Su:2009fz}
have pointed out that the region of parameter space
near the $b$-phobic point will violate perturbativity
or vacuum stability in the limit
where
the $H$ and $A$ fields are quite heavy.
They studied the LHC discovery potential of the LS~2HDM
in the case where the low-energy spectrum only contained
one
light Higgs boson.

\begin{figure}
\vskip -2.5cm
\centerline{\epsfysize=15cm \epsfbox{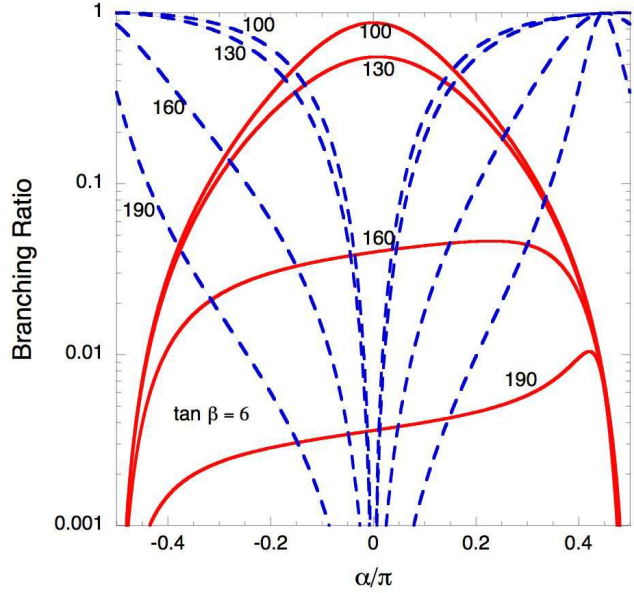} }
\vskip -3cm
\caption{Branching ratios of the light Higgs boson into
fermions in the LS~2HDM with $\tan{\beta} = 6$.
The solid (dashed) lines are the branching ratios into $b\bar{b}$
($\tau^+\tau^-$)
for various values of the
Higgs-boson mass.}
\label{fig:33tb6}
\end{figure}

One can also look at larger values of $\tan\beta$,
where the coupling to $\tau$s will increase substantially.
We have plotted the branching ratios into fermions
for $\tan\beta=6$ in Fig.~\ref{fig:33tb6}.
One sees a dramatic increase in the $\tau^+\tau^-$ mode
away from the $\tau$-phobic point.
In fact,
even for a Higgs mass of 160 GeV,
there is a range of $\alpha$~\cite{Akeroyd:1996di} for which
the $\tau^+\tau^-$ mode
is the {\em dominant} decay.
The fact that the curves do not add up to $100\%$
is due to our not having included the gluon-gluon
and $c\bar{c}$ decay modes in the figure;
these cannot be measured easily at the LHC.
The paper of Arhrib {\it et al.}~\cite{Arhrib:2009hc}
has very similar figures which
(for a specific $h$ mass of 110 GeV)
show these decays as well.
By comparing this figure with Fig.~\ref{fig:33tb1},
one can see the pattern for very large $\tan\beta$.
For $\tan\beta = 100$,
for example,
which is still allowed by perturbation theory
(although, as noted earlier,
may have difficulties with tree-level unitarity),
the $\tau^+\tau^-$ mode would completely dominate the $h$ decays
for all masses
for all masses of $h$,
except very close to $\alpha=0$.
A
comprehensive analysis of the $\tau^+\tau^-$ decay mode at the LHC
can be found in the papers of
Belyaev {\it et al.}~\cite{Belyaev:2009zd,Belyaev:2010bc}.
They
note that if decays to gauge bosons are kinematically forbidden,
and decays to other Higgs bosons are not allowed,
then one can write a simple formula:
\begin{equation}
\mathrm{BR} \left( h\ra \tau^+\tau^- \right) =
\frac{{\displaystyle \frac{\sin^2\alpha}{\cos^2\beta}}\,
\mathrm{BR} \left( h_\mathrm{SM} \ra \tau^+\tau^- \right)}
{\left( {\displaystyle \frac{\sin^2\alpha}{\cos^2\beta} -
\frac{\cos^2\alpha}{\sin^2\beta}} \right)
\mathrm{BR} \left( h_\mathrm{SM} \ra \tau^+\tau^- \right)
+ {\displaystyle \frac{\cos^2\alpha}{\sin^2\beta}} }.
\end{equation}
This equation works well for Higgs masses below 130 GeV.
Above that mass,
$WW^*$ decays can become important.
A recent study of multi-tau-lepton signatures can be found in
the work of Kanemura, {\it et al.}~\cite{arXiv:1111.6089}.

A decay mode of the $h$ that may be important
and is one of the easiest to detect is
$h\ra \mu^+\mu^-$.
The branching ratio is $0.0035$ times that of the $\tau^+\tau^-$ mode.
If the latter does dominate,
then the decay into muons should be clearly detectable.
Discussions of this mode  can be found
in Refs.~\cite{Plehn:2001qg,Han:2002gp,Su:2008bj}.

The previous discussion of $h\ra ZA$ and $h\ra AA$ in the type~I 2HDM
also applies here.
If the coupling to $\tau$s is large,
this will suppress the branching ratios somewhat from the discussion there,
but the general character of the analysis will not change.

\begin{figure}
\vskip -2.5cm
\centerline{\epsfysize=15cm \epsfbox{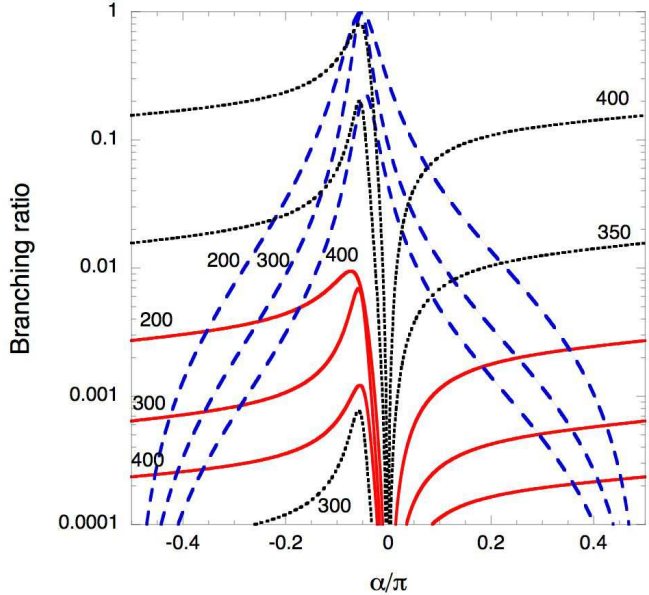} }
\vskip -3cm
\caption{Branching ratios of the heavy Higgs boson into fermions
for $\tan\beta=6$.
The solid
(dashed, dotted) lines,
which are red
(blue, black)
are the branching ratios into $b\bar{b}$
($\tau^+\tau^-$, $t\bar{t}$)
for various values of
the Higgs-boson mass.}
\label{fig:3heavy4}
\end{figure}

What about the other neutral Higgs?
Just as in the type~I 2HDM,
the couplings of the heavier neutral scalar,
$H$,
are identical to that of the lighter Higgs,
$h$,
if one shifts $\alpha$ to $\alpha-\pi/2$.
In Fig.~\ref{fig:3heavy4},
we have plotted the decays of $H$
into top quarks,
bottom quarks,
and tau leptons for $\tan\beta=6$.
We see that for $H$ masses below the top threshold,
the decays into $\tau^+\tau^-$ dominate
as soon as one moves away from the $\tau$-phobic point
(now at $\alpha=\pm\pi/2$ for the $H$ coupling).
For heavier masses,
the $t\bar{t}$ mode dominates the $b\bar{b}$ mode.
For larger $\tan\beta$,
the curves for the $\tau^+\tau^-$ mode
widen and can eventually dominate for
most values of $\alpha$.

The discussion of other decays,
such as $H\ra ZA$ or $H\ra AA$,
is
the same as for the type~I 2HDM.

We finally turn to the decays of the pseudoscalar $A$.
Here,
there are no decays into $WW$ or $ZZ$,
and thus the pseudoscalar decays primarily into fermions.
This provides
a remarkable opportunity for discovery of the pseudoscalar.
The ratio of the branching
fraction
into $\tau^+\tau^-$
to the one into
$b\bar{b}$ is proportional to $\tan^4\beta$,
which can easily exceed unity.
In fact,
including the mass effects,
one can show that
\begin{equation}
\frac{\Gamma \left( A\ra \tau^+\tau^- \right)}
{\Gamma \left( A\ra b\bar{b} \right)}
= \left( \frac{\tan\beta}{1.76} \right)^4,
\end{equation}
and therefore, even for relatively small $\tan\beta$,
the $\tau^+\tau^-$ mode will dominate.
In fact,
for $\tan\beta>3$,
the branching ratio exceeds $90\%$.
This is independent of the $A$ mass,
as long as it is below the
$t \bar{t}$ threshold.
As before,
the branching ratio into $\mu^+\mu^-$ is $0.0035$
times
the one into
$\tau^+\tau^-$,
and if the latter dominates,
this gives a sizable branching ratio into an easily observed signature.

\subsubsection{Higgs decays in the flipped 2HDM}

In the flipped 2HDM,
the RH leptons couple to the same Higgs doublet as the RH up quarks.
As in the lepton-specific model,
the $\tau$-phobic point is different from the $b$-phobic point,
leading to a region of parameter space
in which the $\tau^+\tau^-$ branching ratio
can exceed the $b\bar{b}$ branching ratio.
But unlike the lepton-specific model,
the flipped model cannot have a huge enhancement
of the $\tau$ coupling
to any of the scalars.
This is because any enhancement of the $\tau$ coupling
would also enhance
the top-quark coupling,
and a large enhancement of the latter
would cause serious problems with perturbation theory and unitarity.
One does not expect an enhancement of the $\tau^+\tau^-$ mode
in the flipped model as one changes
$\tan\beta$,
like the one observed in the
lepton-specific model, \textit{cf.}\ Figs.~\ref{fig:33tb1}
and~\ref{fig:33tb6}.

\begin{figure}
\vskip -4cm
\centerline{\epsfysize=12cm \epsfbox{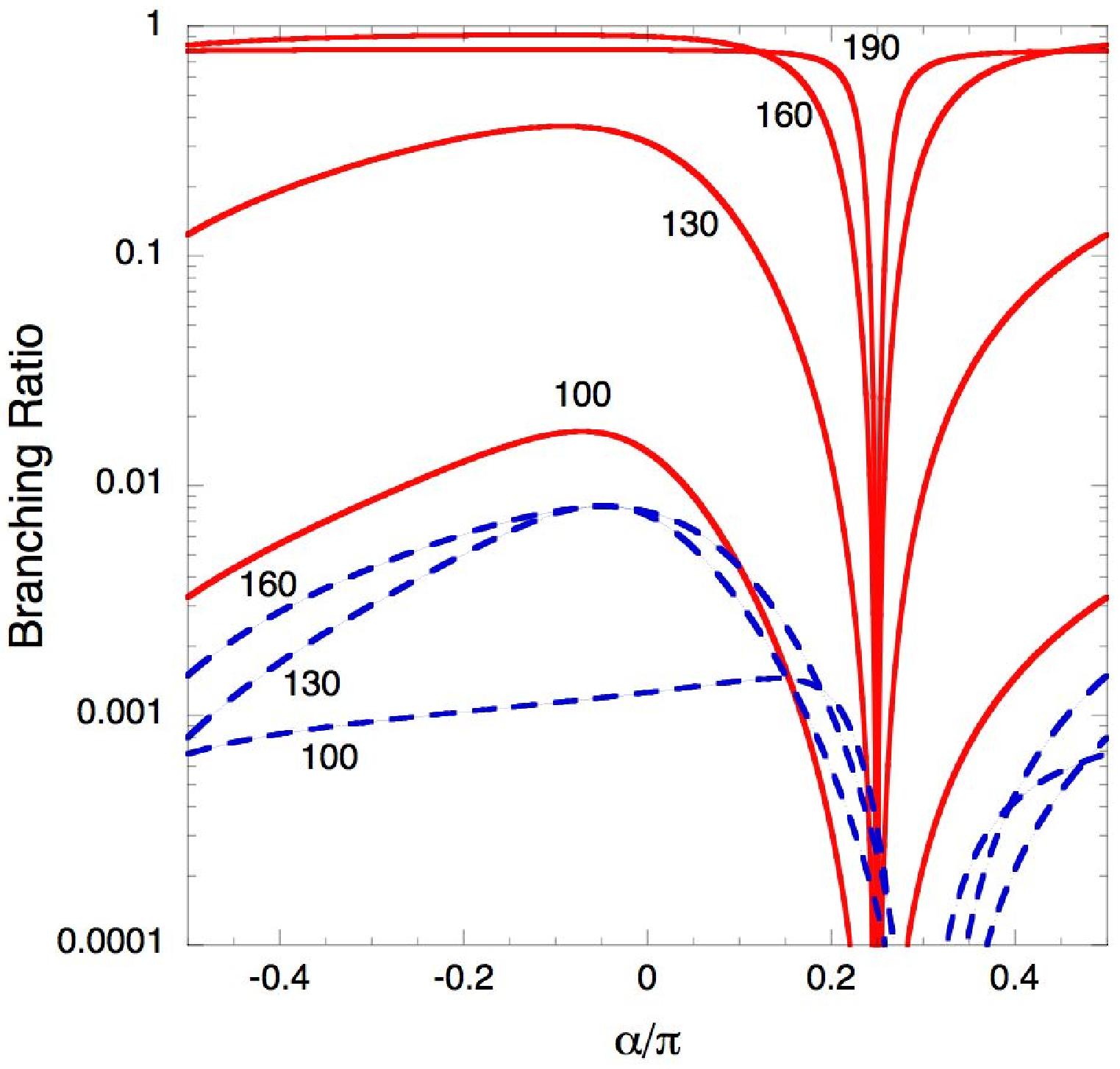}
\epsfysize=12cm \epsfbox{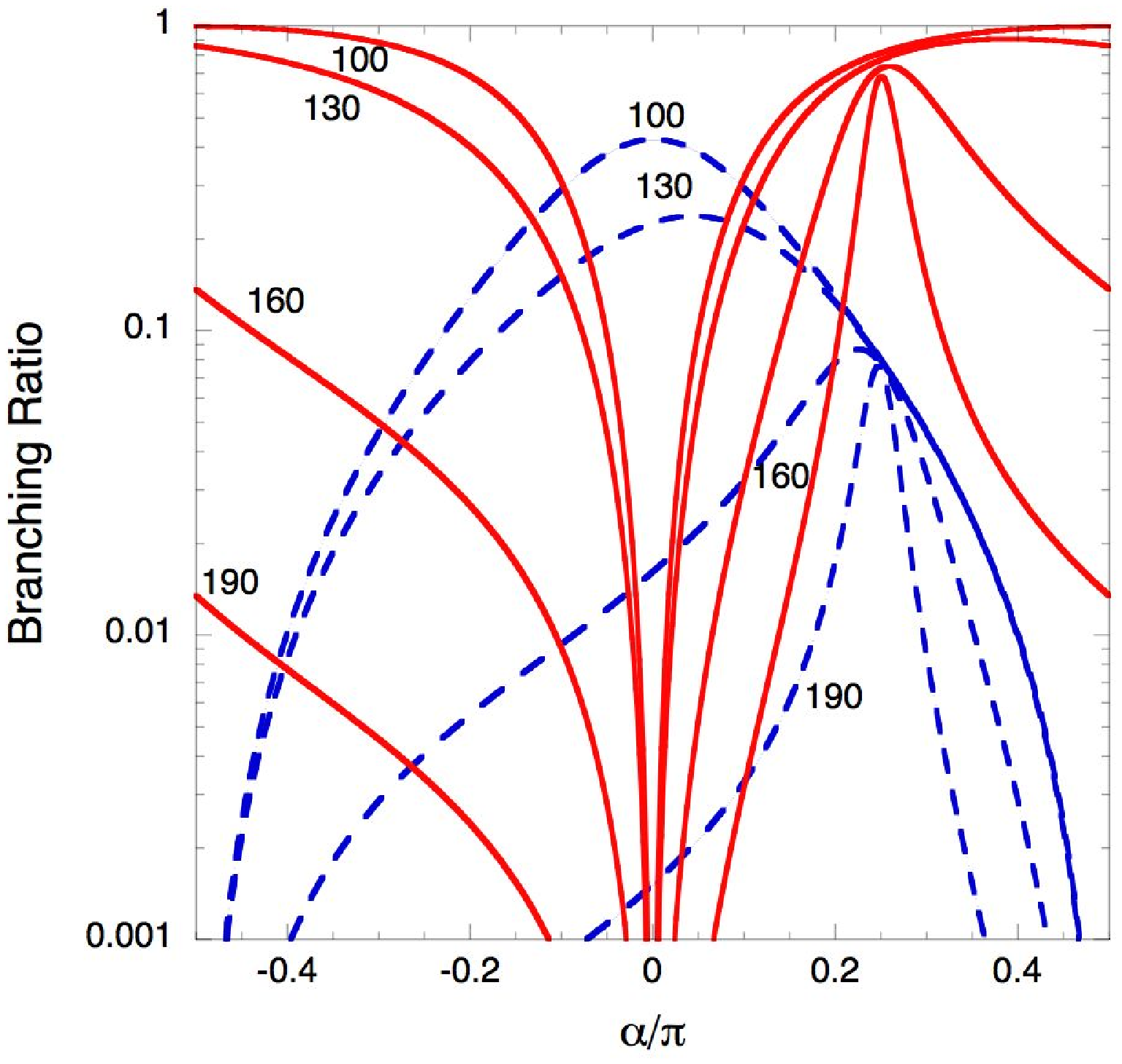} }
\vskip -2cm
\caption{Branching ratios of the light Higgs boson
in the flipped model. We have taken $\tan{\beta} = 1$.
In the left figure,
the solid (dashed) lines are the branching ratios into $WW$,
($\gamma\gamma$); in the right figure,
the solid (dashed) lines are the branching ratios
into $b\bar{b}$
($\tau^+\tau^-$)
for various values of the
Higgs-boson mass.}
\label{fig:34tb1}
\end{figure}

The branching ratios of the light scalar $h$ are presented
in Fig.~\ref{fig:34tb1} for $\tan\beta=1$.
One sees a  region of parameter space
in which the $\tau^+\tau^-$ mode dominates $b\bar{b}$ decays,
but it never reaches branching ratios as high
as in the lepton-specific case
(the remainder of the decays in the $\alpha=0$ region
is a mix of $gg$ and $c\bar{c}$).
All other branching ratios are very similar
to those from the type~II 2HDM.
If $\tan\beta$ is increased,
the region of $\tau^+ \tau^-$ dominance narrows,
and the region in which the $b\bar{b}$ mode dominates grows,
as in the larger $\tan\beta$ curves in the type~II 2HDM.
The $\alpha$ range for which the $\tau^+\tau^-$ decay is dominant
can grow if one takes $\tan\beta < 1$,
but
perturbation theory breaks down
for $\tan\beta < 0.3$.

\begin{figure}[h]
\centerline{\epsfysize=15cm \epsfbox{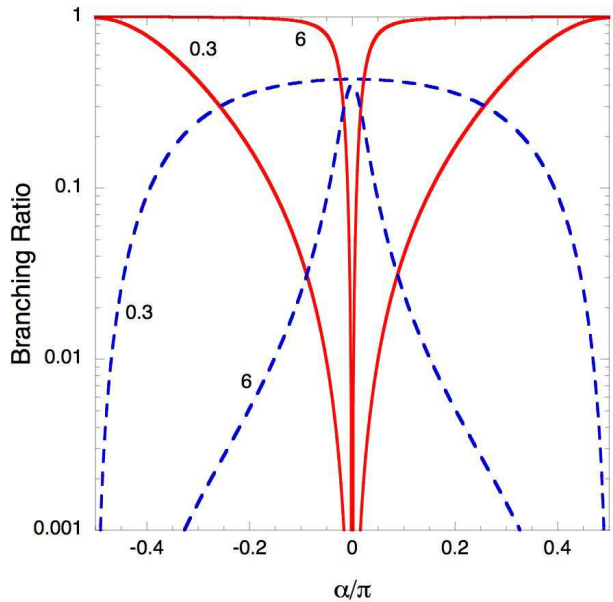} }
\vskip -3cm
\caption{Branching ratios of a
Higgs boson with mass 100 GeV
into fermions for $\tan\beta=0.3$
and $\tan{\beta} = 6$
in the flipped model.
The solid (dashed) lines,
which are red (blue)
are the branching ratios into $b\bar{b}$
($\tau^+\tau^-$).}
\label{fig:34tb036}
\end{figure}

We have plotted the branching ratios of the decays into fermions
for $\tan\beta=0.3$ and $\tan\beta = 6$
and $m_h=100$ GeV
in Fig.~\ref{fig:34tb036}
(for such a low $m_h$
the decays into vector bosons are very small).
As expected,
for $\tan\beta=6$,
the $b\bar{b}$ mode completely dominates
except for a very narrow
region
around the $b$-phobic point.
On the other hand,
for $\tan\beta=0.3$,
$\tau^+\tau^-$ decays dominate
over $b \bar b$ decays
for most values of $\alpha$
(although they never exceed a $40\%$ branching ratio
due to $c\bar{c}$ and gluon-gluon decays).
The fact that in this model
there is a region of parameters
in which the $c\bar{c}$,
gluon-gluon and $\tau^+\tau^-$ decay modes
all contribute equally was first noted in Ref.~\cite{Akeroyd:1996di}.
Although charm pairs will be very difficult to
observe
at either the LHC or Tevatron,
one can see plots of the branching ratios into charm pairs
for this model in Ref.~\cite{Arhrib:2009hc}.
For the other modes,
the discussion is not
much
different from that of
the type~II 2HDM.
In the region where the $\tau^+\tau^-$ decay dominates
one can look at $\mu^+\mu^-$ decays,
which occur at a rate $0.0035$ times the one of $\tau^+\tau^-$.

The couplings of the heavy Higgs, $H$,
are
identical to those of the light Higgs,
when $\alpha$ is shifted by $\pi/2$.
The difference is that the $H$ can,
if heavy enough,
decay into top quarks.
For  $\tan\beta = 0.3$,
the decays into $t \bar t$ and $\tau^+ \tau^-$ are both enhanced,
whereas for $\tan\beta = 6$,
the decay into $b \bar b$
is substantially enhanced,
similar to Fig.~\ref{fig:3heavy2}.
A discussion for fairly light $H$ and $A$ bosons
can be found in Ref.~\cite{Aoki:2009ha}.

The pseudoscalar Higgs does not decay into vector bosons, therefore
its primary decays are into $\tau^+\tau^-$ and $b\bar{b}$.
One has, in the flipped 2HDM,
\begin{equation}
\frac{\Gamma \left( A\ra \tau^+\tau^- \right)}
{\Gamma \left( A\ra b\bar{b} \right)}
= \left(\frac{\cot\beta}{1.76}
\right)^4;
\end{equation}
for $\tan\beta \geq 1$, this is $10\%$ or less.
For $\tan\beta = 0.3$, this would be roughly $10$,
leading to dominance of the $\tau^+\tau^-$ decay,
as in the lepton-specific model,
but for a different region of parameter space.

\subsubsection{Higgs decays in the neutrino-specific 2HDM}

In the neutrino-specific (NS) 2HDM
the RH quarks and charged leptons all couple to $\Phi_2$,
but the right-handed neutrino couples to $\Phi_1$.
In this model,
there are Dirac neutrino masses,
and the vacuum expectation value of $\Phi_1$ must be O(eV).
As a result,
the Yukawa coupling of $\Phi_1$ to neutrinos can be O(1).
The model originally used a $Z_2$ symmetry
which was softly~\cite{Ma:2000cc}
or spontaneously~\cite{Gabriel:2006ns,Wang:2006jy} broken,
in order to have such a small vev,
but this allows for right-handed neutrino masses.
The latter model,
by Gabriel et al~\cite{Gabriel:2006ns} and Wang~\cite{Wang:2006jy},
also has a very light scalar,
with mass of O(eV).
Extending the symmetry to a $U(1)$
and breaking it softly (to avoid a Goldstone boson)
gives the model of Davidson and Logan~\cite{Davidson:2009ha}.
This model is much less fine-tuned,
since it requires that the soft $U(1)$ breaking term
be of the same order as the electron mass
(the small O(eV) vev arises in a see-saw like pattern).
Thus, there are several versions of the model--those with
a $\Phi_1$ vev of O(eV) and a very light scalar, those with a
similar vev and no light scalar, and those that use a see-saw
type mechanism and have a $\Phi_1$ vev of O(MeV).

The first of these models, with a light scalar, which was the original proposal
of Gabriel et al~\cite{Gabriel:2006ns} and Wang, et al~\cite{Wang:2006jy}, has very
recently been excluded.  Following work on the astrophysics of the model~\cite{Sher:2011mx},
Zhou~\cite{Zhou:2011rc} pointed out that there are serious problems with the model.
In particular, the neutrinos emitted by SN1987a would, if there is a light scalar,
interact strongly with the relic neutrino background and would not reach Earth.
In addition, the effects of the neutrinos in the early universe would cause
problems with the WMAP data. Thus, these models are excluded.

Because of the small vev,
mixing between the Higgs doublets is negligible,
and thus one scalar behaves just as the Standard Model Higgs.
In the spontaneously broken version,
they can also decay into the very light scalars,
but these are invisible and thus the Standard Model Higgs
would decay invisibly.
In the Davidson--Logan model~\cite{Davidson:2009ha},
the light  Higgs decays as in the Standard Model.
The phenomenology of the charged Higgs,
which decays into charged leptons and a right-handed neutrino,
is interesting and discussed there.
The phenomenology of the new neutral scalars in the model,
however,
is much less interesting,
since they will only decay into neutrinos,
and thus appear as invisible Higgs decays.

In another version~\cite{Ma:2001mr,Haba:2010zi}, there is relatively
little fine-tuning in the potential, and the
$\Phi_1^\dagger\Phi_2$ term has a coefficient as large as $10$ GeV${}^2$.
Through a see-saw type mechanism, the vev of $\Phi_2$ is O(MeV), and a
further see-saw gives the light neutrino masses.
The vacuum stability of the model is
discussed by Haba and Norita in Ref.~\cite{Haba:2011fn}.
A recent comprehensive analysis of this model,
including precision constraints, can be found in
Ref.~\cite{Haba:2011nb}.
There are also interesting leptogenesis effects, which can
be found in works of Haba and Seto~\cite{Haba:2011ra,Haba:2011yc}

\subsection{Higgs production}

The production of the Higgs boson of the Standard Model at hadron colliders
is very well studied.
Calculations of two-loop corrections,
including next-to-next-to-leading order calculations
and soft-gluon resummations,
have all been carried out in detail.
A very extensive discussion,
with hundreds of references,
is in the review by Djouadi~\cite{Djouadi:2005gi}.

The leading channel at the Tevatron and at the LHC is gluon fusion,
in which
two
gluons and the Higgs are at the vertices of a triangle
with a top quark going round the loop.
The bottom quark plays little role since its coupling to the Higgs is so small
(note that this may not hold in a 2HDM).

The next leading channel, at the LHC, is $W$ fusion or $Z$ fusion.
At the Tevatron,
Higgsstrahlung off a $W$ or a $Z$ is the second leading channel.
Higgsstrahlung off a top quark
might also be observable.

\begin{figure}[h]
\centerline{\epsfysize=15cm \epsfbox{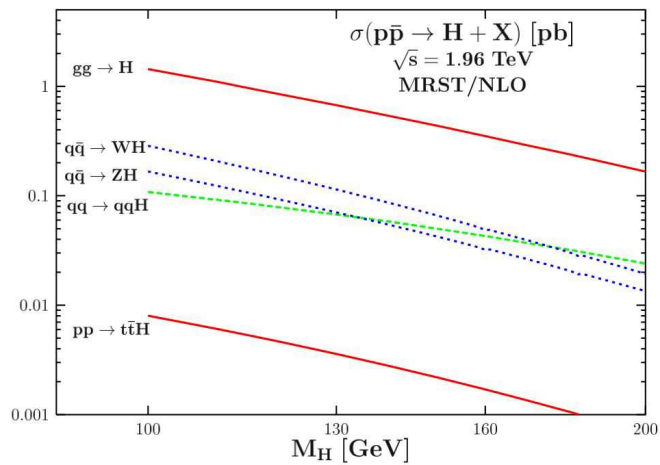}  }
\vskip -8cm
\caption{Production cross sections for the Standard-Model Higgs
at the Tevatron.
This figure is from Djouadi's review article~\cite{Djouadi:2005gj}.}
\label{fig:3djouada}
\end{figure}

\begin{figure}[h]
\centerline{\epsfysize=15cm \epsfbox{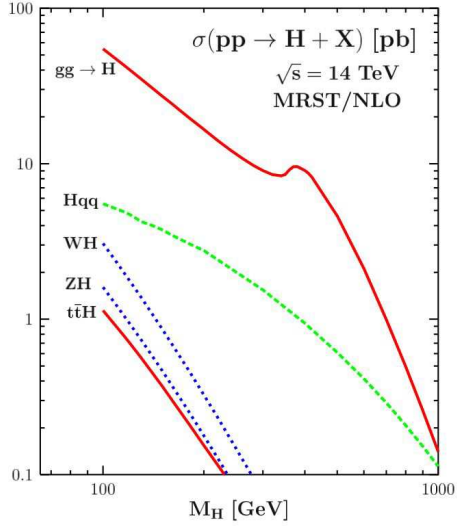}  }
\vskip -7cm
\caption{Production cross sections for the Standard-Model Higgs at the LHC.
This figure is from Djouadi's review article~\cite{Djouadi:2005gj}.}
\label{fig:3djouadb}
\end{figure}

The production cross sections,
from Djouadi's review,
are in Figs.~\ref{fig:3djouada} and~\ref{fig:3djouadb}.
Although gluon fusion dominates,
it may not always provide the best signature.
For example,
in order to detect a Higgs $H$ of mass 120 GeV at the Tevatron.
One must look for decays $H \to b\bar{b}$.
If the Higgs is produced via gluon fusion,
then the background of  $b \bar{b}$ will be much too large.
Instead,
Higgsstrahlung from a $W$ or $Z$ will help tag the $b \bar{b}$
with the decay of the $W$ or $Z$.
Thus,
the searches for a low-mass Higgs
at the Tevatron focus on $WH$ and $ZH$ production,
while the searches for a high-mass Higgs
focus on gluon fusion.
In practice,
of course,
all production modes must be considered.

In the following,
we will look at each of these processes,
and discuss how they are modified in the 2HDMs explored in this Chapter.
In most cases we will determine the ratio of the production cross section
in the 2HDM at hand to the one
in
the Standard Model,
and then one can use Djouadi's figures
to determine the absolute cross section.

\subsubsection{Gluon fusion}

In the Standard Model,
at the parton level,
the cross section for gluon fusion to a Higgs of mass $m_h$
is $m_h^2\,
\delta \left( \hat{s} - m_h^2 \right)
\sigma_o$,
where
\begin{equation}
\sigma_o =
\frac{G_\mu \alpha_s^2}{512 \sqrt{2} \pi}
\left| \sum_q A^h_{1/2} \left( \tau_q \right) \right|^2.
\end{equation}
Here,
$G_\mu$ is the Fermi constant from muon decay,
$\alpha_s$ is the strong coupling constant,
$\tau_q = m_h^2 \left/ \left( 4 m_q^2 \right) \right.$,
and $A_{1/2}^h \left( \tau \right)
= 2 \left[ \tau + \left( \tau - 1 \right) f \left( \tau \right) \right]
\left/ \tau^2 \right.$,
where
\begin{equation}
f \left( \tau \right) = \left\{
\begin{array}{lcl}
\arcsin^2 \left( \sqrt{\tau} \right) &\Leftarrow& \tau \leq 1,
\\
- \frac{1}{4} \left( \log{\frac{1+\sqrt{1-1/\tau}}{1-\sqrt{1-1/\tau}}}
- i \pi \right)^2 &\Leftarrow& \tau > 1.
\end{array}
\right.
\end{equation}
This can be substantially simplified
in the limits of large or small
$\tau$, which apply in most cases that we will consider.
In the limit
$m_q \gg m_h$,
\textit{i.e.}\ when $\tau_q \ll 1$,
$A_{1/2}^h \left( \tau_q \right) \ra 4/3$,
whereas in the limit
$m_q \ll m_h$,
\textit{i.e.}\ when $\tau_q \gg 1$,
$A_{1/2}^h \left( \tau_q \right) \ra
- \left[ \log \left( 4\tau_q \right) - i \pi \right]^2
\left/ \left( 2 \tau_q \right) \right.$.

In the 2HDM,
one's first thought is that
the change in the cross section would be trivial.
If only top-quark loops contribute,
then the only difference is in the coupling of the Higgs to the top quark,
and so one would multiply
the SM cross section by
$\left( \cos{\alpha} / \sin{\beta} \right)^2$ in the case of $h$ production,
or
$\left( \sin{\alpha} / \sin{\beta} \right)^2$ in the case of $H$ production,
in
all four
2HDMs.
In the case of gluon-fusion production of the pseudoscalar $A$,
the form factor is instead $A^A_{1/2}\left( \tau_ q \right)
= 2 f \left( \tau_q \right) \left/ \tau_q \right.$,
and moreover one must multiply
the whole cross section by $\cot^2\beta$.

While
the reasoning above indeed
applies in models
without a substantially enhanced $b$-quark Yukawa coupling,
such as the type~I 2HDM
or the lepton-specific 2HDM,
it does not necessarily apply in the cases of the type~II and flipped 2HDMs
(the behavior of the lepton couplings has no effect on gluon fusion).
This is because the $b$-quark loops can become crucial,
and for large $\tan\beta$ can actually dominate the cross section.

In the type~I or lepton-specific 2HDM,
the production cross section of a light Higgs,
$h$,
through gluon fusion is simply multiplied
by the factor
$\left( \cos{\alpha} / \sin{\beta} \right)^2$.
In the decoupling limit,
\textit{i.e.}\ when $\cos(\alpha-\beta)$ is small,
this is near unity,
but in general
it is quite smaller.
This factor is somewhat discouraging
for the $\tau^+\tau^-$ signature of the lepton-specific model.
As one can see from Fig.~\ref{fig:33tb1},
$h \to \tau^+ \tau^-$ is dominant (when $\tan{\beta} = 1$)
near $\alpha = \pm \pi / 2$,
precisely where the factor
$\left( \cos{\alpha} / \sin{\beta} \right)^2$
strongly suppresses the production rate.
For larger $\tan\beta$,
however,
as seen in Fig.~\ref{fig:33tb6},
the region in which
the $\tau^+ \tau^-$ decays
dominate does extend a substantial distance away from $\alpha=\pm\pi/2$,
and
then
the suppression would not be that severe.

In the type~II or flipped 2HDM,
the contribution of the top quark is also multiplied by
$\left( \cos{\alpha} / \sin{\beta} \right)^2$,
but now the
diagram with a b-quark loop
can contribute for large $\tan\beta$,
since then the $b$ Yukawa coupling becomes large.
The $b$-loop contribution
to the amplitude
is multiplied by $- \tan{\alpha} \tan{\beta}$
relative to the
$t$-loop contribution.
Using the limits of small $\tau_q$ (when $q = t$)
and large $\tau_q$ (when $q = b$) above,
we find that the cross section
increases over the type~I 2HDM cross section
by a factor of
$\left| 1 + \left( 5 - 8 i \right) \tan{\alpha} \tan{\beta}
/ 100 \right|^2$
for $m_h = 100$ GeV,
and that this correction roughly scales as $m_h^{-2}$
for other values of $m_h$.
Note that $\tan\alpha$
can be of either sign in the type~II 2HDM;
in the MSSM it is
always
negative.
Numerically,
for positive $\alpha$,
this is 1.1 (1.4, 2.9, 12.0)
for $\tan{\alpha} \tan{\beta} = 1$
(3, 10, 30).
For negative $\alpha$,
this is 0.9
(0.8, 0.9, 6.0)
for $\tan{\alpha} \tan{\beta} = -1$
($-3$, $-10$, $-30$).
We see that for large $\tan\beta$,
one can get an increase of almost an order of magnitude
in the production rate relative to the type~I 2HDM.
However,
one should keep in mind the discussion of the last section
in which it is noted that values of $\tan\beta$ larger than $6$ are allowed
(by unitarity and perturbation theory)
only for a very small region of parameter space,
so very large enhancements are unlikely.

For the heavy neutral Higgs,
$H$,
the results are similar.
In the type~I or lepton-specific 2HDM,
one must multiply the SM production cross section
by the factor
$\left( \sin{\alpha} / \sin{\beta} \right)^2$.
As before,
this is small in the region where the $\tau^+\tau^-$
branching ratio
is large.
In the type~II or flipped 2HDM,
the factor $- \tan{\alpha} \tan{\beta}$ in the previous paragraph
becomes $\cot{\alpha} \tan{\beta}$.
The results of the previous paragraph are thus qualitatively unchanged.
Note that
when the mass of the $H$ mass rises beyond the mass of the top quark,
the rate of $gg$-fusion production of $H$ drops quickly,
just as in the Standard Model.

For $gg$-fusion production of the pseudoscalar $A$,
the form factor $A_{1/2}^A$ is simply
$2f(\tau)/\tau$,
which in the limit $m_t \gg m_A$ is $2$ instead of $4/3$.
There is also a factor $\xi^u_A = \cot{\beta}$ in the amplitude.
This yields a total factor $\left( 9 / 4 \right) \cot^2{\beta}$
in the production cross section of an $A$ in the type~I 2HDM
compared to the production cross section of an SM Higgs of identical mass.
This will be substantial for small $\tan{\beta}$.
For the type~II or flipped 2HDM,
the ratio of production cross section relative to the type~I 2HDM is,
for $m_A = 100$ GeV,
$\left| 1 - \left( 3.5 - 4 i \right) \tan^2\beta / 100 \right|^2$.
Numerically,
this is 0.93
(0.82, 0.59, 1.44)
for $\tan^2{\beta} = 1$
(3, 10, 30).
The difference of production cross sections
between the various 2HDMs is thus much less dramatic.

\subsubsection{$WH$ and $ZH$ production, $WW$ and $ZZ$ fusion,
and $b\bar{b}H$ and  $t\bar{t}H$ production}

It is straightforward to see how these production rates
are affected in 2HDMs.
First consider the production processes involving vector bosons
($W^\ast \to W H$, $Z^\ast \to Z H$, $WW\ to H$ or $ZZ \to H$ where $H$ is a neutral Higgs ).
In all four 2HDMs,
for the light Higgs $h$
the SM rate is multiplied by $\sin^2{\left( \alpha - \beta \right)}$,
and for the heavy Higgs $H$
it is multiplied by $\cos^2{\left( \alpha - \beta \right)}$.
The pseudoscalar $A$ cannot be produced via this mode,
since there are no $W^+ W^- A$ and $Z Z A$ vertices.

For $t\bar{t}H$ production,
one simply multiplies
the SM rate
by the appropriate coupling-constant factor.
In all four models,
this is $\left( \cos{\alpha} / \sin{\beta} \right)^2$
for the light scalar $h$,
$\left( \sin{\alpha} / \sin{\beta} \right)^2$
for the heavy scalar $H$,
and $\cot^2{\beta}$ for the pseudoscalar.

Contrary to what happens in the Standard Model,
there is in some 2HDMs the possibility
of substantial $b\bar{b}H$ production.
The rate,
well above the threshold,
is,
compared to the SM $t\bar{t}H$ production cross section,
negligible for the type~I and lepton-specific 2HDMs,
but for the type~II and flipped 2HDMs it is
$\left( \sin{\alpha} / \cos{\beta} \right)^2 \left( m_b / m_t \right)^2$
for the light $h$.
A detailed study,
with numerous references,
of this mode in the MSSM
by Schaarschmidt~\cite{Schaarschmidt:2007zz,Aad:2008zzm,Schaarschmidt:2008zz},
showed that an $h$ in the 120--200 GeV mass range
can be discovered at the LHC through this process,
if $\tan{\beta} > 6$,
by looking for the signature $h \to \tau^+ \tau^-$.
In fact,
for a large $\tan\beta$,
which is not unnatural in the MSSM,
this can become the primary discovery mode,
especially for a relatively light Higgs $h$.
One would expect a similar result in the type~II 2HDM
(the flipped model would have less enhancement
in the $h \to \tau^+ \tau^-$ decay mode),
although,
as noted earlier,
large values of $\tan\beta$ are not favored in the type~II 2HDM.

\subsubsection{Other production mechanisms}

With several neutral scalars,
another production mechanism can become important.
One can produce two Higgs bosons
(either the same one or different ones).
(This can also occur in the Standard Model,
of course.)
In any diagram in which a scalar is produced,
that scalar can then convert itself
into two scalars via a trilinear coupling.
In addition,
one can have double-Higgsstrahlung off a $W$ or $Z$,
or one can have a gluon fusion into two scalars
via a box diagram with a heavy quark in the loop;
the latter mechanisms are independent of the trilinear coupling.
An analysis
of all these mechanisms
was done in Ref.~\cite{Plehn:1996wb}.
Total cross sections for the LHC
were given in Djouadi's review~\cite{Djouadi:2005gi},
and are dominated by gluon fusion,
giving cross sections of order 10~fb
over the low-mass range for the scalars.
Although detection is much more difficult
due to large backgrounds~\cite{Baur:2002qd,Baur:2003gp},
analyses~\cite{Moretti:2004dg} indicate that
it might be possible to distinguish the trilinear coupling from zero
with an integrated luminosity of 300~fb$^{-1}$,
for scalar masses in the 150--200~GeV range;
a higher luminosity is needed for lighter scalars.

In the MSSM there are many more possibilities involving neutral scalars.
One can have $q\bar{q} \ra Z^\ast \ra hA, HA$ or,
through triangle and box graphs,
$gg \ra hh, HH, hH, AA, hA, HA$.
In addition,
one can have double Higgsstrahlung.
These are discussed in Refs.~\cite{Djouadi:1999rca,Dawson:1998py}
and summarized in Djouadi's review~\cite{Djouadi:2005gj}.
For the quark annihilation process,
cross sections range from 10 to 100~fb
as $m_A$ varies from 100 to 170~GeV,
and for gluon fusion the cross sections
(for $\tan\beta=30$)
are a little more than an order of magnitude larger.
Detection is difficult,
but possible,
and is summarized in Ref.~\cite{Djouadi:2005gj}.

In the general 2HDM,
one has the same processes as in the MSSM,
but there is now a much larger parameter space.
The quark-annihilation process has precisely the same form as in the MSSM,
but now one need no longer have a small $\cos(\alpha-\beta)$,
and thus the $ZhA$ coupling can be larger than in the MSSM.
This is encouraging and leads to some interesting possibilities.
For example,
in the lepton-specific model and for $\tan\beta > 2$,
the dominant decay of the $A$ is into
$\tau^+ \tau^-$,
and for much of parameter space the decay of the light Higgs
is also into
$\tau^+ \tau^-$.
Thus one might have four-$\tau$ events
with branching ratios as high as tens of femtobarns.
This signature needs further investigation.
A study of pair production of the lightest Higgs bosons in the type II model
was carried out in Refs.~\cite{Moretti:2004wa,Moretti:2007ca,Moretti:2010kc}.
They showed that while pair production in the Standard Model is very difficult to
observe at the LHC, it can be bigger in the type II model, and they also show that
there can be sensitivity to the quartic couplings, which could help distinguish
the model from the MSSM.

For the gluon-initiated process,
triangle diagrams produce a single $h/H/A$,
real or virtual,
which then converts into a pair of scalars.
Alas,
this process is proportional to trilinear scalar couplings and,
while these are known in the MSSM,
they are unknown in the general 2HDM.
The box diagrams which give gluon fusion into two scalars
will be similar to those of the MSSM.
Thus all one can really say is that
the rate could be substantially larger than in the MSSM,
but accurate predictions are impossible.

\subsection{The inert Higgs model}
\label{sec:inert}

The inert Higgs model is a 2HDM with an
unbroken $Z_2$ symmetry under which one of the doublets
transforms non-trivially,
\viz $\Phi_2 \rightarrow -\Phi_2$,
and all other SM fields are invariant.
This `parity' imposes natural flavour conservation.
Initially a similar model~\cite{Deshpande:1977rw} was introduced to
explain neutrino masses.
More recently  such a model was
proposed in the context of radiative neutrino masses \cite{Ma:2006km}
and also to attack the naturalness
problem of the SM by allowing for a larger mass
(between 400 and 600 GeV) for the SM Higgs
while keeping full consistency with
electroweak precision tests \cite{Barbieri:2006dq}, thus
solving the `little hierarchy' problem \cite{Barbieri:2000gf}.
Even more recently, an inert doublet was introduced to allow for
the possibility of several mirror families of fermions \cite{Martinez:2011ua}.

In the inert Higgs model the Higgs doublet
$\Phi_2$---the inert doublet---does not couple
to matter and acquires no vacuum
expectation value, leaving the $Z_2$ symmetry unbroken.
The scalar spectrum consists of the SM-like Higgs
obtained from
$\Phi_1$ and one charged  and two neutral states from  $\Phi_2$.
Since the
$Z_2$ is unbroken the lightest
inert particle will be stable and will contribute to the dark
matter density  \cite{Ma:2006km,Barbieri:2006dq}. This
possibility has been analysed by several
authors
\cite{Ma:2006fn,Majumdar:2006nt,LopezHonorez:2006gr,Sahu:2007uh,Gustafsson:2007pc,Lisanti:2007ec,Hambye:2009pw,Melfo:2011ie}.
The early cosmological evolution of the model has been discussed
by Ginzburg {\em et all} in~\cite{arXiv:1009.4593}.

The scalar potential is
the one in eq.~(2) but with $m_{12}^2 = 0$.
The asymmetric phase, where
\begin{equation}
\left\langle \phi^0_1 \right\rangle = \frac{v}{\sqrt{2}}
\quad \mbox{and} \quad
\left\langle \phi^0_2 \right\rangle = 0,
\label{assym}
\end{equation}
corresponds to a sizeable region of parameter
space~\cite{Ma:2006km,Barbieri:2006dq} and the scalar masses
are given by
\be
\begin{array}{rclrcl}
m^2_h &=& \lambda_{1} v^2, &
m_S^2 &=&  {\displaystyle m_{22}^2 +
\frac{\left( \lambda_3 + \lambda_4 + \lambda_5 \right) v^2}{2},}
\\*[2mm]
m_+^2 &=& {\displaystyle m_{22}^2 + \frac{\lambda_3 v^2}{2},} &
m_A^2 &=& {\displaystyle m_{22}^2 +
\frac{\left( \lambda_3 + \lambda_4 - \lambda_5 \right) v^2}{2},}
\end{array}
\ee
where $h$ is the usual Higgs boson obtained from $\Phi_1$,
$\left( S + i A \right) \left/ \sqrt2 \right.$ is the neutral component
of the inert Higgs doublet,
and $H^+$ is the charged inert Higgs field. The doublet  $\Phi_1$ gives
mass to the gauge bosons $W^\pm$ and $Z^0$. In the
limit of Peccei--Quinn symmetry, $\lambda_5 \rightarrow 0$,
the neutral inert scalars become degenerate. Direct detection of halo
dark matter places a limit on this degeneracy \cite{Akerib:2005kh}.

The inert scalars can be produced at colliders through their
couplings to the electroweak gauge bosons subject to the
constraint of the $Z_2$ symmetry \cite{Cao:2007rm,Lundstrom:2008ai}.
In addition, they also participate in
cubic and quartic Higgs couplings:
\begin{eqnarray}
V_\mathrm{interactions} &=&
\frac{\lambda_{2}}{2}
\left( H^+ H^- + \frac{S^2 + A^2}{2} \right)^2
+ \lambda_{3} \left( vh +  \frac{h^2}{2} \right)
\left( H^+ H^- + \frac{S^2 + A^2}{2} \right)
\no & &
+ \frac{\lambda_4 + \lambda_5}{2}
 \left(  vh +  \frac{h^2}{2} \right) S^2 +
\frac{\lambda_4 - \lambda_5}{2}
\left( vh + \frac{h^2}{2} \right) A^2.
\label{int}
\end{eqnarray}
As pointed out in Ref.~\cite{Cao:2007rm}, assuming the mass hierarchy
$ m^2_+ > m^2_A > m^2_S$, the stable scalar $S$ appears
as missing
energy in the decays of $H^+$ and $A$. Since there is no linear term
in  $A$ or  $S$ in eq.~(\ref{int}), the decay of $A$ must occur
through the gauge interaction
\begin{equation}
\frac{g}{2 c_W}\, Z_{\mu} \left( S \partial^{\mu} A
- A \partial^\mu S \right).
\end{equation}
Hence \cite{Cao:2007rm} the dominant decay of $A$ is into
$S f \bar{f}$, where $f$ is a fermion, either a lepton or a quark,
and $S$ appears as missing energy.
Concerning $H^\pm$, the gauge interactions with $S$ and $A$ are given by:
\begin{equation}
\frac{ig}{2}\, W^-_\mu \left( S \partial^\mu H^+ - H^+ \partial^\mu S \right)
+
\frac{g}{2}\, W^-_\mu \left( A \partial^\mu H^+ - H^+ \partial^\mu A \right)
+ \hc
\end{equation}
Hence the dominant decays of $H^\pm$
are into $W^\pm S$ and  $W^\pm A$,
with, in the second case,  subsequent decay of $A$ into $ZS$.

There are also trilinear gauge interactions among
$H^\pm$, $Z$, and $\gamma$,
as well as the quadrilinear terms required by gauge
invariance.

The LEP I data on the width of the $Z^0$ gauge boson
force the sum of the masses of the $S$ and $A$ to be larger
than the mass of the $Z^0$ \cite{Gustafsson:2007pc,Cao:2007rm},
thus preventing $Z^0 \to A S$.
The electroweak precision tests
put constraints on the inert-scalar mass splittings
as a function of the $h$ mass \cite{Barbieri:2006dq}.
If the $Z_2$-odd scalars are much lighter than the SM Higgs boson,
they may have a great impact on the direct search for the latter,
because $h \to S S$ and $h \to A A$ may become the dominant
decay channels.
In particular,
the $h \to S S$ channel is invisible.
The lower limit $m_h > 114.4\,$GeV,
obtained by LEP II,
was based on a direct search via $h \rightarrow b \bar b$,
which is dominant for a light Higgs boson in the SM.
In the present case this limit can be relaxed down
to about $106\,$GeV,
assuming the mass splitting of the new neutral scalars to
be $m_A - m_S = 10$ GeV,
due to the fact that in this model the invisible decay
of the Higgs boson may be dominant and, as a result, the decay branching
ratio into $b \bar b$ is highly suppressed.

A scan over the parameter space was performed in~\cite{Cao:2007rm},
assuming the charged-Higgs mass to be much larger than $m_h$,
the mass splitting of the new neutral scalars to be $m_A -m_S = 10\,$GeV,
and also taking into account relevant constraints including
the $Z^0$-decay constraint mentioned above and the vacuum stability bound.
The invisible decay mode $h \to S S$ is found to dominate
in the light- and intermediate-mass region of $m_h$,
\ie when $100 < m_h < 160\,$GeV.
The $10\,$GeV mass gap between $A$ and $S$
implies that the contribution of the $SS$ mode
to the decay of the SM-like Higgs boson $h$
is much larger than that of the $AA$ mode.
In particular,
the branching ratio for the invisible mode is
50--65\% in the mass region $100 < m_h < 150\,$GeV
and for $m_S \sim 40$--$60\,$GeV.
As a result,
the usual decay modes of the SM Higgs boson are highly suppressed.
This fact,
together with the strong suppression of the $\gamma \gamma$ mode,
will make it very difficult,
in this framework,
to use this mode
to detect the SM-like Higgs boson at the LHC
in the mass range  $100 < m_h < 150\,$GeV \cite{Cao:2007rm}.
However,
it was also shown there that it is very promising
to look for the SM-like Higgs boson
through its invisible decay in the so-called weak-boson-fusion process
\cite{Eboli:2000ze}.
This process is of the form $q \bar q \to q^\prime q^\prime V V
\rightarrow q^\prime q^\prime h$
($V$ denotes a gauge boson),
with the subsequent decay of $h$ to undetectable particles.
The decay mode $h \rightarrow W^+ W^- $ dominates over the
invisible one for $h$  heavier than $160\,$GeV,
where it starts behaving as in the SM.
The decay pattern and decay branching ratios of the new neutral
scalars have also been examined in  Ref.~\cite{Cao:2007rm}.
They conclude that an $S$ with mass $\sim 50\,$GeV
should be observable at the LHC and,
at the same time,
it would constitute a good dark-matter candidate.
A complementary analysis of collider phenomenology
of the inert Higgs model was performed in Ref.~\cite{Lundstrom:2008ai}.

The possibility of extending the inert Higgs model in order to introduce
CP-violation in the scalar sector was considered in
Refs.~\cite{Grzadkowski:2009bt,Grzadkowski:2010au}.
In another extension of the inert doublet model, Barr and
Kephart~\cite{arXiv:1111.0963} pointed out that a difficulty with generic multi-doublet
models is that many fine-tunings are necessary to alleviate the hierarchy problem, and thus,
if one believes that the hierarchy problem of the Standard Model is solved via anthropic
fine-tuning, then having more than one Higgs doublet becomes exceedingly unlikely.
They looked for the conditions under which an N doublet model only has a single fine-tuning,
and found that if the N doublets formed a representation of a global symmetry group, then only
one fine-tuning is necessary, and classified all possible symmetry groups with $N \leq 6$.
They found that in some models, such as SU(N) with an N-plet, the extra Higgs doublets become
completely inert, and the scalars and pseudoscalars are all degenerate in mass.
Different mass relations arise from different representations.

A different version of the inert Higgs model,
which has recently received increasing attention,
is the Lee--Wick Standard Model
(LWSM)~\cite{Lee:1969fy,Lee:1970iw,Grinstein:2007mp},
whose neutral sector is similar to the inert Higgs model
but whose charged sector is similar to a type II model with $\tan\beta=1$.
In this model,
for every SM field,
a higher-derivative kinetic term is introduced;
for a Higgs scalar,
this term is quartic in the derivatives.
These terms lead to the presence
of additional poles in the propagators.
By introducing auxiliary fields,
the model can be written as one without higher-derivative terms,
but with additional fields corresponding to Lee--Wick partners
(one for each field of the SM).
The signs of the kinetic and mass terms of these partners
are opposite to those of normal particles,
\ie the states have negative norm.

A major attractiveness of the LWSM
is the elimination of quadratic divergences in the Higgs sector.
In supersymmetry,
the minus sign which cancels those divergences
arises from the different statistics of the partners;
in the LWSM,
it arises from a sign difference in the propagator
due to the negative-norm states.
If those states are not stable,
then they do not appear as `out' states in the $S$ matrix
and unitarity is preserved.
While one does have microcausality violation,
no logical paradoxes arise macroscopically~\cite{Lee:1970iw,Grinstein:2008bg}.
There is now an extensive literature on the LWSM~\cite{Rizzo:2007ae,Dulaney:2007dx,Alvarez:2008za,Underwood:2008cr,
Carone:2008bs,Carone:2008iw,Grinstein:2008qq,Carone:2009it,Fornal:2009xc,
Shalaby:2009re,Wu:2008rr,Espinosa:2007ny,Krauss:2007bz} .

A comprehensive analysis of the Higgs sector of the LWSM
was recently carried out by Carone and Primulando~\cite{Carone:2009nu}.
They were motivated by the realization~\cite{Carone:2008bs}
that electroweak precision constraints do not severely constrain
the Lee--Wick mass scale for the Higgs sector.
The Higgs doublet and its Lee-Wick partner form a 2HDM,
but the mixing between the neutral states is symplectic rather than orthogonal.
The neutral-scalar sector is very similar to the one of the inert Higgs model,
but with some differences in signs.
The primary focus of Ref.~\cite{Carone:2009nu}
was on the charged Higgs sector
(which will be discussed in Chapter 4),
but they did find bounds from LEP,
showed that constraints in the charged sector also
constrain the neutral sector,
and have plotted bounds in the neutral-Higgs-boson plane.
The effects of the model on electroweak parameters
and on the $Zb\bar b$ coupling were studied
by Chivukula  {\it et al.}\/~\cite{Chivukula:2010nw}
A more detailed analysis of the neutral sector,
including bounds from direct
searches at LEP and the Tevatron,
as well as prospects at the LHC,
was very recently published by Alvarez {\it et al.}\/~\cite{Alvarez:2011ah};
they find interesting differences
in the usual sum rules in the neutral sector;
for example,
the usual relation $g^2_{h1-V-V}+g^2_{h2-V-V}=1$ has a different
sign between the two terms.
They used the HiggsBounds code to study implications at colliders.

Another 2HDM,
called the ``quasi-inert'' model,
was recently introduced by Cao {\it et al.}\/ ~\cite{Cao:2011yt}.
In that model, motivated by the possible observation of W plus dijets at the
Tevatron~\cite{Aaltonen:2011mk}, there is a second doublet whose tree-level vev
vanishes, and which couples primarily to the first generation of quarks.
They set to zero the $\Phi_1^\dagger\Phi_2 + \hc $ term in the potential, which
can be accomplished by a $Z_2$ symmetry
in which both $\Phi_2$ and $u_R$ change sign.
This symmetry must be weakly broken to allow the up quark to get a mass,
and this results in a small vev for $\Phi_2$.
Cao \etal calculate electroweak precision effects
and flavour constraints on their model.
The primary motivation for the model
is that the Tevatron can produce the charged Higgs
$(p\bar{p}\rightarrow H^\pm)$,
which subsequently decays into $H^0 W^\pm$ or $A^0 W^\pm$,
leading to $\ell^\pm \nu j j$ or to $W^\pm$ plus dijets.
Since the charged Higgs is produced resonantly,
the signal can be large.
Cao \etal show that a reasonable region of parameter space exists
which can explain the recent observation of $W$ plus dijets
at the Tevatron~\cite{Aaltonen:2011mk}.
Even if this observation is not confirmed,
the model is interesting in its own right
and leads to other unique signatures.
Another scenario by Chen {\it et al.}~\cite{Chen:2011wp}
switched the role of the neutral and charged scalars,
and then the dijet comes directly from the charged scalar.
This can be produced non-resonantly by $W^\pm H^\mp$ production,
and can also explain the non-observation of a resonance in $\ell\nu jj$
by CDF (although the bounds on that are much weaker
and a resonance could still exist).
Similar models,
which focus more on explaining the $B_s\rightarrow \mu\mu$ rate
but also discuss the $W$ plus dijet signature,
can be found in Refs.~\cite{Chen:2011wp,Dutta:2011kg}.

\newpage

\section{Models with tree-level flavour-changing neutral currents}
\label{sec:fcnc}

\subsection{The type~III 2HDM}

In the previous chapter
it was shown that one can eliminate
the potentially dangerous tree-level FCNC
through a discrete symmetry.
Suppose,
however,
that we reject any such symmetry.
The tree-level FCNC can certainly be suppressed
by making the neutral scalars extremely heavy,
but scalar masses in the multi-TeV range (or higher) seem unnatural.
In this section,
we examine the constraints from FCNC
and show that a reasonable {\it Ansatz}\/
for the neutral flavour-changing couplings
allows for scalar masses well below the TeV scale.

It is easiest to discuss the tree-level FCNC
in the Higgs basis described in chapter~5.
In that basis,
the scalar doublets are rotated so that
the vev is entirely in the first doublet,
while the second doublet has zero vev.
The general Yukawa couplings can be written as
\ba
{\mathcal L}_\text{Yukawa} &=&
\eta_{ij}^U\bar{Q}_{iL}\tilde{H}_1U_{jR}
+ \eta_{ij}^D\bar{Q}_{iL}{H}_1D_{jR}
+ \eta_{ij}^L\bar{L}_{iL}{H}_1E_{jR}
\no & &
+ \hat{\xi}_{ij}^U\bar{Q}_{iL}\tilde{H}_2U_{jR}
+ \hat{\xi}_{ij}^D\bar{Q}_{iL}{H}_2D_{jR}
+\hat{\xi}_{ij}^L\bar{L}_{iL}{H}_2E_{jR} + \hc,
\ea
 where $H_1$ and $H_2$ are the two scalar doublets.
In the Higgs basis
those doublets have been rotated so that only $H_1$ has a vev,
\ie
\be
\left\langle H_1 \right\rangle_0 = \left(
\begin{array}{c} 0 \\ v \left/ \sqrt{2} \right.
\end{array} \right),
\quad
\left\langle H_2 \right\rangle_0 =
\left( \begin{array}{c} 0 \\ 0  \end{array} \right),
\ee
where $v$ is real.
In this basis,
only the Yukawa couplings of the doublet $H_1$,
\viz the $\eta_{ij}$,
generate fermion masses;
those $\eta_{ij}$ may be bi-diagonalized and do not lead to tree-level FCNC.
When that bi-diagonalization is performed,
the neutral flavour-changing couplings become
\be
{\mathcal L}_\mathrm{FCNC} =
\xi^U_{ij} \bar{U}_{iL} {H_2^0}^\ast U_{jR}
+ \xi^D_{ij} \bar{D}_{iL} H_2^0 D_{jR}
+ \xi^L_{ij} \bar{L}_{iL} H_2^0 L_{jR},
\ee
where
\be
\xi^{U,D,L} = {V_L^{U,D,L}}^\dagger\,
\hat{\xi}^{U,D,L}\, V_R^{U,D,L}.
\ee
Since $V_R$ is completely unknown and the $\hat{\xi}$ are arbitrary,
these $\xi^{U,D,L}$ coefficients are arbitrary;
in order to look at specific processes,
some assumptions must be made about their magnitudes.

One of the earliest papers discussing tree-level FCNC
was the one of Bjorken and Weinberg~\cite{Bjorken:1977vt},
who considered radiative muon decay and chose $\xi^L_{\mu e}$
to be the Yukawa coupling of the muon.
Later,
in 1980,
McWilliams and Li~\cite{McWilliams:1980kj} and Shanker~\cite{Shanker:1981mj}
considered $K$--$\bar K$ mixing,
as well as many processes involving kaon and muon decays.
They argued that the heaviest fermion
sets the scale for the entire Yukawa-coupling matrix.
The flavour-changing vertex should be
the product of the largest Yukawa coupling and a mixing angle factor.
Since they did not know the mixing angle factors,
they set them equal to one.
Thus,
the $\xi^{U,D,L}$ were set equal to the top,
bottom,
and tau Yukawa couplings,
respectively.
The most stringent bound came from $K$--$\bar K$ mixing
and led to a lower bound of $150\,$TeV on the mass of $H_2^0$.
For most of the 1980's,
this led most authors to assume that there must be a discrete symmetry
which prohibits the FCNC,
and attention focused on the type~I and type~II 2HDMs.

Cheng and Sher~\cite{Cheng:1987rs} argued that
this estimate of the lower bound is not reasonable.
They argued that the most conspicuous feature
of the fermion mass structure is its hierarchical structure and that,
therefore,
setting all the flavour-changing couplings to be equal
to the heaviest-fermion Yukawa coupling was not reliable.
They proposed what has since become known as the Cheng--Sher {\it Ansatz}:
that the flavour-changing couplings should be of the order of
the geometric mean of the Yukawa couplings of the two fermions.
In other words,
\be
\label{eq:ansatz}
\xi_{ij} = \lambda_{ij}\, \sqrt{m_im_j}\, \frac{\sqrt{2}}{v},
\ee
where the $\lambda_{ij}$ are of order one.
Since the most severe bounds on FCNC arise from the first two generations
and this {\it Ansatz}\/ especially suppresses
the Yukawa couplings of those generations,
it will reduce the lower bound on the Higgs mass.

More specifically,
Cheng and Sher's argument was as follows.
Consider a  model with $n$ Higgs doublts $\Phi_i$ ($i=1...n$)
and call $\lambda^\prime_i$ the matrix of Yukawa couplings to,
say,
the charge $-1/3$ quarks.
First suppose that the fermion mass matrix
is of the Fritzsch \\
form\footnote{At the time of Cheng and Sher's paper
the Fritzsch {\it Ansatz}\/ gave acceptable mixing angles.
A generalization of it was proposed, and its phenomenology discussed
in detail, in Ref.~\cite{DiazCruz:2004tr}.} :
\be
M=
\left( {\begin{array}{ccc}
 0 & A & 0 \\ A & 0 & B \\ 0 & B & C \\ \end{array}}\right).
\ee
In this case,
calling the eigenvalues $m_1$,
$- m_2$,
and $m_3$,
one has $A \simeq \sqrt{m_1 m_2}$,
$B\simeq \sqrt{m_2 m_3}$,
and $C \simeq m_3$.
Cheng and Sher then simply assumed that
the Yukawa-coupling matrices were given by
\be
\lambda^\prime_i = \frac{\sqrt{2}}{v_i}
\left( {\begin{array}{ccc}
0 & A_i & 0 \\ A_i & 0 & B_i \\ 0 & B_i & C_i \\ \end{array}}\right),
\ee
and that their matrix elements
had the same structure as the full mass matrix,
namely $A_i = a_i \sqrt{m_1 m_2}$,
$B_i = b_i \sqrt{m_2 m_3}$,
and $C_i = c_i m_3$,
with coefficients $a_i, b_i, c_i$ of order unity.
In other words,
the requirements,
obtained by comparing the $\lambda^\prime_i$ with $M$,
\begin{equation}
\sum_i a_i = \sum_i b_i = \sum_i c_i = 1
\end{equation}
are not satisfied through any fine-tuned cancellations
among the different couplings.
Essentially,
the {\it Ansatz}\/ states that
mass matrix zeros are not obtained through any cancellations
among non-zero matrix elements,
and that non-zero mass matrix elements
are also not obtained through precise cancellations among larger terms.
Cheng and Sher then pointed out that
this argument does not apply only to mass matrices with the Fritzsch structure,
but to any other structure in which one requires that
the hierarchy of eigenvalues does not arise through delicate cancellations.
It was later noted~\cite{Antaramian:1992ya} that,
if the hierarchical structure is due to approximate flavour symmetries,
then the Cheng--Sher {\it Ansatz}\/ will be satisfied.
The {\it Ansatz}\/ is thus quite general.

In the type~III~\footnote{The model was first referred to as type III in
Ref.~\cite{Hou:1991un}.}
model the Cheng--Sher {\it Ansatz}\/ is assumed
and its implications are explored.
Many papers focused on a few specific processes,
including $\Delta m_B$~\cite{Gronau:1988qt},
$t  \to  c h$,
and $h \to \bar{t} c + \bar{c} t$~\cite{Hou:1991un},
rare $\mu$,
$\tau$,
and $B$ decays (emphasising $B \to K \mu \tau$)~\cite{Sher:1991km},
$\mu \to e\gamma$ (at two-loop level)~\cite{Chang:1993kw},
$t \to c \gamma$ and $t \to c Z^0$~\cite{Luke:1993cy},
muon--electron conversion~\cite{Kosmas:1993ch},
and $b \to s\gamma$~\cite{Wolfenstein:1994jw}.
In 1996,
an extremely comprehensive analysis of the model by Atwood,
Reina,
and Soni~\cite{Atwood:1996vj} looked at many of these processes
as well as at implications for $Z^0 \to b \bar{b}$
and $Z^0 \to b\bar{s} + s\bar{b}$.

Over the years,
the various bounds have steadily improved.
If the type~III model is correct,
then one would expect the $\lambda_{ij}$ in eq.~\eqref{eq:ansatz}
to be all of order unity,
but this is a fairly loose requirement
since there are unknown mixing angles.
Scalar masses will enter in all specific processes.
In the Higgs basis,
the imaginary part of $H_2^0$ is the usual pseudoscalar,
$A$,
whereas the real part of $H^0_2$ is a linear combination of $h$ and $H$.
Since $H^0_2$ has no vev,
its coupling to $W^+W^-$ vanishes,
and thus the relevant mixing angle
which rotates $(H_2^0,H_1^0)$ into $(h,H)$ is $\alpha - \beta$.
In a model with tree-level scalar exchange
one has an effective mass in the matrix element given by
\be
\frac{1}{m_\mathrm{eff}^2} = \frac{c^2}{m_h^2}  + \frac{s^2}{m_H^2},
\ee
where $s\equiv \sin(\alpha-\beta)$ and $c\equiv \cos(\alpha-\beta)$.
Henceforth,
we shall refer to the scalar mass $m_\mathrm{eff}$
and to the pseudoscalar mass $m_A$
as free parameters.
Some processes,
such as $B_s \to \mu\tau$,
proceed only through pseudoscalar exchange,
while other processes,
such as $B_s \to K\mu\tau$,
proceed only through scalar exchange.
The most stringent bounds in the quark sector
come from meson--antimeson mixing,
to which both scalars and pseudoscalar contribute.

The most recent analysis of $F^0$--$\bar{F}^0$ mixing,
where $F = K$,
$D$,
$B_d$,
or $B_s$,
was by Gupta and Wells~\cite{Gupta:2009wn},
who considered the tree-level exchange of scalars and pseudoscalar
and found that
\ba
\Delta m_F &=& \frac{\xi_{ij}^2}{m_F} \left( \frac{S_F}{m^2_\mathrm{eff}}
+ \frac{P_F}{m^2_A} \right),
\\
S_F &=& \frac{B_F f_F^2 m_F^2}{6}
\left[ 1 + \frac{m^2_F}{\left( m_i + m_j \right)^2} \right],
\\
P_F &=& \frac{B_F f_F^2 m_F^2}{6}
\left[ 1 + \frac{11 m^2_F}{\left( m_i + m_j \right)^2} \right].
\ea
In these expressions,
$m_F$ is the meson mass,
$f_F$ is the pseudoscalar decay constant,
and $B_F$ is the vacuum insertion parameter
defined in Ref.~\cite{Atwood:1996vj}.
Adding in quadrature the theoretical and experimental errors
to the SM prediction,
Gupta and Wells demanded that the sum of the SM value
and the new contribution not exceed the experimental value
by more than two standard deviations for the $B_d$ and $B_s$ systems;
for the $D$ and $K$ systems,
they demanded that the new contribution not exceed the experimental value
by more than two standard deviations.
The results,
using the Cheng--Sher {\it Ansatz}\/
and assuming that $m_\mathrm{eff} = m_A = 120\,$GeV,
are
\be
\left( \lambda_{ds}, \lambda_{uc}, \lambda_{bd}, \lambda_{bs} \right)
\leq (0.1, 0.2, 0.06, 0.06).
\label{vjisw}
\ee
This might seem problematic for the type~III model,
where one expects $\lambda_{ij} \sim 1$.
But,
the pseudoscalar contribution is typically a factor of $7$--$10$ larger than
the scalar contribution,
therefore,
increasing $m_A$ to,
say,
$400\,$GeV increases all the bounds in eq.~\eqref{vjisw}
by more than a factor of three.
This was illustrated more explicitly
by Golowich {\it et al.}\/~\cite{Golowich:2007ka},
who have shown that the bound is increased substantially
as the pseudoscalar mass increases.\footnote{The reader is cautioned that
the parameter $\Delta_{ij}$ in Ref.~\cite{Golowich:2007ka}
differs from our $\lambda_{ij}$ by
$\Delta_{ij} = \lambda_{ij} \left/ \sqrt{2} \right.$.}
In addition,
as pointed out in Ref.~\cite{Atwood:1996vj},
there are additional diagrams,
involving boxes and triangle graphs,
some of them with a charged Higgs,
and a mild cancellation involving these contributions
would also weaken the bounds.
Finally,
it has been argued~\cite{Davidson:2010xv} that
it might be more appropriate to use $\lambda_{ij} \! \left/ \tan\beta \right.$
rather than $\lambda_{ij}$,
which for large $\tan\beta$ would also decrease the bounds.

Golowich {\it et al.}\/~\cite{Golowich:2007ka}
also studied the effects of $\lambda_{ct}$ and $\lambda_{ut}$
on $D$--$\bar{D}$ mixing.
For scalar masses between $100$ and $400\,$GeV,
they found that $\sqrt{\lambda_{ut}\lambda_{ct}}$
must be less than,
approximately,
8--10.
Thus,
bounds on top-quark flavour-changing neutral currents are much weaker.
Direct bounds on $\lambda_{tc}$ can be obtained from $t \to c h$,
which was first discussed by Hou~\cite{Hou:1991un}
and much more recently
in Refs.~\cite{CorderoCid:2004vi,Larios:2006pb,Aranda:2009cd}.
For a Higgs mass of $120\,$GeV,
the branching ratio is roughly $0.005\, \lambda_{ct}^2$.
The fact that the SM decay $t \to b W$
fits its prediction implies that
the branching ratio for $t \to c h$ cannot be very large,
and thus one would expect $\lambda_{ct}$ to be less than 10 or so,
but we know of no current experimental bounds on this process.
The signature for $t\rightarrow ch$
is quite different from that of $t \rightarrow H^+b$, as discussed in
Chapter 4.

Other processes will bound products of the $\lambda_{ij}$
instead of the individual $\lambda_{ij}$.
For example,
the processes $B_s \to \mu^+ \mu^-$,
with tree-level pseudoscalar exchange,
and $B \to K \mu^+\mu^-$,
with tree-level scalar exchange,
will bound the product $\lambda_{bs} \lambda_{\mu\mu}$;
it has been shown~\cite{Gupta:2009wn,Joshipura:2010tz} that
data will be obtained at the Tevatron and LHC
which will bound the product to approximately $0.6$
(for scalar or pseudoscalar masses of $120\,$GeV).
One of the most interesting decays,
since it vanishes in the SM,
is $B \to K \mu \tau$.
This process was searched for by BABAR~\cite{Aubert:2007rn} and,
while the bounds only give $\sqrt{\lambda_{bs}\lambda_{\mu\tau}}$
to be less than $\mathrm{O}(10)$,
the absence of an SM rate makes this promising for future $B$ factories.

In the above we have focused on quarks.
Bounds for leptons in the type~III model
have also been discussed extensively---with large mixing angles
in the leptonic charged weak current,
one imagines that mixing in the neutral current might be large.
The anomalous magnetic moment of the muon,
with neutral-scalar exchange,
is proportional to $\lambda_{\mu\tau}^2$
and does not depend on any other couplings.
This was first used in Ref.~\cite{Nie:1998dg}
to obtain a bound $\lambda_{\mu\tau} < 50$.
Later,
the experimental precision improved and there is now a significant discrepancy,
of approximately $3\sigma$,
between that observable and its SM prediction
(the precise discrepancy depends on the dataset used
in determining hadronic corrections~\cite{deRafael:2009zz,Prades:2009qp}).
Diaz {\it et al.}\/~\cite{Diaz:2002uk,Diaz:2004mk}
assumed that the discrepancy arises from
flavour-changing neutral-Higgs exchange and obtained
a {\em lower} bound on $\lambda_{\mu\tau}$,
with $10 < \lambda_{\mu\tau} < 80$.

In the same work,
Diaz {\it et al.}\/~\cite{Diaz:2002uk} considered $\mu\to e\gamma$,
in which a scalar is exchanged and the internal fermion is a $\tau$.
Taking a range of values for the scalar mass,
and a heavy pseudoscalar,
they obtained $\lambda_{e\tau} \lambda_{\mu\tau} < 0.04$.
This might be marginally acceptable for the type~III model,
similar to the bounds on $F$--$\bar{F}$ mixing discussed above.
In this case,
however,
a very recent analysis by Hou, Lee, and Ma~\cite{Hou:2008yb}
has a different expression for the $\mu\to e\gamma$ width
and they find that $\mu \to e\gamma$ is acceptable
for $\lambda_{e\tau} = \lambda_{\mu\tau} = 1$
if the scalar mass is above $150\,$GeV.
They also note that the contribution
of the scalar and pseudoscalar to $\mu \to e\gamma$ have opposite signs,
hence some cancellation might be possible.
In any event,
failure to observe a signal in $\mu\to e\gamma$
within the next three years at the MEG experiment---which will be sensitive
to an additional two orders of magnitude in the rate---would be
a serious problem for the type~III model.
Diaz {\it et al.}\/ then used the lower bound on $\lambda_{\mu\tau}$
discussed in the previous paragraph to obtain $\lambda_{e\tau} < 10^{-3}$.
Although their work did not explicitly use
the {\it Ansatz}\/ of the type~III model,
their conclusion---if the discrepancy
in the anomalous magnetic moment of the muon
is due to scalar exchange then there must be a substantial hierarchy
between $\lambda_{\mu\tau}$ and $\lambda_{e \tau}$---is robust,
and the size of that hierarchy would rule out the type~III model.

Other bounds may be obtained by mixing leptons and quarks.
In Ref.~\cite{Li:2008xx} the process $\tau \to \mu P$,
where $P$ is a pseudoscalar meson,
was studied.
It was claimed that this process gives $\lambda_{\mu\tau} < 10^{-3}$,
which is stronger than the previous bounds by several orders of magnitude.
However,
in Ref.~\cite{Li:2008xx} it was assumed that $\lambda_{uu}$,
$\lambda_{dd}$,
$\lambda_{ss}$,
$\lambda_{bs}$,
and $\lambda_{bb}$ are all larger than 100,
and that the light scalar and pseudoscalar have mass $120\,$GeV,
leading to this stringent bound.
Subsequently~\cite{Li:2010vf},
in a study of $\tau \to \mu P P$,
these assumptions were relaxed
and $\lambda_{\mu\tau}\lambda_{ss} < 200$ was found.

Other bounds on the $\lambda_{ij}$
do not involve flavour-changing neutral currents.
If one assumes that only $\lambda_{tt}$ and $\lambda_{bb}$ are nonzero,
then neutral scalars play no role,
and one can bound $\lambda_{tt}$ and $\lambda_{bb}$
by considering the charged-Higgs contributions
to $b \to s\gamma$~\cite{BowserChao:1998yp,Xiao:2003ya,Idarraga:2005ia};
one obtains a bound on $\lambda_{tt}$ of approximately 1.7,
for a charged-Higgs mass below $300\,$GeV.
More recently,
Mahmoudi and Stal~\cite{Mahmoudi:2009zx}
analysed $b \to s\gamma$ and $\Delta m_{B_d}$ as well as $B$,
$K$,
and $D_s$ decays,
and obtained bounds on all the second- and third-generation
diagonal $\lambda_{ij}$ (including for the leptons).
Huang and Li~\cite{Huang:2004pk} showed the $\Delta m_{B_s}$
can also be used to bound $\lambda_{cc}$ and $\lambda_{ss}$.
Since we are focusing here on the FCNC,
we shall not discuss these bounds further,
but refer the reader to Ref.~\cite{Mahmoudi:2009zx} for a detailed analysis.
The type III model says nothing about the flavour hierarchy problem.
Blechman {\em et al.}~\cite{Blechman:2010cs}, within a type III model scenario,
looked for a basis independent constraint on the Yukawa couplings that would generate a
mass hierarchy with the Yukawa couplings being of the same order.
Their constraint was that the determinant of the Yukawa couplings of $\Phi_2$ vanish.
They require a fairly large, but not unreasonable, value of $\tan\beta$.
The values of the $\lambda_{ij}$ in their model depend on the pseudoscalar and heavy
scalar masses. The phenomenological constraints are discussed in Ref.~\cite{Blechman:2010cs}.

Over the past twenty years,
bounds on the $\lambda_{ij}$ have steadily improved,
and the type~III model is now facing challenges.
In the leptonic sector,
a negative result in the MEG experiment would essentially rule out the model
(unless there is a cancellation between
the scalar and pseudoscalar contributions).
In the quark sector,
a pseudoscalar in the $100$--$200\,$GeV mass range
would cause problems for $B$--$\bar{B}$ and $B_s$--$\bar{B}_s$ mixing,
and one might expect an observable signal in $B_s \to \mu^+ \mu^-$.
With the large top-quark sample at the LHC,
the most promising signature is likely to be $t\to c h$;
if $h$ is light (as expected)
and the branching ratio is below $10^{-3}$,
then the type~III model would be in serious difficulty.

\subsection{BGL models}

All the tree-level flavour-changing transitions in the SM
are mediated by the charged weak current,
with the flavour mixing controlled by the CKM (quark mixing) matrix.
Branco,
Grimus,
and Lavoura \cite{Branco:1996bq}
have built explicitly a class of 2HDMs
(that we call BGL models)
in which the tree-level flavour-changing couplings of the neutral scalars
are related in an exact way to elements of the CKM matrix.
In BGL models the required suppression of the scalar-mediated FCNC
is obtained through relations involving
the small off-diagonal elements of the CKM matrix $V$.
Some variants of the BGL models fall into a wider category,
which was coined later on \cite{D'Ambrosio:2002ex}
as models of Minimal Flavour Violation (MFV)
\cite{Chivukula:1987py,Buras:2000dm,Blanke:2006ig}.

The Standard Model with three families,
consisting of SU(2) doublets ($Q_L$ and $L_L$)
and SU(2) singlets ($U_R$, $D_R$, and $E_R$)
has a large flavour group of unitary transformations $G_F = U(3)^5$
which commutes with the gauge group.
The gauge group $G_F$ can be decomposed as
\begin{equation}
G_F \equiv SU(3)^3_q \times  SU(3)^2_l \times U(1)_B \times U(1)_L
 \times U(1)_Y \times U(1)_{PQ}   \times U(1)_{E_R},
\end{equation}
where
\begin{eqnarray}
 SU(3)^3_q &=&  SU(3)_{Q_L} \times  SU(3)_{U_R}  \times  SU(3)_{D_R},
\\
  SU(3)^2_l &=&  SU(3)_{L_L} \times  SU(3)_{E_R}.
\end{eqnarray}
The notation is borrowed from Ref.~\cite{D'Ambrosio:2002ex}.
The Yukawa couplings break the flavour group $G_F$.
One can formally recover this flavour invariance
by treating the Yukawa couplings as dimensionless fields (spurions),
transforming in such a way that
the Yukawa interactions become $G_F$-invariant.
The MFV hypothesis consists of assuming that,
even if New Physics exists,
$G_F$ is only broken by the Yukawa couplings,
with dominance of the Yukawa coupling of the top quark.
The most general scalar potential under the MFV hypothesis
was recently derived~\cite{arXiv:1103.2915}.

The relevant question for BGL models is:
under what conditions the neutral-scalar couplings
in 2HDMs are only functions of the CKM matrix $V$?
Namely,
in BGL models these exact functions
result from the imposition of discrete symmetries.
Similar functions had been considered previously
as an {\it ad hoc}\/ assumption
\cite{Antaramian:1992ya,Hall:1993ca,Jo1991,Joshipura:1990pi}.\footnote{Another
proposal for the structure of the scalar couplings to fermions
is the suggestion that
the two Yukawa couplings are aligned in flavour space
\cite{Tuzon:2010vt,Jung:2010ik,Jung:2010ab}.}

Let us write down the Yukawa interactions:
\begin{equation}
{\cal L}_Y = - \overline{Q^0_L} \left( Y^d_1 \Phi_1 + Y^d_2 \Phi_2
\right) d^0_R
- \overline{Q^0_L} \left( Y^u_1 \tilde \Phi_1
+ Y^u_2 \tilde \Phi_2 \right) u^0_R
+ \hc
\label{1e2}
\end{equation}
After spontaneous symmetry breaking the quark mass matrices are
\be
M_d = \frac{1}{\sqrt{2}} \left( v_1  Y^d_1 +
                           v_2 e^{i \alpha} Y^d_2 \right),
\quad
M_u = \frac{1}{\sqrt{2}} \left( v_1  Y^u_1 +
                           v_2 e^{-i \alpha} Y^u_2 \right),
\label{mmmm}
\ee
%
where $\alpha$ a general phase for the vev of $\Phi_2$
and $v_1$ and $v_2$ are, without loss of generality, real.
These matrices are bi-diagonalized as
\ba
U^\dagger_{dL} M_d U_{dR} = D_d \equiv \mbox{diag}
\left( m_d, m_s, m_b \right),
\label{umu}\\
U^\dagger_{uL} M_u U_{uR} = D_u \equiv \mbox{diag}
\left( m_u, m_c, m_t \right).
\label{uct}
\end{eqnarray}
In terms of the quark mass eigenstates $u$ and $d$,
the Yukawa couplings are:
\begin{eqnarray}
{\cal L}_Y &=&
\left[ \frac{\sqrt{2} H^+}{v}\, \bar{u} \left(
V N_d \gamma_R - N^\dagger_u V \gamma_L \right) d +  \hc \right]
\no & &
- \frac{H^0}{v} \left(  \bar{u} D_u u + \bar{d} D_d \ d \right)
\no & &
- \frac{R}{v} \left[ \bar{u} \left( N_u \gamma_R + N^\dagger_u \gamma_L
\right) u + \bar{d} \left( N_d \gamma_R + N^\dagger_d \gamma_L \right) d
\right]
\no & &
+ i\, \frac{I}{v} \left[ \bar{u} \left( N_u \gamma_R - N^\dagger_u \gamma_L
\right) u - \bar{d} \left( N_d \gamma_R - N^\dagger_d \gamma_L
\right) d \right],
\end{eqnarray}
where $v \equiv \sqrt{v_1^2 + v_2^2} = (\sqrt{2} G_F)^{-1/2}
\approx 246\,$GeV ($G_F$ is the Fermi constant),
$\gamma_L = (1 - \gamma_5)/2$,
$\gamma_R = (1 + \gamma_5)/2$,
and
\begin{eqnarray}
H^0 &=& \frac{1}{v} \left( v_1 \rho_1 + v_2 \rho_2 \right),
\\
R &=& \frac{1}{v} \left( v_2 \rho_1 - v_1 \rho_2 \right),
\\
I &=& \frac{1}{v} \left( v_2 \eta_1 - v_1 \eta_2 \right).
\end{eqnarray}
The fields $\rho_j$,
$\eta_j$ ($j = 1, 2$)
arise when one expands  \cite{Lee:1973iz} the neutral scalar fields
around their vevs:
$\Phi^0_j =  \left( e^{i \alpha_j} \left/ \sqrt{2} \right. \right)
\left( v_j + \rho_j + i \eta_j \right)$.
The physical neutral-scalar fields are linear combinations of $H^0$,
$R$,
and $I$.

The flavour-changing neutral currents are controlled
by the matrices $N_d$ and $N_u$,
which are given by:
\begin{eqnarray}
N_d &=& \frac{1}{\sqrt{2}}\, U^\dagger_{dL} \left( v_2 Y^d_1 -
v_1 e^{i \alpha} Y^d_2 \right)  U_{dR},
\label{ndnd} \\
N_u &=& \frac{1}{\sqrt{2}}\,
U^\dagger_{uL} \left( v_2  Y^u_1 -
\label{nunu}  v_1 e^{-i \alpha} Y^u_2 \right) U_{uR}.
\end{eqnarray}
In general these matrices are not diagonal,
hence there are FCNC.
Let us consider,
for example,
$N_d$ and rewrite it as \cite{Lavoura:1994ty}:
\begin{equation}
N_d = \frac{v_2}{v_1} D_d - \frac{v_2}{\sqrt{2}}
\left( \frac{v_2}{v_1} +  \frac{v_1}{v_2}\right)  U^\dagger_{dL}
e^{i \alpha} Y^d_2 U_{dR};
\label{nd}
\end{equation}
clearly,
the first term in the right-hand side conserves flavour,
but the second one leads,
in general,
to FCNC.
The CKM matrix is given by  $V = U^\dagger_{uL}  U_{dL}$,
therefore,
if we want $N_d$ to be entirely controlled by $V$,
we need,
on the one hand,
to get rid of its dependence on $U_{dR}$ and,
on the other hand,
to relate  $U^\dagger_{dL}$ to $V$.
A solution to these two requirements by means of symmetries
was found in Ref.~\cite{Branco:1996bq} and corresponds to the BGL models.

In Ref.~ \cite{Branco:1996bq} a flavour symmetry is imposed which
constrains $U_{uL}$ to be of the form
\begin{equation}
U_{uL}  = \left( \begin{array}{ccc}
\times  & \times & 0 \\
\times & \times & 0  \\
0 & 0 & 1
\end{array}\right),
\label{calm}
\end{equation}
where $\times$ denotes an arbitrary entry.
This leads,
from the definition of $V$,
to:
\begin{equation}
V_{3j} = \left( U_{dL} \right)_{3j}.
\label{3j}
\end{equation}
In addition,
the flavour symmetry imposes the following condition
\begin{equation}
Y^d_2 U_{dR} =  \left( \begin{array}{ccc}
0 & 0 & 0  \\
0 & 0 & 0 \\
\times & \times & \times
\end{array}\right).
\label{bom}
\end{equation}
This guarantees that only the third row of $U_{dL}$ appears in $N_d$,
as can be easily checked from eq.~(\ref{nd}).
This row is exactly the one that coincides with the third row of $V$.

To get rid of $U_{dR}$ in $N_d$,
the flavour symmetry also enforces the relation
\begin{equation}
\frac{v_2 e^{i \alpha}}{\sqrt{2}}\, Y^d_2 = P M_d,
\label{pro}
\end{equation}
where $P$ is a projection matrix.
In order to be consistent with eq.~(\ref{bom}),
the matrix $P$ must be given by:
\begin{equation}
P  = \left( \begin{array}{ccc}
0  & 0 & 0 \\
0 & 0 & 0  \\
0 & 0 & 1
\end{array}\right),
\end{equation}
immediately leading to texture-zeros for $Y^d_2$
such that eq.~(\ref{bom}) is verified.
Furthermore,
the appearance of $M_d$ in eq.~(\ref{pro}) allows to absorb $U_{dR}$.

Branco,
Grimus,
and Lavoura have imposed
the following symmetry $S$ on the 2HDM:
\begin{equation}
Q^0_{L3} \rightarrow e^{i \psi} Q^0_{L3},
\quad
u^0_{R3} \rightarrow e^{2 i \psi} u^0_{R3},
\quad
\Phi_2 \rightarrow e^{i \psi} \Phi_2,
\label{bgl}
\end{equation}
where $\psi \neq 0, \pi$ and all other fields are invariant under $S$.
The Yukawa couplings consistent
with this symmetry have the structure
\be
 Y^d_1 = \left( \begin{array}{ccc}
\times  & \times & \times \\
\times & \times &  \times \\
0 & 0 & 0
\end{array} \right),
\
Y^d_2 = \left( \begin{array}{ccc}
0 & 0 & 0 \\
0 & 0 & 0 \\
\times & \times & \times
\end{array} \right),
\
Y^u_1 = \left( \begin{array}{ccc}
\times & \times & 0 \\
\times & \times & 0 \\
0 & 0 & 0
\end{array} \right),
\
Y^u_2 = \left( \begin{array}{ccc}
0 & 0 & 0 \\
0 & 0 & 0 \\
0 & 0 & \times
\end{array} \right).
\label{del}
\ee
Notice that the matrices $Y^u$ are block diagonal;
this is crucial in order for $U_{uL}$ to be as in eq.~(\ref{calm}).
Furthermore,
these four matrices satisfy
\begin{equation}
P Y^d_{1} = 0, \
P Y^d_{2} = Y^d_{2}, \
P Y^u_{1} = 0, \
P Y^u_{2} = Y^u_{2}.
\label{pppp}
\end{equation}
The structure of zeros
in the matrix $Y^d_2$ leads to the important relation:
\begin{equation}
\left( U^\dagger_{dL} Y^d_2 \right)_{ij}=
\left( U^\dagger_{dL} \right)_{i3} \left( Y^d_2 \right)_{3j}
= V_{i3} \left( Y^d_2 \right)_{3j}.
\end{equation}
Together with eq.~(\ref{pro}) inserted in eq.~(\ref{nd}),
one then obtains \cite{Branco:1996bq}
\begin{equation}
\left( N_d \right)_{ij} =
\frac{v_2}{v_1}\, \left( D_d \right)_{ij}
- \left( \frac{v_2}{v_1} +  \frac{v_1}{v_2} \right)
V^\dagger_{i3} V_{3j} \left( D_d \right)_{jj},
\label{24}
\end{equation}
whereas
\begin{equation}
N_u = - \frac{v_1}{v_2}\, \mbox{diag} \left( 0, 0, m_t \right)
+ \frac{v_2}{v_1}\, \mbox{diag} \left( m_u, m_c, 0 \right).
\label{25}
\end{equation}
In this example there are scalar-mediated FCNC in the down sector
but no FCNC in the up sector.

In general,
BGL models are six different models,
as was emphasised in~\cite{Branco:1996bq}.
Three of these models have FCNC only in the down sector,
and are obtained from the three different projection matrices
similar to the matrix $P$ introduced above,
but,
in each case,
with the diagonal unit entry
in one of the other two alternative positions.
Another three models
are obtained by exchanging the patterns of zeros of the $Y^d_i$
and $Y^u_{i}$ matrices,
\ie by exchanging the up and down quarks.
From eq.~(\ref{bgl}) it is straightforward to
write down the flavour symmetries corresponding to each of the
six cases.
The relations given in eq.~(\ref{pppp}) result from the imposed symmetry.
All BGL models obey relations of this type
for the corresponding projection matrices.
These relations guarantee that the scalar flavour-changing neutral couplings
can be written in terms of quark masses and CKM-matrix entries
\cite{Botella:2009pq}.
The stability of these equations under renormalization
is crucial;
this feature was analysed in~\cite{Botella:2011ne},
with the help of the one-loop renormalization-group equations
for the Yukawa couplings generalized from Ref.~\cite{Ferreira:2010xe}.

In each of the six models,
one can study the phenomenological constraints by
comparison with the type~III model.
For the specific model discussed in the last paragraph,
for example,
one can compare the FCNC coupling $bsH$ in the two models directly:
\begin{equation}
\lambda_{bs}\frac{\sqrt{2m_bm_s}}{v} \leftrightarrow
\frac{m_b}{v}\left(\frac{v_1}{v_2}+\frac{v_2}{v_1}\right)
V_{ts}^\ast V_{tb}
\end{equation}
which numerically gives
$\lambda_{bs}= 0.14 \left(\frac{v_1}{v_2}+\frac{v_2}{v_1}\right)$,
as compared to $\lambda_{bs}=1$ in the type~III model.
Similar expressions can be found for the other FCNC couplings,
as well as for the other five models.
Thus,
the bounds in the previous section can be used
to find the phenomenological bounds in the BGL model
as well~\footnote{In Ref.~\cite{Joshipura:2007sf} a specific analysis of BGL models,
especially their consequences for neutral-meson--antimeson mixing,
has been performed.}.

In the example given here the scalar-mediated FCNC
are suppressed by the matrix elements of the third row of the CKM matrix;
this is due to the fact that
$P$ is the projection matrix with unit (33) entry.
In the alternative cases,
corresponding to a different projection matrix $P$,
the suppression is given
by the row indicated by the non-zero entry in the corresponding
matrix $P$.
In these additional cases the suppression of the scalar-mediated FCNC
is not as strong as in the example above.
From the point of view of the authors of Ref.~\cite{D'Ambrosio:2002ex},
these additional cases do not qualify as MFV models,
since they do not comply with the ingredients imposed by their definition
of MFV.
In the next section we generalize BGL models
and compare them with the MFV definition of Ref.~\cite{D'Ambrosio:2002ex}.

\subsection{MFV generalized}

In Ref.~\cite{Botella:2009pq} the question was addressed of
how to find a general expansion for $N_d^0$ and $N_u^0$
which conforms with the requirement of having all the
flavour-changing couplings of the neutral scalars
related in an exact way to elements of the quark mixing matrix.
Here,
$N_d^0$ and  $N_u^0$ denote $N_d$ and $N_u$ when still in a weak basis:
\begin{eqnarray}
N_d^0 = U_{dL} N_d U^\dagger_{dR} &=&
\frac{1}{\sqrt{2}} \left( v_2  Y^d_1
- v_1 e^{i \alpha} Y^d_2 \right),
\\
N_u^0 = U_{uL} N_u U^\dagger_{uR} &=&
\frac{1}{\sqrt{2}} \left( v_2  Y^u_1
- v_1 e^{-i \alpha} Y^u_2 \right).
\end{eqnarray}
The strategy was imposing that $N_d^0$ and  $N_u^0$ be only functions
of $M_d$ and $M_u$ with no other flavour dependence.
Furthermore,
$N_d^0$ and  $N_u^0$ should transform appropriately under
weak-basis (WB) transformations.

Weak-basis transformations are defined by:
\begin{equation}
Q^0_L \rightarrow W_L Q^0_L, \quad
d^0_R \rightarrow W^d_R d^0_R, \quad
u^0_R \rightarrow  W^u_R u^0_R.
\label{wea}
\end{equation}
Under this transformation,
the quark mass matrices $M_d$ and $M_u$ transform like
\begin{equation}
M_d \rightarrow W^\dagger_L M_d W^d_R. \quad
M_u \rightarrow W^\dagger_L M_u W^u_R,
\end{equation}
and the matrices $U_{dL}$,
$U_{dR}$,
$U_{uL}$,
and $U_{uR}$ defined in eqs.~(\ref{umu}) and~(\ref{uct}) transform as:
\be
U_{dL} \rightarrow W^\dagger_L U_{dL}, \quad
U_{uL} \rightarrow W^\dagger_L U_{uL}, \quad
U_{dR} \rightarrow {W^d_R}^\dagger U_{dR}, \quad
U_{uR} \rightarrow {W^u_R}^\dagger U_{uR}.
\ee
Under a WB transformation $N^0_d$ and $N^0_u$
transform in the same way as $M_d$ and $M_u$,
respectively.
Furthermore,
the Hermitian matrices $H_{d,u} \equiv  M_{d,u} M^\dagger_{d,u}$
transform under a WB as:
\begin{equation}
H_d \rightarrow W^\dagger_L H_d W_L, \quad
H_u \rightarrow W^\dagger_L H_u W_L.
\end{equation}
From eqs.~(\ref{umu}) and~(\ref{uct}) it follows that
\begin{equation}
U^\dagger_{dL} H_d U_{dL} = D^2_d, \quad
U^\dagger_{uL} H_u U_{uL} = D^2_u.
\end{equation}
It is convenient,
for our propose,
to write $H_d$,  $H_u$ in
terms of projection operators \cite{Botella:2004ks}:
\begin{equation}
H_d = \sum_i {m^2_{d}}_i P^{dL}_i,
\end{equation}
where
\begin{equation}
P^{dL}_i =  U_{dL} P_i U^\dagger_{dL},
\quad
\label{bblo}
\left( P_i \right)_{jk} = \delta_{ij} \delta_{ik}.
\end{equation}
Analogous expressions hold for $H_u$.

Based on the above considerations,
the following expansion for $N^0_d$ and $N^0_u$,
with the correct transformation properties under WB transformations,
was proposed \cite{Botella:2009pq}:
\begin{eqnarray}
N^0_d &=& \lambda_1 M_d
+ \lambda_{2i} U_{dL} P_i U^\dagger_{dL} M_d
+ \lambda_{3i} U_{uL} P_i U^\dagger_{uL} M_d
+ \ldots,
\label{amd1} \\
N^0_u &=&
\tau_1 M_u
+ \tau_{2i} U_{uL} P_i U^\dagger_{uL} M_u
+ \tau_{3i} U_{dL} P_i U^\dagger_{dL} M_u
+ \ldots.
\label{amd2}
\end{eqnarray}
In the quark mass eigenstate basis these equations become
\begin{eqnarray}
N_d &=&
\lambda_1 D_d
+ \lambda_{2i} P_i D_d
+ \lambda_{3i} V^\dagger P_i V D_d
+ \ldots,
\label{38} \\
N_u &=&
\tau_1 D_u
+ \tau_{2i} P_i D_u
+ \tau_{3i} V P_i V^\dagger D_u
+ \ldots.
\label{39}
\end{eqnarray}
This meets the requirement of having all the flavour-changing couplings
related in an exact way to the quark mixing matrix,
with no additional flavour dependence.

In  eqs.~(\ref{amd1}) and~(\ref{amd2})
the lambda and tau coefficients are
dimensionless.
The building blocks of the expansion are
given explicitly in eq.~(\ref{bblo}) for the index $d$;
they are of the same form for the index $u$.
Notice that,
in addition,
there is the possibility of having different coefficients
for difference values of the index $i$ (ranging from 1 to 3),
corresponding to different projectors $P_i$.

In Ref.~\cite{D'Ambrosio:2002ex} a different expansion is used.
There,
the building blocks are $Y_d Y^\dagger_d$ and $Y_u Y^\dagger_u$,
where $Y_d$ and $Y_u$ are the Yukawa-coupling matrices.
These matrices are also dimensionless.
Such an expansion can be accommodated in eqs.~(\ref{amd1}) and~(\ref{amd2}),
however,
but in that case the lambdas and taus
would become dimensionless functions of the quark masses.

In theories where the MFV requirement
results from the imposition of a symmetry
there are constraints on the coefficients lambda and tau
appearing in eqs.~(\ref{38}) and~(\ref{39}).
This is the case for the BGL example presented in the previous subsection,
which corresponds to the truncation
\begin{eqnarray}
N^0_d &=& \frac{v_2}{v_1}\, M_d
- \left( \frac{v_2}{v_1} + \frac{v_1}{v_2} \right)
U_{uL} P_3 U^\dagger_{uL} M_d,
\label{bgl1} \\
N^0_u &=& \frac{v_2}{v_1}\, M_u
- \left( \frac{v_2}{v_1} + \frac{v_1}{v_2} \right)
U_{uL} P_3 U^\dagger_{uL} M_u,
\label{bgl2}
\end{eqnarray}
with lambdas and taus fully determined as functions of
$\tan \beta \equiv v_2 / v_1$.

More details about different aspects of this MFV generalization
can be found in Ref.~\cite{Botella:2009pq}.

A study on the effectiveness of the two different hypothesis,
Natural Flavour Conservation and Minimal Flavour Violation,
in suppressing the strength of flavour-changing neutral currents
in models with more than one Higgs doublet was performed
in Ref.~\cite{Buras:2010mh}.

\subsection{Two-Higgs Leptonic Minimal Flavour Violation}

In order to study the phenomenological implications
of models with an extended Higgs sector
it is necessary to specify the lepton sector
in addition to the quark sector.
Furthermore,
the analysis of stability under renormalization
requires the entire set of renormalization-group equations
both in the quark and lepton sectors.
In Ref.~\cite{Botella:2011ne} the extension
to the lepton sector of models of BGL type was considered.
In particular the minimal discrete symmetry required in order
to implement the models in a natural way was given and stability
was analysed.
Different extensions of the MFV principle
to the leptonic sector were considered previously in
Refs.~\cite{Cirigliano:2005ck,Davidson:2006bd,Branco:2006hz}.

The case of Dirac-type neutrinos is straightforward.
The Yukawa couplings for both the quark and the
lepton sectors are given by:
\begin{eqnarray}
\mathcal{L}_{Y} &=&
-\overline{Q_{L}^{0}} \left( Y^d_1 \Phi_1
+ Y^d_{2} \Phi_2 \right) d_{R}^{0}
- \overline{Q_{L}^{0}} \left(
Y^u_1 \tilde \Phi_1 + Y^u_2 \tilde \Phi_2 \right) u_{R}^{0}
\nonumber \\ & &
-\overline{L_{L}^{0}} \left( Y^e_1 \Phi_1 + Y^e_2 \Phi_2 \right) l_{R}^{0}
-\overline{L_{L}^{0}} \left( Y^\nu_1 \tilde \Phi_1
+ Y^\nu_2 \tilde \Phi_2 \right) \nu_{R}^{0}
+ \hc,
\label{YukawaDirac1}
\end{eqnarray}
where $Y^e_j$ and $Y^\nu_j$
denote the couplings of the left-handed leptonic doublets $L^0_L$
to the right-handed charged leptons $l^0_R$ and neutrinos $\nu^0_R$.
If there are no additional Majorana mass terms,
the parallel between  the quark and leptoni sectors
allows to apply similar discussions to both sectors.

In the case of Majorana-type neutrinos
there is no lepton-number conservation.
In the seesaw framework with three right-handed neutrinos,
an additional invariant mass term of Majorana type
for right-handed neutrinos,
\ie $(1/2) {\nu_R^{0}}^T C^{-1} M_R \nu_R^{0}$,
must be included in the Lagrangian.
As a result there will be three light neutrinos
$\nu_i$ and three heavy neutrinos $N_i$.
In order to obtain FCNC in the charged-lepton sector
completely controlled by the lepton mixing (PMNS) matrix,
together with no FCNC in the light-neutrino sector,
a $Z_4$ symmetry is required.
An example is~\cite{Botella:2011ne}:
\begin{equation}
L^0_{L3} \rightarrow e^{i \psi} L^0_{L3},
\quad
\nu^0_{R3} \rightarrow e^{2 i \psi} \nu^0_{R3},
\quad
\Phi_2 \rightarrow e^{i \psi} \Phi_2,
\label{utl}
\end{equation}
with $\psi = \pi/2$;
all other fields are invarant under $Z_4$.
The most general matrices $Y^e_j$,
$Y^\nu_j$,
and $M_R$ consistent with this $Z_4$ symmetry have the following structure:
\ba
& & Y^e_1 = \left( \begin{array}{ccc}
\times  & \times & \times \\
\times & \times &  \times \\
0 & 0 & 0
\end{array} \right), \
Y^e_2   =  \left( \begin{array}{ccc}
0 & 0 & 0  \\
0 & 0 & 0 \\
\times & \times &  \times
\end{array}\right),
\\ & &
Y^\nu_1  = \left( \begin{array}{ccc}
\times  & \times & 0 \\
\times & \times &  0 \\
0 & 0 & 0
\end{array}\right),\
Y^\nu_2   =  \left(\begin{array}{ccc}
0  & 0 & 0 \\
0 & 0 &  0 \\
0 & 0 & \times
\end{array}\right),\
M_R   =   \left( \begin{array}{ccc}
\times  & \times & 0 \\
\times & \times &  0 \\
0 & 0 & \times
\end{array}\right),
\label{mel}
\ea
where $\times$ denotes an arbitrary entry
while the zeros are enforced by the symmetry $Z_4$.
Note that the choice of $Z_4$ is crucial
in order to guarantee $(M_R)_{33}\neq 0$ and,
thus,
a non-vanishing $\det M_R$.

The scalar couplings to the neutrinos
are more involved than those to the charged leptons,
since they include couplings to two light neutrinos,
two heavy neutrinos,
or one light and one heavy neutrino.
They are explicitly given in Ref.~\cite{Botella:2011ne}.

As a result of the $Z_4$ symmetry,
the scalar potential acquires an exact ungauged
accidental continuous symmetry,
which is not,
though,
a symmetry of the full Lagrangian.
The simplest way of avoiding the ensuing pseudo-Goldstone boson
after spontaneous symmetry breaking
is through the introduction of a soft symmetry-breaking term
of the form  $m_{12}^2 \Phi_{1}^\dagger \Phi_{2} + \hc$

\newpage

\section{Charged Higgs bosons}
\label{sec:charg}

One of the most important features of all 2HDMs
is the existence of a charged scalar $H^\pm$.
This state is orthogonal to the longitudinal component $G^\pm$
of the gauge boson $W^\pm$.
The study of the properties of $H^\pm$ will be essential to understand
which 2HDM---if any---has been chosen by nature.
Most of the phenomenological studies of 2HDMs
have indeed focused on the charged scalar.
The $H^\pm$ can be readily pair-produced through Drell--Yan processes
and---unlike the neutral scalars---can never decay ``invisibly".
In addition to direct production,
it can have sizable indirect effects in $B$ physics
and is a major topic of analysis in studies of rare $B$ decays.

The first paper to include the words
``charged Higgs"\footnote{As emphasised by Georgi \cite{Georgi:1994qn},
the phrase ``charged Higgs" is misleading.
If the Higgs is defined to be the field that acquires a vev,
then the Higgs cannot be charged since electromagnetism is unbroken.
If the charged Higgs is defined to be the charged member of an $SU(2)$ doublet
whose neutral component acquires a vev,
then the charged Higgs is the longitudinal component $G^\pm$ of the $W^\pm$.
The phrase ``charged scalar'' is thus preferable,
yet the nomenclature ``charged Higgs'' is by now so common
that we shall adopt it here.} in its title
was the one by Tomozawa \cite{Tomozawa:1977ea},
in which the charged scalar in a vector-like model has been studied.
In the context of a 2HDM,
the phenomenology of charged scalars was first discussed in detail
by Donoghue and Li \cite{Donoghue:1978cj}.
Following that there was an explosion of interest in the charged Higgs.
This interest is sufficiently great that a series of three conferences
studying the properties of charged scalars---CHarged 2006,
CHarged 2008,
and CHarged 2010---have been held in Uppsala;
the proceedings of those conferences are readily available
\cite{charged06,charged08,charged10}
and provide an enormous source of information on
both the theory and phenomenology
of charged Higgs bosons.

For all four models without tree-level FCNC discussed in Chapter 3,
the most general Yukawa couplings were written in~\cite{Barger:1989fj} as
\be
{\mathcal L}_{H^\pm} = - H^+ \left( \frac{\sqrt{2}\,V_{ud}}{v}\,
\bar{u} \left( m_u X P_L + m_d Y P_R \right) d +
\frac{\sqrt{2}\, m_\ell}{v}\,
Z \bar{\nu_L}\ell_R\right)
+ \hc
\label{svhid}
\ee
where $V_{ud}$ is the element of the CKM matrix
corresponding to the charge $2/3$ quark $u$ and the charge $-1/3$ quark $d$.
The values of $X$, $Y$, and $Z$ depend on the particular model
and are given in Table~\ref{tab:6_models}.
\begin{table}[h]
\begin{center}
\begin{tabular}{|c|c|c|c|c|}  \hline
{} & Type I  & Type II & Lepton-specific & Flipped     \\
\hline
$X$ & $\cot{\beta}$ & $\cot{\beta}$ & $\cot{\beta}$ & $\cot{\beta}$ \\
\hline
$Y$ & $\cot{\beta}$ & $- \tan{\beta}$ & $\cot{\beta}$ & $- \tan{\beta}$ \\
\hline
$Z$ & $\cot{\beta}$ & $- \tan{\beta}$ & $- \tan{\beta}$ & $\cot{\beta}$ \\
\hline
\end{tabular}
\caption{The parameters $X$, $Y$, and $Z$ in eq.~(\ref{svhid})
for the four models without FCNC.}
\end{center}
\label{tab:6_models}
\end{table}
In the type~I model the couplings to all fermions
are suppressed if $\tan\beta \gg 1$,
meaning a fermiophobic charged Higgs.
In the same limit $\tan\beta \gg 1$ one has
in the lepton-specific model a quark-phobic but leptophilic model,
which could lead to a huge branching ratio for $H^\pm \to \tau^\pm \nu$.
In both cases,
the quark-phobic nature of the model eliminates constraints
from rare $B$ decays.
The type~II model is the most studied one;
large contributions to rare $B$ decays are possible in it.
The flipped model has only recently been studied.
In the next section we shall review the properties of the charged Higgs
in each of these models.

In models with tree-level FCNC,
discussed in Chapter 3,
it is much more convenient to use the Higgs basis,
in which one doublet has a vev and the other doublet is vev-less
(see the discussion in Chapter 5).
In the models without tree-level FCNC,
this basis is a disadvantage,
since the Yukawa couplings and discrete symmetry adequately specify the basis;
with FCNC this disadvantage vanishes.
This was pointed out by Atwood,
Reina,
and Soni~\cite{Atwood:1996vj}.
As shown there,
in the Higgs basis,
the $H_1$ field (which acquires vev) has diagonal couplings to fermions,
which are identical to those in the Standard Model,
whereas the couplings of $H_2$ to quarks are given by
\be
{\mathcal L} =
\bar Q_{Li} \left( \hat{\xi}^U_{ij} \tilde H_2 u_{Rj}
+ \hat{\xi}^D_{ij} H_2 d_{Rj} \right)
+ \hc
\ee
where the quark fields are mass eigenstates
and the matrices $\hat{\xi}^{U,D}$ are in general not diagonal.
Defining the rotation matrices $V_{L,R}^{U,D}$,
the neutral flavour-changing couplings
are related to the original ones by $V_L$ and $V_R$.
Since the definition of the $\xi_{ij}^{U,D}$ is arbitrary,
one can,
without loss of generality,
replace the rotated couplings by the original ones.
For the charged flavour changing couplings, we then have
\be
\xi^U_{\rm charged} = \xi^U V_{CKM}, \
\xi^D_{\rm charged} = V_{CKM} \xi^D.
\ee
The Cheng--Sher Ansatz discussed in Chapter 3
gives the matrices $\xi_{ij}^{U,D}$ as a constant of ${\rm O} (1)$
times the geometric mean of the respective Yukawa couplings.
Similar results are found for the leptonic sector,
with $V_{CKM}$ replaced by the $V_{PMNS}$ matrix.
In the third section of this Chapter
we shall review the phenomenology of the charged Higgs in these models.

\subsection{Models without tree-level FCNC}

The differences among the various 2HDMs
concern the Yukawa couplings to fermions.
At the recent CHarged 2010 meeting there have been many discussions
of benchmarks for these various 2HDMs~\cite{charged10}.
They were summarized,
in the context of LHC searches,
by Guedes \etall~\cite{Guedes:2011ki}.
More details,
together with numerous plots,
are in the talk by Santos at that meeting \cite{charged10}.
A large group has studied specific benchmarks
for each of the various models;
that work is summarized in the reports
by Krawczyk {\em et al} and by Osland {\em et al} at the meeting.
Some of the models,
other than the type~II model,
have been discussed in the talk by Akeroyd~\cite{charged10}.
A very comprehensive analysis of charged Higgs phenomenology
for all four models can be found in the recent article
of Aoki \etall~\cite{Aoki:2009ha}.
Jung, Pich, and Tuz\'on~\cite{Jung:2010ik}
have considered many processes within their general
``aligned two Higgs doublet model''---in which it is assumed that
the Yukawa-coupling matrices are proportional,
hence FCNC are absent at tree level.
Since that model has many parameters,
we shall not go through their analysis explicitly.
However,
a valuable feature of that analysis is that,
from its results,
one can explore the limits from indirect processes in all four models,
which are all limiting cases of the aligned two Higgs doublet model.
A discussion of the decay of the top quark into $H^+b$ in a very
general 2HDM can be found in Ref.~\cite{hep-ph/9808278}.

We shall summarize in the following the production and decay
of the charged Higgs in each of the four models,
and discuss various constraints.
In the type~II model and in the flipped model
there is a lower bound of about $300\, {\rm GeV}$ for the mass of $H^\pm$
stemming from $b \to s \gamma$;
this bound comes from charged Higgs bosons in the loop.
However,
this bound is only valid if there is no additional New Physics.
If there is New Physics,
even a relatively mild cancelation can weaken the bound substantially.
Therefore,
we shall also examine other,
somewhat weaker bounds.

\subsubsection{The type~II model}

Since supersymmetric and Peccei--Quinn models are all of type~II,
the type~II model is the most studied one.
The Yukawa couplings are given in Table 3;
one can see that the coupling of the charged Higgs
to the top and bottom quarks is governed by
either the bottom-quark mass times $\tan{\beta}$,
which may be large,
or by the top-quark mass times $\cot{\beta}$.
As a result,
one expects potentially large virtual effects in $b$-quark decays and mixing.
In fact,
one of the major motivations for the $B$ factories
was the possibility of New Physics coupling strongly to the third generation;
the type~II 2HDM is the simplest example of this.

A very strong bound on the mass of the charged Higgs
comes from studies of $\bar{B} \to X_s \gamma$.
The charged Higgs appears in the loop.
A nice review with a comprehensive list of references
is in the article of El Kaffas \etall~\cite{WahabElKaffas:2007xd}.
Early analyses can be found in Refs.~\cite{Bertolini:1986th,Deshpande:1987nr,Grinstein:1987pu,Grinstein:1987vj}.
An explosion of interest occurred after the realization that
the top quark was heavy and after QCD corrections
were considered~\cite{Grigjanis:1988iq,Cella:1990sh,Cella:1994px,Cella:1994np,Ciuchini:1993ks,Ciuchini:1993fk}.
These models were shown to be quite
scale dependent~\cite{Buras:1993xp,Ali:1993ct}
and this has led to a calculation of the next-to-leading (NLO) order,
where the scale dependence is substantially
reduced~\cite{Misiak:1992bc,Adel:1993ah,Greub:1996tg,Buras:1997bk}.
Discussions can be found in the reviews of Haisch~\cite{Haisch:2008ar}
and of Hurth and Nakao~\cite{Hurth:2010tk},
and more recent results on all $B$ decays in the recent talks
of Rozanska~\cite{charged10} and Hurth~\cite{Hurth:2011jc}.
The effects of the charged Higgs in the loop
add constructively~\cite{Ciafaloni:1997un,Ciuchini:1997xe,Borzumati:1998tg,Borzumati:2003rr,Bobeth:1999mk}.
The most detailed calculation
is the ${\rm O} ( \alpha_s^2 )$ calculation
by Misiak~\cite{Misiak:2006zs}
and gives $m_{H^\pm} > 295\, {\rm GeV}$
for virtually all values of $\tan{\beta}$
(the bound increases slightly for low values of $\tan{\beta}$,
see~\cite{Haisch:2008ar}).
A more recent measurement of the rate by BELLE~\cite{:2009qg}
does not change this bound.

\begin{figure}[h]
\vskip 0.5cm
\centerline{\epsfysize=8cm \epsfbox{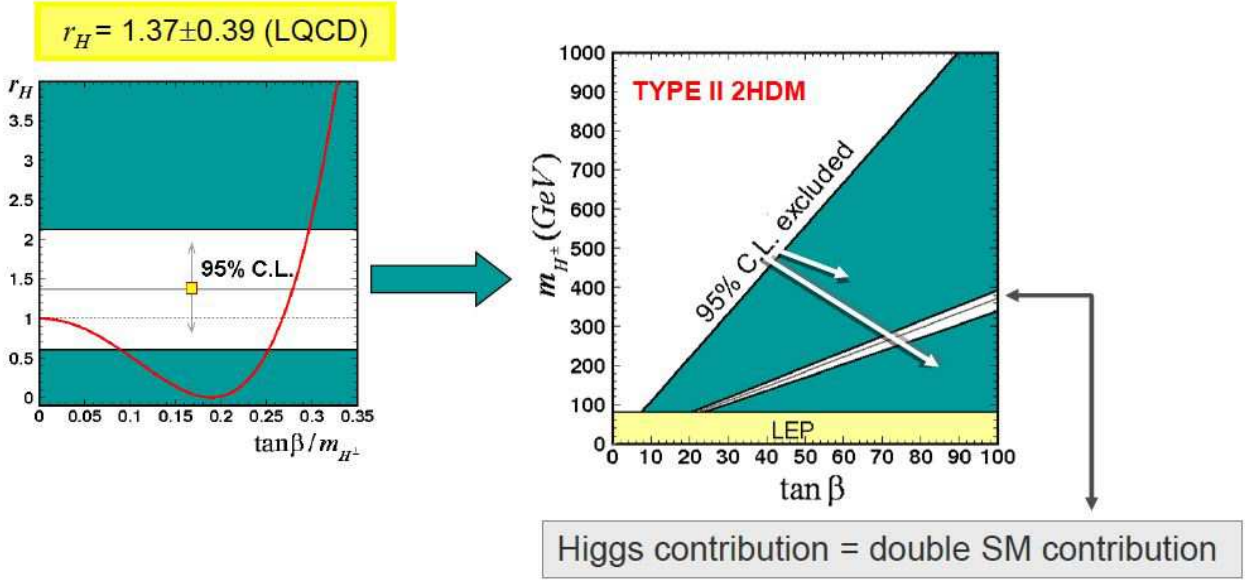} }
\caption{On the left one has the prediction for $r_H$---the ratio
between the value of
${\rm BR} \left( B^+ \to \tau^+ \nu_\tau \right)$
in the type~II 2HDM to that in the Standard Model---as a function of
$\tan{\beta} / m_{H^\pm}$.
Given the experimental value,
one has on the right the ensuing constraints in the
$m_{H^\pm}$--$\tan{\beta}$ plane;
the narrow window corresponds to the allowed region
near $\tan{\beta} / m_{H^\pm} = 0.2$
and the yellow band corresponds to the region of $m_{H^\pm}$
already excluded by direct searches at LEP.
Figure from the talk by Rozanska
at the CHarged 2010 workshop~\cite{charged10}.}
\label{6fig:btaunu}
\end{figure}

Other bounds are even more severe,
but only at large $\tan{\beta}$.
The process $B \to \tau \nu$ has been studied.
This is not a loop process and proceeds instead
through tree-level virtual $H^\pm$ exchange.
The rate is given by~\cite{Hou:1992sy,Grossman:1994ax}
\be
\frac{{\rm BR} \left( B^+ \to \ell^+ \nu_\ell \right)}
{{\rm BR} \left( B^+ \to \ell^+ \nu_\ell \right)_{\rm SM}}
= \left( 1 - \frac{m_B^2 \tan^2{\beta}}{m^2_{H^\pm}}\right)^2.
\end{equation}
Combining the BELLE~\cite{Hara:2010dk,hep-ex/0604018} and
BABAR~\cite{Aubert:2009wt} results,
the Heavy flavour Averaging Group found~\cite{hfag} the branching fraction to be
$\left( 1.64 \pm 0.34 \right) \times 10^{-4}$.
The ratio of this measured value to
the Standard Model prediction~\cite{Czarnecki:1998tn} is $1.37\pm 0.39$.
This gives a lower bound which,
at 95\% confidence level,
excludes a region which rises from $300\, {\rm GeV}$ for $\tan{\beta} = 40$
to $1100\, {\rm GeV}$ for $\tan{\beta} = 100$.
However,
there is a small window in this region which is still allowed.
The reason for that window is shown in Fig.~\ref{6fig:btaunu},
where one can see a narrow region of $\tan{\beta}$ which is still allowed.

That window can be closed by considering
the process $B \to D^{(*)} \tau\nu_\tau$.
In that process the CKM angles are better known;
also,
many of the experimental and theoretical uncertainties
cancel out in the ratio~\cite{Tanaka:2010se,Grossman:1995yp}
in which the $\tau$ is replaced by a lighter lepton---in which case
the contribution from $H^\pm$ exchange is negligible.
It is found (see Ref.~\cite{Tanaka:2010se} for details)
that the window is completely closed by this process---which does have
its own window,
but that is closed by $B^+ \to \tau^+ \nu_\tau$.

For small values of $\tan{\beta}$,
the value of $R_b$,
which is the ratio of $\Gamma \left( Z \to b \bar b \right)$
to the total hadronic width of the $Z$,
can be affected at one loop through the exchange of $H^\pm$.
Haber and Logan found~\cite{Haber:1999zh} that
the ensuing constraints will be more severe
than those from $b \to s \gamma$ for values of $\tan\beta < 1.4$;
radiative corrections to their results are in Ref.~\cite{Degrassi:2010ne}.
Very similar bounds were obtained~\cite{Geng:1988bq}
by considering $\Delta m_B$ and $\Delta m_{B_s}$.
Other bounds,
which tend to be weaker,
can be found from rare $K$,
$D$,
and $\tau$ decays;
a comprehensive analysis is found in the recent paper
of Deschamps \etall~\cite{Deschamps:2009rh}.
Such a comprehensive analysis can be very valuable
since it is possible that New Physics might weaken some,
but not all,
of the bounds.

\begin{figure}[h]
\vskip 0.5cm
\centerline{\epsfysize=8cm \epsfbox{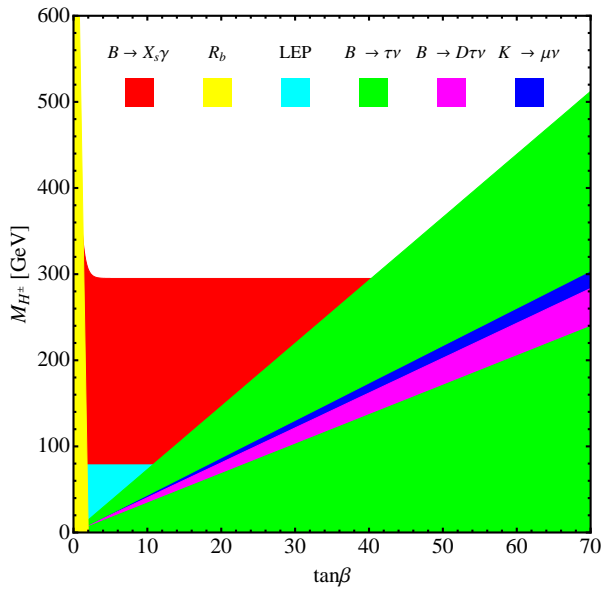} }
\caption{Bounds in the $m_{H^\pm}$--$\tan{\beta}$ plane
from various $B$-physics constraints.
This figure was extracted from the article by Haisch~\cite{Haisch:2008ar}.}
\label{6fig:bbounds}
\end{figure}

The results are shown in Fig.~\ref{6fig:bbounds}.
There one clearly sees the various contributions.
Note that the direct bound from LEP
is substantially weaker than the indirect bounds.

It is still important to look at bounds from direct production of $H^\pm$,
because it is always possible that New Physics weakens the bounds
from $b \to s \gamma$.
In fact,
that is
precisely what occurs in supersymmetric models.
It can be seen that a weakening of this bound would allow,
for smaller values of $\tan{\beta}$,
charged-Higgs masses somewhat below $100\, {\rm GeV}$.

To study direct production bounds,
and to explore prospects for the LHC,
the production rates and branching ratios of the charged Higgs are needed.
The branching ratios have been studied quite thoroughly---Ref.~\cite{Gunion:1989we}
contains an extensive review of the literature.
A recent analysis~\cite{Logan:2010ag} used
the FORTRAN code HDECAY~\cite{Djouadi:1997yw}.
The program includes final-state mass effects,
full one-loop QCD corrections and running masses,
and off-shell decays to $t \bar b$ below threshold.
In principle,
decays to $S^0 W^\pm$ are also possible,
where $S^0$ is a neutral scalar.
However,
the masses of these scalars are unknown and,
moreover,
the branching ratios in most of the parameter space are much smaller than 1\%;
thus,
they are neglected in much of this discussion.
We shall,
however,
discuss these possibilities at the end,
since there could be a substantial rate
in the type~II model---but not in the MSSM---for part of the parameter space,
as emphasised by Kanemura \etall~\cite{Kanemura:2009mk}.
Note that there is no $Z^0 W^\pm H^\mp$ vertex in the 2HDM.

\begin{figure}[h]
\begin{center}
\includegraphics[angle=-90,totalheight=5.5cm]{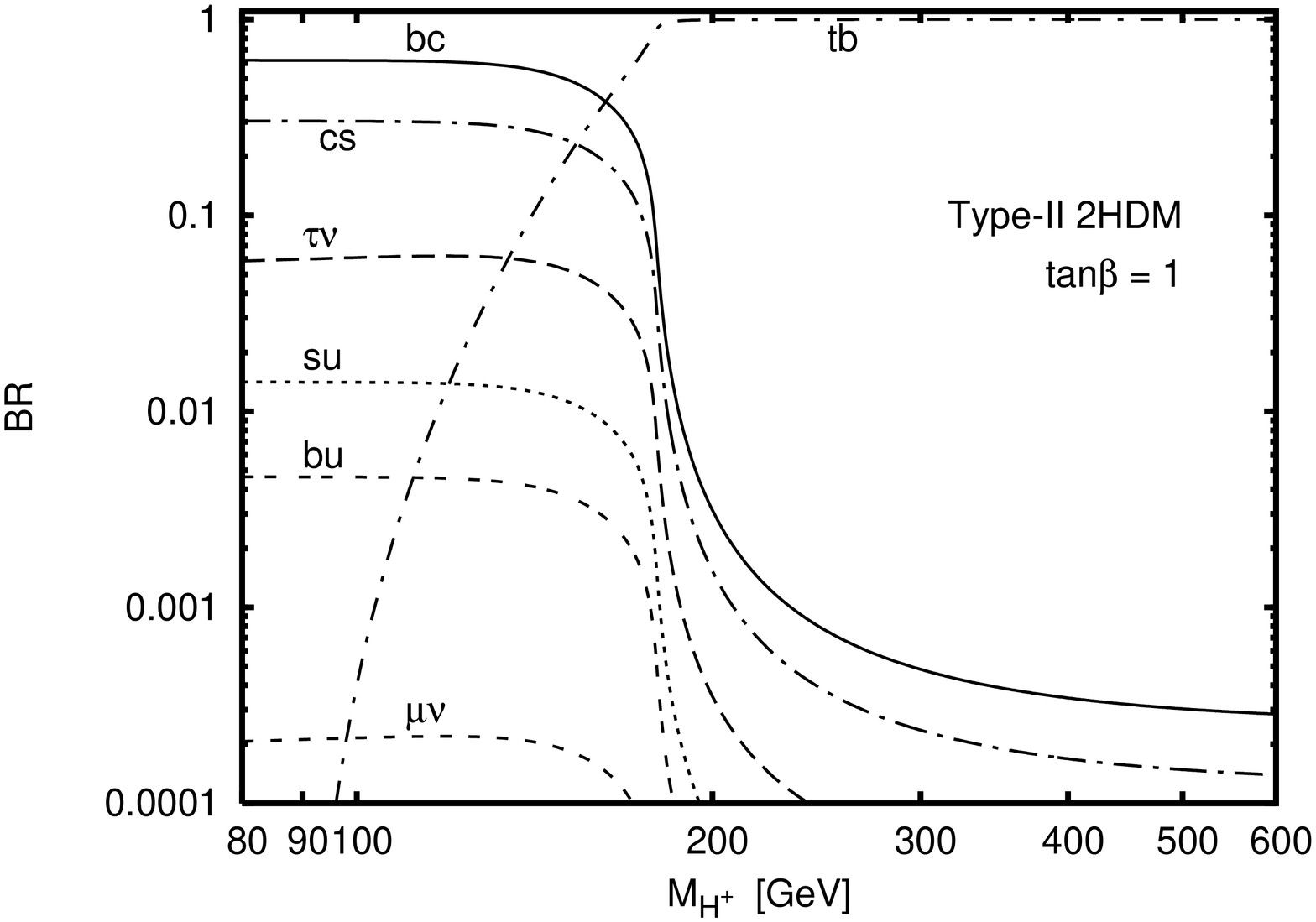}
\includegraphics[angle=-90,totalheight=5.5cm]{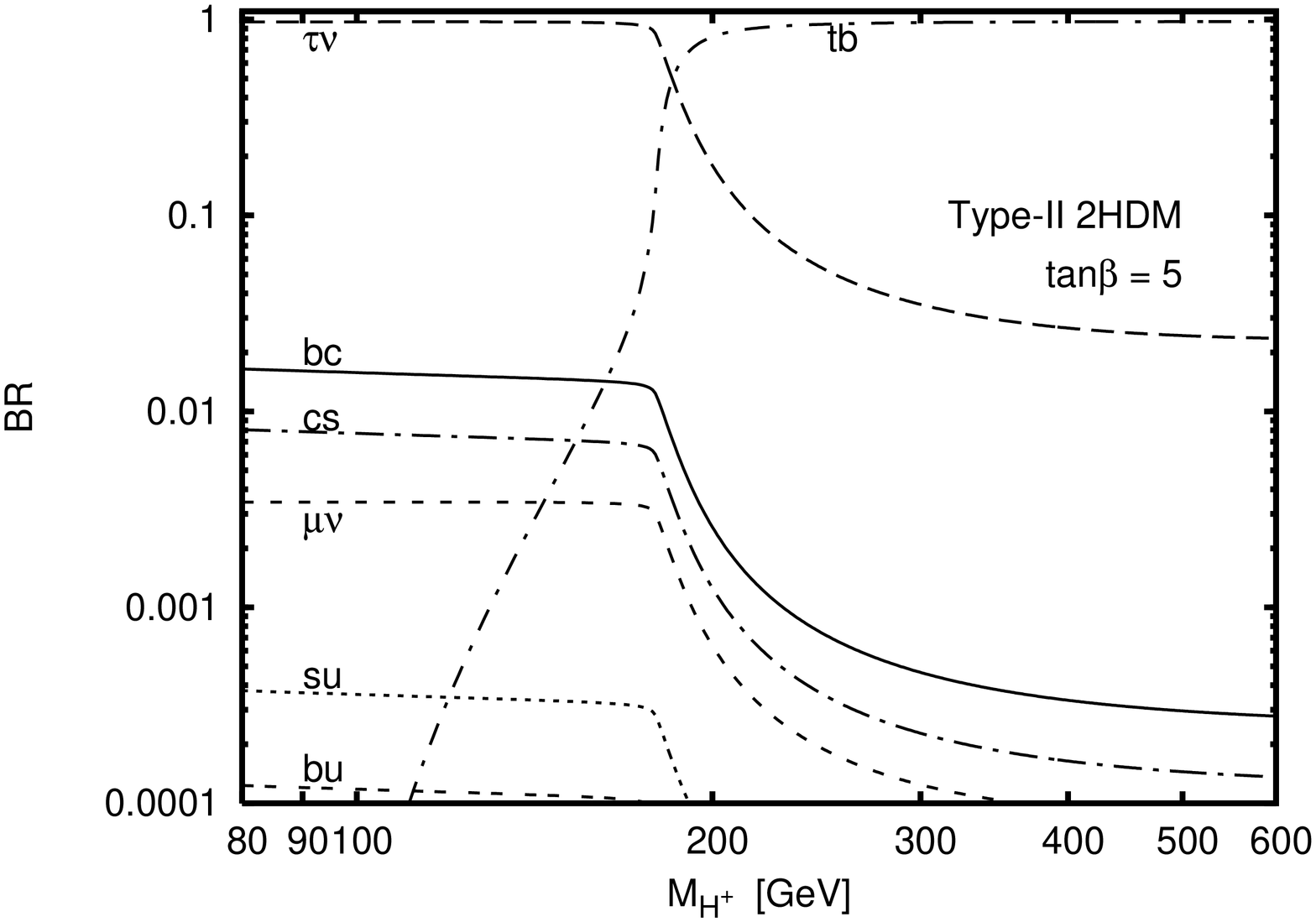}
\end{center}
\caption{The branching ratios of the charged Higgs in the type-II 2HDM.
Decays into $S^0 W^\pm$ ($S^0$ is a neutral Higgs boson) are not included.
This figure is from Ref.~\cite{Logan:2010ag}.}
\label{6fig:bratios}
\end{figure}

The results are shown in Fig.~\ref{6fig:bratios}.
We see that,
above the kinematic limit,
the decays into $t \bar b$ are completely dominant,
but,
for lower masses $m_{H^\pm}$,
the decays into $\tau^+ \nu_\tau$ dominate.
The decay into $c \bar s$ can also be significant
for small values of $\tan{\beta}$.
Now,
all the current direct searches
are sensitive only to charged-Higgs masses
below the kinematic threshold for decays into $t \bar b$.
Therefore,
the main decay modes in the present direct searches
are into $\tau^+ \nu_\tau$ and $c \bar s$.
The LEP combined limit on $m_{H^\pm}$ is $78.6\, {\rm GeV}$
and has been calculated~\cite{:2001xy}
assuming the existence of only those two decay modes
(in the flipped model the decay into $c \bar b$ is more significant
than the decay into $\tau^+ \nu_\tau$~\cite{Logan:2010ag}).

Searches have also been performed at the Tevatron.
The CDF Collaboration~\cite{Aaltonen:2009ke} has assumed that
${\rm BR} \left( H^+ \to  c \bar s \right) = 1$
and has therefrom found bounds on the branching ratio of $t \to H^+ b$.
Logan and MacLennan~\cite{Logan:2010ag} noted that
the CDF analysis also applies to decays into $c \bar b$
and therefore also applies to the flipped model.
However,
in the type~II model the CDF assumption
is valid only for small (less than 1) values of $\tan{\beta}$
and only for charged-Higgs masses near the LEP bound.
The D0 Collaboration~\cite{:2009zh},
on the other hand,
has considered both scenarios
in which ${\rm BR} \left( H^+ \to  c \bar s \right) = 1$
and ${\rm BR} \left( H^+ \to \tau^+ \nu_\tau \right) = 1$;
the latter scenario is certainly valid at large $\tan{\beta}$
in the type~II model.
D0~\cite{arXiv:1107.1268} and CDF~\cite{arXiv:1110.5349} have also
looked for associated production of a charged Higgs with a W.

\begin{figure}[h]
\centerline{\epsfysize=12cm \epsfbox{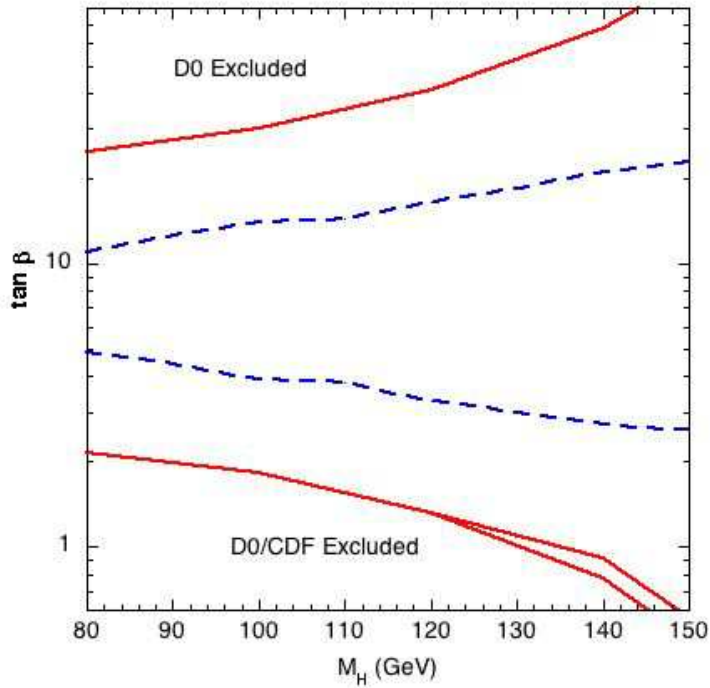} }
\caption{Current Tevatron bounds on the mass $M_H$ of the charged Higgs
as a function of $\tan{\beta}$.
The small sliver near the lower-right corner is excluded by CDF but not by D0.
The dashed lines are the bounds which are projected
for the ATLAS experiment with $30\, {\rm fb}^{-1}$~of data;
the regions that would then be excluded are those outside the dashed lines.}
\label{6fig:lightbounds}
\end{figure}

These papers give upper limits on the branching ratio
of the top quark into a charged Higgs and a bottom quark.
This can be converted into a bound in the $m_{H^+}$--$\tan{\beta}$ plane.
We use the well-known expressions for the branching ratio
(see the appendix of Ref.~\cite{Logan:2010ag} for a simple expression)
and find the results in Fig.~\ref{6fig:lightbounds}.
As expected,
the present bounds only exclude regions of parameter space
for either large or small $\tan{\beta}$;
a charged-Higgs mass of about $100\, {\rm GeV}$ is thus still allowed.
Recall that one is assuming that some New Physics
cancels the contribution to $b \to s \gamma$,
and it is possible that this New Physics
could also lead to alternative production and decay mechanisms.

In the future much of this parameter space will be probed at LHC.
The ATLAS~\cite{Aad:2009wy} and CMS~\cite{Baarmand:2006dm} Collaborations
have studied top-quark pair production
in which one of the tops decays into $H^+  b$.
They have assumed that the charged Higgs $H^+$
subsequently decays into $\tau^+ \nu$,
as is the case in the type~II model except for very small $\tan{\beta}$
of ${\rm O} (1)$.
Techniques to improve these studies have been discussed in
Ref.~\cite{arXiv:1103.1827}.
Also shown in fig.~\ref{6fig:lightbounds} is the expected reach of LHC,
as found in the ATLAS analysis for a collision energy of $14\, {\rm TeV}$
and with $30\, {\rm fb}^{-1}$ of data;
ATLAS gave the expected reach in the $t \to H^+ b$ branching ratio
and we have converted this into a bound in the $m_{H^\pm}$--$\tan{\beta}$ plane.
The CMS analysis was explicitly model dependent by using the MSSM,
and in that particular case it could partially close
the allowed window at intermediate $\tan{\beta}$.
Again,
for intermediate values of $\tan{\beta}$ there is in general
a large window that can be closed neither at the Tevatron nor at the LHC.
The ILC,
on the other hand,
will readily be able to produce charged-Higgs pairs
and thus to cover the entire parameter space for Higgs masses
lower the mass of the top.
There has also been a discussion of the possibility of detecting the
$H^\pm\rightarrow \mu^+\nu_\mu$ after 300 fb${}^{-1}$ in
Ref.~\cite{arXiv:1109.5356}.

For larger Higgs masses the decay of $H^+$ is overwhelmingly into $t \bar b$.
This will be subject to very large backgrounds~\footnote{If the charged Higgs
is produced in associated with a top quark, it has been argued~\cite{arXiv:1111.4530}
that one  might be able to detect the decay into $\bar{t}b$.} and is thus quite challenging.
For moderately large $\tan{\beta}$
the branching ratio of $H^+ \to \tau^+ \nu_\tau$ is approximately $10\%$
and this might be sufficient to pick out the signal.
In fact,
most search strategies must study the latter decay mode.

The primary production mechanisms at the LHC
are single charged-Higgs production and pair production.
The single charged-Higgs production processes
include gluon fusion~\cite{DiazCruz:1992gg},
quark fusion~\cite{Gunion:1986pe,Moretti:1996ra,He:1998ie,Dittmaier:2007uw},
quark-gluon fusion~\cite{Berger:2003sm,Plehn:2002vy},
associated $W^\pm H^\mp$ production
\cite{BarrientosBendezu:1998gd,Moretti:1998xq,Brein:2000cv,Eriksson:2006yt,Hashemi:2010ce,arXiv:1104.0889,arXiv:1112.0086}
and $AH^\pm$, $H H^\pm$, $h H^\pm$
production~\cite{Kanemura:2001hz,Cao:2003tr,Belyaev:2006rf,arXiv:hep-ph/0306045,hep-ph/0412365}.
Most of these works have focused on the MSSM,
although enhancement in the general type~II model
is referred to in Ref.~\cite{Asakawa:2005nx}.
Pair production has also been calculated:
the Drell--Yan and $b \bar{b} \to H^+ H^-$ processes
most recently in~\cite{Alves:2005kr},
and weak-boson fusion in~\cite{Moretti:2001pp}.
Another production mechanism is associated production of a charged
Higgs with a top quark, which can lead to interesting consequences
for the top quark polarization and the angular correlations of its
decay products~\cite{arXiv:1012.0527}, and also in the left-right
asymmetry in the  polarized top quark production cross
section~\cite{arXiv:1109.2420}.
Although strictly a type II model, associated $W^\pm H^\mp$ production
in a model with spontaneous CP violation (i.e. a relative phase between the
vevs) was studied in Ref.~\cite{arXiv:1011.1409}. They found no observable
effects from associated production, but did note that the model allows the
charged Higgs mass to be substantially smaller than the conventional type
II model.


The ATLAS~\cite{Aad:2009wy} and CMS~\cite{Bayatian:2006zz} Collaborations
have used the production cross sections to explore the reach of the LHC.
Their conclusions are that detection of the charged Higgs boson
through its decay into $t \bar b$ will be swamped by large backgrounds,
hence the decay mode $\tau^+ \nu_\tau$ is the most promising one.
Both analyses use various versions of the MSSM
but one would not expect the results in the type-II 2HDM
to differ substantially.
In each case,
the reach on the branching ratio is give,
and we have converted this into a reach
in the $m_{H^\pm}$--$\tan{\beta}$ plane.
The ATLAS Collaboration finds a discovery (\ie, $5 \sigma$) reach
which ranges (for an integrated luminosity of $30\ {\rm fb}^{-1}$)
from $\tan{\beta} = 28$ for $m_{H^\pm} = 200\, {\rm GeV}$
to $\tan{\beta} = 65$ for $m_{H^\pm} = 350\, {\rm GeV}$
(these values of $\tan{\beta}$ are lower bounds).
The CMS Collaboration,
in a slightly different MSSM scenario,
finds a discovery reach,
for the same integrated luminosity,
which varies from $\tan\beta = 28$ for $m_{H^\pm} = 200\, {\rm GeV}$
to $\tan\beta = 65$ for $m_{H^\pm} = 450\, {\rm GeV}$.
The exclusion bounds correspond to much lower values of $\tan{\beta}$,
of course---ATLAS,
for example,
can exclude $\tan{\beta} > 12\ (50)$
for charged-Higgs masses of $200\ (600)\, {\rm GeV}$.
The main production mode is $g b \to t H^-$.
For associated $W^\pm H^\mp$ production,
a signal can be found~\cite{Eriksson:2006yt} for large $\tan{\beta}$
in the 150--300$\, {\rm GeV}$ mass region.

Since most of the above analyses were done in the context of the MSSM,
it is important to focus on the differences in the type~II 2HDM.
As noted in Ref.~\cite{Eriksson:2006yt},
if the mass of a neutral scalar is larger than $m_{H^\pm} + m_W$,
then one can resonantly produce this neutral scalar in the $s$-channel,
leading to a huge enhancement in the cross section for associated production.
It is difficult to make a plot of the expected reach,
though,
since the rate is very sensitive to the neutral-scalar couplings and mass,
but one should keep in mind the possibility that
the rate for associated $H^\pm W^\mp$ production
might be considerably larger than expected.

A detailed analysis of the way in which
the phenomenology of the type~II 2HDM
may differ from the one of the MSSM
was recently carried out by Kanemura \etall~\cite{Kanemura:2009mk}.
Their main point is that
the Higgs masses are very tightly constrained in the MSSM,
therefore certain decays that would otherwise be allowed cannot occur there.
For example,
for much of parameter space in the MSSM the charged Higgs,
the heavy neutral scalar $H$,
and the pseudoscalar $A$ are very close in mass,
therefore decays such as $H^\pm \to W^\pm A$ or $H^\pm \to W^\pm H$
are kinematically forbidden,
and $H^\pm \to W^\pm h$ is suppressed by phase space.
In the general type~II 2HDM,
on the contrary,
those decays are allowed.

To be specific,
Kanemura \etall~\cite{Kanemura:2009mk}
have considered $\tan{\beta} = 1$--3
and $m_{H^\pm} > 250\, {\rm GeV}$---as noted above,
the LHC will be insensitive to such masses in all the MSSM scenarios.
They also consider the case $m_A = m_{H^\pm}$,
$m_H = 150\, {\rm GeV}$,
and $m_h = 50\, {\rm GeV}$ as an illustrative example.
For $\sin{\left( \beta - \alpha \right)} = 0.1$,
they find that,
for $\tan{\beta} = 3$ the dominant decay mode
(over $90\%$ branching ratio)
of the charged Higgs is into $W^\pm h$
and for $\tan{\beta} = 1$ the branching ratio of that mode
rises from $40\%$ to $80\%$
as the charged-Higgs mass goes from $250$ to $600\, {\rm GeV}$.
They obtained similar results for $\sin{\left( \beta - \alpha \right)} = 0.9$
(now with $m_h = 120\, {\rm GeV}$ in order to avoid LEP bounds),
except that the dominant decay mode then is $W^\pm H$.
Note that,
for the charged-Higgs mass region that they consider
the dominant branching ratio is into $t \bar b$ in the MSSM.

This has a huge effect on the phenomenology.
By the time the LHC has been running for a few years,
the neutral Higgs bosons will presumably have been discovered and,
once their masses are known,
the decay $H^\pm \to W^\pm (h,H)$
should be quite straightforward to detect.
Thus,
for a substantial part of the parameter space,
the charged Higgs will be easier to detect in the type~II 2HDM than in the MSSM.

Borzumati and Djouadi~\cite{Borzumati:1998xr} have studied the observation
of the decays $H^\pm \to W^\pm (h,H)$ at LEP and the Tevatron.
It should be noted that their rate can be subject
to large radiative corrections~\cite{Akeroyd:1998uw,Akeroyd:2000xa}.
Searches for the charged Higgs by using the decay mode
$H^\pm \to b \bar b W^\pm$
have been performed at the Tevatron~\cite{Carena:2000yx}
and will be studied at the LHC~\cite{Drees:1999sb, Assamagan:2002ne}.

The production and decay of the charged Higgs in the ``Complex two Higgs doublet
model" was studied in Ref.~\cite{arXiv:1011.1409}. In this model, one violates CP
invariance by making the $m_{12}$ term in Eq. 2 complex. They calculate CP violating
asymmetries for $H^+$ and $H^-$ production and decays, including one loop effects.
They modified the codes for FeynArts and FormCalc to incorporate this model.
It is found that CP violating asymmetries are substantially smaller than in the
MSSM, and do not exceed $3 \%$.

\subsubsection{The type~I model}

Since both supersymmetric models and Peccei--Quinn models require a type~II 2HDM,
there has been substantially less discussion of the type~I model.
Nonetheless,
many of the early works on 2HDMs discuss that model
and it is often considered in more general analyses.
The seminal paper of Barger, Hewett, and Phillips~\cite{Barger:1989fj}
analysed numerous constraints on the charged Higgs in the type~I 2HDM,
but this was before LEP and a relatively light top quark was assumed.
There have been numerous studies since then,
as will be seen in this section.

There are several key features of the type~I model~\cite{aker}.
The relative branching ratios of the charged Higgs
decaying into fermions are independent of $\tan{\beta}$.
Assuming that there are no decays into lighter scalars---such as
$H^+ \to A W^+,\, h W^+,\, H W^+$---the decay into $\tau^+ \nu_\tau$
will have a branching ratio of $70\%$ and the decay into $c \bar s$
will have branching ratio equal to $30\%$,
when $m_{H^\pm}$ is below the top threshold.
Above that threshold,
$H^+$ decays almost always into $t \bar b$.
Since the couplings to fermions are proportional to $\cot{\beta}$,
the charged Higgs becomes fermiophobic in the large $\tan{\beta}$ limit.
This might avoid constraints from flavour physics,
allowing for the possibility of a charged Higgs
in the $100\, {\rm GeV}$ mass range.

If kinematically possible,
one may consider the decays $H^+ \to S W^+$,
where $S$ may be either $h$, $H$, or $A$.
In this case,
the $W^+$ may even be virtual,
since in the large $\tan{\beta}$ (fermiophobic) limit
a three-body decay can still dominate decays
into two fermions~\cite{Akeroyd:1998dt}.
In fact,
it was pointed out in that paper that a three-body decay
can dominate even if $\tan{\beta}$ is not so large,
since the Yukawa couplings of the $H^+$ are small
if its mass is below the one of the top quark.

As in the type-II 2HDM,
we shall begin by considering the constraints from indirect processes.
Since all of these involve Yukawa couplings,
which vanish in the large $\tan{\beta}$ limit,
they will provide bounds only in the low-$\tan{\beta}$ region.
As noted earlier,
a recent comprehensive review by Jung, Pich, and Tuz\'on~\cite{Jung:2010ik}
has considered many processes within their general ``aligned 2HDM''.
One may use their results to explore the limits from indirect processes
in all four 2HDMs,
which are all limiting cases of the general approach of Pich \etall;
they give bounds in each of the four 2HDMs that we discuss explicitly
(the bounds for the type~II model agree with those in the last subsection).

In the type~I 2HDM the strongest bounds
come from the process $Z \to b \bar b$,
from $\epsilon_K$,
and from $\Delta m_{B_s}$.
The value of $R_b$,
the ratio of the width $Z \to b \bar b$
to the total hadronic width of the $Z$,
has been calculated
in 2HDMs~\cite{Ciuchini:1997xe,Haber:1999zh,Field:1997gz},
most recently in Ref.~\cite{Degrassi:2010ne}.
Comparing it with the experimental result
one obtains bounds with vary from $\tan{\beta} > 2$
to $\tan{\beta} > 1$ when the charged-Higgs mass varies
from $80$ to $400\, {\rm GeV}$.
The value of the mass difference between the $B_s$ and the $\bar B_s$
was calculated in~\cite{Urban:1997gw}
(some minor but not insignificant errors in the calculation
were pointed out in~\cite{WahabElKaffas:2007xd}),
and this also provides bounds.
Recent calculations by Buras \etall~\cite{Buras:2008nn,Buras:2010pza}
of $\epsilon_K$ also give constraints.

\begin{figure}[h]
\vskip 0.5cm
\centerline{\epsfysize=8cm \epsfbox{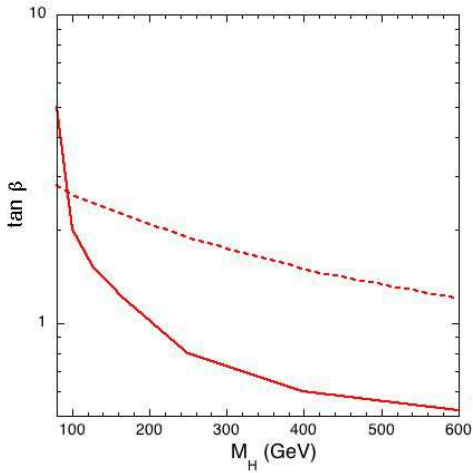} }
\caption{Lower bounds on $\tan{\beta}$ in the type~I 2HDM
as a function of the charged-Higgs mass $M_H$.
The solid line is the bound from Ref.~\cite{Jung:2010ik}
and comes from consideration of $Z \to b \bar b$,
$\epsilon_K$,
and $\Delta m_{B_s}$.
The dashed line is the bound in Refs.~\cite{Mahmoudi:2009zx}
and~\cite{Gupta:2009wn},
which arises from $B \to X_s \gamma$
and is very sensitive to assumptions and to input parameters.}
\label{6fig:type1indirect}
\end{figure}

Putting all of these together leads to the excluded region
in fig.~\ref{6fig:type1indirect}.
There,
the solid line corresponds to the bounds
of Jung, Pich, and Tuz\'on~\cite{Jung:2010ik}
for the processes in the previous paragraph.
As they note,
other bounds from $B \to D \tau \nu$
followed by
$\tau \to \ell \nu$ are weaker.
In Ref.~\cite{Jung:2010ik} the bound
from $\bar B \to X_s \gamma$ is also weaker,
but Ref.~\cite{Mahmoudi:2009zx}---which is primarily concerned
with models with tree-level FCNC and will be discussed
in the next section---obtains a stronger bound
from that process,
shown as the dotted line in fig.~\ref{6fig:type1indirect}.\footnote{A
separate analysis by Gupta and Wells~\cite{Gupta:2009wn}
found results similar to those of Ref.~\cite{Mahmoudi:2009zx}.}
The precise value for radiative $b$ decays
is very sensitive to input parameters and to theoretical errors,
so this is not necessarily a disagreement
between~\cite{Mahmoudi:2009zx} and~\cite{Jung:2010ik}
(one should also note that $\tan{\beta}$ is not directly measurable).
In any event,
one can see that there are no substantive bounds
on $m_{H^\pm}$ for moderate and large $\tan{\beta}$,
which is in sharp contrast to what happens in the type~II 2HDM.

We next proceed to the direct bounds.
For the type~II 2HDM it was necessary to discuss the branching ratios
as a function of $\tan{\beta}$.
That is not needed for the type~I 2HDM,
since the relative fermionic branching ratios
are in this case independent of $\tan{\beta}$ and are,
approximately,
$65\%$ into $\tau^+ \nu_\tau$ and $35\%$ into $c \bar s$
(below the top threshold,
which is the relevant case for the current direct bounds).
The combined LEP bound~\cite{:2001xy} is $78.7\, {\rm GeV}$.
However,
this assumes that there are no non-fermionic decays
such as $H^\pm \to A W^\pm$
(where the $W^\pm$ is virtual).
The DELPHI Collaboration~\cite{Abdallah:2003wd}
also searched for a charged Higgs that decays into either $\tau^+ \nu_\tau$,
$c \bar s$,
or $A W^*$;
it found a bound of $76.7\, {\rm GeV}$
by assuming that the $A$ was heavier than $12\, {\rm GeV}$,
similar results were also reported by the OPAL
Collaboration~\cite{arXiv:0812.0267}.

As for the type~II model,
the Tevatron also has bounds from Drell--Yan pair production of a charged Higgs.
The CDF bounds in Ref.~\cite{Aaltonen:2009ke}
have assumed that ${\rm BR} \left( H^+ \to  c \bar s \right) = 1$,
which is never the case in the type~I 2HDM,
so those bounds do not apply.
The D0 bound~\cite{:2009zh}
has considered the case in which
the sum of ${\rm BR} \left( H^+ \to  c \bar s \right)$
and \\
${\rm BR} \left( H^+ \to  \tau^+ \nu_\tau \right)$ is $1$,
which applies here unless there are light scalars.
The result is the lower line in fig.~\ref{6fig:lightbounds}.
This bound,
however,
is slightly weaker than the one derived by Jung, Pich, and Tuz\'on
from indirect processes.
Thus,
no stronger bounds,
as of yet,
have been obtained by the Tevatron on the type~I 2HDM.

The ATLAS study of a light charged Higgs
possibly
being produced in top-quark decays~\cite{Aad:2009wy}
was explicitly focussed on the type~II 2HDM.
However,
the primary difference between the type~I and type~II models
is in the coupling to $t \bar b$.
In the type~II model that coupling
is large at large $\tan{\beta}$,
whereas it is always small for large $\tan{\beta}$ in the type~I model;
both models give similar results for small $\tan{\beta}$.
As a result,
the discovery bound at the LHC
would correspond to the lower dashed line in Fig.~\ref{6fig:lightbounds}.
For large $\tan{\beta}$,
the top-quark branching ratio into a charged Higgs is too small to detect
in the type~I 2HDM.

A study by Aoki et al.~\cite{arXiv:1104.3178} of the light charged Higgs
at the LHC explicitly focused on the type I and Lepton-Specific models
has very recently appeared.   They analyse production rates from top pair
production (in which one top decays into $H^+b$), single top production,
and direct production through $c s \rightarrow H^+ +$ jet.    They plot
the reach of the LHC for $10$ and $30$ inverse femtobarns at $\sqrt{s}=14$
TeV and find that over the range of charged Higgs masses up to $150$ GeV,
upper bounds on $\tan\beta$ between 6 and 10 can be obtained (recall that
the very large $\tan\beta$ limit is fermiophobic in the type I model).
Charged Higgs pair production is also considered, and it is shown that
constraints are more parameter-dependent, but the process might also be
detected at the LHC.   An analysis at $7$ TeV is currently under
investigation by the same authors.

For larger charged-Higgs masses
the decay into $t \bar b$ will be swamped by background.
The $\tau^+ \nu_\tau$ decay,
which has a branching ratio as high as $10\%$ in the type~II 2HDM,
is much less important in the type~I 2HDM
and typically has a branching ratio of order $m_\tau^2 / m_t^2 \sim 10^{-4}$.
This makes detection at the LHC impossible if the primary decay is fermionic.

However,
there are also neutral scalars $S$ (which may be $h$,
$H$,
or $A$ in the 2HDM) and one may consider $H^\pm \to S W^\pm$.
In the large $\tan{\beta}$ limit the charged Higgs is fermiophobic
and $H^\pm \to S W^\pm$ would be the leading decay mode
if kinematically accessible
(and it might still be the leading decay mode
even if the $W^\pm$ is virtual \cite{Borzumati:1998xr}).
We thus see that in the large $\tan{\beta}$ limit
the most promising decay modes are---unlike
what occurs in the type~II model---into
a neutral scalar and a $W$.
As noted in the last subsection,
searches of a charged Higgs through the decay mode
 $H^\pm \to b \bar b W^\pm$
have been studied at the Tevatron~\cite{Carena:2000yx}
and LHC~\cite{Drees:1999sb, Assamagan:2002ne}.
A plot of the results is premature since there are several additional parameters
(such as $\alpha$ and the scalar mass)
but the decay $H^\pm \to b \bar b W^\pm$ offers the best hope.
An updated analysis of this mode would be welcome.

\subsubsection{The lepton-specific model}

In the lepton-specific (LS) 2HDM
the same Higgs doublet couples to both the up-type and the down-type
(right-handed) quarks,
just as in the type~I 2HDM,
therefore the charged Higgs is quark-phobic for large $\tan{\beta}$.
Indirect bounds arising from hadronic decays
will thus be identical to those in the type~I model.
But,
in contrast to what happens in the type~I 2HDM,
at large $\tan{\beta}$ the charged Higgs becomes strongly leptophilic
in the LS~2HDM,
hence the $\tau^+ \nu_\tau$ decay mode of the $H^+$ can be dominant.
In fact,
as we shall see,
that decay can be dominant {\em even above the $t \bar b$ threshold},
leading to quite dramatic experimental signatures.

Some early discussions of the charged Higgs in the LS 2HDM
can be found in Refs.~\cite{Barger:1989fj,Akeroyd:1994ga,Park:2006gk}.
It has also been discussed recently
in the context of dark-matter models \cite{Goh:2009wg}
and of neutrino mass models \cite{Aoki:2008av}.
The most recent comprehensive analyses
of the phenomenology of the charged Higgs in the LS~2HDM
are the articles by Su and Thomas~\cite{Su:2009fz},
by Aoki \etall~\cite{Aoki:2009ha},
and by Logan and MacLennan~\cite{Logan:2009uf};
we shall follow those analyses closely.
The branching ratio of the charged Higgs
into $\tau^+ \nu_\tau$ is given in Table~\ref{6tab:ls}
(branching ratios below $5\%$ are not shown).
\begin{table}[h]
\begin{center}
\begin{tabular}{|c|c|c|c|c|}
\hline
{Charged Higgs mass} & $\tan\beta = 1$ & $\tan\beta = 5$ &
$\tan\beta = 10$ & $\tan\beta = 20$
\\ \hline
$100$ & $0.70$ & $0.95$ & $0.99$ & $1.00$
\\ \hline
$200$ & $0.05$ & $0.20$ & $0.80$ & $0.97$
\\ \hline
$300$ & $0.00$ & $0.05$ & $0.40$ & $0.92$
\\ \hline
\end{tabular}
\caption{Branching ratio of the charged Higgs into $\tau \nu$
for various values of the charged Higgs mass (in GeV) and of $\tan{\beta}$.
For $m_{H^\pm} = 100\, {\rm GeV}$ and $\tan{\beta} = 1$
the remaining $30\%$ branching ratio is into $c \bar s$;
for all the other entries
the remaining branching ratio is almost entirely into $t \bar b$.}
\end{center}
\label{6tab:ls}
\end{table}
Decays into $W^\pm (h, H, A)$ are not included in the branching ratio computations;
by using MSSM values for the mixing angles and for the light-scalar masses,
Logan and MacLennan have shown that the branching ratios never exceed $10\%$,
rendering irrelevant for the LS 2HDM
some detection strategies involving decays into scalars
which have been discussed for the type~I and type~II 2HDMs.

We see that,
for $\tan{\beta} = 1$,
the branching ratios are identical to those in the type~I model.
But,
the $\tau^+ \nu_\tau$ decay rapidly becomes dominant
as $\tan{\beta}$ increases
and is significant---and sometimes dominant---even above
the $t \bar b$ threshold.
Another important feature is that the total width of the charged Higgs
is much lower than in other models,
remaining below $1\, {\rm GeV}$ over the entire mass range
for $\tan{\beta} < 40$.
Thus one would expect very monochromatic $\tau$s in the decay.

The indirect bounds from $B$ decays
are very similar in the LS and type~I 2HDMs,
since they are only relevant for $\tan{\beta}$ close to 1,
where the Yukawa couplings are identical in both models \cite{Jung:2010ik}.
The decay $B \to X_s \gamma$ gives precisely the same bound as shown in Fig. ~\ref{6fig:type1indirect}
for the type~I 2HDM.
The decays $B \to \ell^+ \ell^-$
and $b \to c \tau \nu$ do not give useful bounds \cite{Logan:2009uf}.
One can get an effect from $B^+ \to \tau^+ \nu_\tau$,
since the charged Higgs can mediate this decay,
but that effect also turns out to be negligible
for charged-Higgs masses which are not yet excluded by the direct searches.
Logan and MacLennan~\cite{Logan:2009uf}
have also considered deviations from flavour universality in $\tau$ decays,
finding an exclusion region for a fairly light charged Higgs
that begins at $\tan{\beta} = 65$ for $m_{H^\pm} = 100\, {\rm GeV}$,
and rises linearly to $\tan{\beta} = 200$ for $m_{H^\pm} = 250\, {\rm GeV}$.

The LEP direct-search limit of $78.6\, {\rm GeV}$ for the type~I 2HDM
still applies here.
OPAL~\cite{Abbiendi:2003ji}
has presented a stronger bound of $92.0\, {\rm GeV}$
under the assumption that the branching ratio
into $\tau^+ \nu_\tau$ is $100\%$.
This assumption is valid for large $\tan{\beta}$ and thus,
over the entire range of $\tan{\beta}$,
the lower bound will vary from $78.6$ to $92.0\, {\rm GeV}$.
At the Tevatron,
the D0 bound discussed for the type~I 2HDM also applies in the LS 2HDM,
but (as in the former case) it is not stronger
than the bounds already imposed by indirect searches.
Similarly,
the ATLAS bound discussed in the previous section,
for a charged Higgs to be produced in top decays,
applies indifferently in the type~I and LS 2HDMs.

There is an additional possibility at the LHC.
In all other 2HDMs,
detection of the charged Higgs above the $t \bar b$ threshold
is very difficult due to the huge backgrounds;
the analyses have had to rely on the small branching fraction
of the $H^+$ into $\tau^+ \nu_\tau$.
In the LS model,
however,
that branching fraction is very large and will
(provided $\tan{\beta}$ is not too small)
dominate.
This sounds promising but,
unfortunately,
the production rate of $H^+$ through $g g \to \bar t b H^+$
scales like $\cot^2{\beta}$ and that will suppress the result.
Still,
a sufficient enhancement at intermediate $\tan{\beta}$ may occur.
As noted by Aoki \etall~\cite{Aoki:2009ha},
one could readily distinguish the type~II and LS 2HDMs
through the decay rates into leptons as opposed to quark,
if the charged Higgs could be detected.

One can also look for pair production of $H^+ H^-$
proceeding to $\tau^+ \nu_\tau \tau^- \bar \nu_\tau$.
A similar phenomenon was studied
by Davidson and Logan~\cite{Davidson:2010sf,Davidson:2009ha}
in the neutrino-specific 2HDM,
to be discussed shortly.
They looked at charged-Higgs pair production at the LHC but,
in their case,
the decays of the charged Higgs studied were into $\mu \nu$ and $e \nu$.
It is clear that $\tau \nu$ will be much more difficult
and the missing energy in the $\tau$ decay
will render irrelevant some of the cuts that Davidson and Logan used.
The possibility of pair production of charged Higgs above the top threshold
with the charged Higgs decaying entirely into $\tau\nu$ had not yet,
to our knowledge,
been studied.

One could look at associated production
of the charged Higgs with a neutral Higgs,
proceeding to three $\tau$s and missing energy.
The only detailed analysis of this possibility
that we are aware of is the one of Aoki \etall~\cite{Aoki:2009ha};
at the LHC with $300\, {\rm fb}^{-1}$ of accumulated luminosity,
they have found---for scalar masses of $150\, {\rm GeV}$---that
the signal is comparable to the background---which
originates primarily from $W^\pm Z$---and they have discussed possible cuts.
They note that $\tau$ misidentification
(hadrons being identified as $\tau$s)
may be a serious problem;
realistic simulations are necessary.
They also study $\mu \mu \tau \nu$ events;
the rate is much smaller,
since the branching ratio of neutral scalars into muons
is much less than the one into $\tau$s,
but the resolution of the muon pair is much better.
If the neutral scalar has already been identified,
this could provide a very useful tag for the charged Higgs.
Once again,
realistic simulations are needed.

Of course,
at an eventual ILC,
signatures will be very clean and backgrounds negligible,
and all of the various 2HDMs
will be readily distinguishable~\cite{Aoki:2009ha}.

\subsubsection{The flipped 2HDM}

In the flipped 2HDM the Yukawa couplings to the quarks
are the same as in the type~II 2HDM
but the Yukawa couplings to the leptons
are proportional to $\cot{\beta}$ instead of $\tan{\beta}$.
As a result,
the various bounds from hadronic processes
such as $B \to X_s \gamma$ are identical in the flipped and type~II 2HDMs.
Without a cancellation due to New Physics,
the lower bound on the charged-Higgs mass is close to $300\, {\rm GeV}$.
Although the flipped model has been discussed
in several recent general studies of the four 2HDMs
\cite{Barger:2009me,Aoki:2009ha,Jung:2010ik,charged10},
the only paper explicitly dedicated to the phenomenology of the charged Higgs
in the flipped 2HDM that we are aware of
is the one by Logan and MacLennan~\cite{Logan:2010ag}.

\begin{figure}[ht]
\centerline{\epsfysize=7cm
\epsfbox{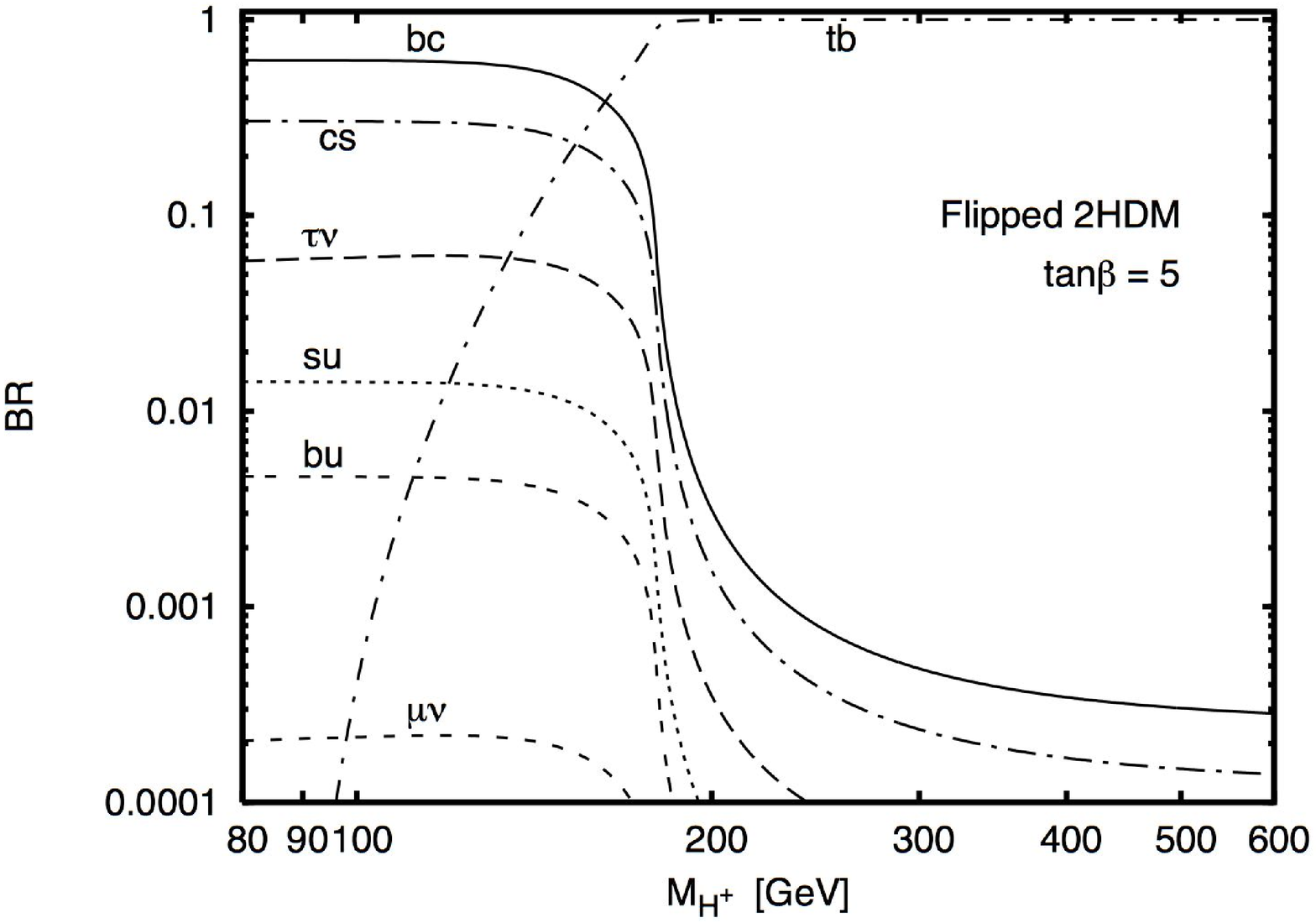}\epsfysize=7cm \epsfbox{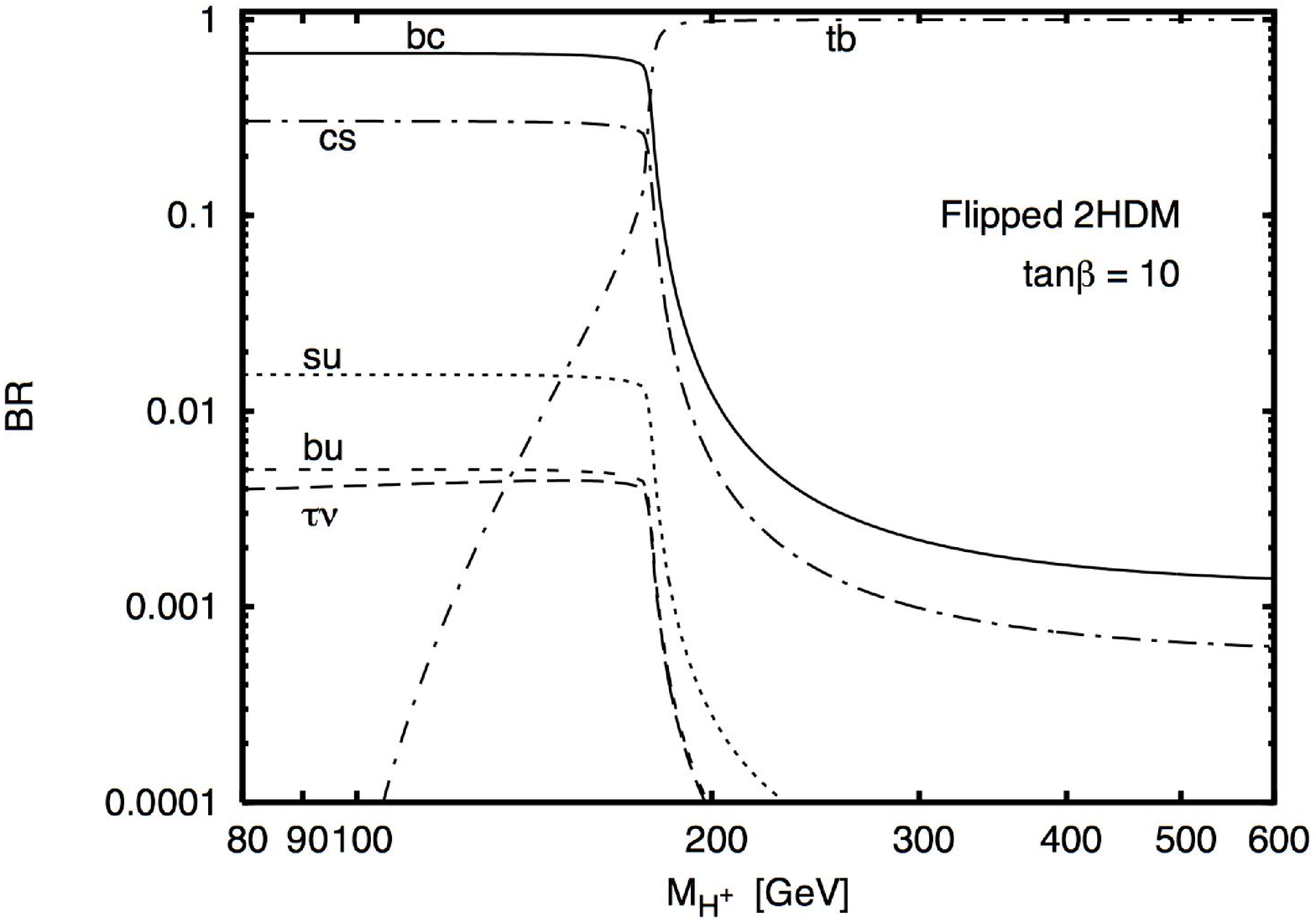} }
\caption{Branching ratios of the charged Higgs in the flipped 2HDM,
from Ref.~\cite{Logan:2010ag}.
Decays into $W^\pm$ and a neutral Higgs boson are not considered.
The ratio of the dominant channels
depends quadratically on the strange quark mass, and thus has a sizable
uncertainty.}
\label{6fig:flip}
\end{figure}

For moderately large $\tan{\beta}$,
branching ratios in the flipped and type~II 2HDMs are quite different.
For $\tan{\beta} = 1$ the branching ratios are identical,
see the left panel of fig.~\ref{6fig:bratios}.
But as $\tan{\beta}$ increases the branching ratios to leptons
become suppressed,
as shown in fig.~\ref{6fig:flip}.
One can see that,
below the $t \bar b$ threshold,
the dominant decay of $H^+$
is into $\tau^+ \nu_\tau$ for $\tan{\beta} \sim 1$
and into $c \bar b$ or $c \bar s$ for larger $\tan{\beta}$.

As stated above,
the indirect bounds on $m_{H^\pm}$ from $b \to s \gamma$,
from $\Delta m_B$,
and from $R_b$ are identical in the flipped and type~II 2HDMs.
The first process yields a lower bound of $295\, {\rm GeV}$ on $m_{H^\pm}$
and the other two processes exclude the region
$\tan{\beta} < 1$ for higher masses---see Jung \etall~\cite{Jung:2010ik}
for plots of these bounds.
As noted earlier,
though,
it is possible that New Physics cancels out
the effects of the charged Higgs in the loop---even a fairly mild cancellation
would weaken the bounds.

The bound on $m_{H^\pm}$ from LEP~\cite{:2001xy}
explicitly assumed that $H^+$ decays into
either $\tau^+ \nu_\tau$ or $c \bar s$.
While this is true in the type~II 2HDM,
it is not true for the flipped 2HDM,
since the decay into $c \bar b$ may dominate
for large or moderate $\tan{\beta}$.
The ALEPH Collaboration~\cite{Heister:2002ev}
produced bounds independent of the quark flavours
and one thus knows that $m_{H^\pm} > 79.3\, {\rm GeV}$.

At the Tevatron,
the bounds discussed earlier also assume that the charged Higgs decays
into either $c \bar s $ or $\tau^+ \nu_\tau$.
Logan and MacLennan~\cite{Logan:2010ag}
have analysed these experiments in the context of the flipped 2HDM.
Their conclusions are somewhat similar to fig.~\ref{6fig:lightbounds},
with a lower bound in the low-$\tan{\beta}$ region
and an upper bound in the high-$\tan{\beta}$ region,
\cf~\cite{Logan:2010ag} for the detailed plots.
Just as in the type~II 2HDM,
there is a large window at intermediate $\tan{\beta}$ where the Tevatron
is unable to produce bounds on $m_{H^\pm}$.

The prospects of detection of the $H^\pm$ at the LHC are less promising than
in the type II 2HDM.   Firstly, suppose that $m_{H^\pm}$ is above the top threshold.
Then one can study associated production of a charged Higgs with a top.  This is
similar to what happens in the type II model.   There, some hope was obtained
by looking at the relatively rare decay $H^+ \rightarrow \tau^+\nu_\tau$.  In
the flipped 2HDM, however, that decay mode is negligible for moderate or large
tan $\beta$.  Thus, there is no advantage to the flipped 2HDM over the type II 2HDM.
Secondly, suppose that $m_{H^\pm}$ is below the top threshold.
The ATLAS and CMS studies looking for top decays into a charged Higgs assumed
that the latter primarily decays into $\tau^+\nu_\tau$.  This is generally
not the case in the flipped 2HDM and it is unlikely that one can get
better bounds than the Tevatron.   However, an ATLAS
study \cite{ATLASstudy,901632} of $t\rightarrow H^+b, H^+\rightarrow c\bar{s}$
shows that the sensitivity (at 1 fb$^{-1}$ at 7 TeV) is better than the
Tevatron.  As shown in Refs.~\cite{Logan:2010ag} and \cite{hep-ph/9509203}, b-tagging
could be used to enhance the detection prospects of
$t\rightarrow H^+b, H^+\rightarrow c\bar{b}$ and distinguish between
$H^+\rightarrow c\bar{b}$ and $H^+\rightarrow c\bar{s}$.  Such a b-tag
would provide sensitivity in the region around the W mass which can't
be probed at the Tevatron due to backgrounds.

What about charged-Higgs pair production?
For $m_{H^\pm} < m_t$ the dominant decay
mode~\cite{Grossman:1994jb,Akeroyd:1994ga} is
(for moderate $\tan{\beta}$) into $c \bar b$,
leading to $c \bar b \bar c b$ final states
which will be impossible to detect due to large QCD backgrounds;
for a heavier charged Higgs
the $t \bar b \bar t b$ signal will also be impossible to pick out.
Thus,
Logan and MacLennan~\cite{Logan:2010ag} concluded that
the prospects for detection of the charged Higgs at LHC are not bright
in the flipped 2HDM
(at the ILC,
of course,
signatures would be easy to pick out).

\subsubsection{Other models}

An interesting model that has a potentially exciting
charged-Higgs phenomenology is the neutrino-specific model
\cite{Davidson:2009ha}.
In it,
the second Higgs doublet only couples to
right-handed
neutrinos and has an extremely small vacuum expectation value,
of order eV.
Then, the only decay modes of the charged Higgs
are into a charged lepton and a right-handed neutrino,
the coupling constant being proportional
to the neutrino mass times the appropriate lepton-mixing-matrix element.
With the normal neutrino mass hierarchy,
the decay of $H^+$ would be primarily into $\mu^+ \nu$ and $\tau^+ \nu$,
with a branching ratio of roughly $50\%$ into each;
for an inverted mass hierarchy,
the decays are primarily into $e^+ \nu$ (roughly $50\%$),
with the remainder into $\mu^+ \nu$ and $\tau^+ \nu$.
Thus,
the most dramatic signatures would be $H^+ \to e^+ \nu$
and $H^+ \to \mu^+ \nu$.

Davidson and Logan~\cite{Davidson:2009ha}
have considered the process
$p p \to H^+ H^- \to \ell^+ \nu {\ell^\prime}^- \bar \nu$.
Backgrounds to this process come from the production of $W^+ W^-$,
$ZZ$,
$Z\gamma$,
or $t \bar t$
(in which the $b$ quarks in the decay of the tops are missed).
For $m_ {H^\pm} = 100\, (300)\, {\rm GeV}$,
the production cross section is $300\, (5)\, {\rm fb}$.
They have looked at various cuts
in order to increase the signal-to-background ratio.
In the $m_ {H^\pm} = 100\, {\rm GeV}$  case,
the luminosity needed for a $5\sigma$ discovery
is $10$ to $70\, {\rm fb}^{-1}$ for normal neutrino mass hierarchy
and $8$ to $15\, {\rm fb}^{-1}$ for inverted hierarchy
(the range depends on the allowed range of neutrino masses and mixings).
In the $m_ {H^\pm} = 300\, {\rm GeV}$  case,
the luminosity in the normal-hierarchy case
is $55$ to $450\, {\rm fb}^{-1}$ and,
in the inverted-hierarchy case,
$25$ to $55\, {\rm fb}^{-1}$
(these calculations assumed a $14\, {\rm TeV}$ LHC).
Thus,
the signatures in this model are dramatic
and within reach of the upgraded LHC.

The primary interest in the Inert Doublet (ID) model is dark matter
and not the charged Higgs,
since a candidate for that matter naturally arises in the model;
most phenomenological studies of the ID model
have focussed on its neutral sector,
and especially on the possibility that the Standard-Model Higgs particle
decays into the dark-matter particle.
Since the charged Higgs in the ID model does not couple to fermions,
it will only decay into $H W^\pm$ or $A W^\pm$,
where the $W^\pm$ is either real or virtual,
depending on the masses.
Cao \etall~\cite{Cao:2007rm} have considered
the associated production of the charged Higgs and a pseudoscalar at LHC
and have found cross sections
(which are very sensitive to the masses of both the charged Higgs and the pseudoscalar)
of several hundred fb for  masses
below $200\, {\rm GeV}$.
They have also considered charged-Higgs pair production
and concluded that,
even for relatively low values of the charged-Higgs mass,
the backgrounds overwhelm the signals.
Later,
Miao \etall~\cite{Miao:2010rg,Dolle:2009ft}
found results for the production cross section similar to those of Cao \etal
and performed a detailed analysis of possible cuts.
They considered eight different benchmark points
and tailored the cuts to each of those points.
For two of the points a $5\sigma$ discovery of the charged Higgs
is possible with a luminosity of $300\, {\rm fb}^{-1}$;
in both of these points $m_{H^\pm}$ is only $110\, {\rm GeV}$---a
heavier charged Higgs probably cannot be discovered.
Huitu \etall~\cite{Huitu:2010uc} have shown that,
if one extends the ID model by including a scalar singlet,
then the charged Higgs could be long-lived,
leading to other detection possibilities.
Finally,
as noted by Ginzburg~\cite{charged10},
the ILC will easily be able to detect and study the charged Higgs
up to its kinematic limit.

As pointed out in Chapter 3, the Lee-Wick Standard Model is a very
unusual 2HDM, in which the second doublet states have wrong sign kinetic
terms and masses and negative norms.   The neutral sector of the theory is
similar to an inert model, but the charged sector is identical~\cite{Carone:2009nu}
to a type II 2HDM with $\tan\beta=1$ and with the sign differences for the
three point Higgs-Higgs-$\gamma,Z$ and the Yukawa couplings, as well as a factor
of $-1$ for each charged Higgs propagator.
Carone and Primulando~\cite{Carone:2009nu} analysed $B_q-\bar{B}_q$ mixing,
$B\rightarrow X_s\gamma$, $R_b$ from Z-decay,    Generally, the effects are of
opposite sign to the conventional 2HDM, but the constraints turn out to be similar.
They found that the lower bound on the charged Higgs mass from $B_d-\bar{B}_d$,
$B_s-\bar{B}_s$ and $b\rightarrow X_s\gamma$ are given by $303$, $354$ and $463$ GeV,
respectively, and that no further bounds can be found from $R_b$.

\subsection{Models with tree-level flavour-changing neutral currents}

\subsubsection{The type~III model}

The most general Yukawa couplings of two Higgs doublets
are given by \cite{Davidson:2005cw,Haber:2006ue}
\be
\bar Q_L \eta_1^U U_R \tilde \Phi_1
+ \bar Q_L \eta_1^D D_R \Phi_1
+ \bar Q_L \eta_2^U U_R \tilde \Phi_2
+ \bar Q_L \eta_2^D D_R \Phi_2
+ \hc,
\ee
where the $\eta_k$ ($k = 1, 2$) are $3 \times 3$ matrices;
we have not included leptonic Yukawa couplings for simplicity.
With our standard definitions and assuming real vevs,
the fermion mass matrices are
\be
M^F = \frac{v}{\sqrt{2}}
\left( \eta_1^F \cos{\beta} + \eta_2^F \sin{\beta} \right),
\ee
with $F= U, D$.
Defining \cite{Mahmoudi:2009zx}
\be
\kappa^F \equiv \eta_1^F \cos{\beta} + \eta_2^F \sin{\beta}
\ee
and the orthogonal combination
\be
\rho^F \equiv - \eta^F_1 \sin{\beta} + \eta^F_2 \cos{\beta},
\ee
one can move to the (``Higgs'') basis where only one doublet,
called $H_1$,
has a vev,
thus generating fermion mass matrices and coupling with $\kappa^F$.
The other doublet,
$H_2$,
has zero vev and couples with $\rho^F$.
The coupling of the charged Higgs boson then is
\be
H^+ \bar U \left( V \rho^D P_R - \rho^U V P_L \right) D + \hc,
\ee
where $V$ is the CKM matrix, and the $\rho^{U,D}$ matrices have been
rotated by the same unitary matrices that diagonalize $\kappa^{U,D}$.
Of course,
all the charged-Higgs couplings are flavour-changing,
but the neutral sector will have tree-level FCNC couplings
unless $\rho^U$ and $\rho^D$ are diagonal.
This occurs if either $\eta_1^U = \eta_1^D = 0$ (the type~I model)
or $\eta_1^D = \eta_2^U = 0$ (the type~II model);
these relations translate into $\rho^F = \kappa^F \cot{\beta}$
and $\rho^F = - \kappa^F \tan{\beta}$,
respectively,
and,
since the $\kappa^F$ are diagonal,
the $\rho^F$ then are diagonal too.

In the type~III 2HDM,
neither of these assumptions is made.
The Cheng--Sher \cite{Cheng:1987rs} {\it Ansatz},
discussed in Chapter 4,
is
\be
\rho^F_{ij} = \lambda^F_{ij}\, \frac{\sqrt{2 m_i m_j}}{v},
\ee
with the $\lambda^F$ of ${\rm O} (1)$.
In Chapter 3 we have shown how this {\it Ansatz}
is being challenged by current experiments.

Other versions of the 2HDM include the aligned 2HDM \cite{Pich:2009sp},
in which it is assumed that the Yukawa-coupling matrices
$\eta^F_{1,2}$ are proportional.
That model has no tree-level FCNC
and was discussed in the last section
(see \cite{Jung:2010ik} for a detailed analysis
of the charged-Higgs phenomenology of the aligned 2HDM).
In another 2HDM,
Mahmoudi and Stal~\cite{Mahmoudi:2009zx}
used the Cheng--Sher parametrization and noted that,
if one assumes that the $\lambda_{ij}$ are diagonal,
then both the aligned 2HDM and the various $Z_2$ models
arise as special cases;
we shall include that model in this section.
One can also find interesting phenomenological possibilities
by considering specific Yukawa textures~\cite{arXiv:1105.4951,arXiv:1106.5035}.

We first discuss direct searches.
The LEP bounds on charged-Higgs masses will still apply,
but the relative branching ratios to $\tau^+ \nu_\tau$ vs.\ $c \bar s$
are parameter-dependent.
Still,
a bound of around $75$ to $80\, {\rm GeV}$ is expected based on the energy scale.
Hadronic colliders can extend the reach.
In 1999,
He and Yuan~\cite{He:1998ie} discussed a new method of detecting charged Higgs
in models with tree-level FCNC.
The idea is that,
if a model has a large $\bar t c H$ coupling ($H$ is a neutral Higgs),
then,
from isospin symmetry,
there will be a large $\bar b c H^+$ coupling
and one can produce a charged Higgs through the $s$ channel,
since the $b$ and $c$ content of the proton is not negligible.
Although this process can exist in standard 2HDMs,
it is suppressed by the small $V_{cb}$ CKM-matrix element.

In the type~III 2HDM,
$( \rho^U V)_{cb} \approx \rho^U_{ct} V_{tb}
+ \rho^U_{cc} V_{cb} \approx \rho^U_{ct}$,
whereas $(V \rho^D)_{cb} \approx \rho^D_{sb}$.
The latter is,
in the Cheng--Sher {\it Ansatz},
much smaller,
therefore the dominant vertex involves $\bar c_R b_L$ and not---as in
the standard Type I or Type II 2HDM---$\bar c_L b_R$.
With the Cheng--Sher {\it Ansatz} this vertex will be quite large,
${\rm O} (\sqrt{m_cm_b/v^2}) \sim 1\%$.
He and Yuan~\cite{He:1998ie}
find that,
for a $14\, {\rm TeV}$ LHC with $100\, {\rm fb}^{-1}$ luminosity,
one would get,
for $m_{H^+} = 300\, (800)\, {\rm GeV}$,
over $130,000\, (380)$ single-top events from the charged-Higgs decay
and
(if kinematically accessible)
about $180,000\, (4,000)$ events $H^+ \to W^+ h \to W^+
(b \bar b, \tau^+ \tau^-$).
With appropriate cuts,
this can by seen over the Standard-Model rate (from $W^\ast$)
up to charged Higgs masses below $350\, {\rm GeV}$, whereas for heavier masses, the signature is not observable.

A study of this process $H^+\to t\bar{b}$
at the Tevatron,
where the rates are of course smaller,
was carried out by the D0 Collaboration~\cite{Abazov:2008rn},
which was however unable to set bounds
on $m_{H^\pm}$
unless the coupling of the charged Higgs to $q_i\bar{q_j}$
was substantially higher than expected in the type~III 2HDM.
Further analysis of the D0 results was carried out
by Cardenas and Rodr\'\i guez~\cite{arXiv:0810.3046}.
They looked at a charged Higgs above the $t \bar b$ threshold and found that,
if $\lambda_{tt} = \lambda_{tc} = 2.8\, (5.0)$,
then $m_{H^\pm}$ must be above $230\, (264)\, {\rm GeV}$.
Below the $t \bar b$ threshold,
they found from top decays that the lower bound on $m_{H^\pm}$
is $145\, (160)\, {\rm GeV}$ for $\lambda_{tt} = \lambda_{tc} = 2.8\, (5.0)$
(the value $2.8$ is the perturbative validity upper bound \cite{Martinez:2002tn}).
For $\lambda_{tt} = \lambda_{tc} = 1$,
 no substantive bound could be found
on $m_{H^\pm}$.
Single-top production at the LHC will provide a good probe of the model.

\begin{figure}
\begin{center}
\includegraphics[width=3.5in]{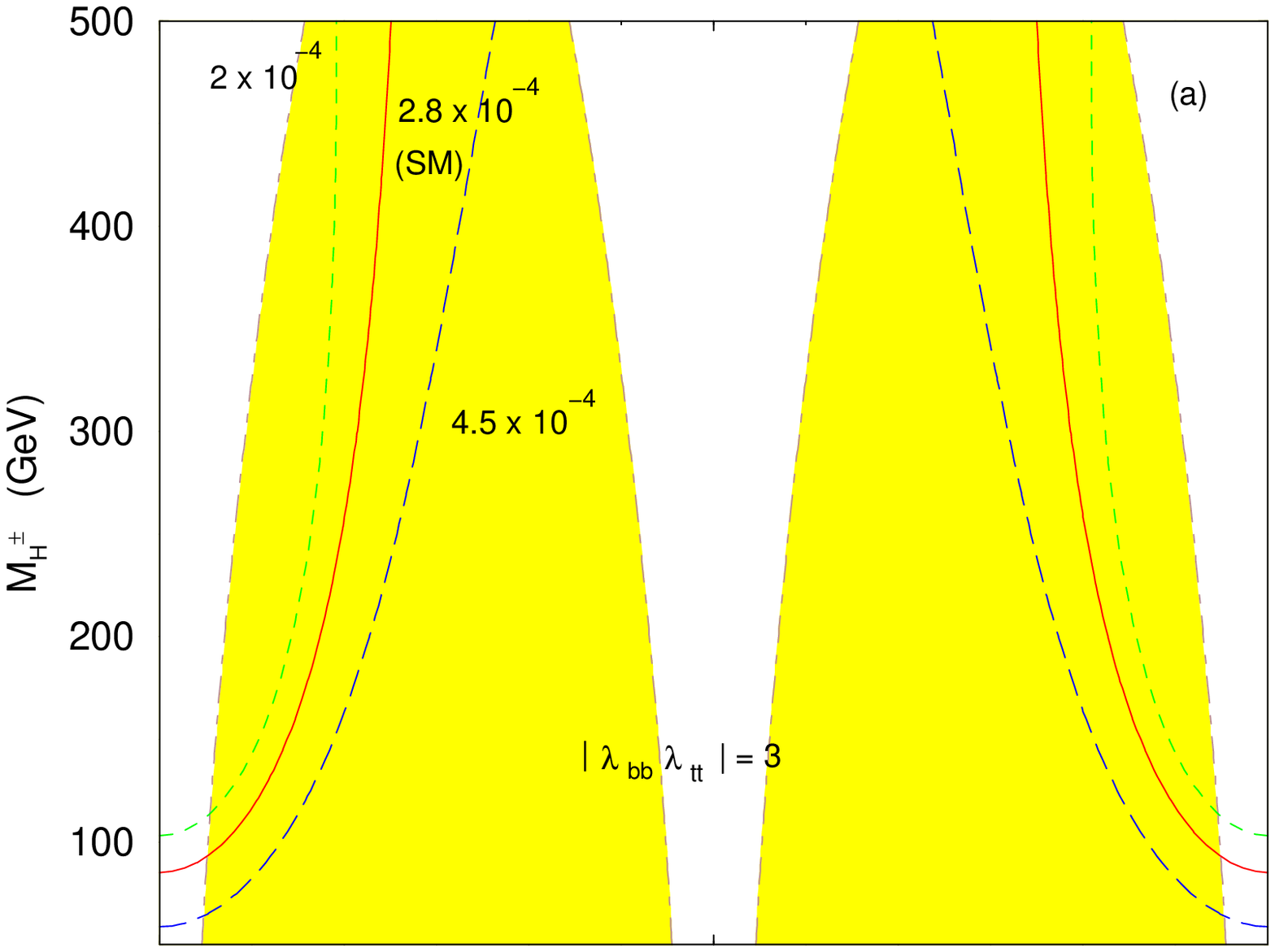}
\includegraphics[width=3.5in]{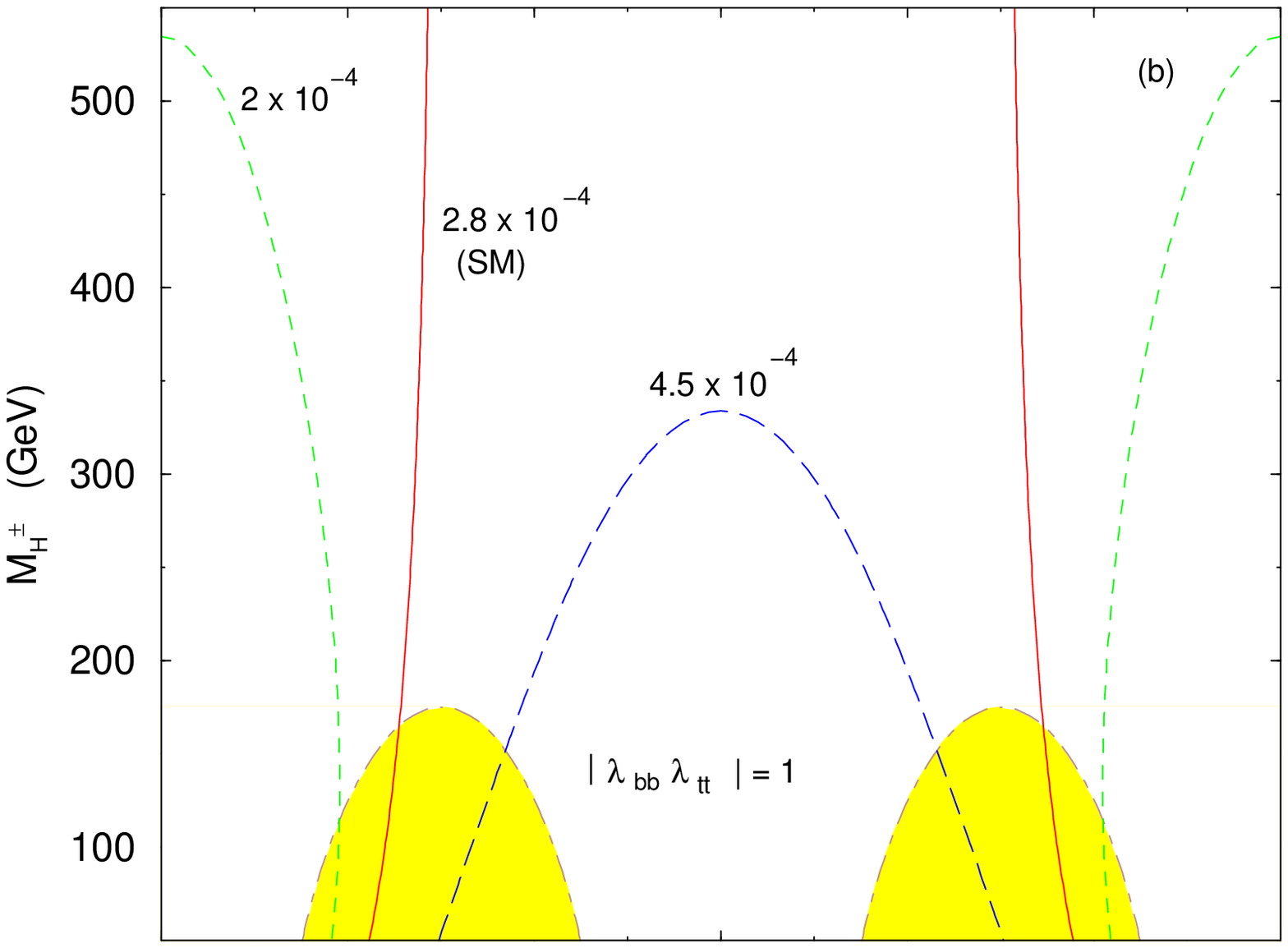}
\includegraphics[width=3.5in]{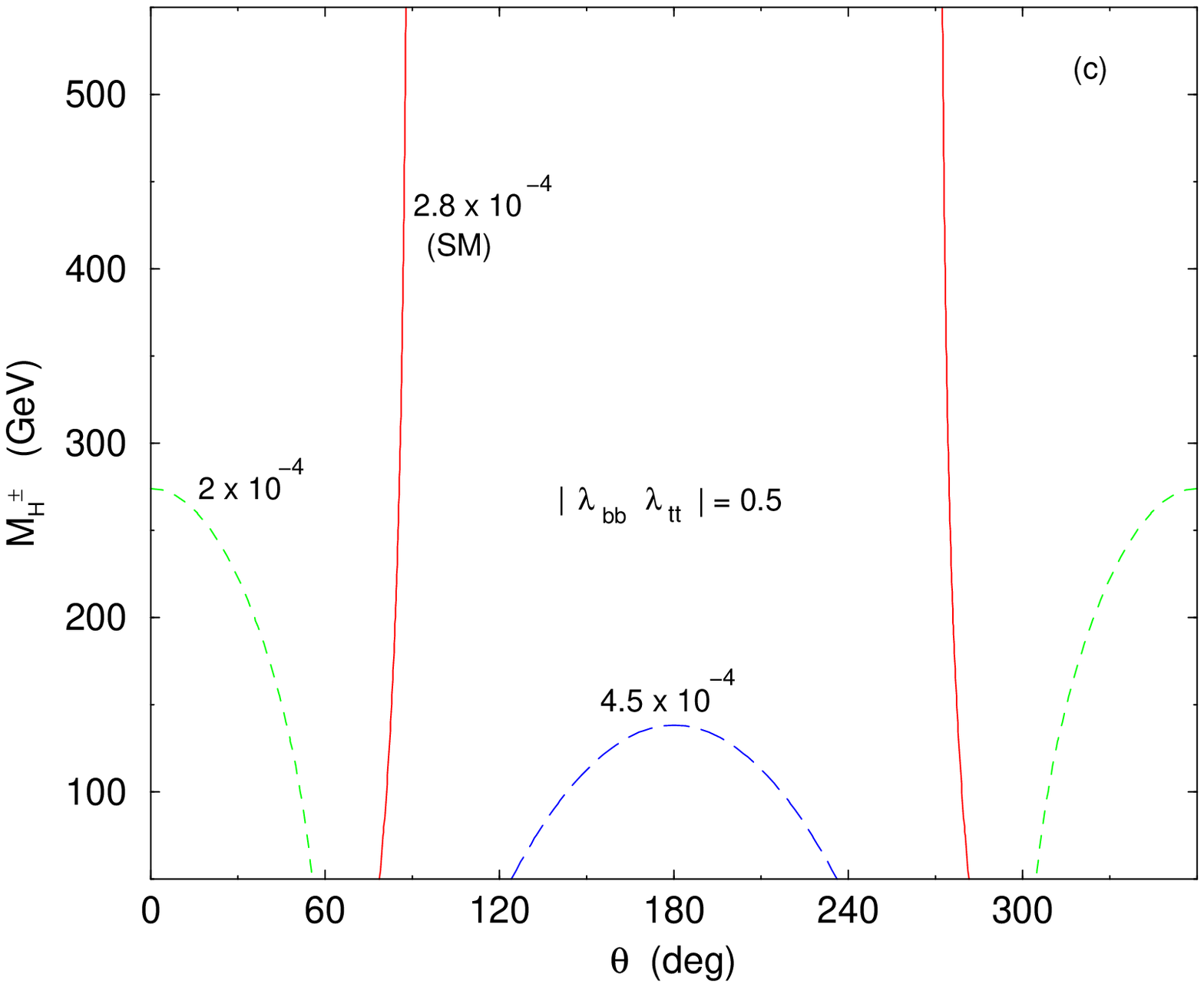}
\end{center}
\caption{Contours of the branching ratio of $b \to s \gamma$
as a function of the charged-Higgs mass
and of the phase $\theta$ between $\lambda_{tt}$ and $\lambda_{bb}$,
for three different values of $\left| \lambda_{tt} \lambda_{bb} \right|$.
The shaded areas are excluded
by the experimental bound on the neutron electric dipole moment.
This figure is from Ref.~\cite{BowserChao:1998yp}.}
\label{6fig:model3}
\end{figure}

As in the type~I and type~II models,
one can set bounds on the parameters of the type~III 2HDM
from indirect processes.
The first discussion of bounds on $m_{H^\pm}$ from $B \to X_s \gamma$
in the type~III 2HDM was Ref.~\cite{Atwood:1996vj},
but this was long before NLO corrections,
which are important,
had been calculated.
A comprehensive study is difficult because of
the large number of parameters in the type~III 2HDM.
A simplified analysis,
in which only $\lambda_{tt}$ and $\lambda_{bb}$ are nonzero,
was performed by Bowser-Chao \etall~\cite{BowserChao:1998yp}.
This has the advantage of a smaller parameter space
and of the fact that
only charged-Higgs loops are relevant---unlike the general case,
where neutral-Higgs loops are important too.
They allowed for the possibility that those two couplings are complex---the
relative phase determines whether the charged-Higgs loops
interfere destructively or constructively,
and can also give rise to a non-zero neutron electric dipole moment (EDM).
Their results are shown in fig.~\ref{6fig:model3}
for three different values of $\left| \lambda_{tt}\lambda_{bb} \right|$.
On the $x$ axis is the phase angle,
which is $180^\circ$ in the type~I and type~II models.
The type~II model corresponds to the middle figure at $\theta = 180^\circ$;
and one can see the bound of roughly $300\, {\rm GeV}$ of that model.
The shaded region is excluded
by the experimental upper bound on the neutron EDM.
Note that,
for a reduction by a factor of two
of $\left| \lambda_{tt} \lambda_{bb} \right|$,
the lower bound on $m_{H^\pm}$ drops
almost to the observed lower bound from LEP
(this work was done in 1999 and the experimental constraints
are much tighter now,
but NLO effects
have been
calculated and lower the curves
only a little,
resulting in similar results).
They also gave bounds from $B$--$\bar B$ mixing and $R_b$,
but the experimental results have changed a bit since then
and the results are rather sensitive to those changes.
The important aspect of the paper by Bowser-Chao \etal is that,
through slightly lower values of $\left| \lambda_{tt} \lambda_{bb} \right|$,
one can substantially lower the bound on $m_{H^\pm}$ in the type~III 2HDM.

The most comprehensive analysis of $B \to X_s\gamma$
and of $B$--$\bar B$ mixing in the type~III 2HDM
was the one by Xiao and Guo~\cite{Xiao:2003ya}
and also,
with a specific focus on $B \to (K^\ast, \rho) \gamma$,
the study of Xiao and Zhuang~\cite{Xiao:2003vq}.
There are numerous plots in these two papers.
From the $B$--$\bar B$ mass difference,
and assuming that only $\lambda_{bb}$ and $\lambda_{tt}$ are nonzero,
they~\cite{Xiao:2003ya,Xiao:2003vq}
find that $\lambda_{tt} < 1.7$ for any $m_{H^\pm} < 300\, {\rm GeV}$;
moreover,
even for such small values of $\lambda_{tt}$,
there is an excluded region:
from 0.75 to 1.6 for $m_{H^\pm} = 200\, {\rm GeV}$,
from 0.95 to 1.4 for $m_{H^\pm} = 300\, {\rm GeV}$,
and from 1.15 to 1.25 for $m_{H^\pm} = 400\, {\rm GeV}$.
They also present results similar to those of Bowser-Chao \etal
for $B \to X_s \gamma$,
although they use a different set of parameters
and also improved,
updated data.
They provide a comprehensive set of formulae
which can easily be employed for other parameter choices.

Of course,
if additional $\lambda_{ij}$ are taken to be non-zero
the parameter space becomes quite large.
Some recent analyses that discuss other $\lambda_{ij}$
include the work of Diaz \etall~\cite{Diaz:2005rv},
in which the effects of non-zero $\lambda_{sb}$ and $\lambda_{tc}$
were discussed
($\lambda_{sb}$ was found to be unbounded,
$\lambda_{tc}$ was found to be smaller than 1),
and of Idarraga \etall~\cite{Idarraga:2008zz},
who also considered bounds from leptonic $B$ decays
but chose very large $\lambda_{\mu\mu}$ and $\lambda_{\tau\tau}$.
Also note that,
if neutral fields are allowed in the loop,
one could get an effect proportional to $\lambda_{bs}^2$,
which is,
however,
negligible.

The possibility of $s$-channel charged-Higgs production at the LHC
may provide the best method of detection,
and detailed simulations would be welcomed.

As discussed in Chapter~\ref{sec:fcnc}, the BGL and MFV couplings can be obtained from
the type III model with the Cheng-Sher ansatz replaced by another ansatz
(which may depend on parameters and which will vary for different models).
Thus, everything in this section would apply to these models, but with
$\lambda_{ij}$ not being of O(1).

\newpage

\section{The scalar sector of the 2HDM}
\label{sec:notation}

The scalar sector of the 2HDM has many interesting features.
In its most general form,
the potential apparently has 14 independent parameters.
However,
the fact that the Higgs doublets $\Phi_1$ and $\Phi_2$
are {\em not\/} physical observables---only the scalar
mass eigenstates are physical particles---means that
we have the freedom to re-define those doublets,
provided we preserve the form of their kinetic terms.
These basis changes of the Higgs doublets
allow one to absorb some of the parameters in the potential
and are essential to understand the number
of physical parameters really present in it.

For several good reasons---the most usual of which
is preventing the occurrence of FCNC in the 2HDM---it is common
to impose a variety of global symmetries on the 2HDM,
thus reducing the number of free parameters.
In a highly non-intuitive result,
it has been proven that there are only {\em six} such symmetries
which have distinct effects on the scalar potential.
The six models resulting from each of those symmetries
have different physical implications:
different spectra of scalars,
different interactions with gauge bosons,
and,
in some cases,
predictions of massless axions or potential dark matter candidates.

It is the scalar potential that determines the vacuum of the 2HDM,
and that vacuum,
unlike what happens in the SM,
is not unique:
with two doublets the possibility arises that
the vacuum of the model spontaneously breaks the CP symmetry---which in fact
is precisely the reason why T.~D.~Lee first proposed the 2HDM
in 1973~\cite{Lee:1973iz}.
However,
for certain values of the parameters in the potential,
vacua which violate the electromagnetic symmetry,
giving mass to the photon,
are also possible.
Those have,
of course,
to be avoided.
Even if one considers only vacua which preserve both CP
and the usual gauge symmetries of the SM,
the 2HDM has a rich vacuum structure:
some of the possible 2HDM potentials can display
so-called ``inert vacua'',
in which one of the
neutral scalars does not couple to gauge bosons at all
(and can easily be made to decouple from fermions as well).
Some other potentials may have two different
electromagnetism-preserving minima,
with different predictions for the masses of the gauge bosons,
for instance.
The 2HDM has however a feature
which distinguishes it from other multi-Higgs models,
such as SUSY,
the Zee model,
or the 3HDM:
its vacua are stable and no tunneling from a neutral,
CP-conserving vacuum to a deeper,
CP- or charge-breaking vaccuum,
is possible.
And vice-versa:
any CP- or charge-breaking minimum that one finds
is guaranteed to be the global minimum of the model.

However,
not all values of the parameters of the 2HDM potential
ensure that there is a stable minimum,
unless one can be sure that the potential is bounded from below.
This basic requirement allows one to impose constraints
on the quartic scalar couplings.
A renormalization-group improvement of those constraints
translates in possibly severe bounds on the masses
of the physical scalar particles.
In the following we shall analyse these questions.
We shall discuss the vacuum structure of the potential,
derive general formulae for the scalar masses,
present bounds on quartic couplings obtained from the
requirement that the potential is bounded from below,
and discuss the  symmetries of the scalar potential
and their extension to the Yukawa sector.

\subsection{The scalar potential, notation 1}

The most general renormalizable,
\ie quartic,
scalar potential may be written~\cite{hep-ph/9409421}
\ba
V_H &=&
m_{11}^2 \Phi_1^\dagger \Phi_1
+ m_{22}^2 \Phi_2^\dagger \Phi_2
- \left( m_{12}^2 \Phi_1^\dagger \Phi_2 + \hc \right)
\no & &
+ \sfrac{1}{2} \lambda_1 \left( \Phi_1^\dagger\Phi_1 \right)^2
+ \sfrac{1}{2} \lambda_2 \left( \Phi_2^\dagger\Phi_2 \right)^2
+ \lambda_3 \left( \Phi_1^\dagger\Phi_1 \right)
\left( \Phi_2^\dagger\Phi_2 \right)
+ \lambda_4 \left( \Phi_1^\dagger\Phi_2 \right)
\left (\Phi_2^\dagger\Phi_1 \right)
\no & &
+ \left[
\sfrac{1}{2} \lambda_5 \left( \Phi_1^\dagger\Phi_2 \right)^2
+ \lambda_6 \left( \Phi_1^\dagger\Phi_1 \right)
\left( \Phi_1^\dagger\Phi_2 \right)
+ \lambda_7 \left( \Phi_2^\dagger\Phi_2 \right)
\left( \Phi_1^\dagger\Phi_2 \right)
+ \hc \right],
\label{2_VH1}
\ea
where ``H.c.''~stands for the Hermitian conjugate.
The parameters $m_{11}^2$,
$m_{22}^2$,
and $\lambda_{1,2,3,4}$ are real.
In general,
$m_{12}^2$ and $\lambda_{5,6,7}$ are complex.
Thus,
the Higgs potential in eq.~(\ref{2_VH1})
depends on six real and four complex parameters,
\ie a total of fourteen degrees of freedom.
However,
as we shall see below,
the freedom to redefine the basis means that in reality
only eleven degrees of freedom are physical.

In eq.~(\ref{2_VH1}) we are following the definitions
of Davidson and Haber~\cite{Davidson:2005cw};
often other definitions are used,
in which the same symbol may be employed for quantities which differ
from ours in sign,
a factor of two,
or complex conjugation.

\subsection{The scalar potential, notation 2}
\label{2_sec:notation}

An alternative notation for the scalar potential,
which has been championed by Botella and Silva~\cite{Botella:1994cs},
is
\ba
V_H &=&
\sum_{a,b = 1}^2 \mu_{ab}\, \Phi_a^\dagger \Phi_b
+ \sfrac{1}{2} \! \sum_{a,b,c,d = 1}^2
\lambda_{ab,cd} \left( \Phi_a^\dagger \Phi_b \right)
\left( \Phi_c^\dagger \Phi_d \right),
\label{2_VH2}
\ea
where,
by definition,
\be
\lambda_{ab,cd} = \lambda_{cd,ab}.
\ee
In eq.~(\ref{2_VH2}) hermiticity implies
\be
\mu_{ab} = \mu_{ba}^\ast \quad \mbox{and} \quad
\lambda_{ab,cd} = \lambda_{ba,dc}^\ast.
\label{2_hermiticity_coefficients}
\ee
The notation of eq.~(\ref{2_VH2}) is useful for
the study of invariants,
basis transformations,
and symmetries.
The correspondence between notations 1 and 2 is given by
\ba
\mu_{11} = m_{11}^2, & & \mu_{22} = m_{22}^2, \no
\mu_{12} = -m_{12}^2, & & \mu_{21} = -{m_{12}^2}^\ast \no
\lambda_{11,11} = \lambda_1, & & \lambda_{22,22} = \lambda_2, \no
\lambda_{11,22} = \lambda_{22,11} = \lambda_3, & &
\lambda_{12,21} = \lambda_{21,12} = \lambda_4,
\label{2_ynum} \label{2_znum} \\
\lambda_{12,12} = \lambda_5, & & \lambda_{21,21} = \lambda_5^\ast, \no
\lambda_{11,12} = \lambda_{12,11} = \lambda_6, & &
\lambda_{11,21} = \lambda_{21,11} = \lambda_6^\ast, \no
\lambda_{22,12} = \lambda_{12,22} = \lambda_7, & &
\lambda_{22,21} = \lambda_{21,22} = \lambda_7^\ast. \nonumber
\ea

Once again,
one must be careful when confronting eq.~(\ref{2_VH2})
to similar equations written in other papers,
since the same symbol may be used in different papers
for quantities which differ in sign,
a factor of two,
or complex conjugation.

\subsection{The scalar potential, notation 3}
\label{3_sec:notation}

The previous two notations
consider the scalar doublets $\Phi_a$ ($a = 1, 2$) individually.
A third notation emphasises the presence of field bilinears
$\Phi_a^\dagger \Phi_b$ in the scalar potential.
An early use of bilinears is due to Velhinho,
Santos,
and Barroso \cite{Velhinho:1994vh};
the notation has later been much employed by Nagel \cite{Nagel:2004sw},
by Maniatis \etal \cite{Maniatis:2006fs,Maniatis:2006jd,Maniatis:2007vn},
by Nishi \cite{Nishi:2006tg,Nishi:2007nh,Nishi:2007dv},
and by Ivanov \cite{Ivanov:2005hg,Ivanov:2006yq,arXiv:0910.4492}.
Following Nishi~\cite{Nishi:2006tg} we write:
\be
V_H = \sum_{\mu = 0}^3 M_\mu r_\mu
+ \sum_{\mu, \nu = 0}^3 \Lambda_{\mu \nu} r_\mu r_\nu,
\label{2_VH3}
\ee
where
\be
\Lambda_{\mu \nu} = \Lambda_{\nu \mu}
\ee
and
\be
\begin{array}{rcl}
r_0 &=&
\frac{1}{2}
\left( \Phi_1^\dagger \Phi_1 + \Phi_2^\dagger \Phi_2 \right),
\\*[2mm]
r_1 &=&
\frac{1}{2}
\left( \Phi_1^\dagger \Phi_2 + \Phi_2^\dagger \Phi_1 \right)
= \mbox{Re}\left( \Phi_1^\dagger \Phi_2 \right),
\\*[2mm]
r_2 &=&
- \frac{i}{2}
\left( \Phi_1^\dagger \Phi_2 - \Phi_2^\dagger \Phi_1 \right)
= \mbox{Im} \left( \Phi_1^\dagger \Phi_2 \right),
\\*[2mm]
r_3 &=&
\frac{1}{2}
\left( \Phi_1^\dagger \Phi_1 - \Phi_2^\dagger \Phi_2 \right).
\end{array}
\label{2_r_Ivanov}
\ee
In eq.~(\ref{2_VH3}) we have adopted an Euclidean metric.
It differs from the notation of Ivanov \cite{Ivanov:2006yq},
who pointed out that $r_\mu$ parametrizes the gauge orbits of the Higgs fields
in a space equipped with a Minkowski metric.

Notation 3 is convenient for studies of features
such as the existence and number of minima
of the scalar potential.
Since the Yukawa couplings involve the Higgs doublets individually
rather than bilinears,
notation 3 cannot be applied for studies of the
full theory with both scalars and fermions.

The correspondence between notations 1 and 3 is given by
\be
M_\mu = \left( m_{11}^2 + m_{22}^2,\ - 2 \mbox{Re}\left( m_{12}^2 \right),\
2 \mbox{Im} \left( m_{12}^2 \right),\ m_{11}^2 - m_{22}^2 \right),
\label{2_M_mu}
\ee
\be
\Lambda_{\mu \nu} = \left( \begin{array}{cccc}
(\lambda_1+\lambda_2)/2 + \lambda_3 &
\mbox{Re}\left( \lambda_6 + \lambda_7 \right) &
- \mbox{Im} \left( \lambda_6 + \lambda_7 \right) &
(\lambda_1 - \lambda_2) / 2 \\
\mbox{Re}\left( \lambda_6 + \lambda_7 \right) &
\lambda_4 + \mbox{Re}\left( \lambda_5 \right) &
- \mbox{Im} \left( \lambda_5 \right) &
\mbox{Re}\left( \lambda_6 - \lambda_7 \right) \\
- \mbox{Im} \left( \lambda_6 + \lambda_7 \right) &
- \mbox{Im} \left( \lambda_5 \right) &
\lambda_4 - \mbox{Re}\left( \lambda_5 \right) &
- \mbox{Im} \left( \lambda_6 - \lambda_7 \right) \\
(\lambda_1 - \lambda_2) / 2 &
\mbox{Re}\left( \lambda_6 - \lambda_7 \right) &
- \mbox{Im} \left( \lambda_6 - \lambda_7 \right) &
(\lambda_1+\lambda_2) / 2 - \lambda_3
\end{array} \right).
\label{2_Lambda_munu}
\ee
The correspondence between notations 2 and 3 is given by
\ba
M_\mu &=& \sum_{a,b = 1}^2 \left( \sigma_\mu \right)_{ab} \mu_{ba},
\label{2_M_vs_Y}
\\
\Lambda_{\mu \nu} &=&
\sfrac{1}{2} \sum_{a,b,c,d = 1}^2
\left( \sigma_\mu \right)_{ab} \left( \sigma_\nu \right)_{cd}
\lambda_{ba,dc},
\label{2_Lambda_vs_Z}
\ea
where $\sigma_0$ is the $2 \times 2$ identity matrix
and $\sigma_{1,2,3}$ are the three Pauli matrices.

\subsection{Basis transformations}
\label{2_sec:BASIS}

The doublets $\Phi_a$ are not physical---only
the scalar mass eigenstates,
corresponding to particles,
are physical.
Thus,
any combination of the doublets which respects the symmetries of the theory
will produce the same physical predictions.
We call any combination of $\left( \Phi_1, \Phi_2 \right)$
a {\it basis} for the doublets.
We may rewrite the potential in terms of new doublets $\Phi^\prime_a$,
obtained from the original ones by a (global) basis transformation
\be
\Phi_a^\prime = \sum_{b = 1}^2 U_{ab} \Phi_b,
\label{2_basis-transf}
\ee
where $U$ is a $2 \times 2$ unitary matrix.
Under this unitary
basis transformation
the gauge-kinetic terms are unchanged,
but the coefficients of the potential in notation 2
transform as
\ba
\mu^\prime_{ab} &=&
\sum_{c,d = 1}^2 U_{ac}\, \mu_{cd}\, U_{bd}^\ast
= \left( U \mu U^\dagger \right)_{ab},
\label{2_Y-transf}
\\
\lambda^\prime_{ab,cd} &=&
\sum_{e,f,g,h = 1}^2
U_{ae}\, U_{cg}\, \lambda_{ef,gh}\, U_{bf}^\ast \, U_{dh}^\ast.
\label{2_Z-transf}
\ea

The basis transformation of eq.~(\ref{2_basis-transf})
induces a transformation of the 4-vector in eq.~(\ref{2_r_Ivanov})
given by
\be
r_0^\prime = r_0, \quad
r_i^\prime = \sum_{j=1}^3 R_{ij} \left( U \right) r_j
\label{2_r_rotated}
\ee
where \cite{Nishi:2006tg}
\be
R_{ij} \left( U \right) = \sfrac{1}{2}\,
\tr \left( U^\dagger \sigma_i U \sigma_j \right).
\label{2_R_matrix}
\ee
The matrix $R$ belongs to SO(3)
and the transformation of $U$ into $R \left( U \right)$ is the
SU(2)$\rightarrow$SO(3)
two-to-one\footnote{Note that
$R \left( U \right) = R \left( - U \right)$.}
mapping.
Under this rotation of the 3-vector $r_i$,
the parameters of the scalar potential in notation 3
transform as
\begin{itemize}
\item scalars,
\be
M_0^\prime = M_0, \quad \Lambda_{00}^\prime = \Lambda_{00},
\ee
\item vectors,
\be
M_i^\prime = \sum_{j=1}^3 R_{ij} M_j,
\quad
\Lambda_{0i}^\prime = \sum_{j=1}^3 R_{ij} \Lambda_{0j},
\label{iudrt}
\ee
\item and a symmetric tensor
\be
\Lambda_{ij}^\prime = \sum_{k,l = 1}^3 R_{ik} R_{jl} \Lambda_{kl}.
\label{nhgui}
\ee
\end{itemize}

We see in eqs.~(\ref{2_Y-transf}) and~(\ref{2_Z-transf})
that the overall phase of $U$
does not impact the change of the parameters of the potential.
As a result,
one may consider $U \in$ SU(2).
We then parametrize
\be
U = \left( \begin{array}{cc}
e^{i \chi} c_\psi & e^{i \left( \chi - \xi \right)} s_\psi \\
- e^{i \left( \xi - \chi \right)} s_\psi & e^{- i \chi} c_\psi
\end{array} \right),
\label{2_U_DH}
\ee
where $c_\psi = \cos{\psi}$ and $s_\psi = \sin{\psi}$;
a similar notation will be used below
for multiples of the angle $\psi$.
Then,
in notation 1,
the parameters $m_{ij}^2$ and $\lambda_i$ of eq.~(\ref{2_VH1}) transform as
\ba
{m_{11}^2}^\prime &=&
m_{11}^2 c_\psi^2 + m_{22}^2 s_\psi^2
- \mbox{Re}\left( m_{12}^2 e^{i \xi} \right) s_{2 \psi},
\label{2_M11}
\\*[2mm]
{m_{22}^2}^\prime &=&
m_{11}^2 s_\psi^2 + m_{22}^2 c_\psi^2
+ \mbox{Re}\left( m_{12}^2 e^{i \xi} \right) s_{2 \psi},
\label{2_M22}
\\*[2mm]
{m_{12}^2}^\prime &=&
e^{i \left( 2 \chi - \xi \right)} \left[
\sfrac{1}{2} \left( m_{11}^2 - m_{22}^2 \right) s_{2 \psi}
+ \mbox{Re}\left( m_{12}^2 e^{i \xi} \right) c_{2 \psi}
+ i \mbox{Im} \left( m_{12}^2 e^{i \xi} \right)
\right],
\label{2_M12}
\\*[2mm]
\lambda_1^\prime &=&
\lambda_1 c_\psi^4 + \lambda_2 s_\psi^4
+ \sfrac{1}{2} \lambda_{345} s_{2 \psi}^2
+ 2 s_{2 \psi} \left[
c_\psi^2 \mbox{Re}\left( \lambda_6 e^{i \xi} \right)
+ s_\psi^2 \mbox{Re}\left( \lambda_7 e^{i \xi} \right)
\right],
\label{2_L1}
\\*[2mm]
\lambda_2^\prime &=&
\lambda_1 s_\psi^4 + \lambda_2 c_\psi^4
+ \sfrac{1}{2} \lambda_{345} s_{2 \psi}^2
- 2 s_{2 \psi} \left[
s_\psi^2 \mbox{Re}\left( \lambda_6 e^{i \xi} \right)
+ c_\psi^2 \mbox{Re}\left( \lambda_7 e^{i \xi} \right)
\right],
\label{2_L2}
\\*[2mm]
\lambda_3^\prime &=& \lambda_3
+ \tfrac{1}{4} s_{2 \psi}^2 \left(
\lambda_1 + \lambda_2 - 2 \lambda_{345} \right)
- s_{2 \psi} c_{2 \psi}
\mbox{Re}\left[ \left( \lambda_6 - \lambda_7 \right) e^{i\xi} \right],
\label{2_L3}
\\*[2mm]
\lambda_4^\prime &=&
\lambda_4
+ \sfrac{1}{4} s_{2 \psi}^2
\left( \lambda_1 + \lambda_2 - 2 \lambda_{345} \right)
- s_{2 \psi} c_{2 \psi}
\mbox{Re}\left[ \left( \lambda_6 - \lambda_7 \right) e^{i \xi} \right],
\label{2_L4}
\\*[2mm]
\lambda_5^\prime &=&
e^{2 i \left( 2 \chi - \xi \right)} \left\{
\sfrac{1}{4} s_{2 \psi}^2
\left( \lambda_1 + \lambda_2 - 2 \lambda_{345} \right)
+ \mbox{Re}\left( \lambda_5 e^{2 i \xi} \right)
+ i c_{2 \psi} \mbox{Im} \left( \lambda_5 e^{ 2 i \xi}\right)
\right. \nonumber \\*[1mm] & & \left.
- s_{2 \psi} c_{2 \psi}
\mbox{Re}\left[ \left( \lambda_6 - \lambda_7 \right) e^{i \xi} \right]
-i s_{2 \psi} \mbox{Im}
\left[ \left( \lambda_6 - \lambda_7 \right) e^{i\xi} \right]
\right\},
\label{2_L5}
\\*[2mm]
\lambda_6^\prime &=&
e^{i \left( 2\chi - \xi \right)} \left\{
- \sfrac{1}{2} s_{2 \psi} \left[
\lambda_1 c_\psi^2 - \lambda_2 s_\psi^2 - \lambda_{345} c_{2 \psi}
- i \mbox{Im} \left( \lambda_5 e^{2 i \xi} \right)
\right]
\right. \nonumber \\*[1mm] & & \left.
+ c_\psi c_{3 \psi} \mbox{Re}\left( \lambda_6 e^{i \xi} \right)
+ s_\psi s_{3 \psi} \mbox{Re}\left( \lambda_7 e^{i \xi}\right)
+ i c_\psi^2 \mbox{Im} \left( \lambda_6 e^{i \xi} \right)
+ i s_\psi^2 \mbox{Im} \left( \lambda_7 e^{i \xi} \right)
\right\},
\label{2_L6}
\\*[2mm]
\lambda_7^\prime &=&
e^{i \left( 2\chi - \xi \right)} \left\{
- \sfrac{1}{2} s_{2 \psi} \left[
\lambda_1 s_\psi^2 - \lambda_2 c_\psi^2 + \lambda_{345} c_{2 \psi}
+ i \mbox{Im} \left( \lambda_5 e^{2 i \xi} \right) \right]
\right. \nonumber \\*[1mm] & & \left.
+ s_\psi s_{3 \psi} \mbox{Re}\left( \lambda_6 e^{i \xi} \right)
+ c_\psi c_{3 \psi} \mbox{Re}\left( \lambda_7 e^{i \xi} \right)
+ i s_\psi^2 \mbox{Im} \left( \lambda_6 e^{i \xi} \right)
+ i c_\psi^2 \mbox{Im} \left( \lambda_7 e^{i \xi} \right)
\right\},
\label{2_L7}
\ea
where
\be
\lambda_{345} := \lambda_3 + \lambda_ 4
+ \mbox{Re}\left( \lambda_5 e^{2 i \xi} \right).
\label{2_L345}
\ee

A basis transformation may be utilized to eliminate
some of the degrees of freedom in the scalar potential.
This implies that not all the parameters in that potential
have physical significance.
Thus,
the three parameters in eq.~(\ref{2_U_DH})
may be used to absorb three out of the 14 parameters in the scalar potential.
As a result,
there are only 11 physical degrees of freedom in the potential and,
thus,
only eleven independent observables.
Note,
though,
that we are still discussing the most general potential;
when one imposes a global symmetry
(see section~\ref{sec:symmetries} ahead)
on the 2HDM,
the number of parameters which may be eliminated
through basis transformations may be less than three.

\subsection{
GCP transformations}
\label{subsec:GCPs}

The standard CP transformation for a Higgs doublet reads
\be
\Phi \left( t, \vec{x} \right) \rightarrow
\Phi^{\textrm{CP}} \left( t, \vec{x} \right)
= \Phi^\ast \left( t, - \vec{x} \right).
\label{2_StandardCP}
\ee
The reference
to the time ($t$) and space ($\vec{x}$) coordinates will henceforth
be suppressed.

However,
in the presence of several identical doublets,
the possibility of arbitrary basis transformations
should be included in the definition of the CP transformation.
Let us illustrate this problem with a simple example.
We start from the basis $\left( \Phi_1, \Phi_2 \right)$
and the usual definition of CP:
\be
\Phi_1^\textrm{CP} = \Phi_1^\ast, \quad
\Phi_2^\textrm{CP} = \Phi_2^\ast.
\label{2_Ex_CP1}
\ee
Now we perform a basis transformation
\be
\begin{array}{rcl}
\Phi_1 &=&
\tfrac{1}{\sqrt{2}} \left( \Phi_1^\prime
+ e^{i \pi/4}\; \Phi_2^\prime \right),
\\*[2mm]
\Phi_2 &=&
\tfrac{1}{\sqrt{2}} \left( - e^{-i \pi/4}\; \Phi_1^\prime
+ \Phi_2^\prime \right).
\end{array}
\label{2_Ex_CP2}
\ee
Substituting eqs.~(\ref{2_Ex_CP2}) into eqs.~(\ref{2_Ex_CP1}),
we obtain
\[
\begin{array}{rcl}
\tfrac{1}{\sqrt{2}} \left[
\left( \Phi_1^\prime \right)^\textrm{CP}
+ e^{i \pi/4} \left( \Phi_2^\prime \right)^\textrm{CP}
\right]
&=&
\tfrac{1}{\sqrt{2}} \left[
\left( \Phi_1^\prime \right)^\ast
+ e^{-i \pi/4} \left( \Phi_2^\prime \right)^\ast
\right]
\\*[2mm]
\tfrac{1}{\sqrt{2}} \left[
- e^{-i \pi/4} \left( \Phi_1^\prime \right)^\textrm{CP}
+ \left( \Phi_2^\prime \right)^\textrm{CP}
\right]
&=&
\tfrac{1}{\sqrt{2}} \left[
- e^{i \pi/4} \left( \Phi_1^\prime \right)^\ast
 + \left( \Phi_2^\prime \right)^\ast
\right],
\end{array}
\]
which leads to
\be
\begin{array}{rcl}
\left( \Phi_1^\prime \right)^\textrm{CP} &=&
{\displaystyle \frac{1+i}{2} \left( \Phi_1^\prime \right)^\ast
- \frac{i}{\sqrt{2}} \left( \Phi_2^\prime \right)^\ast,}
\\*[4mm]
\left( \Phi_2^\prime \right)^\textrm{CP} &=&
{\displaystyle - \frac{i}{\sqrt{2}} \left( \Phi_1^\prime \right)^\ast
+ \frac{1-i}{2} \left( \Phi_2^\prime \right)^\ast.}
\end{array}
\label{hvjwe}
\ee
This is
clearly {\em not}\/
the usual
CP transformation, which
means that eq.~\eqref{2_Ex_CP1}
is too restrictive as a
definition of CP.
Even the usual CP transformation may look
different from eq.~\eqref{2_Ex_CP1} in a different
basis.

As a result, we must consider
a more general version of the CP transformation,
which we denote with the superscript
`GCP'~\footnote{These are known in the literature
by many authors as
Generalized CP
transformations~\cite{Neufeld:1987wa}---~\cite{Ferreira:2010yh}.}:
\be
\begin{array}{rcl}
\Phi_a & \rightarrow &
\Phi^{\textrm{GCP}}_a
= \displaystyle{\sum_{b=1}^2} X_{ab} \Phi_b^\ast,
\\*[3mm]
\Phi_a^\dagger & \rightarrow &
{\Phi^{\textrm{GCP}}}^\dagger_a
= \displaystyle{\sum_{b=1}^2} X_{ab}^\ast \Phi_b^T,
\label{2_GCP}
\end{array}
\ee
where $X$ is an arbitrary unitary matrix.
GCP transformations
were first discussed by Lee and Wick
\cite{Lee:1966ik}.
Their explicit use for quarks is due to
Bernab\'eu, Branco,
and Gronau~\cite{Bernabeu:1986fc}.
GCP transformations in the scalar sector
were developed by the Vienna
group \cite{Ecker:1981wv,Ecker:1983hz,Ecker:1987qp,Neufeld:1987wa}.
Note that,
unlike the standard CP transformation of eq.~\eqref{2_StandardCP},
GCP transformations are such that
the square of the transformation is {\em not},
in general, equal to unity (see section~\ref{2_sec:cpsq})---as we already
see in the practical example of eqs.~(\ref{hvjwe}).

Under the GCP transformation in eq.~(\ref{2_GCP}),
the gauge-kinetic terms stay invariant
but the coefficients $\mu_{ab}$ and $\lambda_{ab,cd}$
transform as
\ba
\mu_{ab}^\textrm{GCP} &=&
\sum_{c,d=1}^2 X_{ac} \mu_{cd}^\ast X_{bd}^\ast
= \left(
X \mu^\ast X^\dagger
\right)_{ab},
\\
\lambda_{ab,cd}^\textrm{GCP} &=&
\sum_{e,f,g,h = 1}^2
X_{ae} X_{cf} \lambda_{eg, fh}^\ast X_{bg}^\ast X_{dh}^\ast.
\label{YZ-CPtransf}
\ea

We next turn to the interplay
between GCP transformations and basis transformations.
If the GCP transformation is given by eq.~(\ref{2_GCP}),
then $\Phi^\prime_a = \sum_b U_{ab} \Phi_b$ has a GCP transformation
given by
\be
\Phi_a^\prime  \rightarrow
\left( \Phi_a^\prime \right)^\textrm{GCP}
= \sum_{b=1}^2 X_{ab}^\prime {\Phi_b^\prime}^\ast,
\ee
where
\be
X^\prime = U X U^T.
\label{2_X-prime}
\ee
The fact that $U^T$,
rather than $U^\dagger$,
appears in eq.~(\ref{2_X-prime}) is crucial.
If one had $U^\dagger$,
then one would be able to find a basis such that $X$ was diagonal.
Because one has $U^T$ in eq.~(\ref{2_X-prime})
\textit{it is not possible to reduce,
through a basis transformation,
all GCP transformations
to the standard CP transformation} of eq.~(\ref{2_StandardCP}).
However,
Ecker,
Grimus,
and Neufeld \cite{Ecker:1987qp} have proved that
for every matrix $X$ there exists a unitary matrix $U$
such that
\be
X^\prime = U X U^T = \left( \begin{array}{cc}
\cos{\theta} & \sin{\theta} \\
- \sin{\theta} & \cos{\theta}
\end{array} \right),
\label{2_GCP-reduced}
\ee
with $0 \leq \theta \leq \pi/2$.
The value of $\theta$ may be determined
through either one of two ways:
(i) the (twice degenerate) eigenvalue of
$\left( X + X^T \right) \left( X^\ast + X^\dagger \right)$ is
$4 \cos^2{\theta}$;
(ii) the eigenvalues of $X X^\ast$ are $e^{\pm 2 i \theta}$.

The GCP transformation in eq.~(\ref{2_GCP})
induces the following transformation
of the four-vector in eq.~(\ref{2_r_Ivanov}):
\be
r_0^\textrm{GCP} = r_0, \quad
r_i^\textrm{GCP} =
\sum_{j=1}^3
\bar R_{ij} \left( X \right)\, r_j
\label{2_r_X_rotated}
\ee
where \cite{Maniatis:2007vn}
\ba
\bar R \left( X \right) &=& R \left( X \right) \bar R_2,
\no
R_{ij} \left( X \right) &=& \sfrac{1}{2}\,
\tr \left( X^\dagger \sigma_i X \sigma_j \right),
\\
\bar R_2 &=& \textrm{diag} \left( 1, -1 , 1 \right).
\nonumber
\ea
Both $\bar R_2$ and $\bar R \left( X \right)$ are improper rotations,
\ie O(3)
matrices with determinant $-1$.
For the simplified form of $X$ in eq.~(\ref{2_GCP-reduced})  one has
\be
\bar R \left( X \right) = \left( \begin{array}{ccc}
\cos{(2 \theta)} & 0 & - \sin{(2 \theta)} \\
0 & - 1 & 0 \\
\sin{(2 \theta)} & 0 & \cos{(2 \theta)}
\end{array} \right).
\label{2_Rbar-reduced}
\ee

\subsection{\label{sec:symmetries}The six classes
of symmetry-constrained scalar potentials}

The large number of free parameters in the scalar potential of the 2HDM
reduces the theory's predictive power.
Any symmetry that we may impose on the 2HDM
to constrain its scalar potential is therefore welcome.
Also,
as we discussed in previous sections,
the 2HDM is in general plagued by flavour-changing neutral currents;
these,
however,
may be eliminated---or strongly suppressed---by imposing
an internal symmetry on the 2HDM.

Symmetries leaving the kinetic terms unchanged
may be of either one of two types:
\begin{enumerate}
\item One may relate $\Phi_a$
to some unitary transformation of $\Phi_b$,
\be
\Phi_a \rightarrow \Phi_a^S = \sum_{b=1}^2 S_{ab} \Phi_b,
\label{2_S-transf-symmetry}
\ee
where $S$ is a unitary matrix.
We then require the potential to be invariant under this
transformation.
As a result of this invariance,
\ba
\mu_{ab} &=& \sum_{c,d = 1}^2 S_{ac} \mu_{cd} S_{bd}^\ast,
\label{2_Y-S}
\\
\lambda_{ab,cd} &=&
\sum_{e,f,g,h = 1}^2
S_{ae} S_{cf} \lambda_{eg,fh} S_{bg}^\ast S_{dh}^\ast.
\label{2_Z-S}
\ea
These are known as Higgs Family (HF) symmetries.\footnote{Notice that
this is \textit{not} the situation considered
in eqs.~(\ref{2_basis-transf})--(\ref{2_Z-transf}).
There,
the coefficients of the Lagrangian \textit{do change}
under the transformation.
In contrast,
eqs.~(\ref{2_S-transf-symmetry})--(\ref{2_Z-S})
imply the existence of a HF symmetry of the scalar potential
because the coefficients of $V_H$ are \textit{unchanged}.}
\item One may relate $\Phi_a$ with some unitary transformation
of $\Phi_b^\ast$:
\be
\Phi_a \rightarrow \Phi^{\textrm{GCP}}_a
= \sum_{b=1}^2 X_{ab} \Phi_b^\ast,
\label{2_GCP_symmetry}
\ee
where $X$ is an arbitrary unitary matrix.
We then require that the potential be invariant under this symmetry:
\ba
\mu_{ab} &=&
\sum_{c,d=1}^2 X_{ac} \mu_{cd}^\ast X_{bd}^\ast,
\no
\lambda_{ab,cd} &=&
\sum_{e,f,g,h = 1}^2
X_{ae} X_{cf} \lambda_{eg,fh}^\ast X_{bg}^\ast X_{dh}^\ast.
\label{2_YZ-CPtransf}
\ea
These are known as GCP symmetries.
\end{enumerate}
Under the basis transformation
$\Phi_a \rightarrow \Phi_a^\prime = U_{ab} \Phi_b$
of eq.~(\ref{2_basis-transf})
the specific forms of the HF and GCP symmetries get altered,
respectively, into:
\begin{eqnarray}
S^\prime &=& U S U^\dagger,
\label{2_S-prime_2}
\\
X^\prime &=& U X U^T.
\label{2_X-prime_2}
\end{eqnarray}
Therefore,
a symmetry relation among the coefficients of the scalar potential will,
in general,
appear as a different relation
if the coefficients of the potential
are written using a different basis
for the Higgs doublets.

One may,
of course,
impose on a theory several HF symmetries and/or GCP symmetries simultaneously.
Ivanov \cite{Ivanov:2006yq} has proved that,
no matter what combination of HF and/or GCP symmetries
one imposes on the scalar potential of the 2HDM,
one always ends up with one of six distinct classes of potentials.
This issue was studied further by Ferreira, Haber, and Silva
\cite{Ferreira:2009wh}.
In table~\ref{2_master1}
\begin{table}[h]
\caption{
The six classes of symmetries (I--VI) of the scalar potential
and a practical example of a symmetry in each of those classes.
The number in the last column is the minimal number of parameters ($n$)
in the scalar potential,
obtainable in a specific basis.
}
\begin{center}
\begin{tabular}
{|cc|ccc|ccccccc|c|}
\hline \hline
class & symmetry &
$m_{11}^2$ & $m_{22}^2$ & $m_{12}^2$ &
$\lambda_1$ & $\lambda_2$ & $\lambda_3$ & $\lambda_4$ &
$\lambda_5$ & $\lambda_6$ & $\lambda_7$ & $n$ \\
\hline
I & U(2) &  & $ m_{11}^2$ & 0 &
   & $\lambda_1$ &  & $\lambda_1 - \lambda_3$ &
0 & 0 & 0 & 3 \\
II & CP3 &  & $m_{11}^2$ & 0 &
   & $\lambda_1$ &  &  &
$\lambda_1 - \lambda_3 - \lambda_4$  & 0 & 0 & 4 \\
III & CP2 &  & $m_{11}^2$ & 0 &
  & $\lambda_1$ &  &  &
   &  & $- \lambda_6$ & 5\\
IV & U(1) &  &  & 0 &
 &  & &  &
0 & 0 & 0 & 6 \\
V & $Z_2$ &   &   & 0 &   &  &  &  &
   & 0 & 0 & 7 \\
VI & CP1 &  &  & real &
 & &  &  &
real & real & real & 8 \\
\hline \hline
\end{tabular}
\end{center}
\label{2_master1}
\end{table}
we present
an example of a symmetry in each of the six classes of symmetries
found by Ivanov,
and the constraints on the parameters of the potential
following from that specific symmetry.
The number of physical parameters in the potential may in general,
within each one of Ivanov's classes,
be further reduced by choosing a specific basis
for the scalar
doublets, much
in the same way as the general
2HDM potential has 14 parameters which may,
however,
be reduced to 11 through a suitable basis
choice;
the number of physical parameters for each class
is given in the last column of table~\ref{2_master1}.

The six specific symmetries given as examples in table~\ref{2_master1}
are the following:
\begin{itemize}
\item U(2)
is the strongest (most general) HF symmetry,
\be
S = \left( \begin{array}{cc}
e^{- i \xi} \cos{\theta} & e^{- i \psi} \sin{\theta} \\
- e^{i \psi} \sin{\theta} & e^{i \xi} \cos{\theta}
\end{array} \right),
\label{2_eq:U2}
\ee
where $\xi$, $\theta$, and $\psi$ are arbitrary.
\item CP3 is a GCP symmetry with
\be
X = \left( \begin{array}{cc}
\cos{\theta} & \sin{\theta} \\
- \sin{\theta} & \cos{\theta}
\end{array} \right),
\label{2_eq:CP3}
\ee
the first-quadrant angle $\theta$ being generic,
\ie different from the two specific values $0$ and $\pi / 2$.
\item CP2 is the GCP symmetry of eq.~(\ref{2_eq:CP3})
but with $\theta = \pi / 2$, {\em i.e.}
\be
X = \left( \begin{array}{cc}
0 & 1 \\
- 1 & 0
\end{array} \right).
\label{2_eq:CP2}
\ee
\item U(1)
is a restricted version of the HF symmetry of eq.~(\ref{2_eq:U2})
with
\be
S = \left( \begin{array}{cc}
e^{- i \xi}  & 0 \\
0 & e^{i \xi}
\end{array} \right),
\label{2_eq:U1}
\ee
where $\xi$ is
arbitrary.\footnote{See, though, the discussion after eq.~\eqref{2_S_2/3}.}
\item
$Z_2$ is the symmetry under $\phi_2 \to - \phi_2$,
\be
S = \left( \begin{array}{cc}
1  & 0 \\
0 & -1
\end{array} \right).
\label{2_eq:Z2}
\ee
\item CP1 is the standard CP symmetry, with
\be
X = \left( \begin{array}{cc}
1 & 0 \\
0 & 1
\end{array} \right).
\label{2_eq:CP1}
\ee
\end{itemize}

Some of these models may be further simplified
by choosing an appropriate basis for the Higgs doublets:
\begin{itemize}
\item A specific basis may be chosen
in the CP2 model
\cite{Davidson:2005cw}
such that $\lambda_5$ is real and $\lambda_6 = \lambda_7 = 0$,
hence that model only has five parameters.
\item In the U(1) model
one may render $\lambda_5$ real
through a rephasing of the doublets,
so that that model only has seven parameters.
\item One may perform a real rotation of the doublets
in the CP1 model
such that $m_{12}^2$ becomes zero \cite{Lee:1973iz},
hence that model only has nine parameters.
\end{itemize}

It should be noted that the scalar potential
of the minimal supersymmetric Standard Model (MSSM)
does {\em not}\/ fall
into any of Ivanov's symmetry classes.
In fact,
in the tree-level scalar potential of the MSSM one has
\cite{Gunion:1989we,Haber:1993an,Maniatis:2006fs}
\be
\lambda_1 = \lambda_2 = \frac{g^2 + {g^\prime}^2}{4},
\quad
\lambda_3 = \frac{g^2 - {g^\prime}^2}{4},
\quad
\lambda_4 = - \frac{g^2}{2},
\quad
\lambda_5 = \lambda_6 = \lambda_ 7
= 0,
\label{2_eq:susylim}
\ee
where $g$ and $g^\prime$ are the gauge coupling constants
of SU(2) and U(1),
respectively.
This is much similar
to the U(2)-symmetric 2HDM,
with the crucial difference that the latter has
$\lambda_3 + \lambda_4 = \lambda_1$,
while in the MSSM $\lambda_3 + \lambda_4 = - \lambda_1$.
It is often stated that the scalar sector of the MSSM
is a particular case of a 2HDM,
but that statement is potentially misleading:
the relations among the quartic couplings in eqs.~\eqref{2_eq:susylim}
are renormalization-group (RG) invariant in the MSSM,
but only due to the presence of
extra particles---namely the gauginos;
analogous relations among the couplings of a 2HDM
are {\em not}\/ RG-protected if the 2HDM is not supersymmetrized
\cite{Haber:1993an,Ferreira:2010jy}.
An RG analysis of the relations between couplings shown in
table~\ref{2_master1} was undertaken in~\cite{arXiv:0909.2855,arXiv:1106.1436}.
The list of possible potential symmetries increases
when one considers field transformations
which do {\em not}\/ leave the gauge-kinetic terms invariant,
as was shown recently by Battye \etal \cite{arXiv:1106.3482}
and by Pilaftsis~\cite{arXiv:1109.3787};
of course,
those extra symmetries lead to relations among the parameters of the potential
which are not RG-invariant.
See also the discussion on custodial symmetry in Appendix~\ref{ap:cust}.

The remainder of this section is devoted to a careful explanation
of table~\ref{2_master1},
following the presentation in~\cite{Ferreira:2009wh}.

\subsubsection{\label{subsec:HFsymmetry}Higgs Family symmetries}

Higgs Family symmetries have a long history in the 2HDM.
Glashow and Weinberg \cite{Glashow:1976nt} and,
separately,
Paschos \cite{Paschos:1976ay} have introduced the discrete $Z_2$ symmetry
\be
Z_2: \quad \Phi_1 \rightarrow \Phi_1, \quad \Phi_2 \rightarrow - \Phi_2,
\label{2_Z2}
\ee
and extended it to the quark sector
in order to avoid flavour-changing neutral currents.
This symmetry enforces $m_{12}^2 = 0$ and $\lambda_6 = \lambda_7 = 0$.

We may consider the $Z_2$ symmetry in a different scalar basis,
\be
\begin{array}{rcl}
\Phi_1^\prime &=& 2^{-1/2} \left( \Phi_1 + \Phi_2 \right),
\\*[1mm]
\Phi_2^\prime &=& 2^{-1/2} \left( \Phi_1 - \Phi_2 \right).
\end{array}
\label{2_from_Z2_to_Pi2}
\ee
obtaining the interchange symmetry
\be
\Pi_2: \quad \Phi^\prime_1 \leftrightarrow \Phi^\prime_2.
\label{2_Pi2}
\ee
This is equivalent to applying eq.~(\ref{2_S-prime_2}) in the form
\be
\sfrac{1}{2}
\left( \begin{array}{cc}
1 & 1 \\ 1 & -1
\end{array} \right)
\left( \begin{array}{cc}
1 & 0 \\ 0 & -1
\end{array} \right)
\left( \begin{array}{cc}
1 & 1 \\ 1 & -1
\end{array} \right)
=
\left( \begin{array}{cc}
0 & 1 \\ 1 & 0
\end{array} \right).
\label{2_Z2ToPi2}
\ee
The $\Pi_2$ symmetry enforces $m_{22}^2 = m_{11}^2$,
$\mbox{Im} \left( m_{12}^2 \right) = 0$,
$\lambda_2 = \lambda_1$,
$\lambda_7 = \lambda_6^\ast$,
and $\mbox{Im} \left( \lambda_5 \right) = 0$.
Thus,
\begin{itemize}
\item the constraints obtained by applying $Z_2$
are apparently different from those obtained by applying $\Pi_2$;
\item however,
the two symmetries are equivalent,
since applying $Z_2$ in a given basis
is the same as applying $\Pi_2$ in a basis obtained from the first one
through the transformation~(\ref{2_from_Z2_to_Pi2});
\item the $Z_2$-symmetric and $\Pi_2$-symmetric potentials
must lead to exactly the same physical predictions---we say that
they are in the same class---because physical observables
cannot depend
on the basis in which we choose to write the Higgs doublets.
\end{itemize}

Equation~(\ref{2_S-prime_2}) constitutes a conjugacy relation
within the group U(2).
Thus,
HF symmetries associated with matrices $S$ and $S^\prime$
which are in the same conjugacy class of U(2)
correspond to the same model.
Moreover,
symmetries $S$ and $S^\prime$ related by
an overall phase transformation
($S^\prime = e^{i \xi} S$)
also lead to the same
physics,
since that overall phase transformation does not affect
the bilinears $\Phi_a^\dagger \Phi_b$.

Ferreira and Silva \cite{Ferreira:2008zy} have shown that
there are only two classes of HF symmetries
generated by one single generator
in the scalar potential of the 2HDM:
the discrete $Z_2$ symmetry
and a continuous U(1) symmetry
\be
\Phi_1 \rightarrow e^{-i \theta} \Phi_1, \quad
\Phi_2 \rightarrow e^{i \theta} \Phi_2,
\label{2_U1}
\ee
for an arbitrary $\theta$.
This U(1) symmetry
(suitably extended to the quark sector)
was first introduced by Peccei and Quinn~\cite{Peccei:1977hh}
in connection with the strong-CP problem.
The Higgs potential invariant under U(1)
has $m_{12}^2 = 0$ and $\lambda_5 = \lambda_6 = \lambda_7 = 0$
and is therefore also invariant under $Z_2$.

It is important to note that,
for instance,
a potential invariant under
\be
S_{2/3} =
\left( \begin{array}{cc}
e^{-i 2 \pi/3} & 0 \\
0 & e^{i 2 \pi/3}
\end{array} \right).
\label{2_S_2/3}
\ee
is automatically invariant
under the full Peccei--Quinn U(1) group.
Even though we only want to enforce a symmetry group
$Z_3 = \{ S_{2/3},\, S_{2/3}^2,\, S_{2/3}^3=1 \}$,
we automatically obtain a potential with full U(1) symmetry.
In fact,
invariance under any $Z_n$ group,
with $n>2$,
will lead us to a U(1)-invariant potential.
Another possibility of obtaining the same result
is to choose an irrational multiple of $\pi$
for the angle $\theta$ in eq.~\eqref{2_U1}.
This is an important point because continuous symmetries,
when broken,
may lead to massless scalars (Goldstone bosons).
An innocent-looking discrete symmetry
may have the same effect on the scalar potential as a continuous symmetry
and therefrom arises the possibility of undesired massless scalars.

We must however point out two caveats
to the discussion in the preceding paragraph.
The first caveat is that we are assuming a renormalizable theory,
from which we exclude all terms in the potential
with dimension larger than four.
If,
however,
we take the reasonable view that the 2HDM
is just the low-energy limit
of a larger theory,
and decide to include effective operators of dimensions five,
six,
or above,
then the equivalence between different symmetries
(such as the $Z_n$ with $n>2$,
all of them leading to the same U(1)-invariant scalar potential)
might no longer be verified.
The second caveat pertains to the fermionic sector:
given a specific symmetry of the scalar sector,
there are in general many ways of
extending that symmetry to the fermion sector,
often with completely different effects on the Yukawa terms.
We shall return to this issue in more detail in section~\ref{sec:fermions}.

One may also impose a symmetry with multiple generators
on the scalar potential.
For example,
the scalar potential invariant under both $Z_2$ and $\Pi_2$
{\em in the same basis} has
$m_{11}^2 = m_{22}^2$,
$m_{12}^2 = 0$,
$\lambda_1 = \lambda_2$,
and $\lambda_ 6 = \lambda_7 = \mbox{Im} \left( \lambda_5 \right) = 0$.
Thus,
the potential invariant under $Z_2 \times \Pi_2$
only has five parameters
($m_{11}^2$, $\lambda_1$, $\lambda_3$, $\lambda_4$,
and $\mbox{Re}\left( \lambda_5 \right)$) and,
indeed,
one may show \cite{Davidson:2005cw, Ferreira:2009wh} that
it is equivalent,
in a different basis,
to a potential with CP2 symmetry:
\be
Z_2 \times \Pi_2 \Leftrightarrow \textrm{CP2} \Leftrightarrow
\textrm{Class III}.
\label{2_equiv-CP2}
\ee
The U(2)-invariant potential
may similarly be obtained \cite{Ferreira:2009wh} through
the imposition of the CP3 and $U(1)$ symmetries in the same basis,
as can easily be seen in table~\ref{2_master1}.

One can also prove \cite{Ferreira:2010hy} that
the existence of either the $Z_2$
(or, equivalently, $\Pi_2$),
U(1), or U(2) symmetries
is sufficient to guarantee the existence of a basis choice in which
all the parameters of the scalar potential are real.
That is,
the corresponding scalar Higgs sectors are explicitly
CP-conserving. Therefore, all models belonging to the
classes in table~\ref{2_master1}
have CP-conserving scalar potentials.

\subsubsection{\label{sec:GCP}CP symmetries}

We now want to discuss the potentials obtained
by imposing one single GCP symmetry.
Writing $X$ as in eq.~(\ref{2_GCP-reduced}),
Ferreira, Haber, and Silva \cite{Ferreira:2009wh} have shown that
there are only three classes of potentials obtainable by imposing
a single GCP symmetry:
CP1 (class VI),
CP2 (class III),
and CP3 (class II).
The potential CP1 results from applying the GCP symmetry
with the matrix $X$ in eq.~(\ref{2_GCP-reduced}) with $\theta = 0$:
\be
\Phi_1 \to \Phi_1^\ast, \quad \Phi_2 \to \Phi_2^\ast;
\label{eq:cp1}
\ee
this is the standard CP symmetry,
which forces all coefficients in the potential to be real.
The potential CP2 results from applying the GCP symmetry
with the matrix $X$ in eq.~(\ref{2_GCP-reduced}) with $\theta = \pi / 2$:
\be
\Phi_1 \to \Phi_2^\ast, \quad \Phi_2 \to - \Phi_1^\ast.
\ee
The potential CP3 results from applying the GCP symmetry
with any other (arbitrary) angle $\theta \neq 0, \pi/2$.
The theories with symmetry CP2 and CP3 are
(of course) CP conserving,
but they have potentials more restrictive than CP1.

In the CP3 symmetry,
any single angle $\theta$ different from $0$ or $\pi/2$
in eq.~\eqref{2_GCP-reduced} leads to the same potential.
However,
if one wants to extend the CP symmetry to the Yukawa sector,
different values of $\theta$ will have
different consequences for the quark masses---only $\theta = \pi/3$
allows for six massive quarks \cite{Ferreira:2010bm}.

As mentioned regarding eq.~(\ref{2_equiv-CP2}),
we may reach class III of 2HDM scalar potentials
either by requiring symmetry under the GCP transformation
of eq.~\eqref{2_eq:CP2} or,
alternatively,
by requiring joint symmetry under $Z_2$ and $\Pi_2$.
There are,
indeed,
many other ways to obtain the class III scalar potential.
Similarly \cite{Ferreira:2009wh}
\be
U(1) \times \Pi_2 \equiv \textrm{CP3} \equiv \textrm{Class IV}.
\label{2_equiv-CP3}
\ee
In general,
there are many possible symmetries
leading into any of the six classes of symmetry-constrained 2HDM potentials.
The different symmetries are equivalent
with respect to the scalar potential,
but they may differ when one tries to extend them to the Yukawa sector.
Ferreira, Haber, and Silva \cite{Ferreira:2009wh} have proved that
i) except for the class VI potential,
all other five classes of potentials can be
obtained through multiple applications of HF symmetries;
ii)  one can obtain all six classes of scalar potentials
through multiple applications of the standard CP symmetry in different bases.
An interesting geometric interpretation
of these properties is presented in~\cite{Ferreira:2010yh}.

\subsubsection{The square of the GCP transformation}
\label{2_sec:cpsq}

Applying the GCP transformation twice to the scalar fields,
\be
\left( \Phi_a^{\textrm{GCP}} \right)^{\textrm{GCP}} =
\sum_{b=1}^2 X_{ab} \left( \Phi_b^{\textrm{GCP}} \right)^\ast =
\sum_{b,c=1}^2X_{ab} X_{bc}^\ast \Phi_c,
\ee
one obtains a HF symmetry with $S = X X^\ast$.
Thus,
$({\rm GCP})^2$ provides a distinction among the three GCP symmetries:
\begin{eqnarray}
(CP1)^2 &=& \boldsymbol{1},
\nonumber\\
(CP2)^2 &=& - \boldsymbol{1},
\nonumber\\
(CP3)^2 &=&
\left(
\begin{array}{cc}
  \phantom{-} \cos{2\theta} & \quad \sin{2\theta}\\
   - \sin{2\theta} & \quad \cos{2\theta}
\end{array}
\right).
\end{eqnarray}
While $\left( {\rm CP2} \right)^2$ is reduced to the identity transformation
through a global hypercharge transformation \cite{Maniatis:2007vn},
$\left( {\rm CP3} \right)^2$ is a non-trivial HF symmetry
of the class II scalar potential \cite{Ferreira:2009wh}.

\subsubsection{\label{2_sym_bilin}Symmetries and bilinears}

Ivanov's description of the possible classes
of scalar potentials in the 2HDM
is most conveniently summarized by looking at the corresponding
vectors and tensor of eqs.~(\ref{iudrt}) and~(\ref{nhgui}).
We start by looking at the implications of the symmetries we have
studied so far on the vector $\vec{r} = \{ r_1, r_2, r_3\}$,
whose components were introduced in eq.~(\ref{2_r_Ivanov}).
Notice that a unitary transformation $U$ on
the fields $\Phi_a$ induces an orthogonal
transformation $R$ on the vector of bilinears
$\vec{r}$,
given by eq.~(\ref{2_R_matrix}).
For every pair of unitary transformations $\pm U$
of $SU(2)$,
one can find some corresponding transformation $R$
of $SO(3)$,
in a two-to-one correspondence.
We then see what these symmetries imply
for the coefficients of eq.~(\ref{2_VH3})
(recall the $\Lambda_{\mu \nu}$ is a symmetric matrix).
Below, we list the transformation of $\vec{r}$ under which the
scalar potential is invariant, followed by the
corresponding constraints on the
quadratic and quartic scalar potential parameters, $M_\mu$ and
$\Lambda_{\mu\nu}$.

\begin{itemize}
\item The class I symmetry implies
\begin{equation}
\vec{r} \rightarrow
R\, \vec{r},
\hspace{10mm}
\left[
\begin{array}{c}
M_0\\
0\\
0\\
0
\end{array}
\right],
\hspace{4ex}
\left[
\begin{array}{cccc}
\Lambda_{00} &\,\,\, 0 & \,\,\,0 & \,\,\,0\\
 0 &\,\,\, \Lambda_{11} &\,\,\, 0 &\,\,\, 0\\
 0 & \,\,\,0 & \,\,\,\Lambda_{11} & \,\,\,0\\
0 & \,\,\,0 & \,\,\,0 &\,\,\, \Lambda_{11}
\end{array}
\right],
\label{Iv-SO3}
\end{equation}
where $R$ is an arbitrary $3\times 3$ orthogonal
matrix of unit determinant.
Thus the vectors $M_i$ and $\Lambda_{0i}$ vanish
while the tensor $\Lambda_{ij}$ is proportional to the unit matrix.
It is evident that class I has only three parameters
in the scalar potential.
\item For class II, the vectors $M_i$ and $\Lambda_{0i}$ once again vanish
while the tensor $\Lambda_{ij}$ has two---instead of three,
as in class I---degenerate eigenvalues.
For instance, if one applies the GCP transformation of
eq.~\eqref{2_eq:CP3}
\begin{equation}
\vec{r} \rightarrow
\left[
\begin{array}{ccc}
\phantom{-}c_2 & \phantom{-}0 & \phantom{-}s_2\\
\phantom{-}0 & -1 & \phantom{-}0\\
-s_2 & \phantom{-}0 & \phantom{-}c_2
\end{array}
\right]\ \vec{r},
\hspace{10mm}
\left[
\begin{array}{c}
M_0\\
0\\
0\\
0
\end{array}
\right],
\hspace{4ex}
\left[
\begin{array}{cccc}
\Lambda_{00} &\,\,\, 0 & \,\,\,0 & 0\,\,\,\\
0 &\,\,\, \Lambda_{11} &\,\,\, 0 &\,\,\, 0\\
0 &\,\,\, 0 &\,\,\, \Lambda_{22} &\,\,\, 0\\
0 &\,\,\, 0 &\,\,\, 0 & \,\,\,\Lambda_{11}
\end{array}
\right].
\label{Iv-CP3}
\end{equation}
It is evident that for this symmetry class there are four parameters
in the scalar potential.
\item In class III, the vectors $M_i$ and $\Lambda_{0i}$ vanish
and the eigenvalues of the tensor $\Lambda_{ij}$ are all different.
For instance, if one applies the GCP transformation of
eq.~\eqref{2_eq:CP2}
\begin{equation}
\vec{r} \rightarrow
\left[
\begin{array}{c}
-r_1\\
-r_2\\
-r_3
\end{array}
\right],
\hspace{10mm}
\left[
\begin{array}{c}
M_0\\
0\\
0\\
0
\end{array}
\right],
\hspace{4ex}
\left[
\begin{array}{cccc}
\Lambda_{00} &\,\,\, 0 &\,\,\, 0 &\,\,\, 0\\
0 &\,\,\, \Lambda_{11}\,\,\, &\,\,\, \Lambda_{12} & \,\,\,\Lambda_{13}\\
0 &\,\,\, \Lambda_{12} &\,\,\, \Lambda_{22} &\,\,\, \Lambda_{23}\\
0 &\,\,\, \Lambda_{13} & \,\,\,\Lambda_{23} & \,\,\,\Lambda_{33}
\end{array}
\right],
\label{Iv-CP2}
\end{equation}
but the $3 \times 3$ real symmetric matrix
\[
\left( \begin{array}{ccc}
\Lambda_{11} & \Lambda_{12} & \Lambda_{13} \\
\Lambda_{12} & \Lambda_{22} & \Lambda_{23} \\
\Lambda_{13} & \Lambda_{23} & \Lambda_{33}
\end{array} \right)
\]
may be diagonalized and has in general three distinct eigenvalues.
Therefore,
in this symmetry class the scalar potential has five parameters.
\item For class IV, the vectors $M_i$ and $\Lambda_{0i}$ are parallel
and,
moreover,
in the subspace orthogonal to those vectors the tensor $\Lambda_{ij}$
has degenerate eigenvalues.
For instance,
with the $U(1)$ symmetry of eq.~\eqref{2_eq:U1}, we have
\begin{equation}
\vec{r} \rightarrow
\left[
\begin{array}{ccc}
c_2 & -s_2 & \phantom{-}0\\
s_2 &\phantom{-}c_2 & \phantom{-}0\\
0 & \phantom{-}0 & \phantom{-}1
\end{array}
\right]\ \vec{r},
\hspace{10mm}
\left[
\begin{array}{c}
M_0\\
0\\
0\\
M_3
\end{array}
\right],
\hspace{4ex}
\left[
\begin{array}{cccc}
\Lambda_{00} &\,\,\, 0 & \,\,\,0 & \,\,\,\Lambda_{03}\\
 0 &\,\,\, \Lambda_{11} &\,\,\, 0 &\,\,\, 0\\
 0 & \,\,\,0 & \,\,\,\Lambda_{11} & \,\,\,0\\
\Lambda_{03} & \,\,\,0 & \,\,\,0 &\,\,\, \Lambda_{33}
\end{array}
\right],
\label{Iv-U1}
\end{equation}
It is evident that there are six independent parameters
in this scalar potential.
\item For class V, the vectors $M_i$ and $\Lambda_{0i}$ are parallel
and the tensor $\Lambda_{ij}$ has three non-degenerate eigenvalues.
For instance,
with the $Z_2$ symmetry of eq.~\eqref{2_eq:Z2},
\begin{equation}
\vec{r} \rightarrow
\left[
\begin{array}{c}
-r_1\\
-r_2\\
\phantom{-}r_3
\end{array}
\right],
\hspace{10mm}
\left[
\begin{array}{c}
M_0\\
0\\
0\\
M_3
\end{array}
\right],
\hspace{4ex}
\left[
\begin{array}{cccc}
\Lambda_{00} &\,\,\, 0 &\,\,\, 0 &\,\,\, \Lambda_{03}\\
 0 &\,\,\, \Lambda_{11} &\,\,\, \Lambda_{12} &\,\,\, 0\\
 0 & \,\,\,\Lambda_{12} &\,\,\, \Lambda_{22} &\,\,\, 0\\
\Lambda_{03} &\,\,\, 0 &\,\,\, 0 &\,\,\, \Lambda_{33}
\end{array}
\right],
\label{Iv-Z2}
\end{equation}
but the $2 \times 2$ real symmetric matrix
\[
\left( \begin{array}{cc}
\Lambda_{11} & \Lambda_{12} \\
\Lambda_{12} & \Lambda_{22}
\end{array} \right)
\]
may be diagonalized and has in general distinct eigenvalues.
Therefore,
the class V scalar potential has seven parameters.
\item For class VI, the vectors $M_i$ and $\Lambda_{0i}$ lie on the same plane
and the tensor $\Lambda$ has one eigenvector orthogonal to that plane.
For instance,
with the CP1 symmetry of eq.~\eqref{2_eq:CP1},
\begin{equation}
\vec{r} \rightarrow
\left[
\begin{array}{c}
\phantom{-}r_1\\
-r_2\\
\phantom{-}r_3
\end{array}
\right],
\hspace{10mm}
\left[
\begin{array}{c}
M_0\\
M_1\\
0\\
M_3
\end{array}
\right],
\hspace{4ex}
\left[
\begin{array}{cccc}
\Lambda_{00} & \,\,\,\Lambda_{01}  & \,\,\, 0 & \,\,\, \Lambda_{03}\\
\Lambda_{01} & \,\,\, \Lambda_{11} & \,\,\, 0 & \,\,\, \Lambda_{13}\\
 0 & \,\,\, 0 & \,\,\, \Lambda_{22} &  \,\,\,0\\
\Lambda_{03} &  \,\,\,\Lambda_{13} &  \,\,\,0 & \,\,\, \Lambda_{33}
\end{array}
\right],
\label{Iv-CP1}
\end{equation}
but the $2 \times 2$ real symmetric matrix
\[
\left( \begin{array}{cc}
\Lambda_{11} & \Lambda_{13} \\
\Lambda_{13} & \Lambda_{33}
\end{array} \right)
\]
may be diagonalized and has in general distinct eigenvalues.
Therefore,
in this symmetry class the scalar potential has nine parameters.
\end{itemize}

Since each unitary transformation of the fields $\Phi_a$
induces an $SO(3)$ transformation on the vector of bilinears $\vec{r}$,
and since the standard CP transformation
corresponds to an inversion of $r_2$,
\ie to a $Z_2$ transformation on the vector $\vec{r}$,
Ivanov \cite{Ivanov:2006yq}
actually considered all possible $O(3)$ transformations
acting on $\vec{r}$.
He has identified the following six classes of transformations:
(i) $Z_2$;
(ii) $(Z_2)^2$;
(iii) $(Z_2)^3$;
(iv) $O(2)$;
(v) $O(2) \otimes Z_2$;
and (vi) full $O(3)$.
No other independent symmetry transformations are possible.

\subsection{Bounded from below limits}
\label{2_sec:ufb}

Stability of the 2HDM potential requires that it be bounded from below,
{\em i.e.} that there is no direction in field space along which the
potential tends to minus infinity. This is a basic requirement for any
physical theory - the existence of a stable minimum, around which one can
perform perturbative calculations - which is satisfied by the scalar potential
of the SM
through the trivial condition
$\lambda > 0$, where $\lambda$ is the quartic coupling of the SM scalar
potential.

The 2HDM scalar potential of eq.~\eqref{2_VH1} is much more
complicated than the SM's, and ensuring its stability  requires that one studies all
possible directions along which the fields $\Phi_1$ and $\Phi_2$ (or rather,
their respective eight component fields) tend to arbitrarily large values. In
general, the existence of a non-trivial minimum - by which we mean the fields
$\Phi_i$ acquiring non-zero vacuum expectation values - implies two conditions
on the potential's parameters. They have to be such that:
the quartic part of the scalar potential, $V_4$, is positive for
arbitrarily large values of the component fields, but
the quadratic part of the scalar potential, $V_2$, can take negative
values for at least some values of the fields.

The restrictions on $V_4$ need to be handled carefully:
\begin{itemize}
\item Demanding that $V_4 > 0$ for all $\Phi_i \rightarrow \infty$ is a
{\em strong stability} requirement. This may, however, be too strong, since
several interesting models are excluded by it.
For instance, in tree-level SUSY potentials there is a direction
($\langle \Phi_1 \rangle = \langle \Phi_2 \rangle$) for which $V_4 = 0$.
\item We can also demand stability in a {\em marginal} sense, by requiring $V_4 \geq 0$,
for any direction in field
space tending to infinity.
\item The equality in the marginal stability
bound comes at a price: if there is a given direction in field space such that
$V_4 \rightarrow 0$, it is necessary to demand that, along {\em that} specific
direction, one has $V_2 \geq 0$.
\end{itemize}
A simple way to obtain {\em necessary} conditions on the quartic parameters of the
potential is to study its behaviour along specific field directions. Considering for instance
the direction $|\Phi_1| \rightarrow \infty$ and $|\Phi_2| = 0$, the
expression~\eqref{2_VH1} for the potential renders it obvious that one
can have positive values for $V_4$ if and only if $\lambda_1 \geq 0$.
Likewise, the direction $|\Phi_1| = 0$ and $|\Phi_2| \rightarrow \infty$ gives us
the condition $\lambda_2 \geq 0$. And if one takes $|\Phi_1|^2 = r\cos\theta$, $|\Phi_2|^2 = r\sin\theta$
(with $0 < \theta < \pi/2$ and $r \rightarrow +\infty$) but such that $\Phi_1^\dagger\Phi_2 = 0$
(for instance considering only non-zero upper components for $\Phi_1$ and lower ones
for $\Phi_2$), the bounded-from-below condition becomes
\be
\lim_{r\rightarrow + \infty}
r^2\left(\frac{\lambda_1}{2} \cos^2\theta + \frac{\lambda_2}{2} \sin^2\theta +
\lambda_3 \sin\theta\cos \theta\right) = r^2 f(\theta) \geq 0\;,\; \forall_\theta .
\ee
Minimizing $f(\theta)$ with respect to $\theta$ to obtain its smallest value and demanding
that it be larger or equal to zero, one finds that the coefficients $\lambda_i$ need to obey
$\lambda_3 \geq -\sqrt{\lambda_1 \lambda_2}$.
By studying several such directions, it is possible to reach other conditions on the couplings,
and we can gather all as
\ba
\lambda_1 \geq 0 & , &  \lambda_2 \geq 0 \; ,\nonumber \\
\lambda_3 \geq -\sqrt{\lambda_1 \lambda_2} & , &
\lambda_3 + \lambda_4 - |\lambda_5| \geq -\sqrt{\lambda_1 \lambda_2} \;,
\label{2_eq:ufb}
\ea
where we've taken $\lambda_5$ to be real. In~\cite{Klimenko:1984qx,Maniatis:2006fs} it was proven that, in potentials where
one has $\lambda_6 = \lambda_7 = 0$, these are actually necessary and sufficient conditions to ensure
the positivity of the quartic potential along all directions. As we can see from the discussion in
section~\ref{sec:symmetries},
most of the possible symmetry-constrained 2HDM scalar potentials fall
unto this category - there is at least a basis where
$\lambda_6 = \lambda_7 = 0$ holds, for the $Z_2$, $U(1)$, CP2, CP3 and $U(2)$ models. For the remaining
possibilities - a model with CP1 symmetry, or a model with no symmetry at all (other than the gauge
ones), one can find necessary and sufficient conditions for boundedness, involving $\lambda_6$ and
$\lambda_7$ in the work of~\cite{Maniatis:2006fs} - unfortunately, they do not have a simple analytical expression,
like those of eq.~\eqref{2_eq:ufb}, but they can be handled numerically. In~\cite{Maniatis:2006fs}, all possible
cases - strong stability, marginal stability, analysis of the quadratic terms - were considered. Equivalent conditions, for the strong stability requirements, were found in~\cite{Ivanov:2006yq}. Again,
they are not easily translated into analytical bounds. It is not difficult, though, to find
{\em necessary} conditions involving $\lambda_6$ and $\lambda_7$. For the case they are
real~\cite{Ferreira:2004yd, Ferreira:2009jb},
one finds
\be
2\,|\lambda_6\,+\,\lambda_7|\;<\;\frac{\lambda_1\,+\,\lambda_2}{2}\,+\,
\lambda_3\,+\,\lambda_4\,+\,\lambda_5\;\;\; . \label{2_eq:unb2}
\ee

As mentioned earlier, the requirement of strong stability is too constraining, in the sense that it
actually excludes potentially interesting models. The obvious example is the SUSY potential, for which
the quartic couplings are related to the gauge coupling constants, and they are given by eq.~\eqref{2_eq:susylim}.
With these couplings, the last bound
in eq.~\eqref{2_eq:ufb} is saturated. Likewise, in the CP3 model, for which $\lambda_5 = \lambda_3 +
\lambda_4 - \lambda_1$ and $\lambda_1 = \lambda_2$, the last bound in eq.~\eqref{2_eq:ufb} gives us
$\lambda_1 + \lambda_3 + \lambda_4 \geq 0$ and $\lambda_1 \geq \lambda_1$, the latter clearly demanding the equality, lest the model be excluded.

For most cases, the conditions~\eqref{2_eq:ufb} are all we need. They become specially important if one wishes to analyse the stability of the potential including higher order corrections. Clearly they were obtained through tree-level analysis, requiring that the tree-level potential always be convex. But
one may wonder whether those conditions ensure that the one-loop corrected effective potential
is also bounded from below. The one-loop corrections to the tree-level potential are of the form
\be
V(\phi_i) = V_{tree}(\phi_i) \,+\,\frac{1}{64\pi^2} \sum_{\alpha} m^4_\alpha(\phi_i) \left[\log
\left(\frac{m^2_\alpha(\phi_i)}{\mu^2}\right) - \frac{3}{2}\right],
\ee
where the sum runs over all helicity states of the particles of masses $m_\alpha$
present in the theory, and $\mu$ is the renormalization scale. We have kept the dependence on
 the fields $\phi_i$ explicit. Unbounded from below limits are found analysing the behaviour
of the potential for very large values of the fields $\phi_i$, at which point one should worry
about the appearance of potentially large contributions from the logarithms in the expression
above. A renormalization group (RG) improvement of the bounds then amounts to considering only
the tree-level expressions we have already discussed, but considering the values of the couplings
which appear in those expressions at different renormalization scales. In other words, we
take the bounds from Eqs.~\eqref{2_eq:ufb} and run the couplings therein, using the
$\beta$-functions of the model (see Appendix~\ref{2_sec:rge}), along a range of scales $\mu$ -
from the weak scale $M_Z$ to an upper scale $\Lambda$;
at all scales in the interval chosen, the bounds must hold, and in this way combinations
of parameters which at one scale might be acceptable would violate the bounds at another
scale.

This type of analysis was performed in the
SM~\cite{Lindner:1985uk,Sher:1988mj,Sher:1993mf,Casas:1994qy,Casas:1996aq,Espinosa:1995se}.
There, the Higgs potential quartic
coupling $\lambda$ has a $\beta$-function with a sizeable negative contribution
from the top quark Yukawa. The top being so heavy, this term tends to decrease the value
of $\lambda$ at higher renormalization scales. Thus, if the starting value (at the weak
scale, say) of $\lambda$ is  small, the coupling may well become negative at some higher
scale, and the potential would suddenly be unbounded from below. In this manner we can thus
put a {\em lower} bound on $\lambda$ and thus on the Higgs mass. On the other hand, if the
starting value of $\lambda$ is too large, the RG evolution of the coupling will increase
its value immensely and eventually the theory becomes non-perturbative. We will
return to this when we consider another class of theoretical bounds, in which one requires
unitarity of all 2HDM processes involving scalars, which we'll look into in
Appendix~\ref{2_sec:uni}. Thus, the RG
analysis allows us to impose both higher and lower bound on the mass of the Higgs particles.

In the 2HDM the same type of phenomena can occur, and was treated
in, for instance,
\cite{Sher:1988mj,Kreyerhoff:1989fa,Freund:1992yd,Kastening:1992by,Nie:1998yn,Kanemura:1999xf,Ferreira:2009jb}.
If for instance the $\Phi_1$ is made to
couple to the up quarks, the $\beta$-function for the $\lambda_1$ quartic coupling will
have a large negative top Yukawa contribution, and a similar analysis to the SM case
will hold. Now, however, many other quartic couplings are present and more bounds
need to be obeyed. Nonetheless, the main conclusions hold: smaller values for some
of the $\lambda_i$ at the weak scale are disfavoured as they lead to unbounded from
below potentials at higher scales; and large values of those couplings lead to
Landau poles at high scales, thus a breakdown of perturbation theory. These translate into
bounds on the several Higgs masses. Several observations are in order:
\begin{itemize}
\item Clearly the bounds obtained will depend on what the upper renormalization
scale $\Lambda$ is. Usually this is taken to be the gauge unification scale,
$\sim 10^{16}$ GeV. Varying this scale will change (mostly) the upper bounds on
the masses - the upper bound on the lightest CP-even scalar can change from about
300 to about 100 GeV, if one varies $\Lambda$ from $10^3$ to $10^{16}$
GeV~\cite{Kanemura:1999xf}.
\item The precise values for the bounds have a noticeable
dependence on the value of the top pole mass. This is to expected since the
top quark Yukawa is what drives most of the quartic coupling RG evolution.
Typically, at most some of the bounds can change by roughly $\sim 10$ GeV
for a 5 GeV change in the top pole mass~\cite{Kanemura:1999xf,Ferreira:2009jb}.
\item Most of the analysis performed considered a 2HDM with an intact $Z_2$
symmetry. However, as shown in~\cite{Kanemura:1999xf}, the soft breaking
term $m_{12}$ has a crucial importance in the bounds obtained, which are increasingly
relaxed the more the magnitude of $m_{12}$ increases. Roughly speaking,
with $\Lambda = 10^{16}$ GeV, the
exact $Z_2$ model gives us an upper bound on the lightest CP-even Higgs mass
of roughly 100 GeV, which is easily raised to about 185 GeV for the soft
broken model~\cite{Kanemura:1999xf,Ferreira:2009jb}. For some cases, like
the charged Higgs mass, large values of the soft breaking term eliminate
the upper bound obtained for the exact symmetry (roughly 160 GeV).
\item In~\cite{Ferreira:2009jb} other 2HDM theories were considered,
namely the inert vacua (see section~\ref{2_sec:ssb}) of the $Z_2$
potential, the Peccei-Quinn $U(1)$ model
and the model without $Z_2$ symmetry (dubbed the $CP1$ model in
section~\ref{sec:symmetries}. The Peccei-Quinn and CP1 model bounds obtained
do not differ significantly from the $Z_2$ case. The inert model has quite
restrictive bounds on the two CP-even scalars.
\item In~\cite{Ferreira:2009jb} these bounds were also applied to
theories in which the vaccum of the theory spontaneously breaks
CP (again, see section~\ref{2_sec:ssb}). Such theories always have
the $m_{12}$ term, which can be quite large. However, it was found that
the lightest Higgs scalar has a very low upper bound, roughly 85 GeV.
\item These analyses only take into account the top Yukawa coupling,
coupled to only one of doublets. Even for a theory with
flavour changing neutral interactions, this can be considered a good
approximation.
\end{itemize}

The general conclusion is that requiring that the potential be bounded
from below in the range of scales from $M_Z$ to $\Lambda$ can severely limit
the parameters of the theory, namely the masses of the scalar eigenstates.
However, such bounds are heavily model-dependent - they depend on whether
the 2HDM considered has a symmetry, or if that symmetry is softly broken,
and they depend immensely on what the upper scale $\Lambda$ is taken to be.
Notice, for instance, that we simply do not know what the value of $\Lambda$
should be; if one thinks of the 2HDM as an effective theory
$\Lambda$ should be taken as the scale above which new fields have to be
considered, and it could be as low as 1 TeV.  Hence, great care must be
exercised when applying such bounds, lest one
exclude regions of parameter space which may well be important.

\subsection{Spontaneous symmetry breaking}
\label{2_sec:ssb}

If the scalar potential of the 2HDM is bounded from below, being a quartic polynomial function
it will certainly have a global minimum somewhere. This same argument applies to the SM, but there
we can only have two types of minima: the ``trivial" one, for which the Higgs acquires zero vevs,
and the usual one, where electroweak symmetry breaking occurs, away from the origin, for
$\langle \Phi \rangle = v/\sqrt{2}$.
In particular, vacua which break electric charge or CP conservation are impossible in the SM. In what follows,
we consider a vacuum any stationary point of the potential, regardless of whether it is a
minimum or not.

In the 2HDM, the vacuum structure is much richer. We can have three types of vacua (other than the trivial
case, $\langle \Phi_1 \rangle = \langle \Phi_2 \rangle = 0$):
\begin{itemize}
\item ``Normal" (N) vacua, with vevs which do not have any complex relative phase and can thus be
trivially rendered real:
\be
\langle\Phi_1 \rangle_N = \begin{pmatrix} 0 \\ {\displaystyle\frac{v_1}{\sqrt{2}}} \end{pmatrix} \; , \;
\langle\Phi_2 \rangle_N = \begin{pmatrix} 0 \\ {\displaystyle\frac{v_2}{\sqrt{2}}} \end{pmatrix} ,
\label{2_eq:vacn}
\ee
where $v = \sqrt{v_1^2 + v_2^2} = $ 246 GeV~\footnote{Notice, however, that certain 2HDM potentials can
have more than one solution of this type, with different values for $v$. See the discussion for
eq.~\eqref{2_eq:vnvn}.} and one defines $\tan\beta = v_2/v_1$. This solution,
of course, is the 2HDM equivalent of the SM vacuum. We can distinguish a special case here,
in which the minimization conditions allow for one of the vevs ${v_1 , v_2}$ to be zero. These are
called ``inert models", already discussed in section~\ref{sec:inert}. Notice that, unlike the passage
to the Higgs basis (in which only one of the doublets has a vev, as well), the inert vacua are found in the
basis where a $Z_2$ (or $U(1)$, or other) symmetry is manifest.

\item CP breaking vacua, where the vevs do have a relative complex phase, that is
\be
\langle\Phi_1 \rangle_{CP} = \begin{pmatrix} 0 \\ {\displaystyle\frac{\bar{v}_1}{\sqrt{2}}}
 e^{i\theta} \end{pmatrix} \; , \;
\langle\Phi_2 \rangle_{CP} = \begin{pmatrix} 0 \\ {\displaystyle\frac{\bar{v}_2}{\sqrt{2}}}
\end{pmatrix} ,
\label{2_eq:vaccp}
\ee
with real values for $\bar{v}_1$ and $\bar{v}_2$. The moniker ``CP breaking" is not the most
appropriate, since such vacua are possible even in potentials where the CP symmetry is not defined
(due to it being explicitly broken) - see for instance~\cite{Barroso:2007rr}. Also, the presence of a phase
in~\eqref{2_eq:vaccp} is not a guarantee of spontaneous CP breaking (see section~\ref{sec:cpv}).
\item Charge breaking (CB) vacua, in which one of the vevs carries electric charge,
\be
\langle\Phi_1 \rangle_{CB} = \begin{pmatrix} {\displaystyle\frac{\alpha}{\sqrt{2}}} \\
{\displaystyle\frac{v^\prime_1}{\sqrt{2}}} \end{pmatrix} \; , \;
\langle\Phi_2 \rangle_{CB} = \begin{pmatrix} 0 \\ {\displaystyle\frac{v^\prime_2}{\sqrt{2}}}
\end{pmatrix} ,
\label{2_eq:vaccb}
\ee
with real numbers $v^\prime_1$, $v^\prime_2$, $\alpha$. Due to the presence of a non-zero vev in an
upper component (charged) of the fields, this vacuum breaks electrical charge conservation,
causing the photon to acquire a mass. Thus, they are to be avoided at all costs.
\end{itemize}
Given our definition for the charge
(such that the lower components of fields are neutral),
the vacuum will break the charge and lead
to a massive photon if and only if
\be
\langle \varphi_1^{ +} \rangle\
\langle \varphi_2^{ 0} \rangle \,=\,\alpha v^\prime_2
\neq 0.
\ee
One can give a definition for a charge breaking vacuum
which does not depend on our definition of charge
or on our basis choice~\cite{Barroso:2006pa},
\be
\left| \langle \varphi_1^{ +} \rangle\
\langle \varphi_2^{ 0} \rangle\ \right|^2
+
\left| \langle \varphi_1^{ 0} \rangle\
\langle \varphi_2^{ +} \rangle\ \right|^2
-
2 \textrm{Re} \left(
\langle \varphi_1^{ +} \rangle\
\langle \varphi_2^{ 0} \rangle\\,
\langle \varphi_1^{ 0 \ast} \rangle
\langle \varphi_2^{ + \ast} \rangle
\right)
\neq 0.
\ee
Charge preserving vacua (many times called `` aligned vacua") for the
2HDM and for models with additional singlets or triplets were
studied in Ref.~\cite{DiazCruz:1992uw}.

That all possible vacua in the 2HDM reduce to one of the three forms of eqs.~\eqref{2_eq:vacn},
\eqref{2_eq:vaccp} and \eqref{2_eq:vaccb} can be seen using the freedom
to choose a particular gauge in $SU(2)_L\times U(1)_Y$. Let us write the two doublets in the following
simplified manner:
\be
\Phi_1 = \begin{pmatrix} |\varphi_1^+|\,e^{i\theta_1^+} \\
|\varphi_1^0|\,e^{i\theta_1^0} \end{pmatrix} \;, \;
\Phi_2 = \begin{pmatrix} |\varphi_2^+|\,e^{i\theta_2^+} \\
|\varphi_2^0|\,e^{i\theta_2^0} \end{pmatrix}.
\label{2_eq:dub1}
\ee
The complex phases above will, in general, be functions of
the space-time coordinates.
Then, the local gauge transformation $U_1$, given by the $SU(2)_L$ matrix
\be
U_1\;=\; \begin{pmatrix} u_{11} & u_{12} \\ -u_{12}^* & u_{11}^* \end{pmatrix}
\ee
with
\ba
u_{11} &=& \frac{|\varphi_2^0|}{\sqrt{|\varphi_2^0|^2 + |\varphi_2^+|^2}} \nonumber \\
u_{12} &=& - \frac{|\varphi_2^+|}{\sqrt{|\varphi_2^0|^2 + |\varphi_2^+|^2}}\,
e^{i(\theta_2^+ - \theta_2^0)} ,
\ea
eliminates the upper components of $\Phi_2$. A combined hypercharge and $SU(2)_L$ gauge transformation
can then be used to eliminate the phases of the upper and lower components of $\Phi_1$,
through the matrix
\be
U_2\;=\; \begin{pmatrix} e^{-i\theta_1^+} & 0 \\0 & e^{-i\theta_1^0}\end{pmatrix}.
\ee
Of course, these phases $\theta_1^+$ and $\theta_1^0$ are not the same that appear in
eq.~\eqref{2_eq:dub1} (they have been changed by the gauge transformation $U_1$),
but that is irrelevant for the argument. The final form of the doublets is thus
\be
\Phi_1 = \begin{pmatrix} \mbox{Re}(\varphi_1^+) \\ \mbox{Re}(\varphi_1^0) \end{pmatrix} \; , \;
\Phi_2 = \begin{pmatrix} 0 \\ \varphi_2^0 \end{pmatrix}.
\ee
This is almost  of the form of eq.~\eqref{2_eq:vaccb}, but it has a
complex lower component in $\Phi_2$~\cite{Sher:1988mj}. However, that phase is physically
irrelevant, since it can always be absorbed through a trivial basis transformation -
a re-phasing of $\Phi_2$, $\Phi_2 \rightarrow e^{-i\theta_2^0} \Phi_2$. As such, the form of the
vevs in eq.~\eqref{2_eq:vaccb} is indeed the most general one we need: there exists always a basis
for which the most generic vacuum will have that form. This conclusion could also have been reached
through a series of basis changes, plus a gauge transformation~\cite{Barroso:2006pa}. That
method has the advantage of being easily generalized for an N-Higgs doublet model.

Let us now look in more detail at the solutions of the minimization conditions.
Writing the potential
in terms of the vevs $\tilde{v}_i$ (for any of the three sets~\eqref{2_eq:vacn}
to~\eqref{2_eq:vaccb}), a stationary
point of the potential is found if the set of equations
$\partial V/\partial \tilde{v}_i = 0$ has solutions. In terms
of the notation introduced in section~\ref{2_sec:notation}, and for
completely general vacuum
expectation values such that
$\langle \Phi_a \rangle = \tilde{v}_a/\sqrt{2}$~\footnote{
Notice that we are
using $\tilde{v}_a/\sqrt{2}$ for the complex vev $\langle \Phi_a \rangle$, while
$v_a = |\tilde{v}_a|$ is real.}, the extremum conditions
may be written as
\begin{equation}
\sum_{b=1}^2 \left[ \mu_{ab}
+
\tfrac{1}{4} \sum_{c,d=1}^2 \lambda_{ab,cd}\, \tilde{v}_d^\ast \tilde{v}_c \right]\
\tilde{v}_b = 0
\hspace{3cm}(\textrm{for\ } a = 1,2).
\label{2_stationarity_conditions}
\end{equation}
Multiplying by $\tilde{v}_a^\ast$ leads to
\begin{equation}
\sum_{a,b=1}^2 \mu_{ab} (\tilde{v}_a^\ast \tilde{v}_b) =
- \tfrac{1}{4} \sum_{a,b,c,d=1}^2\lambda_{ab,cd}\,
(\tilde{v}_a^\ast \tilde{v}_b)\, (\tilde{v}_d^\ast \tilde{v}_c).
\label{2_aux_1}
\end{equation}
If one performs a basis transformation such as the one presented
in~\eqref{2_basis-transf}, the vevs are transformed as
\begin{equation}
\tilde{v}_a \rightarrow \tilde{v}_a^\prime = \sum_{b=1}^2 U_{a b} \tilde{v}_b.
\label{2_vev-transf}
\end{equation}
And, for a GCP transformation like in~\eqref{2_GCP},
the vevs transform as
\begin{equation}
\tilde{v}_a \rightarrow \tilde{v}_a^\textrm{GCP} = X_{a \alpha}
\sum_{\alpha =1}^2 \tilde{v}_\alpha^\ast.
\label{2_vev-CPtransf}
\end{equation}

The different CB and CP stationary points are determined by a set of three equations,
a normal one by only two. In fact, since the 2HDM potential depends on eight real component
fields, any stationary point ought to be the solution of a set of eight equations on eight
unknowns which arise from~\eqref{2_stationarity_conditions}. However, given that one can always
choose the simplified forms of the vevs in
eqs.~\eqref{2_eq:vacn} to~\eqref{2_eq:vaccb}, most of those equations are trivially satisfied.

As was shown in~\cite{Ferreira:2004yd} (and later in~\cite{Maniatis:2006fs,Ivanov:2006yq,Nishi:2007nh}),
the CB vevs can always be obtained analytically, and are given by
\be
\begin{pmatrix} {v^\prime}^2_1 + \alpha^2 \\ {v^\prime}^2_2 \\ v^\prime_1 v^\prime_2 \end{pmatrix} =
2\,\begin{pmatrix} \lambda_1 & \lambda_3 & 2\mbox{Re}(\lambda_6) \\
\lambda_3 & \lambda_2 & 2\mbox{Re}(\lambda_7) \\
2\mbox{Re}(\lambda_6) & 2\mbox{Re}(\lambda_7) & 2(\lambda_4 + \mbox{Re}(\lambda_5)) \end{pmatrix}^{-1}
\begin{pmatrix} m^2_{11} \\ m^2_{22} \\ -2 \mbox{Re}(m^2_{12}) \end{pmatrix} .
\label{2_eq:cbsol}
\ee
This expression has an important consequence: if eq.~\eqref{2_eq:cbsol} admits a solution, it is unique, up to
trivial sign changes ($\alpha \rightarrow -\alpha$, $v^\prime_1 \rightarrow -v^\prime_1$ and
$v^\prime_2 \rightarrow -v^\prime_2$) with no physical impact. Charge breaking is in fact impossible in several
symmetry-constrained 2HDM.

Likewise, the CP vacua vevs can always be obtained analytically in terms of the potential's parameters. Restricting
ourselves to potentials where the CP symmetry is defined - {\em i.e.}, where it isn't explicitly broken, see
section~\ref{sec:cpv} - we obtain
\be
\begin{pmatrix} \bar{v}^2_1 \\ \bar{v}^2_2 \\ \bar{v}_1 \bar{v}_2 \cos\theta \end{pmatrix} =
2\,\begin{pmatrix} \lambda_1 & \lambda_3 + \lambda_4 - \mbox{Re}(\lambda_5)) & 2\mbox{Re}(\lambda_6) \\
\lambda_3 + \lambda_4 - \mbox{Re}(\lambda_5)& \lambda_2 & 2\mbox{Re}(\lambda_7) \\
2\mbox{Re}(\lambda_6) & 2\mbox{Re}(\lambda_7) & 4\mbox{Re}(\lambda_5) \end{pmatrix}^{-1}
\begin{pmatrix} m^2_{11} \\ m^2_{22} \\ -2 \mbox{Re}(m^2_{12}) \end{pmatrix} .
\label{2_eq:cpsol}
\ee
Again, up to physically irrelevant sign changes, the CP vacuum is {\em unique}.

The normal vacuum turns out to be the most difficult to solve. In fact, for many potentials, the minimization
conditions cannot be solved analytically. The equations $\partial V/\partial v_1 = 0$ and
$\partial V/\partial v_2 = 0$ give, for the most general 2HDM potential,
\ba
m_{11}^2 v_1 - \mbox{Re}(m_{12}^2) v_2 + \frac{\lambda_1}{2} v_1^3 + \frac{\lambda_{345}}{2} v_1 v_2^2 +
\frac{1}{2}\left[3\mbox{Re}(\lambda_6)v_1^2 v_2 + \mbox{Re}(\lambda_7)v_2^3\right] & = 0 \nonumber \\
m_{22}^2 v_2 - \mbox{Re}(m_{12}^2) v_1 + \frac{\lambda_2}{2} v_2^3 + \frac{\lambda_{345}}{2} v_2 v_1^2 +
\frac{1}{2}\left[\mbox{Re}(\lambda_6)v_1^3 + 3\mbox{Re}(\lambda_7)v_2 v_1^2\right] & = 0 ,
\label{2_eq:nsol}
\ea
where $\lambda_{345} = \lambda_3 + \lambda_4 + \mbox{Re}(\lambda_5)$.
A few important observations about these equations:
\begin{itemize}
\item For some models (unbroken $Z_2$, $U(1)$, CP2, CP3 and $U(2)$) these equations can be solved analytically.
However, the presence of soft breaking terms may prevent that.
\item For any models in which $m^2_{12} = \lambda_6 = \lambda_7 = 0$ (for instance, models with a $Z_2$ symmetry
or higher), eqs.~\eqref{2_eq:nsol} may admit
solutions of the form ${v_1 \neq 0, v_2 = 0}$ and ${v_1 = 0, v_2 \neq 0}$. In the first case we have
\be
v_1^2 = -\frac{2 m^2_{11}}{\lambda_1} ,
\label{2_eq:in1}
\ee
in the second
\be
v_2^2 = -\frac{2 m^2_{22}}{\lambda_2} ,
\label{2_eq:in2}
\ee
as long that, of course, $m^2_{11}<0$ or $m^2_{22}<0$ . These lead to the so-called
Inert models~\cite{Deshpande:1977rw,Barbieri:2006dq}.
The name derives from the fact that, unlike solutions
of eq.~\eqref{2_eq:nsol} which have ${v_1 \neq 0, v_2 \neq 0}$,
these vacua lead to scalar particles which do not couple to gauge bosons, and can easily
be made to decouple from fermions.
As such, these models provide excellent candidates for dark matter.
\item The equations~\eqref{2_eq:nsol} do not, in general, have a {\em unique} solution. Even
when they do not admit inert vacua they can lead, depending on the values of the parameters,
to several sets of vevs $\{v_1 , v_2\}$ which are not related by trivial
sign changes~\cite{Barroso:2007rr}. In~\cite{Ivanov:2007de} it was however proven that there
can be no more than {\em two} such solutions which
are simultaneously minima of the potential.
\end{itemize}

\subsection{Vacuum stability}
\label{2_sec:vacstab}

In the SM there is only {\em one} possible type of vacuum, other than the trivial
one. Indeed, in that theory the scalar potential is such that one
can only have one minimum. In theories with more than one scalar, however, there
is the possibility that minima of different natures occur, and thus that the theory
may allow for tunneling from one minimum to another. An example of this behaviour
occurs in SUSY models, where the existence of many charged and/or coloured scalar
fields gives rise to possible charge and/or colour breaking minima~\cite{Frere:1983ag}.
These minima would imply massive photons and/or gluons, and as such one wishes to
avoid them. Hence, it is desirable to impose bounds on the theory's
parameters to ensure that the {\em global} minimum preserves the SM's symmetries.

Colour breaking is impossible in the 2HDM, but we have already seen, in
section~\ref{2_sec:ssb}, that charged vevs are possible. Also, one may have CP breaking
vevs. Thus, the question arises: can these vacua of different natures coexist with one
another? Could one tunnel, for instance, from a normal minimum to a deeper charge-breaking
one? In other words, given a minimum in the 2HDM, is it stable? The limited
number of scalars, and the inexistence of cubic terms in the potential, in the 2HDM
allows us to treat this question in a fully analytical way. It has been possible to
show that~\cite{Ferreira:2004yd,Barroso:2005sm,Barroso:2007rr}:
\begin{itemize}
\item For a potential where a normal stationary point and a charge breaking
one exist, with vevs as given by eqs.~\eqref{2_eq:nsol} and~\eqref{2_eq:cbsol},
the difference in the values of the scalar potential at both those vacua (respectively
$V_N$ and $V_{CB}$) is given by
\be
V_{CB} - V_N\,=\,\left(\frac{M^2_{H^\pm}}{4 v^2}\right)_N\,
\left[(v^\prime_1 v_2 - v^\prime_2 v_1)^2 + \alpha^2 v_2^2\right]\, ,
\label{2_eq:vcbvn}
\ee
where $(M^2_{H^\pm}/4 v^2)_N$ is the ratio of the squared mass of the charged scalar
to the sum of the square of vevs, $v^2 = v_1^2 + v_2^2$, as computed in the normal
stationary point.
\end{itemize}
The significance of~\eqref{2_eq:vcbvn} is plain: {\em if} the normal stationary
point is a {\em minimum} (which implies that $M^2_{H^\pm} > 0$) then one will
necessarily have $V_{CB} - V_N > 0$. That is, if there is a normal minimum, any
CB stationary point will lie {\em above} it - the normal minimum is stable against
charge breaking. In~\cite{Ferreira:2004yd} it was also proven that in that case
the CB stationary point is necessarily a saddle point. Thus, normal and CB minima
cannot coexist in the 2HDM. Of course, it is possible to choose sets of parameters
of the potential such that the global minimum of the potential breaks charge  -
but in that case no normal minima will exist.
\begin{itemize}
\item For a potential where a normal stationary point and a CP breaking
one exist, with vevs as given by eqs.~\eqref{2_eq:nsol} and~\eqref{2_eq:cpsol},
the difference in the values of the scalar potential at both those vacua (respectively
$V_N$ and $V_{CP}$) is given by
\be
V_{CP} - V_N\,=\,\left(\frac{M^2_A}{4 v^2}\right)_N\,
\left[(\bar{v}_1 v_2 \cos\theta - \bar{v}_2 v_1)^2 + \bar{v}_1^2 v_2^2 \sin^2\theta\right]\, ,
\label{2_eq:vcpvn}
\ee
where $(M^2_A/4 v^2)_N$ is the ratio of the squared mass of the pseudoscalar
to the sum of the square of vevs, $v^2 = v_1^2 + v_2^2$, as computed in the normal
stationary point.
\end{itemize}
The significance of~\eqref{2_eq:vcpvn} is plain: {\em if} the normal stationary
point is a {\em minimum} (which implies that $M^2_A > 0$) then one will
necessarily have $V_{CP} - V_N > 0$. That is, if there is a normal minimum, any
CP stationary point will lie {\em above} it - the normal minimum is stable against
CP breaking. In~\cite{Ivanov:2007de} it was also proven that in that case
the CP stationary point is necessarily a saddle point. Thus, normal and CP minima
cannot coexist in the 2HDM. Of course, it is possible to choose sets of parameters
of the potential such that the global minimum of the potential breaks CP  -
but in that case no normal minima will exist.
\begin{itemize}
\item No CB and CP minima can coexist either. This derives from the fact that for
the CP vacuum the square of the charged Higgs mass is given by
\be
(M^2_{H^\pm})_{CP} = -\frac{1}{2}[\lambda_4 - \mbox{Re}(\lambda_5)]
(\bar{v}_1^2 + \bar{v}_2^2),
\ee
whereas in a CB vacuum one of the squared mass
matrix eigenvalues is
\be
M^2_{CB} = \frac{1}{2}[\lambda_4 - \mbox{Re}(\lambda_5)]
({v^\prime_1}^2 + {v^\prime_2}^2 + \alpha^2).
\ee
As we see, the sign of
$\lambda_4 - \mbox{Re}(\lambda_5)$ determines that both these vacua cannot be
simultaneously minima. Thus, if a CP minimum exists the (unique) CB stationary point,
if it exists, {\em cannot} be a minimum as well, and vice-versa.
\item Unlike the CB and CP cases, the normal minimization conditions allow for
multiple solutions, so that one can have an $N_1$ vacuum with vevs
$\{v_{1,1}\, , \, v_{2,1}\}$ and an $N_2$ vacuum with different vevs
$\{v_{1,2}\, , \, v_{2,2}\}$. In that case, the difference in the values of
the potential in those two vacua (respectively, $V_{N_1}$ and $V_{N_2}$) is
given by
\be
V_{N_2} - V_{N_1}\,=\,\frac{1}{4}\,\left[\left(\frac{M^2_{H^\pm}}{v^2}\right)_{N_1}
- \left(\frac{M^2_{H^\pm}}{v^2}\right)_{N_2}\right]\,
(v_{1,1} v_{2,2} - v_{2,1} v_{1,2})^2\, ,
\label{2_eq:vnvn}
\ee
where $(M^2_{H^\pm}/v^2)_{N_1}$ is the ratio of the squared mass of the charged scalar
to the sum of the square of vevs, $(v^2)_{N_1} = {v_{1,1}}^2 + {v_{2,1}}^2$, as computed in the $N_1$
stationary point, and analogously for $(M^2_{H^\pm}/v^2)_{N_2}$.
\end{itemize}
Equation~\eqref{2_eq:vnvn} shows us that there is nothing favouring $N_1$ over $N_2$, the
deepest stationary point will be determined by the values of the parameters - as it should be,
since both vacua have the same symmetries.

In Ref.~\cite{Ivanov:2007de} it was proven that it is possible to have two coexisting normal
minima. Numerical examples of this were found in~\cite{Barroso:2007rr} for the particular case
of a softly broken $U(1)$ model, where it was shown that one may have a curious situation: a
minimum $N_1$ with $\sqrt{v_{1,1}^2 + v_{2,1}^2} =$ 246 GeV, and all particles having
their known masses; and a {\em deeper} $N_2$ minimum, for which $
\sqrt{v_{1,2}^2 + v_{2,2}^2} \neq$ 246 GeV (possibly much larger or smaller). The $N_2$
minimum would have the same unbroken symmetries as $N_1$ {\em but} with a completely different
mass spectrum of scalars, fermions and gauge bosons. This was seen to happen only for a
very small portion of the parameter space. Of course, it is very easy to pick sets of parameters
for which $N_1$ would be the global minimum, with $N_2$ above it or not even existing.

The main consequences of this vacuum analysis are:
\begin{itemize}
\item Minima of different natures cannot coexist in the 2HDM.
\item Whenever a normal minimum exists in the 2HDM, the global minimum of the potential is normal.
No tunneling to a deeper CB or CP minimum is possible.
\item If a CP (CB) violating minimum exists, it is the global minimum of the theory,
and thouroughly stable. No tunnelling to a deeper normal or CB (CP)
minimum can occur.
\end{itemize}

\subsection{Mass matrices for neutral minima}
\label{2_sec:mat}

In order to determine whether a given stationary point is a minimum, one needs
to analyse the second derivatives of the potential, meaning the scalar mass
matrices. For the sake of completeness, we include here the expressions for these matrices
for neutral minima. The mass matrices for CB stationary points can be found
in~\cite{Ferreira:2004yd}.
A discussion in the MSSM with explicit CP violation, which approaches the 2HDM in a given
limit, can be found in Ref.~\cite{hep-ph/9902371}.

\begin{itemize}
\item \bf{Normal minima}
\end{itemize}

With vevs given by~\eqref{2_eq:nsol} and determined by~\eqref{2_eq:vacn},
the squared mass for the charged scalar is given by the eigenvalues of a
$2\times 2$ matrix whose entries are
\be
[M^2_{H^\pm}]_{ij} \,=\,
\frac{\partial^2 V_H}{\partial\varphi^+_i \partial \varphi^-_j}.
\ee
This matrix has a zero eigenvalue (corresponding to the Goldstone boson which gives
mass to the W) so that the charged scalar squared mass is
\be
M^2_{H^\pm}\,=\,-\frac{v^2}{2 v_1 v_2}\,\bar{V},
\ee
where we have defined the quantity
\be
\bar{V}\,=\, -2\mbox{Re}(m_{12}^2) + [\lambda_4 + \mbox{Re}(\lambda_5)] v_1 v_2 +
\mbox{Re}(\lambda_6) v_1^2 + \mbox{Re}(\lambda_7) v_2^2.
\ee
Let us now assume for a moment the potential is explicitly CP conserving
({\em i.e.} we will work in a basis without imaginary couplings).
The pseudoscalar mass matrix is the $2\times 2$ matrix of the second
derivatives of the imaginary parts of the neutral components,
\be
[M^2_A]_{ij} \,=\,\frac{1}{2}\,
\frac{\partial^2 V_H}{\partial\mbox{Im}(\varphi^0_i)\partial\mbox{Im}(\varphi^0_j)}\,=\,
\frac{v_1 v_2}{v^2}\,M^2_A\,\begin{pmatrix} \displaystyle{\frac{v_2}{v_1}} & -1 \\
-1 & \displaystyle{\frac{v_1}{v_2}} \end{pmatrix},
\label{2_eq:mpseu}
\ee
where the entries of this matrix have been simplified through the minimization conditions.
It has one zero eigenvalue (corresponding to the Goldstone boson which gives
mass to the Z), so that the pseudoscalar squared mass is found to be
\be
M^2_A\,=\,M^2_{H^\pm}\,+\,\frac{1}{2} [\lambda_4 - \mbox{Re}(\lambda_5)]\,v^2.
\label{eq:ma}
\ee
As for the CP-even scalars, they are the eigenvalues of the symmetric $2\times 2$ matrix,
given by
\be
[M^2_h]_{ij} \,=\,\frac{1}{2}\,
\frac{\partial^2 V_H}{\partial\mbox{Re}(\varphi^0_i)\partial\mbox{Re}(\varphi^0_j)}\,=\,
\begin{pmatrix} A & C \\ C & B \end{pmatrix},
\label{2_eq:mneut}
\ee
where the matrix's entries are given by
\ba
A &=& m_{11}^2 + \frac{3 \lambda_1}{2}  v_1^2 + \frac{\lambda_{345}}{2} v_2^2 +
3 \mbox{Re}(\lambda_6) v_1 v_2, \nonumber \\
B &=& m_{22}^2 + \frac{3 \lambda_2}{2}  v_2^2 + \frac{\lambda_{345}}{2} v_1^2 +
3 \mbox{Re}(\lambda_7) v_1 v_2, \nonumber \\
C &=& - \mbox{Re}(m_{12}^2) + \frac{3}{2}\left[\mbox{Re}(\lambda_6) v_1^2 +
\mbox{Re}(\lambda_7) v_2^2\right] + \lambda_{345} v_1 v_2,
\ea
where we have defined $\lambda_{345} = \lambda_3 +\lambda_4 + \mbox{Re}(\lambda_5)$.
The mass eigenstates of this matrix are traditionally represented as $h$ and $H$,
respectively the lightest and heaviest state. The diagonalization angle $\alpha$
of the matrix~\eqref{2_eq:mneut} is defined~\footnote{Clearly this is a basis-dependent
definition, as is that of $\tan\beta$. None of these angles are basis invariant
quantities, though their difference is~\cite{Davidson:2005cw}.} as
\ba
H &=& -\cos\alpha \,\mbox{Re}(\varphi^0_1) - \sin\alpha \,\mbox{Re}(\varphi^0_2) \nonumber \\
h &=& \sin\alpha \,\mbox{Re}(\varphi^0_1) - \cos\alpha \,\mbox{Re}(\varphi^0_2)
\ea
so that one gets, after trivial calculations,
\be
\tan 2\alpha = \frac{2 C}{A - B}.
\ee

If we are dealing with the most general 2HDM potential with complex couplings,
then there will be mixing between the CP even and odd scalar particles. The neutral
scalars will be the eigenvalues of a $4\times 4$ matrix,
\be
[M^2_N]_{ij} \,=\,\begin{pmatrix} [M^2_h] & [M^2_I] \\ [M^2_I]^T & [M^2_A] \end{pmatrix},
\ee
composed of three $2\times 2$ blocks. $[M^2_h]$ and $[M^2_A]$ are as given in
eqs.~\eqref{2_eq:mneut} and~\eqref{2_eq:mpseu}, respectively, whereas $[M^2_I]$
is given by the matrix
\be
[M^2_I] \,=\,
\begin{pmatrix} A_I & B_I \\ C_I & D_I \end{pmatrix},
\ee
whose entries are
\ba
A_I &=& \frac{1}{2} v_2 \left[\mbox{Im}(\lambda_5) v_2 +
2 \mbox{Im}(\lambda_6) v_1\right] , \nonumber \\
B_I &=& \mbox{Im}(m_{12}^2) - \mbox{Im}(\lambda_5) v_1 v_2 -
\frac{3}{2} \mbox{Im}(\lambda_6) v_1^2 -
\frac{1}{2} \mbox{Im}(\lambda_7) v_2^2, \nonumber \\
C_I &=& - \mbox{Im}(m_{12}^2) + \mbox{Im}(\lambda_5) v_1 v_2 +
\frac{1}{2} \mbox{Im}(\lambda_6) v_1^2 -
\frac{3}{2} \mbox{Im}(\lambda_7) v_2^2, \nonumber \\
D_I &=& - \frac{1}{2} v_1 \left[\mbox{Im}(\lambda_5) v_1 +
2 \mbox{Im}(\lambda_7) v_2\right].
\ea
Obviously, one of the eigenvalues of $[M^2_N]$ will be zero.

\begin{itemize}
\item \bf{Inert minima}
\end{itemize}

Recall that one needs $m^2_{12} = \lambda_6 = \lambda_7 = 0$ to obtain inert vacua, and
they obey the minimization conditions~\eqref{2_eq:in1} or~\eqref{2_eq:in2}. Considering,
for example, the case $v_2 = 0$ and $v_1 = v/\sqrt{2}$, the scalar mass spectrum is greatly
simplified. The CP-even mass matrix is diagonal (so that in this basis one may consider $\alpha =
\beta = 0$), and the expressions for the masses are:
\ba
M^2_{H^\pm} &=& m_{22}^2 \,+\, \frac{1}{2} \lambda_3 v^2 \nonumber \\
M^2_A &=& M^2_{H^\pm} \,+\,\frac{1}{2} [\lambda_4 - \mbox{Re}(\lambda_5)]\,v^2 \nonumber \\
M^2_H &=& M^2_A \,+\, \mbox{Re}(\lambda_5)\,v^2\nonumber \\
M^2_h &=& \lambda_1 v^2.
\ea
In inert models where we further have $\lambda_5 = 0$ (those arising from a $U(1)$
symmetry, for instance) $A$ and $H$ will be degenerate. Also, notice that though
we maintained the notation $h$ and $H$, it is now not guaranteed that they correspond to the
lightest and heaviest CP-even states - that will depend on the specific values for the
parameters. In fact, notice that the state `$h$' can be made much heavier or much
lighter than the remaining three - it is the only one depending on the coupling $\lambda_1$.
It can be shown that the $H$ and $A$ states do not couple to the $Z$ or to the fermions,
hence the name ``inert".

\begin{itemize}
\item \bf{CP breaking minima}
\end{itemize}

It only makes sense to speak of minima with spontaneous CP violation if that
symmetry is defined - {\em i.e.} if it is not explicitly broken by the potential.
As explained in section~\ref{sec:cpv}, that corresponds to the existence
of a basis where all parameters are real. We now write all masses in such a basis,
for a vacuum with vevs such as~\eqref{2_eq:vaccp}. The charged scalar mass is now, as was
mentioned before,
\be
M^2_{H^\pm} \,=\,-\frac{1}{2} [\lambda_4 - \lambda_5](\bar{v}_1^2 + \bar{v}_2^2).
\ee
Given that the vevs now have an imaginary component, there will be a mixing between
the real and imaginary components of $\varphi_i^0$. The neutral Higgs squared masses
are thus the eigenvalues of a $4\times 4$ symmetric matrix $M^2_{CP}$. Using
the minimization conditions~\eqref{2_eq:cpsol} to simplify, the entries of this matrix
are given by:
\ba
M^2_{CP}(1,1) &=& \lambda_1 \bar{v}_1^2 \cos^2\theta +
2\lambda_6 \bar{v}_1 \bar{v}_2 \cos\theta + \lambda_5 \bar{v}_2^2 \nonumber  \\
M^2_{CP}(1,2) &=& \lambda_6 \bar{v}_1^2 \cos^2\theta +
(\lambda_3 + \lambda_4)\bar{v}_1 \bar{v}_2 \cos\theta + \lambda_7 \bar{v}_2^2 \nonumber  \\
M^2_{CP}(1,3) &=& (\lambda_6 \bar{v}_2 +
\lambda_1 \bar{v}_1 \cos\theta) \bar{v}_1 \sin\theta \nonumber \\
M^2_{CP}(1,4) &=& (\lambda_6 \bar{v}_1 \cos\theta + \lambda_5 \bar{v}_2)
\bar{v}_1 \sin\theta \nonumber \\
M^2_{CP}(2,2) &=& \lambda_5 \bar{v}_1^2 \cos^2\theta +
2\lambda_7 \bar{v}_1 \bar{v}_2 \cos\theta + \lambda_2 \bar{v}_2^2 \nonumber \\
M^2_{CP}(2,3) &=& [\lambda_7 \bar{v}_1 \cos\theta +
(\lambda_3 + \lambda_4 + \lambda_5) \bar{v}_2]
\bar{v}_1 \sin\theta \nonumber \\
M^2_{CP}(2,4) &=& (\lambda_7 \bar{v}_2 + \lambda_5 \bar{v}_1 \cos\theta)
\bar{v}_1 \sin\theta \nonumber \\
M^2_{CP}(3,3) &=& \lambda_1 \bar{v}_1^2 \sin^2\theta \nonumber \\
M^2_{CP}(3,4) &=& \lambda_6 \bar{v}_1^2 \sin^2\theta \nonumber \\
M^2_{CP}(3,4) &=& \lambda_5 \bar{v}_1^2 \sin^2\theta .
\ea
This matrix has a zero eigenvalue, corresponding to the $Z$ Goldstone boson.

\subsection{\label{subsec:HiggsBasis}The Higgs basis}

After spontaneous electroweak symmetry breaking with neutral vacua,
the fields acquire the vacuum expectation values
$v_1/\sqrt{2}$ and $v_2 e^{i \delta}/\sqrt{2}$;
where $v_1$ and $v_2$ are real,
without loss of generality.
It is convenient to rotate into
a new basis of scalar fields such that the vev is all
in the first field,
while the second field has no vev.
This is known as the Higgs basis $\{H_1, H_2 \}$,
obtained through $H_a = \sum_{b=1}^2 U_{ab} \Phi_b$,
where \cite{Lavoura:1994fv}
\ba
U
&=&
\frac{1}{v}
\left[
\begin{array}{cc}
v_1 & v_2\; e^{-i \delta}\\
- v_2 & v_1\; e^{-i \delta}
\end{array}
\right]
\label{2_eq:HBT}
\\
&=&
\frac{e^{-i \delta/2}}{v}
\left[
\begin{array}{cc}
v_1\; e^{i \delta/2}& v_2\; e^{-i\delta/2}\\
- v_2\; e^{i \delta/2}& v_1\; e^{-i\delta/2}
\end{array}
\right],
\label{2_eq:HBT2}
\ea
is unitary, and $v = \sqrt{v_1^2 + v_2^2} = (\sqrt{2} G_F)^{-1/2} = 246$ GeV.
This rotates the vev into $H_1$,
allowing us to parametrize
\be
H_1
=
\left[
\begin{array}{c}
G^+\\
(v + H + i G^0)/\sqrt{2}
\end{array}
\right],
\ \ \
H_2
=
\left[
\begin{array}{c}
H^+\\
(R + i I)/\sqrt{2}
\end{array}
\right],
\label{2_eq:higbas}
\ee
where $ G^+ $ and $ G^0 $ are the Goldstone bosons,
which,
in the unitary gauge,
become the longitudinal components of the $ W^+ $ and of the $ Z^0 $,
and $ H$, $ R $ and $ I $ are real neutral fields.

Notice that there are infinitely many Higgs basis.
Indeed,
we may change the phase of $H_2$,
\be
H_2 \rightarrow e^{i \xi} H_2
\label{jvuep}
\ee
while keeping the vev in $H_1$.
Under this phase transformation,
the fields $R$ and $I$ are rotated by
\be
\left(
\begin{array}{c}
R\\
I
\end{array}
\right)
\rightarrow
\left(
\begin{array}{cc}
\cos{\xi} & \sin{\xi}\\
-\sin{\xi} & \cos{\xi}
\end{array}
\right)
\left(
\begin{array}{c}
R\\
I
\end{array}
\right).
\label{yuesn}
\ee

In going from a generic basis to the Higgs basis,
the couplings in the scalar potential get rotated.
Comparing eq.~(\ref{2_U_DH}) with
eq.~(\ref{2_eq:HBT2}) without the overall phase
$e^{-i\delta/2}$,
we find
\ba
\tan{\beta} &=& \frac{v_2}{v_1},
\nonumber\\
\chi &=& \delta/2,
\nonumber\\
\xi &=& \delta.
\label{2_param_HB}
\ea
Thus,
the potential coefficients in the Higgs basis are obtained from those
in the generic basis,
through Eqs.~(\ref{2_M11})--(\ref{2_L7}),
with $\beta$, $\chi$, and $\xi$ as defined
in eq.~(\ref{2_param_HB}).
The quadratic (quartic) coefficients of the scalar potential in the Higgs basis
are denoted by $\bar m_{ij}^2$ ($\bar \lambda_{i}$).

In the Higgs basis,
the stationarity conditions are simply given by
\ba
\bar m_{11}^2 &=& - \tfrac{1}{2} \bar \lambda_1 v^2,
\nonumber\\
\bar m_{12}^2 &=& \tfrac{1}{2} \bar \lambda_6 v^2.
\label{2_stationarity_HB}
\ea
Let us count the parameters in the Higgs basis:
the complex parameter $\bar m_{12}^2$ is determined by
$\bar m_{11}^2$, $\bar \lambda_1$, and $\bar \lambda_6$.
Thus, we would seem to have 12 parameters for the most general
2HDM potential. However,
the fact that one may rephase $H_2$ implies that
only the relative phases of the complex parameters
$\bar \lambda_5$, $\bar \lambda_6$, and $\bar \lambda_7$ have
physical significance.
We are thus left with 11 physical parameters,
as expected (we can view $m_{11}^2$ as determining $v^2$).
However, as observed earlier, potentials to which symmetries have been
imposed will display, in the Higgs basis,
a smaller number of parameters.
As first pointed out by Lavoura~\cite{Lavoura:1994yu}
and subsequently greatly expanded by
Davidson and Haber \cite{Davidson:2005cw},
the coefficients of the potential in the Higgs basis
are observable,
up to the overall phase of the complex parameters.
The only physically meaningful phases are
$\textrm{Im}(\bar \lambda_5^\ast \bar \lambda_6^2)$,
$\textrm{Im}(\bar \lambda_5^\ast \bar \lambda_7^2)$,
and $\textrm{Im}(\bar \lambda_6 \bar \lambda_7^\ast)$.
These are proportional to the quantities
$J_1$, $J_2$, and $J_3$ (respectively),
introduced by Lavoura and Silva \cite{Lavoura:1994fv}
as basis-invariant
signals of the CP violation in the 2HDM present
after spontaneous electroweak symmetry breaking.
Notice that only two are independent;
for example, $J_1$ and $J_3$.
However,
since in a given theory $\bar \lambda_6$ may vanish
while $\textrm{Im}(\bar \lambda_5^\ast \bar \lambda_7^2)$ does not,
all three must be considered when searching for CP violation.

\subsection{Yukawa couplings in the Higgs basis}
\label{sec:fermions}

The most generic Yukawa interactions that one can write
with two doublets and the fermionic content of the SM
are given by
\be
\mathcal{L}_Y = - \sum_{j=1}^{2} \left[
\bar Q_L \left( \Phi_j Y^d_j n_R
+ \tilde \Phi_j Y^u_j p_R
\right)
+ \bar L_L \Phi_j Y^e_j \ell_R
\right] + \mathrm{H.c.}
\label{uvhfw}
\ee
In this equation, $\tilde \Phi_j = i \tau_2 \Phi_j^\ast$;
$Q_L$, $L_L$, $n_R$, $p_R$ and $\ell_R$ are 3-vectors in
flavour space - the $n_R$ correspond to the negative-charged
quarks, and $p_R$ to the positive-charged ones~\footnote{Which,
after diagonalization, will yield the down and up type quarks.}; $L_L$ and
$\ell_R$ are the leptonic fields; and $Y^d_j$,
$Y^u_j$ and $Y^e_j$ are generic $3\times 3$ complex matrices
containing the Yukawa couplings for, respectively, the down, up and
leptonic sector.
In the initial basis $\left( \Phi_1, \Phi_2 \right)$,
the doublet $\Phi_1$ has vacuum expectation value (vev)
$\tilde{v}_1 \left/ \sqrt{2} \right.$
and the doublet $\Phi_2$ has vev $\tilde{v}_2 \left/ \sqrt{2} \right.$,
where $\tilde{v}_1$ and $\tilde{v}_2$ are allowed to be \emph{complex}.
We define
\be
v = \sqrt{\left| \tilde{v}_1 \right|^2 + \left| \tilde{v}_2 \right|^2}.
\ee
Experimentally,
$v \approx 246$ GeV.
Notice that $v$ is,
by definition,
real and positive.

Let us define the Higgs basis $\left( H_1, H_2 \right)$
by $H_1$ having vev
$v \left/ \sqrt{2} \right.$ while $H_2$ has vanishing
vev~\footnote{The definition used in eq.~\eqref{2_eq:HBT} was
useful for comparison with the basis-change formulae of
eqs.~\eqref{2_M11}--~\eqref{2_L7}. This definition differs
from the first by an irrelevant phase such as in
eq.~\eqref{jvuep}.}.
The transformation from one basis to the other is
\ba
\Phi_1 &=& \frac{1}{v} \left( \tilde{v}_1 H_1 + \tilde{v}_2^\ast  H_2 \right),
\\
\Phi_2 &=& \frac{1}{v} \left( \tilde{v}_2 H_1 - \tilde{v}_1^\ast H_2 \right).
\ea

Let us define the matrices
\ba
M_n &=& \frac{1}{\sqrt{2}} \left( \tilde{v}_1 Y^d_1 + \tilde{v}_2 Y^d_2 \right),
\label{hvgsq} \\
N_n &=& \frac{1}{\sqrt{2}} \left( \tilde{v}_2^\ast Y^d_1 - \tilde{v}_1^\ast Y^d_2 \right).
\label{jvusa}
\ea
Then
\be
\sum_{j=1}^2 \Phi_j Y^d_j =
\frac{\sqrt{2}}{v} \left( M_n H_1 + N_n H_2 \right).
\ee
The matrix $M_n$ in equation~(\ref{hvgsq})
is the mass matrix of the down-type quarks.
Notice that $N_n$ in equation~(\ref{jvusa}) may be written as
\ba
N_n
&=& \frac{\tilde{v}_2^\ast}{\tilde{v}_1}\, M_n - \frac{v^2}{\sqrt{2} \tilde{v}_1}\, Y^d_2
\no
\\
&=& - \frac{\tilde{v}_1^\ast}{\tilde{v}_2}\, M_n + \frac{v^2}{\sqrt{2} \tilde{v}_2}\, Y^d_1.
\ea
It is convenient to pass to the mass basis of the quarks, in
which the mass matrices are diagonal. To do so,
we bi-diagonalize $M_n$, via a simultaneous rotation
on the left-handed and right-handed quark fields:
\ba
{U_L}^\dagger M_n U_R^n &=& M_d,
\\
{U_L}^\dagger N_n U_R^n &=& N_d,
\label{2_bidd}
\ea
where $M_d = \mathrm{diag} \left( m_d, m_s, m_b \right)$
is diagonal with real and positive diagonal elements. In the sector of the up-type quarks,
\ba
M_p &=& \frac{1}{\sqrt{2}} \left( \tilde{v}_1^\ast Y^u_1 + \tilde{v}_2^\ast Y^u_2 \right),
\\
N_p &=& \frac{1}{\sqrt{2}} \left( \tilde{v}_2 Y^u_1 - \tilde{v}_1 Y^u_2 \right),
\ea
and
\be
\sum_{j=1}^2 \tilde \Phi_j Y^u_j =
\frac{\sqrt{2}}{v} \left[ M_p \left( i \tau_2 H_1^\ast \right)
+ N_p \left( i \tau_2 H_2^\ast \right) \right].
\ee
The bi-diagonalization proceeds as
\ba
{U_L}^\dagger M_p U_R^p &=& M_u,
\\
{U_L}^\dagger N_p U_R^p &=& N_u,
\label{2_bidu}
\ea
where $M_u = \mathrm{diag} \left( m_u, m_c, m_t \right)$
is diagonal with real and positive diagonal elements.
Notice that the $U_L$ matrix has to be the same in
eqs.~\eqref{2_bidd} and~\eqref{2_bidu}.

If, after the bi-diagonalization, the matrix $N_d$ ($N_u$)
is {\em not} diagonal, then there are scalar tree-level
flavour-changing neutral interactions in the down (up)
sector, and the FCNC couplings for those interactions
are obtained from the entries of $N_d$ ($N_u$).
In the generic basis of eq.~\eqref{uvhfw},
the condition for non-existence of FCNC is also quite
simple: if the matrices $Y_1^d$ and $Y_2^d$ ($Y_1^u$ and $Y_2^u$)
commute, there is no tree-level FCNC in the down-quark
(up-quark) sector~\cite{Ecker:1987qp}. This, of course,
is trivial if, for instance, $Y_2^x$ is zero, as is
obtained in models with Natural Flavour Conservation,
which were discussed in Chapter~\ref{sec:nfc}.

\subsection{Basis transformations and Yukawa couplings}

We start from the Lagrangian
\be
\mathcal{L} = \mathcal{L}_\mathrm{H} + \mathcal{L}_\mathrm{Y},
\label{full_L}
\ee
where
\ba
- \mathcal{L}_\mathrm{H} =
&=&
\sum_{a,b=1}^2 \mu_{ab} (\Phi_a^\dagger \Phi_b) +
\tfrac{1}{2}
\sum_{a,b,c,d=1}^2 \lambda_{ab,cd} (\Phi_a^\dagger \Phi_b) (\Phi_c^\dagger \Phi_d),
\label{VH2}
\\
\label{yuk}
- \mathcal{L}_\mathrm{Y}
&=&
\bar q_L \left[
\left( Y^d_1 \Phi_1 + Y^d_2 \Phi_2 \right) n_R
+
\left( Y^u_1 \tilde \Phi_1 + Y^u_2 \tilde \Phi_2 \right) p_R
\right]
+ \mathrm{H.c.},
\ea

The Lagrangian can be rewritten in terms of new fields
obtained from the original ones by simple basis transformations
\begin{eqnarray}
\Phi_a
& \rightarrow &
\Phi_a^\prime = \sum_{\alpha =1}^2 U_{a \alpha}\ \Phi_\alpha,
\nonumber\\
q_L
& \rightarrow &
q^\prime_L = U_L\ q_L,
\nonumber\\
n_R
& \rightarrow &
n^\prime_R = U_{nR}\ n_R,
\nonumber\\
p_R
& \rightarrow &
p^\prime_R = U_{pR}\ p_R,
\label{basis-transf}
\end{eqnarray}
where $U\in U(2)$ is a $2 \times 2$ unitary matrix,
while $\left\{ U_L, U_{nR}, U_{pR} \right\} \in U(3)$ are $3 \times 3$ unitary matrices.
Under these unitary basis transformations,
the gauge-kinetic terms are unchanged,
but the coefficients $\mu_{ab}$ and $\lambda_{ab,cd}$ are transformed as
\begin{eqnarray}
\mu_{ab} & \rightarrow &
\mu^\prime_{ab} = \sum_{\alpha,\beta =1}^2
U_{a \alpha}\ \mu_{\alpha \beta}\ U_{b \beta}^\ast ,
\label{Y-transf}
\\
\lambda_{ab,cd} & \rightarrow &
\lambda^\prime_{ab,cd} = \sum_{\alpha,\beta,\gamma,\delta =1}^2
U_{a\alpha}\, U_{c \gamma}\
\lambda_{\alpha \beta,\gamma \delta}\ U_{b \beta}^\ast \, U_{d \delta}^\ast,
\label{Z-transf}
\end{eqnarray}
while the Yukawa matrices change as
\begin{eqnarray}
Y^d_a
& \rightarrow &
Y^{d\, \prime}_a = \sum_{\alpha =1}^2
U_L\ Y^d_\alpha\ U_{nR}^\dagger\ \left( U^\dagger \right)_{ \alpha a}
\nonumber\\
Y^u_a
& \rightarrow &
Y^{u\, \prime}_a = \sum_{\alpha =1}^2
U_L\ Y^u_\alpha\ U_{pR}^\dagger\ \left( U^\top \right)_{ \alpha a}.
\end{eqnarray}
Notice that we have kept the notation of showing explicitly
the indices in scalar-space,
while using matrix formulation for the quark flavour
spaces.
The basis transformations may be utilized in order to absorb
some of the degrees of freedom of
$\mu$, $\lambda$, $Y^d$, and/or $Y^u$,
which implies that not all parameters in the Lagrangian
have physical significance.

\subsection{Symmetries and Yukawa couplings}

The symmetries we have discussed in section~\ref{sec:symmetries}
were imposed on the scalar sector, but they have to
be extended to the full lagrangian. By their definition, they
leave the gauge kinetic terms invariant, but will affect
the Yukawa terms, where the scalars are coupled to the
fermions. As such, one needs to consider how the fermion
fields transform under such symmetries, and whether or not
their impact on the Yukawa sector leads to viable models.

\subsubsection{Family symmetries}

We will now assume that the Lagrangian in eq.~\eqref{full_L} is invariant under the
symmetry
\begin{eqnarray}
\Phi_a
& \rightarrow &
\Phi_a^S
= \sum_{\alpha =1}^2  S_{a \alpha}\ \Phi_\alpha,
\nonumber\\*[1mm]
q_L
& \rightarrow &
q_L^S
= S_L\ q_L,
\nonumber\\*[1mm]
n_R
& \rightarrow &
n_R^S
= S_{nR}\ n_R,
\nonumber\\*[1mm]
p_R
& \rightarrow &
p_R^S
= S_{pR}\ p_R,
\label{S-transf-symmetry}
\end{eqnarray}
where $S \in U(2)$,
while $\left\{ S_L, S_{nR}, S_{pR} \right\} \in U(3)$.
As a result of this symmetry,
the parameters in the lagrangian have to obey the following equations:
\begin{eqnarray}
\mu_{a b} & = &
\sum_{\alpha,\beta =1}^2
S_{a \alpha}\ \mu_{\alpha \beta}\ S_{b \beta}^\ast ,
\label{Y-S}
\\
\lambda_{ab,cd} & = &
\sum_{\alpha,\beta,\gamma,\delta =1}^2
S_{a \alpha}\, S_{c \gamma}\
\lambda_{\alpha \beta, \gamma \delta}\ S_{b \beta}^\ast \, S_{d \delta}^\ast ,
\label{Z-S}
\\
Y^d_a & = &
\sum_{\alpha =1}^2
S_L\ Y^d_\alpha\ S_{nR}^\dagger\ \left( S^\dagger \right)_{\alpha a} ,
\label{Gamma-S}
\\
Y^u_a & = &
\sum_{\alpha =1}^2
S_L\ Y^u_\alpha\ S_{pR}^\dagger\ \left( S^\top \right)_{\alpha a}.
\label{Delta-S}
\end{eqnarray}

Under the basis transformation of eq.~(\ref{basis-transf}),
the specific form of the symmetry
in eq.~\eqref{S-transf-symmetry} is altered as
\begin{eqnarray}
S^\prime &=& U\ S\ U^\dagger ,
\label{S-prime}
\\
S^\prime_L &=& U_L\ S_L\ U^\dagger_L ,
\label{S_L-prime}
\\
S^\prime_{nR} &=& U_{nR}\ S_{nR}\ U^\dagger_{nR} ,
\label{S_nR-prime}
\\
S^\prime_{pR} &=& U_{pR}\ S_{pR}\ U^\dagger_{pR} .
\label{S_pR-prime}
\end{eqnarray}

\subsubsection{CP symmetries}

We will now assume that the Lagrangian in eq.~\eqref{full_L} is invariant under the
CP symmetry
\begin{eqnarray}
\Phi_a
& \rightarrow & \sum_{\alpha =1}^2
X_{a\alpha}\ \Phi_\alpha^\ast,
\nonumber\\*[1mm]
q_L
& \rightarrow &
X_L\ \gamma^0 C \ q_L^\ast,
\nonumber\\*[1mm]
n_R
& \rightarrow &
X_{nR}\ \gamma^0 C \ n_R^\ast,
\nonumber\\*[1mm]
p_R
& \rightarrow &
X_{pR}\ \gamma^0 C \ p_R^\ast,
\label{X-transf-symmetry}
\end{eqnarray}
As a result of this symmetry~\footnote{Eq.~\eqref{Y-X}
can be written in Higgs-family matrix form
as $\mu^\ast = X^\dagger\, \mu\, X$,
in an obvious notation.
This is equivalent to
$\mu^\ast = X^\ast\, \mu\, X^\top$.
Similar rewritings are also possible for
eqs.~(\ref{Z-X})-(\ref{Delta-X}),
sometimes complicating comparisons.}, the parameters of the
lagrangian need to obey
\begin{eqnarray}
\mu_{ab}^\ast
& = &
\sum_{\alpha,\beta =1}^2
X_{\alpha a}^\ast \mu_{\alpha \beta} X_{\beta b},
\label{Y-X}
\\
\lambda_{ab,cd}^\ast
& = &
\sum_{\alpha,\beta,\gamma,\delta =1}^2
X_{\alpha a}^\ast X_{\gamma c}^\ast
\lambda_{\alpha \beta, \gamma \delta} X_{\beta b} X_{\delta d},
\label{Z-X}
\\
Y^{d\, \ast}_a
& = &
\sum_{\alpha =1}^2
X_{a \alpha}
X_L\ Y^d_\alpha\ X_{nR}^\dagger,
\label{Gamma-X}
\\
Y^{u\, \ast}_a
& = &
\sum_{\alpha =1}^2
X_{a \alpha}^\ast
X_L\ Y^u_\alpha\ X_{pR}^\dagger.
\label{Delta-X}
\end{eqnarray}

Under the basis transformation of eq.~(\ref{basis-transf}),
the specific form of the symmetry
in eq.~\eqref{X-transf-symmetry} is altered as
\begin{eqnarray}
X^\prime &=& U\ X\ U^\top ,
\label{X-prime}
\\
X^\prime_L &=& U_L\ X_L\ U^\top_L ,
\label{X_L-prime}
\\
X^\prime_{nR} &=& U_{nR}\ X_{nR}\ U^\top_{nR} ,
\label{X_nR-prime}
\\
X^\prime_{pR} &=& U_{pR}\ X_{pR}\ U^\top_{pR} .
\label{X_pR-prime}
\end{eqnarray}
%

\subsubsection{Symmetries of the scalar-scalar and scalar-fermion interactions}
\label{2_sec:yuksym}

We may now ask whether the six symmetry classes of the Higgs potential
shown in table~\ref{2_master1} can be extended to the fermion sector
in a way consistent with experiment.
This issue is complicated by the fact that the fermion fields
can transform, for a given scalar symmetry, under infinitely many
ways, as detailed in eqs.~\eqref{Gamma-S} and~\eqref{Delta-S} for
Higgs-family symmetries, and eqs.~\eqref{Gamma-X} and~\eqref{Delta-X}
for GCP symmetries. Let us deal with those two types of symmetries
separately.

\begin{itemize}
\item Higgs family symmetries extended to the Yukawa sector
\end{itemize}

As an example of how complex this issue can become, consider that
even for a simple $Z_2$ symmetry in the scalars, for which
the $\Phi_2$ doublet flips its sign, we can choose an
extremely elaborate transformation law for the quark fields, with
arbitrary unitary 3$\times$3 matrices $S_L$, $S_{nR}$ and $S_{pR}$.
To further complicate matters, the field transformation laws can
correspond to an Abelian symmetry (with a single generator, or
a set of generators which commute amongst themselves), or a more
general and complex non-Abelian one (with several
non-commuting generators). In the scalar sector, all but
the class I models (with a full U(2) symmetry) can be
obtained via an Abelian symmetry.

As such, extensions of scalar symmetries to the Yukawa sector
are usually specific examples, where one chooses a particular
form for the fermion transformation matrices. There are many such
examples. For instance, the various implementations of the $Z_2$ symmetry
(or the U(1) one) which preclude the occurrence of tree-level
FCNC~\cite{Glashow:1976nt,Paschos:1976ay} - the so-called
type I, II, X (lepton specific) and Y (flipped) 2HDMs.
Such models are said to have ``Natural Flavour Conservation'',
and their phenomenology was already discussed in
chapter~\ref{sec:nfc}.
Another example would be the BGL model~\cite{Branco:1996bq}, which
does contain FCNC which are naturally small due to a
flavour-dependent fermionic symmetry. A third example would
be a recent
application of a $Z_3$ symmetry to the 2HDM lagrangian, with
interesting consequences regarding the origin of CP
violation~\cite{Ferreira:2011xc}.

The extension of generic Abelian symmetries into the quark sector
was only fully mapped recently by Ferreira and
Silva~\cite{Ferreira:2010ir}. That calculation was greatly
simplified by the usage of the full freedom of choosing a
basis of both scalar and fermionic fields, as explained in
the previous sections: it turns out that it is {\em always}
possible, for a transformation involving a single
transformation matrix $S_X$, to go to a basis of fields
for which {\em all} such matrices are diagonal, containing,
in general, only complex phases. Meaning, it is always possible
to choose a basis of fields such that the fermionic transformation
matrices of eqs.~\eqref{Gamma-S} and~\eqref{Delta-S} are reduced,
through transformations like~\eqref{S_L-prime}--~\eqref{S_pR-prime},
to the form
\ba
S_L &=& \mbox{diag}(e^{i \alpha_1}, e^{i \alpha_2}, e^{i \alpha_3})\,,
\nonumber \\
S_{nR} &=& \mbox{diag}(e^{i \beta_1}, e^{i \beta_2}, e^{i \beta_3})\,,
\nonumber \\
S_{pR} &=& \mbox{diag}(e^{i \gamma_1}, e^{i \gamma_2}, e^{i \gamma_3})\,.
\ea
Thus, all the freedom in choosing the fermions' transformation laws
is reduced to the choice of the arbitrary (real) phases $\alpha_i$,
$\beta_i$ and $\gamma_i$. Even this immense simplification, though,
yields  millions of possible specific symmetry implementations -
that is, of possible different models.

However, as explained in detail in~\cite{Ferreira:2010ir}, it turns
out that the simultaneous requirements of six massive
quarks~\footnote{Or three massive charged leptons, for the argument
is trivially extended to the leptonic sector.} and an acceptable CKM
matrix are extremely powerful and curtail immensely the number
of allowed models. The reason for that is that the effect of the
phases $\alpha_i$ on the CKM matrix, or in the quark squared-mass
matrices, is tantamount to setting many of their entries to zero.
Very easily, for arbitrary choices of the $\alpha_i$, one obtains
a line or column of zeros in the mass matrices, or a diagonal block
on the CKM matrix. Surprisingly, then, the transformation laws
of the quark fields are extremely constrained.
The authors of~\cite{Ferreira:2010ir}
have shown that there are only
$246$ possible forms for the Yukawa matrices for both the
up and down quarks. Up to physically unimportant permutations,
these involve only 34 forms of Yukawa matrices. As such, there is
effectively a maximum of only 34 possible ways of extending
the Abelian Higgs-family symmetries (that is, the models
like classes IV and V, with symmetries like $Z_2$ or the
Peccei-Quinn U(1)) to the quark sector. Most of these
symmetry-constrained Yukawa matrices lead to tree-level
scalar FCNC. However
some of them have already been shown to have naturally
small FCNC, which are CKM suppressed, the aforementioned
BGL models; or to be such that one
can easily find values of parameters which satisfy all
experimental constraints, such as the mass differences of
neutral kaons and $B$-mesons, which typically are
difficult to accommodate when tree-level FCNC are present.

In addition, the general analysis of these Abelian
symmetries leads to some wide-reaching conclusions:
\begin{itemize}
\item Imposing $Z_2$ on the scalars does not imply a continuous
symmetry in the Higgs sector, but it may or not imply a continuous
symmetry in the Yukawa sector, depending on how the
symmetry is extended into the fermions.
\item Imposing $Z_3$ on the scalars does imply a continuous
symmetry in the Higgs sector, but it may or not imply a continuous
symmetry in the Yukawa sector,depending on how the
symmetry is extended into the fermions.
\item Imposing $Z_n$, with $n \geq 4$, on the scalars implies
always a continuous symmetry, both in the Higgs sector and in the
Yukawa sector.
\end{itemize}
This analysis also permits us to qualify the statement we made in
section~\ref{subsec:HFsymmetry}, following eq.~\eqref{2_S_2/3}:
the imposition on the scalar potential of any discrete symmetry
$Z_n$, with $n>2$, always leads to the same potential, the
Peccei-Quinn one of class IV. However, as we mentioned there,
the extension of those symmetries to the fermion sector might be
able to ``lift the degeneracy" of these symmetries - and in fact it does.
The results of~\cite{Ferreira:2010ir} show very clearly that: (a)
imposing $Z_2$ on the Yukawa terms is different from imposing
$Z_n$, with $n >2$; (b) imposing $Z_3$ on the Yukawa terms is
different from imposing any other $Z_n$, $n > 3$; and (c), that imposing
any $Z_n$, with $n \geq 4$, {\em always} leads to the same form of
Yukawa matrices. As such, in terms of $Z_n$ symmetries, the 2HDM
{\em lagrangian} falls under three classes: lagrangians with a $Z_2$
symmetry; lagrangians with a $Z_3$ symmetry; and lagrangians with any
other $Z_n$, $n \geq 4$, which always lead to the same symmetry
constraints, regardless of the value of $n$.

Also, notice that, since any finite discrete group has an abelian
sub-group, the classification of \cite{Ferreira:2010ir}
is important even when considering non-Abelian family symmetries.
The discussion above concerns extensions of the $Z_2$ and U(1) scalar
symmetries to the fermion sector. That leaves out, of the three possible
Higgs-family symmetries, the models invariant under the full U(2) group -
dubbed class I in section~\ref{sec:symmetries}. However, up until now
it has been impossible to extend the U(2) symmetry in a satisfactory
way to the Yukawa sector - all attempts to do so have lead to zero mass
quarks, for instance. A proof of impossibility has not yet been obtained,
and the question remains open.

\begin{itemize}
\item GCP symmetries extended to the Yukawa sector
\end{itemize}

By comparison with Higgs-family symmetries, it is much simpler
to extend the three possible scalar GCP symmetries to the Yukawa
sector. In fact, as we discussed in section~\ref{sec:GCP}, any
GCP transformation on the doublets can be reduced to a simple
rotation matrix of the form of eq.~\eqref{2_GCP-reduced}. As it turns
out, the Vienna group has also shown~\cite{Ecker:1987qp} that a similar
result is attained for the generic $3\times 3$ transformation
matrices of the quark fields in eqs.~\eqref{Gamma-X} and~\eqref{Delta-X}.
That is, it is always possible, through a judicious choice of basis of
quark fields, to reduce the transformation matrix of the left doublets
to the form
\be
X_L\, \Longrightarrow\, \left( \begin{array}{ccc}
\cos{\alpha} & \sin{\alpha} & 0\\
- \sin{\alpha} & \cos{\alpha} & 0 \\
 0 & 0 & 1
\end{array} \right),
\label{2_eq:fred}
\ee
with some angle $0 \leq \alpha \leq \pi/2$. A similar form is
obtained for the matrices $X_{nR}$ and $X_{pR}$ in
eqs.~\eqref{Gamma-X} and~\eqref{Delta-X}, with independent angles
$\beta$ and $\gamma$ in the same range as $\alpha$. The extremely
simple form of the fermion transformation matrices imposes
severe constraints on the Yukawa couplings. In fact, the
constraints are so serious that no ambiguity occurs in the
fermionic sector when one extends the scalar GCP symmetries
to it: each of the three GCP models have only one possible
implementation on the fermion sector. Recalling that the
three GCP scalar models can be parameterized in terms of the
angle $\theta$ in the simplified GCP transformation of
eq.~\eqref{2_GCP-reduced}, it was concluded
that~\cite{Ferreira:2010bm}:

\begin{itemize}
\item For the CP1 symmetry (class VI), with $\theta = 0$,
there is only one way to extend the scalar symmetry to the
fermion sector which does not entail massless quarks or charged
leptons: by forcing all Yukawa couplings to be real. We are thus
left with a lagrangian with generic real matrices $Y_1^d$, $Y_2^d$,
$Y_1^u$ and $Y_2^u$ - as such the model has tree-level scalar
FCNC, which are not in any way ``naturally suppressed". In this model,
CP violation must needs arise spontaneously, through a relative phase between
the two vevs.
\item For the CP2 symmetry (class III), with $\theta = \pi/2$, there is no
way to extend the symmetry to the Yukawa sector without obtaining
at least one massless charged fermion. As such, the CP2 model
may be considered ruled out by experiment. However, one might
also take the point of view that the CP2 symmetry is an
{\em approximate} one, broken by some manner of mechanism, and
as such the massless fermions it predicts will gain a (small) mass
somehow, corresponding to the first generations of particles.
The phenomenology of such models was explored in great detail
in~\cite{Maniatis:2007de,Maniatis:2009vp,Maniatis:2009by,arXiv:1009.1869}.
\item For the CP3 model (class II), with {\em any} $0 < \theta < \pi/2$,
a remarkable thing happens: all values of $\theta \neq \pi/3$ lead to a
massless quark or charged lepton. Only $\theta = \pi/3$ leads to an
acceptable fermion mass spectrum. The Yukawa matrices which result from
such a symmetry are extremely constrained - the quark sector ends up
depending only on ten independent parameters (seven moduli and three
phases). Nonetheless, this model is capable of fitting all quark masses
and the elements of the CKM matrix with relative ease. The model does possess
tree-level FCNC, but they end up being quite suppressed,
in a ``natural" way. The model also possesses a unique feature, in the
sense that CP violation arises in a completely novel way - we will return to
this point in section~\ref{sec:orig_cpv}. However, the value of the Jarlskog
invariant predicted by this model is several orders of magnitude below its
SM value, which leads to values of the unitarity triangle angles $\alpha$
and $\beta$ practically equal - a prediction of the model in contradiction
with the most recent experimental data~\cite{hfag}.
\end{itemize}

In conclusion, when one extends the three GCP scalar symmetries to the
Yukawa sector, one obtains: arbitrary FCNC for the CP1 case; massless
quarks and charged leptons for the CP2 case; a single CP3 symmetry leading
to three massive generations of fermions, with naturally small FCNC
but predictions for heavy meson phenomenology which do not agree with
experiment.

\newpage

\section{CP violation}
\label{sec:cpv}

\subsection{CP invariance and CP violation at the Lagrangian level:
Scalar potential}

\subsubsection{Two Higgs doublets}
\label{5_2HDM}

The scalar potential for $n_d$ SU(2) doublets
is the most general renormalizable polynomial
consistent with the gauge invariance and
can be written
\begin{equation}
V =
\sum_{a,b = 1}^{n_d}
\mu_{ab} \Phi _{a}^{\dagger }\Phi _{b}+
{\textstyle \frac{1}{2}} \! \sum_{a,b,c,d = 1}^{n_d}
\lambda_{ab,cd}
\left( \Phi _{a}^{\dagger }\Phi _{b}\right) \left( \Phi _{c}^{\dagger }\Phi
_{d}\right).
\label{laghiggs}
\end{equation}
Hermiticity of $V$ implies:
\begin{equation}
\begin{array}{ccc}
\mu_{ab}^\ast = \mu_{ba},
\quad
\lambda_{ab,cd}^\ast = \lambda_{ba,dc}.
\end{array}
\label{hermzy}
\end{equation}
One may redefine the $n_d$ doublets
through unitary transformations without changing the physics.
Those transformations are called Higgs-basis transformations (HBT),
defined by:
\begin{equation}
\Phi _{a} \overset{\mathrm{HBT}}{\longrightarrow }
\Phi _{a}^\prime = \sum_{b=1}^{n_d} V_{ab} \Phi_b,
\quad
\Phi_a^\dagger
\overset{\mathrm{HBT}}{\longrightarrow }
{\Phi^\prime_a}^\dagger = \sum_{b=1}^{n_d} V_{ab}^\ast
\Phi_b^\dagger,
\label{trafowb}
\end{equation}
where $V$ is an $n_{d}\times n_{d}$ unitary matrix
acting in the space of the Higgs doublets.
Under a HBT the couplings $\mu$ and $\lambda$ transform as:
\ba
\mu_{ab} &\overset{\mathrm{HBT}}{\longrightarrow }&
\mu_{ab}^{\prime } = \sum_{m,n=1}^{n_d} V_{am} \mu_{mn} V^\dagger_{nb},
\no
\lambda_{ab,cd} &\overset{\mathrm{HBT}}{\longrightarrow }&
\lambda_{ab,cd}^{\prime }= \sum_{m,n,p,q=1}^{n_d} V_{am} V_{cp}
\lambda_{mn,pq} V^\dagger_{nb} V^\dagger_{qd}.
\label{wbyz2}
\ea
The most general CP transformation that leaves the kinetic energy invariant is:
\begin{equation}
\begin{array}{ccc}
\Phi_a \overset{\mathrm{CP}}{\longrightarrow}
{\displaystyle \sum_{b=1}^{n_d}}
U_{ab} \Phi_b^\ast,
\quad
\Phi_a^\dagger \overset{\mathrm{CP}}{\longrightarrow}
{\displaystyle \sum_{b=1}^{n_d}}
U_{ab}^\ast \Phi_b^T,
\end{array}
\label{trafocp}
\end{equation}
where $U$ is an $n_{d}\times n_{d}$ unitary matrix operating in
the space of the Higgs doublets.
This is the definition of a CP transformation for the scalar doublets
in models with several such doublets.
It combines what would be the CP transformation for a single Higgs doublet
with a Higgs basis transformation,
due to the existence of several doublets with the same quantum numbers.

For two Higgs doublets the most general Higgs potential
is explicitly written in eq.~\eqref{2_VH1}.
Hermiticity only allows for four of the coefficients---$m_{12}^2$,
$\lambda_5$,
$\lambda_6$,
and $\lambda_7$---to be complex.
However,
that potential contains an excess of parameters.
There is no loss of generality in redefining the two Higgs doublets
in such a way that the quadratic terms are diagonal,
thus eliminating $m_{12}^2$.
Furthermore,
one of the remaining three phases
can still be eliminated through a rephasing of one of the doublets.
Thus,
there are (at most) only two independent CP-violating phases
in the potential of a 2HDM.

In this section we address the question of what are the
necessary and sufficient conditions for the potential $V$
to be CP-invariant.
At this stage we analyse the potential prior to gauge-symmetry breaking.
We want to derive HBT-invariant conditions
following the general method proposed in~\cite{Bernabeu:1986fc}
and described in detail in~\cite{Branco:1999fs}.
We investigate what restrictions for the couplings $\mu$ and $\lambda$
are implied by CP invariance \cite{Branco:2005em}.

From the potential of eq.~(\ref{laghiggs})
and the definition of the CP transformation given by eq.~(\ref{trafocp}),
it is clear that the necessary and sufficient condition for
$V$ to conserve CP is the existence of
an $n_{d} \times n_d$ unitary matrix $U$ satisfying the
following relations:
\begin{equation}
\mu^\ast_{ab}=\ {\displaystyle \sum_{m,n=1}^{n_d}}
U_{am}^\dagger \mu_{mn} U_{nb},
\quad
\lambda^\ast_{ab,cd}=
{\displaystyle\sum_{m,n,p,q=1}^{n_d}}
U_{am}^\dagger U_{cp}^\dagger \lambda_{mn,pq} U_{nb} U_{qd}.
\label{cpzy}
\end{equation}
From eq.~(\ref{cpzy}), using the the property of invariance of the trace
under similarity transformations, one can derive
necessary conditions for CP invariance fully
expressed in terms of the couplings $\mu$ and $\lambda$.
Examples of such relations are:
\begin{eqnarray}
I_{1}\equiv \mathrm{Tr}
\left( \mu Z_{Y} \widehat{Z} - \widehat{Z} Z_{Y} \mu \right) &=& 0,
\label{inv1}\\
I_{2}\equiv \mathrm{Tr}
\left( \mu Z_{2} \widetilde{Z} - \widetilde{Z} Z_{2} \mu \right) &=& 0,
\label{inv2}
\end{eqnarray}
where we have introduced the following  $n_{d}\times n_{d}$ Hermitian matrices:
\ba
\widehat{Z}_{ab} &\equiv& \sum_{m=1}^{n_d} \lambda_{ab,mm},
\\
\widetilde{Z}_{ab} &\equiv& \sum_{m=1}^{n_d} \lambda_{am,mb},
\\
\left( Z_{Y} \right)_{ab} &\equiv& \sum_{m,n=1}^{n_d}
\lambda_{ab,nm} \mu_{mn},
\\
\left( Z_{2} \right)_{ab} &\equiv& \sum_{m,n=1}^{n_d}
\lambda_{ap,nm} \lambda_{mn,pb}.
\label{contrzy}
\ea

It is clear that the  eqs.~(\ref{inv1})
and~(\ref{inv2}) are HBT invariant.
These two conditions have the remarkable property
of being necessary and sufficient conditions
for $V$ to conserve CP in the case of two Higgs doublets,
barring the consideration of special isolated points such as
$m_{11}^2 = m_{22}^2$ in the basis where $m_{12}^2 = 0$,
or the special isolated point where $\lambda_6 = - \lambda_7$
with generic $\lambda_1$ and
$\lambda_2$~\cite{Branco:2005em}.
It should be emphasised that these isolated points
are of measure zero and are unstable under renormalization,
since they do not correspond to any
symmetry.
In order to check that these two conditions are sufficient
let us express $I_1$ and $I_2$
in terms of the parameters of the potential after
diagonalization of the quadratic terms,
\ie when $m_{12}^2 = 0$:
\begin{eqnarray}
I_{1} &=& \frac{i}{2}
\left( m_{11}^2 - m_{22}^2 \right)^{2}\,
\mathrm{Im} \left( \lambda_6 \lambda_7^\ast \right),
\label{cond1}
\\
I_{2} &=& i \left( m_{11}^2 - m_{22}^2 \right)\,
\mathrm{Im} \left[
\lambda_5 {\lambda_6^\ast}^2
+ \lambda_5 {\lambda_7^\ast}^2
+ 2 \lambda_5 \lambda_6^\ast \lambda_7^\ast
+ \lambda_6^\ast \lambda_7 \left( \lambda_2 - \lambda_1 \right)
\right].
\label{cond2}
\end{eqnarray}
Choosing $\lambda_5$ to be real,
which can be done without loss of generality,
from $I_1 = 0$ we obtain that $\lambda_6$ and $\lambda_7$
have either equal phases of phases differing by $\pi$.
Then the condition $I_2 = 0$ implies
\begin{equation}
\mathrm{Im} \left[ \lambda_5 \left( \lambda_6^\ast + \lambda_7^\ast
\right)^2 \right] = 0.
\label{simp}
\end{equation}
In the case where $\lambda_6$ and $\lambda_7$ have equal phases,
we find from eq.~(\ref{simp}) that that phase is either $0$
or $\pi/2$.
The case $\arg{\lambda_6} = \arg{\lambda_7} = 0 $
obviously corresponds to a CP-invariant $V$,
with the matrix $U$ in eq.~(\ref{trafocp})
being the $2 \times 2$ identity matrix.
It can be easily checked that the case
$\arg{\lambda_6} = \arg{\lambda_7} = \pi/2 $
also corresponds to a CP-invariant $V$ with
\begin{equation}
U = \left(
\begin{array}{cc}
1 & 0 \\
0 & -1
\end{array}
\right).
\label{uuuu}
\end{equation}
Notice that at this stage we are assuming $m_{11}^2 \neq m_{22}^2$
and $\left| \lambda_6 \right| \neq \left| \lambda_7 \right|$.

Let us now consider the singular points
which are unstable under renormalization and,
therefore,
of limited interest.
It is clear from the explicit form of $I_1$
and $I_2$ that in the special cases $m_{11}^2 = m_{22}^2$
or $\lambda_6 = - \lambda_7$
both conditions $I_1 = I_2 = 0$ are trivially
verified irrespective of the values of the phases.
Yet,
it is still possible to have
CP violation in this region of parameters.
In these cases,
barring again cases of special isolated points,
the following basis-invariant necessary condition for CP conservation
is useful:
\begin{equation}
I_{3} \equiv \mathrm{Tr}
\left( Z_2 Z_3 \widehat{Z} - \widehat{Z} Z_3 Z_2 \right) = 0,
\label{inv3}
\end{equation}
where $Z_3$ is one further $n_{d}\times n_{d}$ Hermitian matrix given by
\begin{equation}
\left( Z_{3} \right)_{ab} \equiv
{\displaystyle \sum_{m,n,p,r,s=1}^{n_d}}
\lambda_{am,rp} \lambda_{mr,ns} \lambda_{pn,sb}.
\label{z3}
\end{equation}
An important characteristic of $I_3$ is the fact that,
unlike $I_{1}$ and $I_{2}$,
it is built exclusively from the quartic couplings $\lambda$.
Therefore,
non-vanishing of this invariant necessarily signals hard CP violation.

Let us consider the case where $m_{12}^2 = 0$,
$\lambda_1 = \lambda_2$,
and $\lambda_6 = \lambda_7$.
In that case the quartic couplings by themselves conserve CP,
provided one chooses
\begin{equation}
U = \left(
\begin{array}{cc}
0 & 1 \\
1 & 0
\end{array}
\right).
\label{antd}
\end{equation}
This CP symmetry is only broken by $m_{11}^2 \neq m_{22}^2$,
\ie by the quadratic terms,
and therefore one may say that this potential corresponds to ``hidden"
soft CP breaking.
In this example,
both $I_1$ and $I_3$ vanish identically,
whilst $I_2 = 4 i
\left( m_{11}^2 - m_{22}^2 \right)
\mathrm{Im} \left( \lambda_5 {\lambda_6^\ast}^2 \right)$.

Another interesting example of soft symmetry breaking
with two Higgs doublets is obtained by taking
$\lambda_1=\lambda_2$ and $\lambda_6 = - \lambda_7$.
In this case the quartic couplings by themselves conserve CP,
provided one now chooses $U$
\begin{equation}
U = \left(
\begin{array}{cc}
0 & 1 \\
- 1 & 0
\end{array}
\right).
\label{-11}
\end{equation}
Once again we have an example of hidden soft CP breaking for
$m_{11}^2 \neq m_{22}^2$.
In this example all three CP-odd invariants defined above vanish
and an additional CP-odd invariant is necessary.
In Ref.~\cite{Gunion:2005ja} the following condition was provided:
\begin{equation}
I_4 \equiv \mathrm{Im}
{\displaystyle \sum_{a,b,\ldots i=1}^2}
\left(\lambda_{ac,bd} \lambda_{ce,dg} \lambda_{eh,fi}
\mu_{ga} \mu_{hb} \mu_{if} \right).
\label{i4}
\end{equation}
It was pointed out in~\cite{Davidson:2005cw} that,
under Higgs basis transformations,
the given relations $\lambda_1 = \lambda_2$ and $\lambda_6 = - \lambda_7$
remain invariant,
this case being a special isolated point
in the 2HDM scalar-potential parameter space.
In the language of Ref.~\cite{Ivanov:2005hg}
this corresponds to the absence of the triplet
in the decomposition of the quartic Higgs potential
into irreducible representations of the SU(2) Higgs basis transformation
for two Higgs doublets.

More examples of hidden soft symmetry breaking can be written,
based on different $U$ matrices,
such as
\begin{equation}
U = \left(
\begin{array}{cc}
0 & i \\
 1 & 0
\end{array}
\right)
\quad \mathrm{or} \quad
U = \left(
\begin{array}{cc}
0 & -i \\
1 & 0
\end{array}
\right).
\end{equation}
In both these cases $I_1$ does not automatically vanish.

Under a HBT the specific form of $U$ for a given CP transformation
changes in the following way:
\begin{equation}
U^\prime = V U V^T .
\label{lkj}
\end{equation}
Invariance under CP of the 2HDM potential for $U$ given
by eq.~(\ref{antd}) will therefore look different in a different
Higgs basis. Taking $V$ as
\begin{equation}
V = \frac{1}{\sqrt{2}}\left(
\begin{array}{cc}
1 & 1 \\
1 & -1
\end{array}
\right),
\label{411}
\end{equation}
one obtains:
\begin{equation}
U^\prime = \left(
\begin{array}{cc}
1 & 0 \\
0 & -1
\end{array}
\right).
\label{dif}
\end{equation}
%
This two Higgs doublet model with the reflection symmetry
$\Phi_2 \rightarrow - \Phi_2 $ softly broken by the quadratic terms
proportional to $m^2_{12}$ in eq.~\eqref{2_VH1} was
considered in Refs. \cite{Branco:1985aq}, \cite{Weinberg:1990me},
\cite{Ginzburg:2004vp}.

\subsubsection{Three Higgs Doublets}

In this subsection we briefly mention the case of three Higgs
doublets. The total number of independent CP-violating phases
in the scalar potential is given by \cite{Branco:2005em}:
\begin{equation}
N_{\mathrm{phases}} = \frac{1}{4} \left[ n_d^2 ( n_d^2 -1) \right]
- (n_d -1),
\label{cont}
\end{equation}
the second term corresponds to the number of phases one can eliminate
by rephasing the Higgs fields. In general there are sixteen independent
phases for $n_d = 3$. Models with three Higgs doublets are much more
involved than those with only two. Three Higgs doublets were considered
\cite{Weinberg:1976hu} in an attempt to introduce CP violation in
an extension of the SM with NFC~\cite{Glashow:1976nt,Paschos:1976ay}.
Natural flavour
conservation means that the Higgs neutral currents conserve all
quark flavours for all values of the parameters of the theory, i.e.,
this conservation is a consequence of the group structure and the
representation content of the theory and not of a special choice of
parameters. NFC with several Higgs doublets was implemented in
Ref.~\cite{Weinberg:1976hu} by requiring invariance of the Lagrangian
under separate reflections under which any one of the
doublets (and perhaps some fermions) change sign. Discrete
symmetries of this kind were used to insure that only one Higgs couples to
the right-handed up quarks and another to the right-handed down quarks.
It was pointed out in Ref.~\cite{Weinberg:1976hu} that for three
or more Higgs doublets the scalar potential that is invariant
under these reflections need not conserve CP. On the other hand
it was shown in \cite{Branco:1979pv,Branco:1980sz,Branco:1985pf}
that with three Higgs doublets it is possible to
violate CP spontaneously while having NFC.
In the three Higgs doublets model proposed by
Weinberg \cite{Weinberg:1976hu}, a
$Z_{2}\times Z_{2}\times Z_{2}$
symmetry, under separate reflections of the Higgs doublets of the form $\phi
_{i}\rightarrow -\phi _{i}$, together with an appropriately chosen
transformation for the quark fields, ensures NFC and leads
to a strong reduction in the number of parameters. The Higgs
potential is given by:
%
\begin{equation}
\begin{array}{r}
V= {\displaystyle \sum_{i=1}^3} \left[m_{i}\ \Phi _{i}^{\dagger }\Phi _{i}+a_{ii}\ \left(
\Phi _{i}^{\dagger }\Phi _{i}\right) ^{2}\right] +{\displaystyle \sum_{i<j} } \ \Bigg\{2b_{ij}\
\left( \Phi _{i}^{\dagger }\Phi _{i}\right) \ \left( \Phi _{j}^{\dagger
}\Phi _{j}\right) + \\
\\
+2c_{ij}\ \left( \Phi _{i}^{\dagger }\Phi _{j}\right) \ \left( \Phi
_{j}^{\dagger }\Phi _{i}\right) +\left[ d_{ij}\ e^{i\theta _{ij}}\ \left(
\Phi _{i}^{\dagger }\Phi _{j}\right) ^{2}+h.c.\right] \Bigg\}.
\end{array}%
\   \label{weinpot}
\end{equation}
There are three different $d_{ij}\ e^{i\theta _{ij}}$ terms,
and only these can be complex. It was pointed out by Weinberg
that in general one cannot rotate away simultaneously the three
phases $\theta_{ij}$. A relevant CP-odd invariant relevant to this
model is:
\begin{equation}
I_{2}^{W}=\mathrm{Im}{\displaystyle \sum_{a,\ldots i =1}^{n_d}}
[Z_{abcd}\ Y_{be}\ Z_{efgh}\ \widehat{Z}
_{fi}\ Z_{icha}] \ .  \label{weininv1}
\end{equation}
Its explicit form is:
\begin{equation}
\begin{array}{r}
I_{2}^{W}=d_{12}d_{13}d_{23}\ [(\ m_{3}-m_{2})(a_{11}-b_{23})-(\
m_{3}-m_{1})(a_{22}-b_{13})+ \\
\\
+(\ m_{2}-m_{1})(a_{33}-b_{12})]\sin (\theta _{12}-\theta _{13}+
\theta _{23}) \ .
\end{array}
\label{weinin2}
\end{equation}
Assuming non-degenerate values for the $m_i$, a non-vanishing $I_{2}^{W}$
indicates a non-vanishing $(\theta _{12}-\theta _{13}+\theta _{23})$.

In the general case of three Higgs doublets it is possible to build
relevant simpler CP-odd invariants \cite{Branco:2005em},
which are irrelevant for the case of
two Higgs, since they trivially vanish in that case.

The softly broken three Higgs doublets model of Weinberg and relevant
CP-odd invariant in this case is also discussed in
Ref.~\cite{Branco:2005em}.

\subsection{CP Violation after Spontaneous Symmetry Breaking}

\subsubsection{Spontaneous CP violation}

The idea of spontaneous CP breaking was suggested by T. D. Lee
\cite{Lee:1973iz} in the early stages of unified gauge theories.
In order to have spontaneous CP violation, one must have
a Lagrangian which is CP invariant but after spontaneous
gauge symmetry breaking the vacuum is not CP invariant.
One has to be careful in correctly identifying a vacuum
as CP violating. This has to do with the fact that often a
CP invariant Lagrangian allows not only for a single CP
transformation, but for a whole class of CP transformations.
In order to have a genuine spontaneous CP violation,
the following two conditions have to be satisfied: \\

\noindent
(i) The Lagrangian is invariant under a CP transformation
which may be physically interpreted as CP. \\
(ii) There is no transformation which can be physically
interpreted as CP which leaves both the vacuum and the
Lagrangian invariant. \\

\noindent
In the Standard Model there is only one Higgs doublet. Hermiticity
requires that the parameters of the scalar potential be real and, as
a result the scalar potential of the Standard Model cannot
violate CP. Furthermore, spontaneous CP violation is also ruled out,
in this case, due to the possibility of using a U(1) gauge
transformation to make the vacuum expectation value of the
neutral Higgs real and positive.

Since we are working in the framework of relativistic quantum
field theory, the CPT theorem
applies~\cite{Luders:1954zz,Pauli,Jost:1957zz,Streater:1989vi},
so spontaneous CP
breaking also implies spontaneous T breaking and vice-versa.

For definiteness, let us consider an extension of the SM where
n SU(2)$\times$U(1) scalar doublets are introduced. In order
to include the possible existence of symmetries of the
Lagrangian under which the scalar doublets transform
non-trivially, one has to consider the most general CP
transformation which leaves invariant the kinetic energy
terms of the scalar potential. Thus, we consider the following
CP transformation for the scalar doublets:
\begin{equation}
CP \Phi_{i}(CP)^\dagger = \sum_{j=1}^{n} U_{ij} \Phi^\ast_{j}
\label{cpp}
\end{equation}
corresponding to eq.~(\ref{trafocp}).
Assuming that the vacuum is CP invariant, meaning that:
\begin{equation}
CP|0 \rangle = |0 \rangle
\label{c00}
\end{equation}
one can readily derive from  eqs.~(\ref{cpp}) and (\ref{c00})
the following relation \cite{Branco:1983tn}:
\begin{equation}
 \sum_{j=1}^{n} U_{ij} \langle 0|\Phi_{j}|0 \rangle^\ast =
\langle0 |\Phi_{i}|0 \rangle
\label{rmi}
\end{equation}
If the vacuum is such that none of the CP symmetries allowed by the
Lagrangian satisfy eq.~(\ref{rmi}), then this means that the vacuum is
not CP invariant and we say that CP is spontaneously broken.

From the previous discussion, one concludes that in the presence
of extra symmetries imposed on the Lagrangian, one has to be specially
careful in analysing whether a particular vacuum violates
CP or not. The point is that the presence of an
extra symmetry in the Lagrangian may allow for non-trivial possibilities
for the matrix $U$ in eq.~(\ref{cpp}), which may satisfy  eq.~(\ref{rmi}),
even in the case of complex minimae. This can be best understood
through a simple example: \\

\subsubsection{An example}

Let us consider an extension of the SM with two Higgs doublets,
where a $Z_2$ symmetry is introduced, with the scalars
transforming as:
\begin{equation}
\Phi _{1}{\longrightarrow } - \Phi _{1}
\quad ,\quad  \Phi _{2} {\longrightarrow }
\Phi _{2}
\label{z2z2}
\end{equation}
A possible motivation for the introduction of a  $Z_2$ symmetry, is the
requirement of natural flavour conservation (NFC) in the Higgs
sector \cite{Glashow:1976nt,Paschos:1976ay}.
Indeed, if the right-handed down quarks are odd under $Z_2$,
\begin{equation}
d_R \longrightarrow - d_R
\end{equation}
while all other fields are even, then down quarks receive mass only
from $\Phi _{1}$, while up quarks receive mass only from $\Phi _{2}$,
thus satisfying the NFC principle.

The most general gauge symmetry invariant Higgs potential,
consistent with the $Z_2$ symmetry, can be written as
%
\begin{equation}
V = V_0 + \left[ \lambda_5 \left( \Phi _{1}^{\dagger }\Phi _{2}\right)
\left( \Phi _{1}^{\dagger }\Phi _{2} \right) +  \mathrm{h.c.} \right]
\end{equation}
where $V_0$ denotes the part of the potential which does not depend
on the relative phase of the $ \Phi _{i}$. It is clear that for
$ \lambda_5 > 0$, the minimum of the potential is at:
\begin{equation}
\langle  \phi ^0_{1} \rangle = v_1 e^{i \frac{\pi}{2}};
\qquad \langle  \phi ^0_{2} \rangle = v_2
\label{left<}
\end{equation}
One could be tempted to think that this vacuum violates CP
``maximally''. This is  \underline{not the case}, as it can be seen from
eq.~(\ref{rmi}). Indeed, the presence of the  $Z_2$ symmetry allows for various
choices of matrix $U$ which defines the CP properties of $ \Phi _{1}$,
 $\Phi _{2}$. Apart from the trivial one, one may choose:
\begin{equation}
U = \left(
\begin{array}{cc}
-1 & 0 \\
0 & 1
\end{array}
\right).
\end{equation}
It is clear that with this choice of $U$ , eq.~(\ref{rmi}) is satisfied,
\begin{equation}
 \left(
\begin{array}{cc}
-1 & 0 \\
0 & 1
\end{array}
\right)  \  \left[ \begin{array}{c}  v_1 e^{i \frac{\pi}{2}} \\
v_2 \end{array}  \right]^\ast =  \left[ \begin{array}{c}
v_1 e^{i \frac{\pi}{2}} \\ v_2 \end{array}  \right]
\end{equation}
thus proving that the vacuum of  eq.~(\ref{left<}) is CP invariant.
At this stage, it is worth noting that one encounters an entirely
analogous situation in the  Minimal Supersymmetric Standard Model
(MSSM) where one of the possible minima also has a relative phase
of $\pi/2$, which does not lead to spontaneous CP violation.
It has been shown \cite{Branco:2000dq,Branco:2006pj}
that one may achieve spontaneous
CP violation in an extension of the MSSM where one introduces a singlet
scalar. \\

Coming back to the previous discussion, one may wonder whether this
is a generic feature of vacua with ``calculable'' vacuum phases.
This question was addressed in detail by Branco, G\'erard and
Grimus in Ref.~\cite{Branco:1983tn}. It was shown that indeed this
is the case for most vacua with calculable phases
arising from the presence of extra symmetries in the Higgs potential.
Another interesting example is the case of $S_3$ symmetry introduced in
a Higgs potential with three Higgs doublets, transforming as a
three dimensional reducible representation of $S_3$. The most
general renormalizable Higgs potential can be written \cite{Derman:1978rx}:
\begin{eqnarray}
V & = &  V_{0} + \lambda_1 \left( \Phi _{i}^{\dagger }\Phi _{j}\right)
+  \lambda_2 \left[ \ \left( \Phi _{i}^{\dagger }
\Phi _{i}\right)  \ \left( \Phi
_{i}^{\dagger }\Phi _{j}  + \mathrm{h.c.} \right) \right] \nonumber \\
 & & +  \lambda_3  \left[ \ \left( \Phi _{i}^{\dagger }
\Phi _{i}\right)  \ \left( \Phi
_{j}^{\dagger }\Phi _{k}  + \mathrm{h.c.} \right) \right] +
 \lambda_4 \left[ \ \left( \Phi _{i}^{\dagger }
\Phi _{j}\right)  \ \left( \Phi
_{i}^{\dagger }\Phi _{k}\right)  + \mathrm{h.c.}  \right] +  \\
 & & + \lambda_5  \left[ \ \left( \Phi _{i}^{\dagger }
\Phi _{j}\right)  \ \left( \Phi
_{k}^{\dagger }\Phi _{i}\right)  + \mathrm{h.c.}  \right] +
\lambda_6  \left[ \left( \Phi_{i}^{\dagger }\Phi _{j}\right) ^{2}
 + \mathrm{h.c.}  \right] \nonumber
\end{eqnarray}
where $V_0$  denotes the part of the potential with no phase
dependence. In each square bracket a sum is understood over all
independent permutation of $i$, $j$, $k$ with $i \neq j \neq k$.
It can be readily verified that there is a region of parameter
space where the vacuum has the following phase structure:
\begin{equation}
\left<  \phi ^0_{i} \right> = v \exp \left[ i \frac{2\pi}{3}
(k-1) \right]; \qquad k=1,2,3
\qquad \left<  \phi ^0_{2} \right> = v_2 \ .
\label{k=123}
\end{equation}
It can be easily shown using an argument analogous to the one
used in the $Z_2$ example that, contrary to na\" ive intuition,
the vacuum of eq.~(\ref{k=123}) is CP and T invariant. One may
wonder whether this is a universal feature. Namely, one may ask whether
calculability of vacuua phases necessarily implies CP
invariant vacuua. In Ref.~\cite{Branco:1983tn} it was shown
that this is not the case. A counterexample was found
\cite{Branco:1983tn} based on the group $\Delta(27)$
which is a dyhedral-like subgroup of SU(3) with 27 elements.
In this example, a CP-violating vacuum with calculable
phases was found. \\

\subsubsection{A Survey of Models with Spontaneous CP Violation}

At this stage, the following question is in order: What is the minimal
extension of the SM where one may achieve
spontaneous CP violation while at the same time not entering in conflict
with experiment? In order for a given model to
be a candidate for a realistic example of spontaneous CP violation,
it should satisfy the following conditions:

i) The CP violating phase arising from the vacuum should be able to create
a complex CKM matrix, leading to CP violation through W-mediated weak
currents.

ii) The model should be able to avoid in a plausible way, i.e. without
unreasonable fine tuning, the stringent experimental constraints arising
from FCNC processes, as well as from the knowledge of the location of the
upper vertex of the Unitarity Triangle \cite{Nakamura:2010zzi}.

First let us explain why the model should satisfy the condition (i).
At present, there is strong evidence for a complex CKM matrix
even if one allows for the  presence of New Physics beyond
the SM \cite{Botella:2005fc}.
The best  evidence arises from the non-vanishing of the rephasing
invariant angle $\gamma$ \cite{Nakamura:2010zzi} which  does not
receive important contributions
from New Physics. It is non-trivial for a model with
spontaneous CP violation to satisfy at the same time
both constraints (i) and (ii). The reason has to do with the fact that
in order to have spontaneous  CP violation, the Lagrangian has to be CP
invariant,  which requires real Yukawa couplings. Then the phase arising
from the vacuum expectation value must give rise to complex
quark mass matrices $M_u$, $M_d$, in such  a way that the weak basis
invariant $I_{CP}$ defined by
\begin{equation}
I_{CP} \equiv \mathrm{Tr} \left[ H_u, H_d \right]^3
\label{icp}
\end{equation}
does not vanish \cite{Bernabeu:1986fc}, with $H_u = M_u M_u^\dagger$ and
$H_d = M_d M_d^\dagger$. The non-vanishing of $I_{CP}$ has
to be achieved without
generating too large FCNC, so that condition (ii) is satisfied. This is the
source of the difficulty in constructing realistic models of spontaneous
CP violation. In the sequel we shall present a simple model which
satisfies both conditions in an elegant way. We consider next some of
the minimal models with spontaneous CP violation which have been
considered in the literature. \\

\noindent
{\bf A. Lee's Model} \cite{Lee:1973iz} \\
As previously mentioned, this is the first model of spontaneous CP violation
proposed in the literature. Lee introduced two doublets
but no extra symmetry in the Higgs potential. In this case it was shown by
Lee that there is a non-singular region of the Higgs
parameter space where the vacuum conserves electric charge but violates CP.
At the time Lee suggested his model, there were only
two incomplete generations, since charm had not been discovered. Therefore in
Lee's model with only two generations CP arises
exclusively from Higgs exchange. If we implement Lee's model in the framework
of three generations, it turns out that although
there is only one phase ($\theta$) arising from the vacuum, one can generate
a non-trivial CP violating phase in $V_{CKM}$. This can be seen
by noting that the two quark mass matrices can be written
\begin{eqnarray}
M_d = \frac{1}{\sqrt{2}} ( v_1  Y^d_1 +
                           v_2 e^{i \alpha} Y^d_2 ), \quad
M_u = \frac{1}{\sqrt{2}} ( v_1  Y^u_1 +
\label{lee}                           v_2 e^{-i \alpha} Y^u_2 ),
\end{eqnarray}
where $Y^d_i$, $Y^u_i$ are real matrices.
From eq.~(\ref{lee}) it follows that one generates a complex $H_d$,
with its imaginary part given by:
\begin{equation}
\mathrm{Im}H_d = v_1 v_2 \left[  Y^d_2 {Y^d}^\dagger_1 -
 Y^d_1 {Y^d}^\dagger_2  \right] \ \sin \theta
\label{imhd}
\end{equation}
with an analogous expression for $H_u$.  It is clear from
eq.~(\ref{imhd}) that in spite of having only one phase $\theta$
arising from the vacuum, the structure of $H_d$, $H_u$ is such that the
invariant $I_{CP}$, given in eq.~(\ref{icp}), does not vanish, which is
sufficient to have a non-trivially complex CKM matrix. One concludes
that in Lee's model with three generations, there are two sources of
CP violation, Higgs exchange plus CKM mechanism. Therefore,
the Lee model satisfies condition (i). However, the model has
difficulty in satisfying condition (ii), since it leads, in general,
to too large FCNC, unless one assumes very large Higgs masses,
in the range 1--10 TeV, or invokes some suppression
mechanism \cite{Branco:1985aq}. \\

\noindent
{\bf B. Models with Natural Flavour Conservation in the Higgs sector} \\
We have seen in the previous section that in the case of two Higgs
doublets, the imposition of NFC in the Higgs sector eliminates
the possibility of generating spontaneous CP violation. However,
this is no longer true for three Higgs doublets, where it was shown
\cite{Weinberg:1976hu} that one can generate genuine spontaneous
CP violation. However, it was also shown \cite{Branco:1979pv}
that in this model $V_{CKM}$ is real, which is in disagreement with
present experiment \cite{Nakamura:2010zzi,Botella:2005fc}. \\

\noindent
{\bf C. A Minimal Realistic Model with Spontaneous CP Violation} \\
Now we present what we consider to be the minimal extension of the
SM where one can generate spontaneous CP violation, leading to a complex
CKM matrix, having no conflict with the experimentally suppressed
FCNC processes.

Let us consider an extension of the SM which consists of the
addition of a vector-like singlet quark $D$ and a complex
scalar singlet $S$. The vector-like quark may be the down-type or
the up-type. For definiteness, we consider that it is of the down-type
with electrical charge $Q= -1/3$. For simplicity, we introduce
a $Z_2$ symmetry, under which the new fields are odd,
\begin{equation}
Z_2: \ \  D_L \longrightarrow - D_L,  \ \ D_R \longrightarrow - D_R, \ \
S \longrightarrow - S
\end{equation}
while all the SM fields are even. Strictly speaking, the introduction
of the $Z_2$ symmetry is not necessary. However, its presence in the model
provides a simple solution \cite{Bento:1991ez}  to the strong
CP
problem
\cite{'tHooft:1976up,'tHooft:1976fv,Nelson:1983zb,Nelson:1984hg,Barr:1984qx,Peccei:2006as}.
As a result of the $Z_2$ symmetry, the couplings
$\overline{d_{Li}} D_R \Phi$ are forbidden but the mass term and
couplings:
\begin{equation}
M {\overline {D_L}} D_R + ({f_j} S + {f_j}^{\prime} S^\ast )
{\overline {D_L}} d_R^j + h. c.
\end{equation}
are allowed by gauge and $Z_2$ invariance. As a result the $4 \times 4$
quark mass matrix has the form:
\begin{equation}
{\cal M}_d= \left(\begin{array}{cc}
m_d & 0 \\
M_D & M \end{array}\right),
\end{equation}
where $m_d$ stands for the $3 \times 3$ mass matrix connecting
standard quarks, the zero reflects the presence of the $Z_2$ symmetry
and $M_D$ is a $1 \times 3$ matrix given by:
\begin{equation}
(M_D)_j = \left( f_j V \exp{(i \alpha)} +  {f_j}^{\prime} V
\exp{(- i \alpha)} \right).
\end{equation}
We have assumed that CP is broken by the vacuum with:
\begin{equation}
\langle S \rangle = V \exp (i \alpha )
\label{vev}
\end{equation}
It can be shown \cite{Bento:1990wv} that  the presence in the scalar
potential  of terms like $(S^2 + S^{\ast 2})$,  $(S^4 + S^{\ast 4})$,
implies that there is a region of parameters where the
minimum is at the $\langle S \rangle$ of eq.~(\ref{vev}), with
$\alpha$ a non-trivial phase. This vacuum breaks CP
spontaneously. Note that although one has only one phase
$\alpha$ arising from the vacuum, due to the arbitrariness
of the real couplings $f_j$, $ {f_j}^{\prime}$, $M_d$ is an arbitrary
$1 \times 3$ complex matrix. The matrix ${\cal M}_d$ is diagonalized
by the usual bi-unitary transformation:
\begin{equation}
U^\dagger_L {\cal M}_d U_R = \left( \begin{array}{cc}
d & 0 \\
0 & D \end{array} \right)
\end{equation}
where $d=$ diag.($m_d$, $m_s$, $m_b$) and $D$ denotes the mass of the
physical $Q = -1/3$ vector-like quark. The unitary matrix $U_L$
can be written in block form:
\begin{equation}
 U_L =  \left(\begin{array}{cc}
K & R \\
S & T \end{array}\right)
\end{equation}
The matrix $K$ stands for the usual $3 \times 3$ CKM matrix, which is
determined by the relation:
\begin{equation}
K^{-1} (m_{eff} m^\dagger_{meff}) \ K = d^2
\label{vckm}
\end{equation}
where
\begin{equation}
m_{eff} m^\dagger_{meff} = m_d {m_d}^\dagger -
\frac{m_d {M_D}^\dagger M_D \  {m_d}^\dagger}{ \tilde{M} ^2}
\label{mem}
\end{equation}
where $\tilde{M} ^2 = M_D {M_D}^\dagger + M^2$. The  crucial point is that
the two terms contributing to $m_{eff} m^\dagger_{meff}$ are of the same
order of magnitude, since both $M_D$ and $M$ are $SU(2) \times U(1)$
invariant mass terms which are expected to be of the same large
mass scale. As a result, a nontrivial complex CKM matrix is generated
and the CP violating phase is not suppressed by the large scale
of $M_D$ and $M$ . Of course, this can be explicitly checked by evaluating
$ \mathrm{Tr} \left[ H_u, m_{eff} m^\dagger_{meff} \right]^3$.
It is worth summarizing the main features of this class of models,
where spontaneous CP violation is achieved through the introduction
of at least one vector-like isosinglet quark and a complex singlet scalar:\\

\noindent
(i) They provide a simple framework for having spontaneous CP
violation, while at the same time generating a complex CKM matrix. \\

\noindent
(ii) The $3 \times 3$ quark mixing matrix connecting standard
quarks is not unitary. However, deviations from unitarity are naturally
suppressed by the ratio $m^2/M^2$ where $m$ and $M$  denote the mass
of the standard quarks and the mass of the heavy isosinglet quark(s),
respectively. These deviations from unitarity lead to
Z--mediated flavour changing neutral currents, (FCNC)
which are again naturally suppressed by the ratio  $m^2/M^2$.
This is a general feature of models with vector-like quarks
\cite{delAguila:1985mk,Branco:1986my,delAguila:1985ne,Nir:1990yq,
Silverman:1991fi,Choong:1993gq,Barger:1995dd,Gronau:1996rv,delAguila:1997vn,
Branco:1992wr}.
If the mass of the new quark is of order 1 TeV, the FCNC are sufficiently
suppressed so that they do not enter in conflict with the stringent limits on
$\Delta S=2$ tree level transitions. Yet, one may have significant
contributions to $B_d$--$\overline{B_d}$
mixing~\cite{Barenboim:1997qx,Barenboim:2000zz,Barenboim:2001fd,
AguilarSaavedra:2002kr,AguilarSaavedra:2004mt,Botella:2006va,Botella:2008qm}
and $B_s$--$\overline{B_s}$ mixing which could be detected at LHCb
and in super--B factories.

\subsection{CP-violating quantities from the scalar potential}

As shown in section~\ref{subsec:HiggsBasis},
in the 2HDM one may define the `Higgs basis'
as a basis $\left( H_1, H_2 \right)$ for the scalar $\sud$ doublets
such that the neutral component of $H_1$
has real and positive vacuum expectation value $v \left/ \sqrt{2} \right.$
while $H_2$ has vanishing vev.
In the Higgs basis the doublets are given by eqs.~(\ref{2_eq:higbas}),
where $G^\pm$ and $G^0$ are the three `would-be Goldstone bosons'
while $H^\pm$ are the two physical charged scalars.
The three physical neutral scalars $S_{1,2,3}$
are the linear combinations of $H$,
$R$,
and $I$ given by eq.~(\ref{juaoe}),
where $T$ is a matrix of $\sot$.

However,
as was also stressed in section~\ref{subsec:HiggsBasis},
the Higgs basis is {\em not} uniquely defined,
because,
when one rotates $H_2$ as in eq.~(\ref{jvuep}),
the conditions for the Higgs basis,
\viz the vev of $H_2$ being zero and the vev of $H_1$ being real and positive,
remain satisfied.
The rotation~(\ref{jvuep}) implies
\be
\left( \begin{array}{c} R \\ I \end{array} \right) \to
O \left( \begin{array}{c} R \\ I \end{array} \right),
\label{uixms}
\ee
where $O \in \sod$,
\cf eq.~(\ref{yuesn}).

Under a CP transformation,
besides the change in the sign of the space coordinates,
$H_1$ transforms to its complex conjugate\footnote{In
second-quantized field theory one must employ,
instead of the complex conjugate,
the transpose of the Hermitian conjugate.}
while $H_2$ transforms to its complex conjugate
apart from an arbitrary phase:
\be
H_1 \left( t, \vec r \right)
\stackrel{\mathrm{CP}}{\rightarrow}
H_1^\ast \left( t, - \vec{r} \right),
\quad
H_2 \left( t, \vec r \right)
\stackrel{\mathrm{CP}}{\rightarrow}
e^{i \zeta} H_2^\ast \left( t, - \vec r \right).
\ee
Thus,
in a CP transformation,
\be
\left( \begin{array}{c} R \\ I \end{array} \right)
\stackrel{\mathrm{CP}}{\rightarrow}
O^\prime \left( \begin{array}{c} R \\ I \end{array} \right),
\label{mghir}
\ee
where $O^\prime \in \od$ but $\det{O^\prime} = -1$.

Our task in this section consists in finding quantities
which depend solely on the scalar potential
and {\em are invariant under basis transformations}\/
but {\em change sign under CP}.
The solution to this problem \cite{Lavoura:1994fv}  hinges on the matrix
\be
\epsilon = \left( \begin{array}{cc} 0 & 1 \\ -1 & 0 \end{array} \right),
\label{svhge}
\ee
which has the property
\be
O\, \epsilon\, O^T = \epsilon \det{O}
\ee
for any $O \in \od$.

The scalar potential of a general 2HDM in the Higgs basis
is in eq.~(\ref{potH}).
The quantities $\bar m_{11}^2$ and $\bar m_{12}^2$
are functions of the vev $v$
and of the quartic couplings $\bar \lambda_1$ and $\bar \lambda_6$,
respectively,
through eqs.~(\ref{2_stationarity_HB}).
When one expands the scalar potential
as a function of the component fields of $H_1$ and $H_2$,
one finds terms of the following forms:\footnote{We
do not consider terms containing $G^\pm H^\mp$,
which are trickier to handle.}
\begin{description}
\item $a^S S$,
where $a^S$ is a real coefficient
and $S$ is a neutral combination of fields which is invariant
under both the basis transformation of eq.~(\ref{uixms})
and the CP transformation in eq.~(\ref{mghir}).
For instance,
$S$ may be $H^2$,
$H^3$,
${G^0}^2$,
$H^- H^+$,
$G^- G^+$,
$R^2 + I^2$,
and so on.
\item $\left( b_1^S R + b_2^S I \right) S$,
where $b_1^S$ and $b_2^S$ are real coefficients.
\item $\left[ c_{11}^S R^2 + c_{22}^S I^2
+ \left( c_{12}^S + c_{21}^S \right) R I \right] S$,
where the $c_{ab}^S$ ($a, b = 1, 2$) are real coefficients and,
without loss of generality,
$c_{21}^S = c_{12}^S$.
\end{description}
Considering for instance $V_2$ in eqs.~(\ref{ibhje})--(\ref{ibhje2}),
we see that
$b_{1}^H = v^2\, \real\, \bar \lambda_6$,
$b_{2}^H = - v^2\, \imag\, \bar \lambda_6$,
and that
$c_{11} = - c_{22} = \left( v^2 / 4 \right) \real\, \bar \lambda_5$,
$c_{12} = c_{21} = - \left( v^2 / 4 \right) \imag\, \bar \lambda_5$
(in $V_2$ there are also terms
$m_+^2 H^- H^+$,
$\left( v^2 / 2 \right) \bar \lambda_1 H^2$,
and
$\left( m_+^2 / 2 + v^2 \bar \lambda_4 / 4 \right) \left( R^2 + I^2 \right)$,
which are of the form $a^S S$).
If we also consider $V_3$ in eq.~(\ref{sbuir}),
we find for instance
$b_1^{H^-H^+} = v\, \real\, \bar \lambda_7$
and $b_2^{H^-H^+} = - v\, \imag\, \bar \lambda_7$.

It is clear that,
under the basis transformation of eq.~(\ref{uixms}),
\be
\left( \begin{array}{c} b_1^S \\ b_2^S \end{array} \right)
\to O
\left( \begin{array}{c} b_1^S \\ b_2^S \end{array} \right),
\quad
\left( \begin{array}{cc} c_{11}^S & c_{12}^S \\
c_{21}^S & c_{22}^S \end{array} \right)
\to O
\left( \begin{array}{cc} c_{11}^S & c_{12}^S \\
c_{21}^S & c_{22}^S \end{array} \right)
O^T,
\ee
and similarly,
under the CP transformation of eq.~(\ref{mghir}),
\be
\left( \begin{array}{c} b_1^S \\ b_2^S \end{array} \right)
\stackrel{\mathrm{CP}}{\rightarrow} O^\prime
\left( \begin{array}{c} b_1^S \\ b_2^S \end{array} \right),
\quad
\left( \begin{array}{cc} c_{11}^S & c_{12}^S \\
c_{21}^S & c_{22}^S \end{array} \right)
\stackrel{\mathrm{CP}}{\rightarrow} O^\prime
\left( \begin{array}{cc} c_{11}^S & c_{12}^S \\
c_{21}^S & c_{22}^S \end{array} \right)
{O^\prime}^T.
\ee
Using the matrix $\epsilon$ of eq.~(\ref{svhge}),
it is then easy to construct basis-invariant,
CP-violating quantities like
\[
\left( \begin{array}{cc} b_1^S, & b_2^S \end{array} \right)
\epsilon
\left( \begin{array}{c} b_1^{S^\prime} \\ b_2^{S^\prime} \end{array} \right)
\quad \mathrm{or} \quad
\left( \begin{array}{cc} b_1^S, & b_2^S \end{array} \right)
\left( \begin{array}{cc}
c_{11}^{S^\prime} & c_{12}^{S^\prime} \\
c_{21}^{S^\prime} & c_{22}^{S^\prime}
 \end{array} \right)
\epsilon
\left( \begin{array}{c} b_1^S \\ b_2^S \end{array} \right).
\]
For instance,
from the tensors explicitly given above one obtains
\ba
\left( \begin{array}{cc} b_1^H, & b_2^H \end{array} \right)
\left( \begin{array}{cc}
c_{11} & c_{12} \\
c_{21} & c_{22}
 \end{array} \right)
\epsilon
\left( \begin{array}{c} b_1^H \\ b_2^H \end{array} \right)
&=& - \frac{v^6}{4}\,
\imag
\left( \bar \lambda_6^2 \bar \lambda_5^\ast \right),
\label{hdior} \\
\left( \begin{array}{cc} b_1^{H^- H^+}, & b_2^{H^- H^+} \end{array} \right)
\left( \begin{array}{cc}
c_{11} & c_{12} \\
c_{21} & c_{22}
 \end{array} \right)
\epsilon
\left( \begin{array}{c} b_1^{H^- H^+} \\ b_2^{H^- H^+} \end{array} \right)
&=& - \frac{v^4}{4}\,
\imag \left( \bar \lambda_7^2 \bar \lambda_5^\ast \right),
\\
\left( \begin{array}{cc} b_1^H, & b_2^H \end{array} \right)
\epsilon
\left( \begin{array}{c} b_1^{H^- H^+} \\ b_2^{H^- H^+} \end{array} \right)
&=&
v^3\,
\imag \left( \bar \lambda_6 \bar \lambda_7^\ast \right).
\ea
One easily sees that,
indeed,
\be
J_1 \propto
\imag \left( \bar \lambda_6^2 \bar \lambda_5^\ast \right),
\quad
J_2 \propto
\imag \left( \bar \lambda_7^2 \bar \lambda_5^\ast \right),
\quad
J_3 \propto
\imag \left( \bar \lambda_6 \bar \lambda_7^\ast \right),
\label{J1J2J3}
\ee
are the only basis-invariant CP-violating quantities
in the potential of eq.~(\ref{potH}).

By taking into account eq.~(\ref{jvhgt}) and $T \in \sot$,
the quantity in eq.~(\ref{hdior}) may be written \cite{Mendez:1991gp}
\ba
- \frac{v^6}{2}\,
\mathrm{Im}
\left( \bar \lambda_6^2 \bar \lambda_5^\ast \right)
&=&
M_{23} \left[
\left( M_{13} \right)^2 - \left( M_{12} \right)^2
\right] + \left( M_{22} - M_{33} \right) M_{12} M_{13}
\no &=&
\left( m_1^2 - m_2^2 \right) \left( m_1^2 - m_3^2 \right)
\left( m_2^2 - m_3^2 \right) T_{11} T_{12} T_{13}.
\ea
This shows that there is CP violation in the 2HDM if
all the matrix elements in the first row of the mixing matrix $T$ are nonzero
and,
moreover,
the three physical neutral scalars are non-degenerate.\footnote{If
two of the physical neutral spin-0 fields,
say $S_j$ and $S_k$,
are degenerate,
then the matrix $T$ may be redefined in such a way that
either $T_{1j}$ or $T_{1k}$ becomes zero,
which is a CP-conserving situation.}
This can be confirmed by considering,
for instance,
the interactions in eq.~(\ref{kjcui}).
They tell us that,
for $j = 1, 2, 3$,
if $T_{1j} \neq 0$,
then $S_j$ is a scalar;
therefore,
$T_{11} T_{12} T_{13} \neq 0$
implies that all three physical neutral spin-0 fields are scalars,
contrary to our knowledge that
in the CP-conserving 2HDM one of those fields must be a pseudoscalar.

If we then consider the interactions in line~(\ref{nhydp})
and take into account eq.~(\ref{mbkpq}),
we find that
\be
v^2 \mathrm{Im} \left( \bar \lambda_6 \bar \lambda_7^\ast \right)
= \sum_{j,k,l = 1}^3 \epsilon_{jkl} m_k^2 T_{1k} c_j T_{1l}.
\ee
Indeed,
the interactions in line~(\ref{nhydp}) indicate that,
if $c_j \neq 0$,
then $S_j$ is a scalar.
CP conservation would then require either $S_k$ or $S_l$
($j \neq k \neq l \neq j$) to be a pseudoscalar;
correspondingly,
either $T_{1j}$ or $T_{1k}$,
respectively,
ought to vanish.

One should note that
in the generic 2HDM $J_1$,
$J_2$,
and $J_3$ are not all independent.
One might in principle choose only $J_1$ and $J_2$
as the two independent basis-invariant signals of CP violation.
However,
$J_1 = J_2 = 0$ in a particular 2HDM with $\bar \lambda_5=0$,
yet there might still be CP violation through $J_3 \neq 0$.
Different particular cases of the 2HDM
may require different choices for a minimum set of independent
$J$-invariant and,
to cover all the particular cases,
we need $J_1$,
$J_2$,
and $J_3$,
even though they are not all independent.

An important point made in Refs.~\cite{Gun2004,Branco:2005em,Gunion:2005ja}
concerns spontaneous symmetry breaking (SSB).
Quantities like $J_1$,
$J_2$,
and $J_3$ involve the vev $v$
and therefore refer to CP violation {\em after} SSB;
the Lagrangian before SSB does not involve $v$.
Of course,
all laboratory CP-violating observables concern CP violation after SSB,
but early Universe phenomena,
such as leptogenesis,
involve CP violation {\em before} SSB.
The quantities $J_1$,
$J_2$,
and $J_3$ are the only ones
needed to study CP violation in the scalar potential {\em after} SSB.
In an earlier section,
we found that one needs the four invariants $I_{1,2,3,4}$
in order to study CP violation {\em before} SSB.
Interestingly,
four quantities are needed in order to study
CP violation in the generic 2HDM before SSB,
but only three quantities are required after SSB.

The comparison between the $I$'s and the $J$'s is also of theoretical
interest.
If some $I$ is non-vanishing,
then there is CP violation at the Lagrangian level.
If all the $I$'s vanish but some $J$ do not,
then there is spontaneous CP violation.
The theory is CP conserving only when all the $I$'s and $J$'s vanish.
For discussions see, for 
instance,~\cite{Nishi:2006tg,Maniatis:2007vn,Sokolowska:2008bt,Ferreira:2010hy}.

We end this section with an open problem.
Through the minimization conditions,
one can determine
(at least implicitly,
or numerically)
the vevs in terms of the couplings of the scalar potential.
Therefore,
one should be able to write $J_{1,2,3}$
in terms of $I_{1,2,3,4}$ together with,
possibly,
some CP-conserving quantities.
As far as we know,
this has not yet been achieved.

\subsection{CP-violating quantities with scalars and fermions}

\subsubsection{The general method}

In this section we discuss a systematic method
for the construction of basis-invariant quantities
which was developed by Botella and Silva \cite{Botella:1994cs}.
The main focus of their work was on basis-invariant signals of CP violation,
which we designate by $J$-invariants.
Yet,
they pointed out that their strategy applies to any other property;
for example,
their method has later been applied \cite{Davidson:1996cc}
to $R$-parity in supersymmetric models.
The method applies with any gauge group $G$
and can also be used in an effective field theory
including nonrenormalizable interactions.

To illustrate the main idea,
we start with a generic Lagrangian of the form
\be
\ls_I = \left( \sum_{i,j} g_{ij} \alpha_i \beta_j +
\sum_{k,l} h_{kl} \alpha_k \gamma_l \right) \Phi + \hc,
\ee
where the $g_{i,j}$ and $h_{k,l}$ are coupling constants
and the $\alpha$,
$\beta$,
and $\gamma$ are field operators with their respective $U(\alpha)$,
$U(\beta)$,
and $U(\gamma)$ flavour spaces,
and also transforming like some multiplet of the gauge group $G$.
As an example,
in the SM we have,
after SSB,
\be
\ls_I =
\sum_{i,j} \left( \begin{array}{cc}
\bar u_L, & \bar d_L \end{array} \right)_i
\left[
\left( M_u \right)_{ij}
\left( \begin{array}{c} 1 \\ 0 \end{array} \right)
u_{Rj}
+
\left( M_d \right)_{ij}
\left( \begin{array}{c} 0 \\ 1 \end{array} \right)
d_{Rj}
\right] + \hc,
\ee
and the flavour spaces are U(3)$_L$,
U(3)$_{uR}$,
and U(3)$_{dR}$,
respectively.
In perturbation theory, one can generate interactions
mediated by any power of ${\cal L}_I$. For example, to second order
in perturbation theory, we will find interactions mediated by
\be
(g_{ij} \alpha_i \beta_j \Phi)\
(h_{kl} \alpha_k \gamma_l \Phi)\ .
\ee
Hence, a given property of the theory (say CP violation) may show up at
some order of perturbation theory as a suitable product of couplings.

Under a basis transformation the couplings transform as,
%
\ba
g_{ij} & \rightarrow &
{\displaystyle \sum_{kl}}
U(\alpha)_{ki}\  g_{kl}\  U(\beta)_{lj}\ ,
\nonumber\\
h_{ij} & \rightarrow &
{\displaystyle \sum_{kl}}
U(\alpha)_{ki}\  h_{kl}\  U(\gamma)_{lj}\ .
\ea
The strategy in looking for basis invariant quantities consists in taking
products of couplings (as in the perturbative expansion),
contracting over the internal flavour spaces and taking a trace at the end.
For example, the quantities
\be
H_u = M_u M_u^\dagger\ \ ,\ \ H_d = M_d M_d^\dagger\ \ ,
\ \ H_u H_d
\ee
are tensors in the U(3)$_L$ space
and their traces are weak-basis invariants.
The same is true for the trace of the U(3)$_{uR}$ tensor
$M_u^\dagger M_u$.

In so doing, we have already traced over the basis transformations
that could lead to the spurious phases brought about by basis transformations.
Therefore, the imaginary parts of such traces are unequivocal
signs of CP violation \cite{roldan}.
For example,
the $J$-invariant of the three-family SM is
\be
J \propto \imag \left[ \textrm{tr}
\left( H_u H_d H_u^2 H_d^2 \right) \right].
\label{Jarl}
\ee

A detail concerns spontaneous symmetry breaking (SSB).
After SSB,
the physical degrees of freedom of the neutral scalars
are shifted fields $\eta_i$,
related to the original ones ($\phi_i$) by the vevs ($v_i$) as
\be
\phi_i  = v_i + \eta_i.
\ee
This reparametrizes a Lagrangian term;
for instance,
\be
\mu_{ij} \phi_i^\dagger \phi_j
= \mu_{ij} v_i^\ast v_j
+ \mu_{ij} v_i^\ast \eta_j
+ \mu_{ij} \eta_i^\dagger v_j
+ \mu_{ij} \eta_i^\dagger \eta_j.
\ee
In this way,
$v_i$ becomes an integral part of new couplings like
$\left( \mu_{ij} v_i^\ast \right) \eta_j$.
Thus,
in the construction of basis invariants involving the scalar sector,
one must consider combinations of couplings both
with and without vevs.
This greatly simplifies the study of the scalar sector over the
analysis in the previous section.
In addition, the minimization conditions provide relations between the
couplings in the scalar potential which must be used in identifying the
correct number of independent CP-violating invariants.

This discussion motivates the following prescription for the construction
of $J$-invariants:
\begin{itemize}
\item identify all the scalar and fermion flavour spaces in the theory;
\item make a list of all the couplings according to their transformation
properties under weak basis transformations, including the vacuum
expectation values (which transform as vectors under the scalar basis
change);
make use of the stationarity conditions of the scalar
potential to reduce the number of parameters;
\item construct invariants by contracting over internal flavour spaces
in all possible ways, taking traces at the end
(in order to be systematic
it is advisable to do this firstly in the fermion sector,
using this to define new scalar tensors,
and then perform the analysis of the scalar sector);
\item take the imaginary part to obtain a basis-invariant signal of
CP violation.
\end{itemize}

In general,
a minimal set of CP-violating quantities is not easy to find,
since one could in principle go to arbitrary order in perturbation theory.
The identification of the number of independent $J$-invariants may,
at best,
be done in a case-by-case way
through a careful study of the sources of CP violation in the model.
Moreover, different particular cases of a model may require different
choices for the minimum set of fundamental $J$-invariants.

\subsubsection{Invariants with scalars}

We can use the technique of the previous section
to reproduce the CP-violating invariants of the scalar sector after SSB.
One finds~\cite{Botella:1994cs,Gunion:2005ja},
\ba
J_1 &\propto& \imag \left(
\sum_{a,\ldots f = 1}^2 v_a^\ast\ Y_{ae}\ v_b^\ast\ Y_{bf}\
Z_{ecfd}\ v_c\ v_d \right),
\label{J1}
\\
J_2
& \propto &
\mathrm{Im} \left(
\sum_{a,\ldots h = 1}^2 v_b^\ast\ v_c^\ast\
Z_{bgge}\ Z_{chhf}\ Z_{eafd}\ v_a\ v_d
\right),
\\
J_3 &\propto& \imag \left(
\sum_{a,b,c,d = 1}^2 v_a^\ast Y_{ab}\ Z_{bd,dc}\ v_c \right).
\label{J3}
\ea
These, it may be shown, are equivalent to the $J$ invariants defined
in eqs.~\eqref{J1J2J3}.

\subsubsection{Invariants with scalars and fermions}

In looking for invariants probing CP violation
and involving both scalars and fermions
we follow Botella and Silva \cite{Botella:1994cs}.
We start from the Yukawa Lagrangian in eq.~\eqref{uvhfw}.
The Yukawa-coupling matrices $Y_a^d$ ($a = 1, 2$)
are $3 \times 3$ complex matrices.
Their rows (columns) are acted upon by unitary U(3)$_L$ (U(3)$_{dR}$)
transformations on the space of left-handed quark doublets
(right-handed down-type quark singlets).
Similarly,
the Yukawa matrices $Y_a^u$ are $3 \times 3$ complex matrices.
Its rows (columns) are acted upon by unitary U(3)$_L$ (U(3)$_{uR}$)
transformations on the space of left-handed quark doublets
(right-handed, up-type singlets).
Combinations such as $Y_a^d {Y_b^d}^\dagger$ and
$Y_a^u {Y_b^u}^\dagger$ have both indices in the left-handed-doublet space.
As a result,
in
\bs
\ba
T^d_{ab} &=& \textrm{tr} \left( Y_a^d {Y_b^d}^\dagger \right),
\\
T^u_{ab} &=& \textrm{tr} \left( Y_a^u {Y_b^u}^\dagger \right),
\ea
\es
all quark spaces have been traced over.
These quantities depend only on the scalar indices $a$ and $b$.
As a result,
they can be combined with the $v_a$,
$Y_{ab}$,
and $Z_{abcd}$ to construct weak-basis invariants
depending on both scalars and fermions.
The lowest order basis-invariant measures of CP violation involving
both scalars and fermions are \cite{Botella:1994cs}
\bs
\ba
J^d &=& \imag \left(
\sum_{a,b,c = 1}^2 v_a v_b^\ast Y_{bc} T^d_{ca}
\right)
\\
J^u &=& \imag \left(
\sum_{a,b,c = 1}^2 v_a v_b^\ast Y_{bc} T^u_{ca}
\right).
\ea
\es
Expressing these invariants in the quark mass basis,
we find
\bs
\ba
J^d &\propto&
\imag \left[ m_{12}^2
\sum_{i=1}^{n_G} m_{di} \left( N_d \right)_{ii}^\ast \right],
\\
J^u &\propto&
\imag \left[ m_{12}^2
\sum_{i=1}^{n_G} m_{ui} \left( N_u \right)_{ii}^\ast \right],
\ea
\es
where $n_G$ is the number of generations
and $m_{di}$ ($m_{ui}$) is the mass of the $i$-th down-type (up-type) quark.
In the simplest case of $n_G=1$,
there is no CP violation in the CKM matrix,
there are only two independent
CP-violating invariants---which we may take to be $J_1$ and $J_3$---in
the scalar sector,
and there are only two independent invariants---$J^d$ and $J^u$---involving
both the scalars and the fermions.
For $n_G=2$,
there is still no CP violation in the CKM matrix,
$J_1$ and $J_3$ apply to the scalar sector,
and there are a total of eight invariants
in the scalar--fermion interactions,
which were explicitly constructed in Ref.~\cite{Botella:1994cs}.
For $n_G=3$,
$J_1$ and $J_3$ apply to the scalar sector,
there are 18 invariants in the scalar--fermion interactions,
and there is now also one CP-violating invariant in the CKM matrix,
given by eq.~\eqref{Jarl}.

\subsection{CP basis invariants and the bilinear formalism}
\label{sec:cpvbil}

The basis invariant quantities of
eqs.~\eqref{inv1},~\eqref{inv2},~\eqref{inv3} and~\eqref{i4} determine
whether or not a given 2HDM scalar potential is explicitly CP-conserving.
They have extremely simplified expressions in terms of the bilinear
formalism introduced in section~\ref{3_sec:notation}, as was shown in
refs.~\cite{Nishi:2006tg,Maniatis:2007vn}. We follow the notation
of the Heidelberg group~\cite{Nagel:2004sw,Maniatis:2006fs,Maniatis:2007vn}
and introduce the vectors $\xi$ and $\eta$ and the matrix $E$, given, in terms
of the parameters of the 2HDM potential defined in eq.~\eqref{2_VH1}, as
\ba
\tvec{\xi}&=&\frac{1}{2}
\left(
\begin{array}{c}
- 2 \textrm{Re}(m_{12}^2)\\
2 \textrm{Im}(m_{12}^2)\\
 m_{11}^2-m_{22}^2
\end{array}
\right)
\quad\quad , \quad\quad
\tvec{\eta}=\frac{1}{4}
\left(
\begin{array}{c}
\textrm{Re}(\lambda_6+\lambda_7)\\
-\textrm{Im}(\lambda_6+\lambda_7)\\
\frac{1}{2}(\lambda_1 - \lambda_2)
\end{array}
\right),
\nonumber \\*[3mm]
E &=& \frac{1}{4}
\left(
\begin{array}{ccc}
\lambda_4 + \textrm{Re}(\lambda_5) &
-\textrm{Im}(\lambda_5) &
\textrm{Re}(\lambda_6-\lambda_7) \\
-\textrm{Im}(\lambda_5) &
\lambda_4 - \textrm{Re}(\lambda_5) &
-\textrm{Im}(\lambda_6-\lambda_7) \\
\textrm{Re}(\lambda_6-\lambda_7) &
-\textrm{Im}(\lambda_6 -\lambda_7) &
\frac{1}{2}(\lambda_1 + \lambda_2) - \lambda_3
\end{array}
\right).
\ea
Then, the $I$ invariants of section~\ref{5_2HDM} may be written
as~\footnote{In fact, these four invariants are linear combinations
of those in section~\ref{5_2HDM}, but their usage is equivalent.}
\begin{eqnarray}
I_1 &=& \left(\tvec{\xi} \times \tvec{\eta} \right)^\trans\ . \ E \tvec{\xi},
\label{I1}
\\
I_2 &=& \left(\tvec{\xi} \times \tvec{\eta} \right)^\trans\ . \ E \tvec{\eta},
\label{I2}
\\
I_3 &=& \left(\tvec{\xi} \times (E \tvec{\xi}) \right)^\trans\ . \ E^2 \tvec{\xi},
\label{I3}
\\
I_4 &=& \left(\tvec{\eta} \times (E \tvec{\eta}) \right)^\trans\ . \ E^2 \tvec{\eta}.
\label{Ii4}
\end{eqnarray}
The extremely simple form of these equations, and the appearance of external
products between vectors in them, leads to meaningful geometrical interpretations
for CP-violating/preserving potentials~\cite{Nishi:2006tg,Maniatis:2007vn}.
See also~\cite{Ivanov:2006yq,Ivanov:2007de}.

As for the basis invariant quantities $J_i$ of eqs.~\eqref{J1}--~\eqref{J3},
which determine whether a given vacuum preserves or breaks CP, they too
can be written, in an extremely simplified manner, in terms of the
bilinear formalism. Let us introduce the vector
$\tvec{r}\,=\,(r_1 , r_2 , r_3)^T$, with the $r_i$ defined in
eq.~\eqref{2_r_Ivanov}.  Then~\cite{Maniatis:2007vn},
\begin{eqnarray}
J_1 &=& \left(\tvec{\xi} \times \tvec{\eta} \right)^\trans\ . \ \langle  \tvec{r} \rangle,
\label{J1b}
\\
J_2 &=& \left(\tvec{\xi} \times (E \tvec{\xi}) \right)^\trans\ . \  \langle \tvec{r} \rangle,
\label{J2b}
\\
J_3 &=& \left(\tvec{\eta} \times (E \tvec{\eta}) \right)^\trans\ . \ \langle  \tvec{r} \rangle,
\label{J3b}
\end{eqnarray}
where $\langle \tvec{r} \rangle$ corresponds to the vector $\tvec{r}$ evaluated at some
stationary point of the theory~\footnote{These, too, are linear combinations of the
invariants of~\eqref{J1}--~\eqref{J3}.}.

\subsection{CP violation and symmetries}

As discussed in section~\ref{sec:symmetries}, there are six classes
of 2HDM symmetry-constrained potentials. Those symmetries, and their
impact on the parameters of the potential, were shown in
Table~\ref{2_master1}. Leaving aside the extension of these symmetries
to the Yukawa sector - where, as we have seen in section~\ref{2_sec:yuksym},
each symmetry has very different consequences - each of those models corresponds
to a very specific and different type of scalar physics. For instance, a model
with a Peccei-Quinn~\cite{Peccei:1977hh} U(1) symmetry (class IV) can
have a vacuum with a massless scalar, an axion. That is not possible,
whatever the vacuum, for models with the $Z_2$ or CP1 symmetries
(classes V and VI).

Another aspect for which the six classes behave very differently concerns
the possibility of CP breaking - explicitly  or spontaneously - for each
potential. This question is better handled using the bilinear formalism
formulae presented in section~\ref{sec:cpvbil}. A systematic analysis of all
possible potentials was carried out in~\cite{Ferreira:2010hy}. Briefly,
this consists of:
\begin{itemize}
\item One may wish to consider models with an {\em exact} symmetry, out of the
six considered, or to softly-break that symmetry via the inclusion of
generic dimension-two terms (real or, in the case of $m_{12}^2$, even complex).
\item Prior to spontaneous symmetry breaking, it is necessary to determine
whether CP is a valid symmetry of the potential, or if it is explicitly
broken. This is best achieved computing the four $I$ invariants of
eqs.~\eqref{I1}--~\eqref{Ii4}.
\item For any given model, the extremum conditions need to be solved, to
prove that a certain vacuum, which might break CP spontaneously, is
possible.
\item Having proved that a given sets of vevs is a possible solution of
the extremum conditions, one must verify whether that vacuum effectively
breaks CP. This is best achieved through the calculation of the three
invariant quantities of eqs.~\eqref{J1b}--~\eqref{J3b}.
\end{itemize}
If any of the $I_i$ invariants is different from zero, the potential is
not CP-conserving, and CP is not a defined symmetry of the potential
(for {\em all} possible CP definitions of the form of eq.~\eqref{2_GCP}).
If all $I_i = 0$ then the potential is CP-conserving; a given vacuum is
CP-conserving if and only if all invariants $J_i = 0$. A CP-conserving
scalar sector has some very distinct physical properties: there is a
well-defined pseudoscalar state $A$, and there are two well-defined
CP-even states $h$ and $H$; as a result, though triple vertices of the form
$Z Z h$ and $Z Z H$ are possible, no vertex like $Z Z A$ is allowed.

In~\cite{Ferreira:2010hy} the bilinear formalism of
refs.~\cite{Maniatis:2006fs,Maniatis:2006jd,Maniatis:2007vn} was used
to compute all invariant quantities. As shown in the previous section,
the formulae for the $I$ and $J$ CP basis invariants are extremely simple
in the bilinear formalism, and that allowed a general analysis of all possible
models, without even an
explicit calculation of the vevs (a major simplification, since solving
the extremum conditions can be analytically impossible in some models).
The conclusions of the study of~\cite{Ferreira:2010hy} are summarised
in Table~\ref{tab:CPV}.
\begin{table*}[ht!]
\caption{\em CP properties of the six symmetry-constrained
classes of 2HDM scalar potentials. We consider both the case of exact
symmetries and their soft breaking via dimension-two terms.
In this table, ``Yes'' means that it is possible to choose the parameters of
the potential such as to enable that particular form of CP violation.}
\begin{center}
\begin{tabular}{|c|c|c|c|c|}
\hline
\hline
  & \multicolumn{2}{|c|}{exact} & \multicolumn{2}{|c|}{softly-broken} \\
\cline{2-5}
 symmetry & explicit & spontaneous & explicit & spontaneous  \\
    class & CPV & CPV & CPV & CPV  \\
\hline
I - U(2) & -- & -- & -- & -- \\
II - CP3 & -- & -- & -- & -- \\
III - CP2 & -- & -- & Yes & Yes \\
IV - U(1) & -- & -- & -- & -- \\
V - $Z_2$ &  -- & -- & Yes & Yes \\
VI - CP1 & -- & Yes & Yes & Yes \\
\hline
\hline
\end{tabular}
\end{center}
\label{tab:CPV}
\end{table*}
Some obvious observations are drawn from this table:
\begin{itemize}
\item Any scalar potential with an exact symmetry is CP-conserving.
\item If the symmetry of the potential is {\em continuous} (i.e.,
classes I, II and IV) no CP violation is possible, be it explicitly
or spontaneously, even with generic soft-breaking terms.
\item Discrete symmetries allow for the possibility of spontaneous
symmetry breaking; however, with the exception of the class VI
model, that is only possible via the inclusion of a soft-breaking
term~\footnote{For the potential with a $Z_2$ symmetry this was
well-known for a long time~\cite{Branco:1985aq}.}.
\end{itemize}
Though a general analysis of CP breaking in the scalar sector is
achievable, the study of the CP properties of the theory requires
that one takes into
account the Yukawa terms. And some of the models of
Table~\ref{tab:CPV} may well end up having CP conserved in the
scalar sector but violated by the fermion-scalar interactions,
much like the SM.

\subsection{Two models with an original source of CP violation}
\label{sec:orig_cpv}

In the SM,
the source of CP violation is {\em explicit breaking}, via
Yukawa matrices which are complex. This leads to a complex CKM
matrix, and in turn to a non-zero Jarlskog invariant. In such models,
then, CP is not defined {\em a priori}, since it is not a
symmetry of the lagrangian - it is broken by hard, dimension
four, Yukawa terms.
The MSSM can be another example: in~\cite{Pilaftsis:1999qt}
it was shown that radiative Higgs-sector CP violation can be quite large in the
MSSM. In effect, after quantum effects are included, one obtains, as a
limit, the potential of a general CP-violating 2HDM.

In the
2HDM, as has been explained in previous sections, there is the
possibility of {\em spontaneous breaking} of CP. A complex phase
can appear in the vacuum of the theory and lead to CP breaking. The
presence of such a phase, however, is not sufficient to guarantee
CP violation - one needs to calculate the $J$ invariants of
eqs.~\eqref{J1}--~\eqref{J3} and verify whether at least one of them is
non-zero. An example of a model where this happens is the classic
paper by Branco and Rebelo~\cite{Branco:1985aq} - they considered a
scalar potential with a $Z_2$ symmetry, extended to the fermion
sector in a particular way, and also required CP conservation at
the lagrangian level. Hence, all of the model's parameters are real.
Since the exact symmetry forbids any
CP-breaking vacuum, they added a {\em real} soft breaking term
to the scalar potential, and thus generated a complex CKM matrix.
In such a model some (or all) of the $J_i$ invariants are
non-zero, and there is CP violation in the scalar-scalar interactions,
as well as in the fermion sector.

In this section we will briefly describe two versions of the 2HDM
in which CP violation arises in ways which are different from the
two usual ones described above.

\subsubsection{The CP3 model}

In~\cite{Ferreira:2010bm} a 2HDM was built with the CP3
symmetry extended to the fermion sector. We recall (see eq.~\eqref{2_eq:CP3})
that this corresponds to a transformation on the scalar fields of
the form
\be
\left( \begin{array}{c}
\Phi_1 \\ \Phi_2
\end{array} \right)
= \left( \begin{array}{cc}
\cos{\theta} & \sin{\theta} \\
- \sin{\theta} & \cos{\theta}
\end{array} \right)\
\left( \begin{array}{c}
\Phi_1^* \\ \Phi_2^*
\end{array} \right).
\ee
In~\cite{Ferreira:2010bm} it was shown that the only value
of the angle $\theta$ which leads to six massive quarks (and three
massive charged leptons) is $\theta = \pi/3$. Also, the transformation
laws of the quark fields under this symmetry are uniquely determined
(again, by the requirement of six massive quarks). One finds that the
Yukawa coupling matrices for the down quarks have a very simple form,
\be
Y^d_1 =
\left[
\begin{array}{ccc}
a_{11} & a_{12} & a_{13}\\
a_{12} & - a_{11} & a_{23}\\
a_{31} & a_{32} & 0\\
\end{array}
\right]
\quad , \quad
Y^d_2 =
\left[
\begin{array}{ccc}
a_{12} & -a_{11} & -a_{23}\\
-a_{11} & -a_{12} & a_{13}\\
-a_{32} & a_{31} & 0\\
\end{array}
\right],
\label{special_gamma}
\ee
in a special basis where all the $a_{ij}$ coefficients are real.
Indeed, in that basis {\em all} the parameters of the potential are real.
An analogous form is found for the $Y^u$ matrices for the up quarks, with
different coefficients $b_{ij}$.

The scalar potential of the CP3
model (class II of table~\ref{2_master1}) has a continuous symmetry
which will be broken if both scalar fields acquire a vev - which then
implies the appearance of a massless axion. As such, one needs to add
{\em real} soft breaking terms to the potential, which give mass to the
would-be axion. One then finds that, with such a soft breaking term,
a vacuum with a complex relative phase between the vevs is possible -
of the form $\langle \Phi_1 \rangle = v_1$,
$\langle \Phi_2 \rangle = v_1 e^{i \delta}$.
However, as shown in table~\ref{tab:CPV}, even such a vacuum does not
provoke CP breaking in the scalar sector - all $J_i$ invariants are
equal to zero, even if $\delta \neq 0$. Still, it was shown
in~\cite{Ferreira:2010bm} that the Jarlskog invariant is directly
proportional to $\sin\delta$. As such, in this model a vacuum with a complex
relative phase is possible and does lead to CP violation, even if its scalar
sector preserves CP. This then, is a new type of CP violation:
\begin{itemize}
\item The lagrangian does preserve CP because there is a basis for which
all of its parameters are real. Thus, no explicit CP breaking occurs, as
in the SM.
\item The scalar sector preserves CP because, even for a vacuum with a complex
phase $\delta$, all of the basis-invariant quantities $J_i$ which measure
CP violation in the scalar-scalar interactions are zero.
\item However, CP breaking {\em does} occur, and it is spontaneous, since
$\delta \neq 0$ implies a non-zero Jarlskog invariant.
\end{itemize}
It is interesting that the scalar sector
does the deed (spontaneously break the symmetry)
but it is the fermion sector which pays the
consequences (providing CP violation).
To the best of our knowledge, this type of CP violation is
unheard of in the literature - it arises spontaneously, but
the scalar sector remains CP-conserving.

The model's interest is increased by the limited number of free
parameters it contains - 12 independent Yukawa couplings, and 6
scalar potential parameters (counting the soft breaking terms).
It is easy to fit the six quark masses and a quartet of moduli
of CKM matrix elements (in theory all one needs to fit the entire
CKM matrix~\cite{Branco:1999fs}). This model has FCNC (the matrices
$Y^d_1$ and $Y^d_2$ do not in general commute with one another),
so care must be taken to ensure that it is in agreement with
the stringent FCNC bounds which arise from meson physics.
Remarkably, the model does manage to fit the mass differences of
the $K$, $B_s$ and $B_d$ mesons, as well as the $\epsilon_K$
parameters and the unitarity triangle angle $\beta$, as obtained from
$B$ meson decays. However, the fits performed lead to a
value of the Jarlskog invariant at least three orders of magnitude
smaller than the SM value; this would seem to indicate a relation
between the unitarity triangle angles, $\alpha \simeq \beta$, which
goes against observational data~\cite{hfag}.

\subsubsection{A specific $Z_3$ model}

Recently~\cite{Ferreira:2011xc} a model was proposed wherein one extended a
$Z_3$ symmetry in the scalar sector to the Yukawa terms, in a particular
way. The scalar sector therein resulting is identical to that of the
Peccei-Quinn model; the Yukawa matrices one obtains are extremely simple,
having the form
\[
Y^d_1, Y^u_1 \sim \left( \begin{array}{ccc}
0 & 0 & 0 \\ 0 & 0 & \times \\ \times & \times & 0
\end{array} \right),
\quad
Y^d_2 \sim \left( \begin{array}{ccc}
\times & \times & 0 \\ 0 & 0 & 0 \\ 0 & 0 & \times
\end{array} \right),
\quad
Y^u_2 \sim \left( \begin{array}{ccc}
0 & 0 & \times \\ \times & \times & 0 \\ 0 & 0 & 0
\end{array} \right),
\]
where the symbol $\times$ denotes
a non-zero matrix entry, in general complex. These Yukawa
matrices were first found in~\cite{Ferreira:2010ir}.

If, on top of the
$Z_3$ symmetry, we apply the standard CP transformation, all
terms in the lagrangian are forced to be real. Again, since
the matrices $Y^d_1$ and $Y^d_2$ (and $Y^u_1$ and $Y^u_2$)
do not in general commute, this model has FCNC.

Once more, the Peccei-Quinn model has an axion if both scalar fields
acquire a non-zero vev. Thus, it is necessary to add a soft-breaking
$m_{12}^2$ term to the potential. However, unlike what happened
for the CP3 model of the previous section, it is impossible to obtain
a vacuum of the form $\langle \Phi_1 \rangle = v_1$,
$\langle \Phi_2 \rangle = v_1 e^{i \delta}$, with $\delta \neq 0$,
{\em unless} the soft breaking term $m_{12}^2$ is itself complex (and
has phase $-\delta$)~\footnote{The reason for this drastic difference
in behaviour is the fact that in the CP3 model the $\lambda_5$
quartic coupling is non-zero, unlike what happens for the Peccei-Quinn
potential.}. However, even despite the introduction of a complex soft
breaking term, the scalar potential remains CP-conserving, {\em before
of after} spontaneous symmetry breaking - check table~\ref{tab:CPV}.
As such, this too is an unusual source of CP violation:
\begin{itemize}
\item The model has a scalar sector which is CP-conserving.
\item CP violation occurs due to an {\em explicit} breaking
of the CP symmetry, via a complex coefficient in the
scalar potential.
\item However, unlike the case of the SM, this CP breaking is
{\em soft}, not {\em hard}, since the complex coefficient is
a dimension-two term of the lagrangian.
\end{itemize}
The FCNC which arise in this model have an extra surprise: it is easy
to show that all FCNC couplings are  real, so that no CP violation
occurs in FCNC interactions. Indeed, this model reproduces perfectly
the type of CP violation one obtains in the SM - all quark masses,
CKM matrix elements, meson mass differences and CP-violating quantities
can be fitted with the model's eleven parameters
(see~\cite{Ferreira:2011xc}). Remarkably, one is capable of
fitting all observables with some of the scalar masses as low as $\sim$ 150 GeV,
even in the presence of FCNC. Further, the model's scalar sector
also satisfies constraints on New Physics arising from the oblique
parameters of section~\ref{2_sec:ob}, as well as the LEP2 constraints
on the lightest Higgs mass~\cite{Nakamura:2010zzi}.

As such, this model suggests that even though the CKM mechanism for CP
violation is well established experimentally, the origin for a complex
CKM matrix need not necessarily be that of the SM - a hard breaking
of CP through dimension-four terms. In the 2HDM with a $Z_3$ symmetry,
the possibility exists that the origin of all CP violation is a soft,
dimension-two, term.

\newpage

\section{Recent results from the LHC}

\begin{figure}[ht]
\vskip 0.5cm
\centerline{\epsfysize=12cm \epsfbox{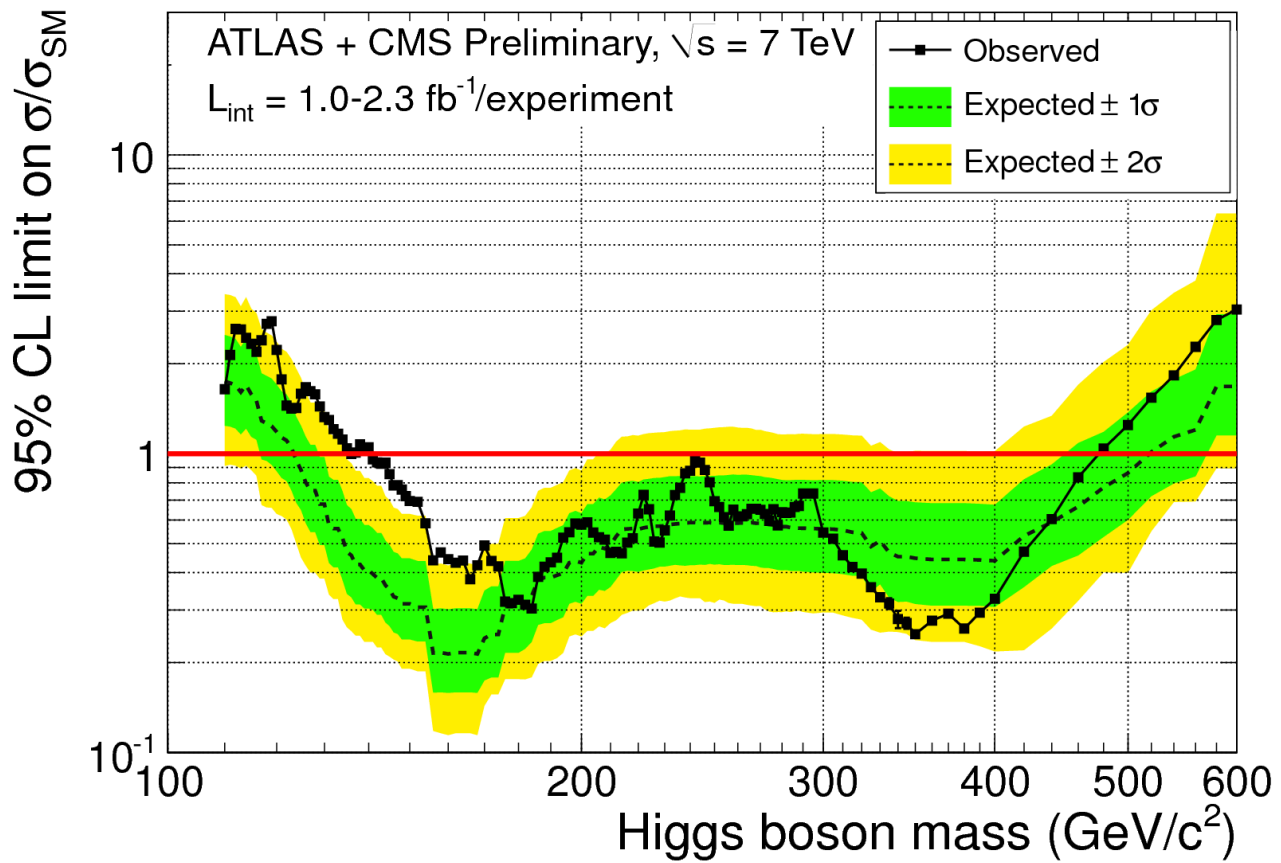} }
\caption{Combined results of the ATLAS and CMS Collaborations
presented at the Hadron Collider Workshop in November 2011.
The solid line is the experimental upper bound,
at 95\% confidence level,
on the cross section for Higgs boson production
divided by the cross section of the Standard Model.
The dashed line is the expected 95\% exclusion bound,
and the green and yellow bands
are the one- and two-standard deviation bands
around that expected bound.}
\label{updatelhc}
\end{figure}

The 35 inverse picobarns of the 2010 Large Hadron Collider (LHC) run
were not useful in constraining the Higgs sector.
Over the summer of 2011,
the LHC collected more than one inverse femtobarn of data,
and at the Hadron Collider Workshop in November 2011
the ATLAS and CMS Collaborations presented~\cite{brazil}
the combined results for that summer run.
They presented upper bounds on the cross section
for Standard Model Higgs boson production
as a function of the Higgs boson mass.
The main plot,
the so-called ``Brazil bands'' plot,
is shown in Fig.~\ref{updatelhc}.
The solid line in this plot gives the upper bound,
at 95\% confidence level,
on the measured cross section for Higgs boson production
relative to the Standard Model cross section,
{\em assuming Standard Model decay signatures}.
One can't quickly draw conclusions about 2HDMs from this plot,
which is meaningful only for Standard Model Higgs boson decays.
As seen throughout this report,
in many 2HDMs,
and for a substantial region of the parameter space,
one expects very different branching ratios from those of the Standard Model.

Nonetheless,
there is one circumstance in which the plot
can directly lead to information about 2HDMs.
That is when the primary decay modes of the $h$ and $H$
of the (CP-conserving) 2HDM
are into $W^+W^-$ and $ZZ$.
In the Standard Model,
these decays dominate for Higgs boson masses above $130~\mathrm{GeV}$,
and thus the branching ratios into $W^+W^-$ and $ZZ$ would be the same,
since the total branching ratio is 100\%
and the only difference between the $W^+W^-$ and $ZZ$ branching ratios
would be the SU(2) couplings and phase space.

If the primary decays are into $W^+W^-$ and $ZZ$,
then the only differences in the event rate
between the Standard Model Higgs boson
and the $h$ and $H$ scalars of the 2HDM
would occur in the production cross section,
which proceeds through top-quark loops.
For the $h$ ($H$),
this is equal to the Standard Model cross section
times $\cos^2\alpha/\sin^2\beta$ ($\sin^2\alpha/\sin^2\beta$).
Clearly,
one can adjust $\alpha$ such as to make
either one or the other of the neutral scalars $h$ or $H$ invisible,
but the sum of the cross sections of $h$ and $H$ is independent of $\alpha$.
Note that the one exception would be in the type II model
with large $\tan{\beta}$,
since then bottom-quark loops can affect the production cross section
of $h$ and $H$.
Including this possibility would add an additional parameter and,
for illustrative purposes,
we shall not include those loops here.
One should keep in mind that,
as noted in Chapter 2,
large $\tan{\beta}$ has difficulties with perturbativity and unitarity,
and requires fine-tuning.

Consider the case $\tan{\beta} = 1$.
Then,
the sum of the cross sections for $h$ and $H$ production,
where both the $h$ and $H$ decay into $W^+W^-$ and $ZZ$,
is twice the cross section for the Higgs boson of the Standard Model.
That means that at least one of those cross sections
must be larger than the one of the Standard Model.
From the plot in Fig.~\ref{updatelhc},
one can see that if {\em both} the $h$ and $H$
have masses in between $140~\mathrm{GeV}$ and $480~\mathrm{GeV}$,
then the production cross section for each of them is below the Standard Model;
therefore,
this case is experimentally excluded.
Suppose instead that $\tan{\beta}$ is very large.
Then,
$\sin^2\beta \sim 1$ and the sum of the cross sections
for $h$ and $H$ production
is similar to the one
of the Standard Model.

\begin{figure}[ht]
\centerline{\epsfysize=12cm \epsfbox{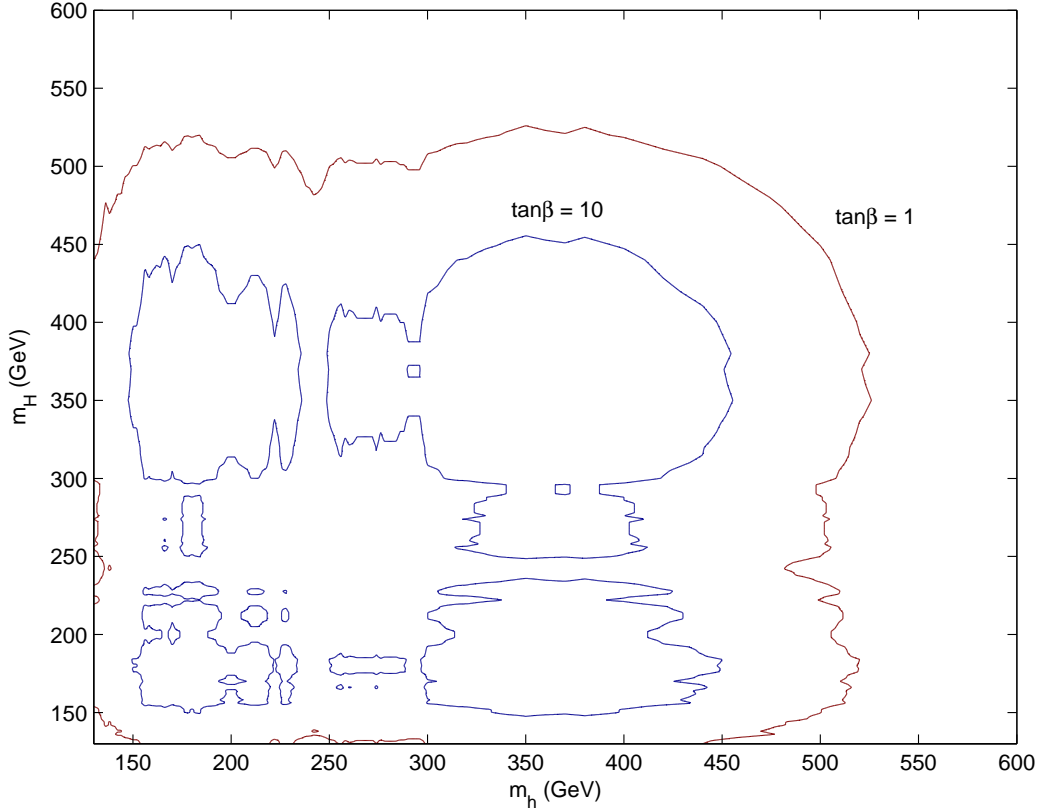} }
\caption{For $\tan{\beta} = 1$ and $\tan{\beta} = 10$,
and assuming that
{\em both $h$ and $H$ decay primarily into $W^+W^-$ and $ZZ$}.
the excluded regions are those inside the contours.}
\label{contours}
\end{figure}

We can thus exclude the regions shown in Fig.~\ref{contours}.
For $\tan\beta=1$,
the entire parameter space in between $140~\mathrm{GeV}$ and $480~\mathrm{GeV}$
is excluded,
as well as some regions beyond.
For $\tan\beta=10$ a much smaller region is excluded.
It is a rather complicated shape due to the complex structure of the lines
in Fig.~\ref{updatelhc}.
As more data is collected,
the islands are expected to gradually merge and move outwards,
eventually excluding most of the region---unless,
of course,
a signal is found,
in which case a region will remain.
It must be once again emphasized that {\em these exclusion regions
are only valid if the primary decay of both the $h$ and $H$
are into $W^+W^-$ and $ZZ$}---and,
as can be seen from Chapter 2,
this is not the case for a substantial part of the parameter space
in various 2HDMs.

In general,
for each of the four models described in Chapter 2
it is possible to find bounds
analogous to those depicted in Fig.~\ref{contours},
but for various values of $\alpha$.
However,
these bounds will often be quite weak,
since a value of $\alpha$ can always be found that makes
either $h$ or $H$ gauge-phobic,
thereby eliminating all LHC bounds.

All of this may have changed  due to LHC experimenters
presenting,
in December 2011~\cite{newatlas,newcms},
evidence for a peak in the $\gamma\gamma$ channel
at a mass of approximately $125~\mathrm{GeV}$.
If this evidence is confirmed,
then Fig.~\ref{contours} will become irrelevant.
What would such a discovery mean for 2HDMs?

The cross section for $pp \rightarrow h \rightarrow \gamma\gamma$
is consistent with the Standard Model,
but substantial discrepancies are still allowed.
In the Standard Model,
a Higgs boson with mass $125~\mathrm{GeV}$
has a branching ratio of approximately 0.002
to $\gamma \gamma$.
In the 2HDM,
the branching ratios can differ,
as seen in many of the plots in Chapter 2,
although there is usually a region of parameter space
in which the branching ratio is similar.
The production cross sections will also be different.
It is clear that one will usually be able to find a set of parameters
in which the recent results can be accommodated.
If the branching ratios into $W^+W^-$,
$ZZ$,
$b \bar b$,
and $\tau^+ \tau^-$ can also be observed,
then these will correspond to other regions of parameter space,
which might be mutually exclusive.
Thus it might be possible to rule out some of the 2HDMs in Chapter 2---or else,
if the Standard Model predictions are not verified,
these 2HDMs might constitute viable alternatives.
This possibility has recently been discussed~\cite{fs^3}.
Note that failure to find a heavy Higgs boson at the LHC
is not necessarily a problem;
the more Standard-Model-like is the $h$ field,
the more gauge-phobic will the $H$ be,
and thus it could easily evade detection.

Clearly,
one will need information about the other decay modes
of this $125~\mathrm{GeV}$ state,
and such information should be forthcoming shortly.

\newpage

\section{Conclusions}
\label{sec:conc}

The Large Hadron Collider has begun taking data, and a flood of information is imminent.
Within a short time, the deepest question in particle physics in a generation, the nature
of electroweak symmetry breaking, will definitively be addressed.   In the Standard Model,
one assumes the simplest possible scalar sector, and the properties of the Higgs boson in
that model have been studied extensively.

The simplest extension of the electroweak Higgs sector is the addition of another scalar doublet.
The purpose of this review article is to discuss this extension in detail.
The phenomenology of Two-Higgs-Doublet models is extremely rich, since it contains a charged
Higgs, a pseudoscalar and two neutral scalars, flavour-changing neutral currents, and more
possibilities for CP violation and baryogenesis.

The most general 2HDM has tree level flavour changing neutral currents, which can be
phenomenologically problematic.  The most studied versions of the 2HDM use a discrete
symmetry to avoid tree level FCNCs, and the phenomenology of the neutral sector of
these versions was studied in Chapter 2, focusing on the bounds from the Tevatron and
expectations from the LHC.    In Chapter 3, we studied models that do contain FCNC at tree level.
In Chapter 4, the phenomenology of the charged Higgs in these models was analysed.

The theoretical structure of 2HDMs is quite complex. One can have CP-violating,
CP-conserving and charge breaking minima, there are several bases that one can choose,
as well as a number of invariants. The theoretical structure is discussed in Chapter 5,
including renormalization group analyses, vacuum stability bounds, symmetry-constrained
lagrangians, etc.
In Chapter 6, CP violation in the Higgs sector is studied in detail.

Many important topics have not been included.  The most important is supersymmetry,
which automatically requires at least two Higgs doublets. There are
extremely comprehensive reviews of the Higgs sector of supersymmetric models, and we referred
the reader to those reviews.  Other than a brief discussion in Chapter 6, we have not included
models with three or more Higgs doublets, and we have, for simplicity, not considered models with
singlets.  In addition, the phenomenological focus has been on the Tevatron and the LHC.
While a linear collider would be enormously helpful in a detailed study of the scalar sector,
it is sufficiently far in the future that we did not discuss it here; the questions that a
linear collider would answer will likely be very different within a couple of years.

While the LHC will help elucidate the nature of the Higgs sector, it will be several years
before the various couplings can be measured precisely enough to test many of the models in
this review, although some will be tested sooner. Of course, paraphrasing Sidney Coleman,
it is possible that the entire structure will be swept into the dustbin of history by a
thunderbolt from Geneva. We may know very shortly.

\newpage

\paragraph{Acknowledgements:}
The work of G.C.B., L.L., M.N.R., and J.P.S.\
was partially supported by the Portuguese
\textit{Funda\c c\~ao para a Ci\^encia e a Tecnologia}
(FCT)
through the projects
CERN/FP/116328/2010,
PTDC/FIS/098188/2008, and CFTP-FCT Unit 777 which are partially funded
through POCTI (FEDER) and by Marie Curie Initial Training Network
`UNILHC' PITN-GA-2009-237920.
The work of P.M.F. was supported in part by FCT
under contract PTDC/FIS/70156/2006. P.M.F. thanks the kind
hospitality of the College of William and Mary during
part of this work.
The work of M.S.\ is funded by the National Science Foundation
grant NSF-PHY-0757481.

\newpage

\appendix

\section{Unitarity limits}
\label{2_sec:uni}

The theoretical bounds discussed in section~\ref{2_sec:ufb}
arise from the condition that the potential must have a minimum,
\viz they prevent the existence of directions in field space
along which the potential is unbounded from below.
There are other theoretical bounds that one must impose on the potential,
namely,
all the (tree-level) scalar--scalar scattering amplitudes
must respect unitarity.
This is equivalent to requiring that the $J = 0$ partial waves
(usually denoted $a_0$)
for scalar--scalar scattering satisfy $\left| a_0 \right| < 1/2$
in the high-energy limit.\footnote{The same condition must be satisfied
by the amplitudes for gauge boson--scalar scattering,
but we need not worry about those, since they yield no new bounds.}
In the SM,
this requirement is equivalent to ensuring that
the quartic coupling in the scalar potential is not too large;
in the original work by Lee, Quigg, and Thacker~\cite{Lee:1977yc,Lee:1977eg}
the bound $m_H < \sqrt{8 \pi \sqrt{2} / \left( 3 G_F \right)}
\simeq 1\, {\rm TeV}$ was thus obtained.
Extending this bound to the 2HDM is complicated,
due to the richer scalar spectrum and,
consequently,
to the need to take into account many scattering amplitudes;
the existence of many quartic couplings also complicates matters.
This leads to an analysis of the eigenvalues of the $S$ matrix
in the scalar sector.

Early work on the unitarity bounds in the 2HDM
was undertaken in~\cite{Casalbuoni:1986hy,Casalbuoni:1987eg,Maalampi:1991fb}.
A comprehensive study of all the scattering amplitudes and of their relevance
for a CP-conserving scalar potential with a $Z_2$ symmetry
was presented in~\cite{Kanemura:1993hm}.
That work was later generalized in~\cite{Akeroyd:2000wc,Horejsi:2005da}
by allowing the presence the following quantities:
\ba
L_1 &=&
\frac{1}{2 v^2} \left[
m^2_H \cos^2{\alpha}
+ m^2_h \sin^2{\alpha}
+ \frac{\sin{2 \alpha}}{2 \tan{\beta}}
\left (m^2_h - m^2_H \right)
\right]
\no & &
+ \frac{m^2_{12}}{v^2 \sin{2 \beta}} \left( 1 - \tan^2{\beta} \right),
\no
L_2 &=&
\frac{1}{2 v^2} \left[
m^2_H \sin^2{\alpha}
+ m^2_h \cos^2{\alpha}
+ \frac{\sin{2 \alpha} \tan{\beta}}{2}
\left (m^2_h - m^2_H \right)
\right]
\no & &
+ \frac{m^2_{12}}{v^2 \sin{2 \beta} \tan^2{\beta}}
\left( \tan^2{\beta} - 1 \right),
\no
L_3 &=& \frac{1}{v^2 \sin{2 \beta}}
\left[\frac{\sin{2 \alpha}}{2} \left( m^2_H - m^2_h \right) - m^2_{12} \right],
\no
L_4 &=& \frac{2 m^2_+}{v^2} \quad, \quad
L_5 \ =\ \frac{m^2_{12}}{v^2 \sin 2\beta}
\quad, \quad
L_6 \ =\ \frac{2 M^2_A}{v^2},
\ea
which are functions of the masses of the physical scalars,
of the angles $\alpha$ and $\beta$,
and the soft-breaking parameter $m^2_{12}$
(notice that we are assuming a CP-conserving vacuum).
The unitarity bounds can be,
and usually are,
expressed in terms of these quantities.
We shall however,
for consistency,
express them in the notation adopted in this review.
The $L_i$ are written,
in terms of the couplings of the scalar potential in Eq.~\eqref{2_VH1},
as
\ba
\lambda_1 &=& 2 (L_1 + L_3),
\\
\lambda_2 &=& 2 (L_1 + L_3),
\\
\lambda_3 &=& 2 (L_3 + L_4),
\\
\lambda_4 &=& \frac{1}{2}\, (L_5 + L_6 - 2 L_4),
\\
\lambda_5 &=& \frac{1}{2}\, (L_5 - L_6 ).
\ea
The computation of the $S$ matrix
for scalar--scalar scattering amplitudes
allows the determination of its eigenvalues;
the relevant ones are given
by~\footnote{For a CP-violating potential,
in which the $\lambda_5$
coupling is complex, just replace $\lambda_5$ by
 $|\lambda_5|$~\cite{Ginzburg:2005dt}.}
\ba
a_\pm &=&
\frac{3}{2} \left( \lambda_1 + \lambda_2 \right)
\pm \sqrt{\frac{9}{4} \left( \lambda_1 - \lambda_2 \right)^2
+ \left( 2 \lambda_3 + \lambda_4 \right)^2},
\\
b_\pm &=&
\frac{1}{2} \left( \lambda_1 + \lambda_2 \right)
\pm \frac{1}{2}\, \sqrt{
\left( \lambda_1 - \lambda_2 \right)^2 + 4 \lambda_4^2},
\\
c_\pm &=&
\frac{1}{2} \left( \lambda_1 + \lambda_2 \right)
\pm \frac{1}{2}\, \sqrt{
\left( \lambda_1 - \lambda_2 \right)^2 + 4 \lambda_5^2},
\\
e_1 &=& \lambda_3 + 2 \lambda_4 - 3\lambda_5
\\
e_2 &=& \lambda_3 - \lambda_5,
\\
f_+ &=& \lambda_3 + 2 \lambda_4 + 3\lambda_5,
\\
f_- &=& \lambda_3 + \lambda_5,
\\
f_1 &=& \lambda_3 + \lambda_4,
\\
p_1 &=& \lambda_3 - \lambda_4.
\ea
The requirement of tree-level perturbative unitarity translates as
\be
\left| a_\pm \right|,\
\left| b_\pm \right|,\
\left| c_\pm \right|,\
\left| f_\pm \right|,\
\left| e_{1,2} \right|,\
\left| f_1 \right|,\
\left| p_1 \right|
< 8 \pi.
\ee
From all these conditions,
the one on $|a_\pm|$ is the most restrictive one,
but all others also contribute to place severe upper bounds
on the scalar masses.
Under the assumption that the $Z_2$ symmetry is unbroken
and for small values of $\tan\beta \simeq 0.5$,
Akeroyd \etal have thus found that
$m_+ < 691\, {\rm GeV}$,
$m_A < 695\, {\rm GeV}$,
$m_h < 435\, {\rm GeV}$,
and $m_H < 638\, {\rm GeV}$;
for larger values of $\tan\beta$
the bound on $m_h$ becomes quite stronger,
dropping below $100\, {\rm GeV}$ for $\tan\beta \simeq 6$.
Despite the constraints from LEP data~\cite{Nakamura:2010zzi},
such low values for $m_h$ are not forbidden, since the
production cross section of a $Z h$ pair at LEP is
suppressed by a factor of $\sin^2(\alpha - \beta)$.

However,
the presence of the soft-breaking term $m_{12}^2$
greatly relaxes these bounds---for large enough values of $m^2_{12}$
the bound on $m_h$ becomes independent of $\beta$
and approximately equal to $670\, {\rm GeV}$.
The unitarity bounds of the 2HDM thus are
quite dependent on the values of some parameters,
but may in some cases be quite constraining,
even ruling out entire sections of parameter space
due to conflicts with experimental findings.

\section{Gauge interactions}

In the Higgs basis
of Eq.~\eqref{2_eq:higbas},
the charged field $H^+$ is physical and has mass $m_+$,
the neutral fields $H$, $R$, $I$ are linear combinations
of the physical neutral fields $S_1$, $S_2$, $S_3$:
\be
\left( \begin{array}{c} H \\ R \\ I \end{array} \right)
= T \left( \begin{array}{c} S_1 \\ S_2 \\ S_3 \end{array} \right),
\label{juaoe}
\ee
where the $3 \times 3$ matrix $T$ is orthogonal.
Without loss of generality we shall assume $T$ to have determinant $+1$.
The field $S_j$ ($j = 1, 2, 3)$ has mass $m_j$.

The gauge-kinetic Lagrangian
\be
\mathcal{L}_\mathrm{g} =
\sum_{k=1}^2 \left( D^\mu H_k \right)^\dagger \left( D_\mu H_k \right)
\ee
may be developed as~\cite{Grimus:2007if}
\begin{subequations}
\ba
\mathcal{L}_\mathrm{g} &=&
( \partial^\mu G^- ) ( \partial_\mu G^+ )
+ ( \partial^\mu H^- ) ( \partial_\mu H^+ )
+ \frac{( \partial^\mu G^0 ) ( \partial_\mu G^0 )}{2}
+ \sum_{j=1}^3
\frac{( \partial^\mu S_j ) ( \partial_\mu S_j )}{2}
\\  & &
\left. \frac{}{} \right.
+ m_W^2 W^{\mu -} W_\mu^+
+ m_Z^2\, \frac{Z^\mu Z_\mu}{2}
\\ & &
+ m_Z Z_\mu \partial^\mu G^0
+ i m_W ( W_\mu^- \partial^\mu G^+ - W_\mu^+ \partial^\mu G^- )
\\ & &
 \left. \frac{}{} \right.
- \left( e m_W A^\mu + g s_W^2 m_Z Z^\mu \right)
\left( W_\mu^- G^+ + W_\mu^+ G^- \right)
\\  & &
+ i \left[ e A_\mu
+ \frac{g \left( s_W^2 - c_W^2 \right)}{2 c_W}\, Z_\mu \right]
\left[ ( G^+ \partial^\mu G^- - G^- \partial^\mu G^+ )
\right. \no & & \left.
+ ( H^+ \partial^\mu H^- - H^- \partial^\mu H^+ ) \right]
\\  & &
+ \frac{g}{2 c_W}\, Z_\mu \Bigg[
\sum_{j=1}^3 T_{1j} \left( S_j \partial^\mu G^0 - G^0 \partial^\mu S_j \right)
+ T_{13} \left( S_1 \partial^\mu S_2 - S_2 \partial^\mu S_1 \right)
\nonumber
\\  & &
\quad \quad \quad \quad + T_{12} \left( S_3 \partial^\mu S_1 - S_1 \partial^\mu S_3 \right)
+ T_{11} \left( S_2 \partial^\mu S_3 - S_3 \partial^\mu S_2 \right)
\Bigg]
\\  & &
+ i\, \frac{g}{2}
\Bigg\{
W_\mu^+ \sum_{j=1}^3 T_{1j}
\left( G^- \partial^\mu S_j - S_j \partial^\mu G^- \right)
+ W_\mu^- \sum_{j=1}^3 T_{1j}
\left( S_j \partial^\mu G^+ - G^+ \partial^\mu S_j \right)
\nonumber
\\  & &
 \left. \frac{}{} \right.
\quad \quad + i W_\mu^+ \left( G^- \partial^\mu G^0 - G^0 \partial^\mu G^- \right)
+ i W_\mu^- \left( G^0 \partial^\mu G^+ - G^+ \partial^\mu G^0 \right)
\nonumber
\\  & &
\quad \quad \quad  + W_\mu^+ \sum_{j=1}^3 \left( T_{2j} + i T_{3j} \right)
\left( H^- \partial^\mu S_j - S_j \partial^\mu H^- \right)
\no  & &
\quad \quad \quad + W_\mu^- \sum_{j=1}^3 \left( T_{2j} - i T_{3j} \right)
\left( S_j \partial^\mu H^+ - H^+ \partial^\mu S_j  \right)
\Bigg\}
\\  & &
+ g \left( m_W W_\mu^+ W^{\mu -}
+ \frac{m_Z}{c_W}\, \frac{Z_\mu Z^\mu}{2} \right)
\sum_{j=1}^3 T_{1j} S_j
\label{kjcui}
\\  & &
- \left( \frac{eg}{2}\, A^\mu + \frac{g^2 s_W^2}{2 c_W}\, Z^\mu \right)
\left[
\left( W_\mu^+ G^- + W_\mu^- G^+ \right) \sum_{j=1}^3 T_{1j} S_j
+ i \left( W_\mu^+ G^- - W_\mu^- G^+ \right) G^0
\right. \no & & \left.
\quad \quad \quad + W_\mu^+ H^- \sum_{j=1}^3 ( T_{2j} + i T_{3j} ) S_j
+ W_\mu^- H^+ \sum_{j=1}^3 ( T_{2j} - i T_{3j} ) S_j
\right]
\\  & &
+ \left( \frac{g^2}{2}\, W^{\mu -} W_\mu^+
+ \frac{g^2}{2 c_W^2}\, \frac{Z^\mu Z_\mu}{2} \right)
\left[ \frac{\left( G^0 \right)^2}{2} + \sum_{j=1}^3 \frac{S_j^2}{2} \right]
\label{int2}
\\  & &
+ \left[
\frac{g^2}{2}\, W^{\mu -} W_\mu^+
+ 2 e^2\, \frac{A^\mu A_\mu}{2}
+ \frac{g^2 \left( s_W^2 - c_W^2 \right)^2}{2 c_W^2}\, \frac{Z^\mu Z_\mu}{2}
\right. \no  & & \left.
\quad \quad \quad \quad
+ \frac{e g}{c_W \left( s_W^2 - c_W^2 \right)}\, A^\mu Z_\mu
\right] \left( G^- G^+ + H^- H^+ \right).
\label{int4}
\ea
\end{subequations}

\section{Scalar--scalar interactions}

The scalar potential
in the Higgs basis is given by
\ba
V &=&
\bar m_{11}^2 H_1^\dagger H_1
+ \bar m_{22}^2 H_2^\dagger H_2
- \left( \bar m_{12}^2 H_1^\dagger H_2 + \mathrm{H.c.} \right)
\no & &
+ \frac{\bar \lambda_1}{2} \left( H_1^\dagger H_1 \right)^2
+ \frac{\bar \lambda_2}{2} \left( H_2^\dagger H_2 \right)^2
+ \bar \lambda_3\, H_1^\dagger H_1\, H_2^\dagger H_2
+ \bar \lambda_4\, H_1^\dagger H_2\, H_2^\dagger H_1
\no & &
+ \left[ \frac{\bar \lambda_5}{2} \left( H_1^\dagger H_2 \right)^2
+ \left( \bar \lambda_6\, H_1^\dagger H_1
+ \bar \lambda_7\, H_2^\dagger H_2 \right) H_1^\dagger H_2
+ \mathrm{H.c.} \right],
\label{potH}
\ea
where, as before, the parameters in the Higgs basis are denoted by
$\bar m_{ij}$ and $\bar \lambda_i$.
The vacuum expectation value of the potential is
\be
V_0 \equiv \left\langle 0 \left| V \right| 0 \right\rangle
= \bar m_{11}\, \frac{v^2}{4}
= - \bar \lambda_1\, \frac{v^4}{8}.
\ee
We now present general expressions for masses and couplings
written in the Higgs basis, for any neutral vacuum. These
are, of course, written in a different basis from the
results of section~\ref{2_sec:mat} and as such cannot
be trivially compared.
The mass terms for the scalars are given by
the part of the potential which is bilinear in the fields:
\be
V_2 = m_+^2 H^- H^+ + \frac{1}{2}
\left( \begin{array}{ccc} H & R & I \end{array} \right)
M
\left( \begin{array}{c} H \\ R \\ I \end{array} \right).
\label{ibhje}
\ee
The mass of the charged Higgs is given by
\be
m_+^2 = \bar m_{22}^2 + \bar \lambda_3\, \frac{v^2}{2},
\label{ibhje1}
\ee
while
\be
M = \left( \begin{array}{ccc}
\bar \lambda_1\, v^2 &
\mathrm{Re}\, \bar \lambda_6\, v^2 &
- \mathrm{Im}\, \bar \lambda_6\, v^2 \\
\mathrm{Re}\, \bar \lambda_6\, v^2 &
m_+^2 + \left( \bar \lambda_4 + \mathrm{Re}\, \bar \lambda_5 \right)
v^2 / \, 2 &
- \mathrm{Im}\, \bar \lambda_5\, v^2 / \, 2 \\
- \mathrm{Im}\, \bar \lambda_6\, v^2 &
- \mathrm{Im}\, \bar \lambda_5\, v^2 / \, 2 &
m_+^2 + \left( \bar \lambda_4 - \mathrm{Re}\, \bar \lambda_5 \right)
v^2 / \, 2
\end{array} \right).
\label{ibhje2}
\ee
The symmetric matrix $M$ is diagonalized by the orthogonal matrix $T$:
\be
T^T \! M T = \mathrm{diag} \left( m_1^2, m_2^2, m_3^2 \right),
\label{jvhgt}
\ee
so that
\be
V_2 = m_+^2 H^- H^+ + \sum_{j=1}^3 m_j^2\, \frac{S_j^2}{2}.
\ee
Like we said before,
we assume $\det{T} = +1$ without loss of generality.

The part of $V$ which is trilinear in the fields may be written
\bs
\ba
V_3 &=&
\left[ G^- G^+ + \frac{\left( G^0 \right)^2}{2} \right]
\sum_{j=1}^3 S_j T_{1j} \frac{m_j^2}{v}
\\ & &
+ G^+ H^- \sum_{j=1}^3 S_j \left( T_{2j} + i T_{3j} \right)
\frac{m_j^2 - m_+^2}{v}
\\ & &
+ G^- H^+ \sum_{j=1}^3 S_j \left( T_{2j} - i T_{3j} \right)
\frac{m_j^2 - m_+^2}{v}
\\ & &
+ G^0 \left(
S_1 S_2 T_{13}\, \frac{m_1^2 - m_2^2}{v}
+ S_3 S_1 T_{12}\, \frac{m_3^2 - m_1^2}{v}
\right. \no & & \left.
+ S_2 S_3 T_{11}\, \frac{m_2^2 - m_3^2}{v}
\right)
\\ & &
+ v H^- H^+ \sum_{j=1}^3 c_j S_j
\label{nhydp} \\
& & + \sum_{j, k, l = 1}^3 S_j S_k S_l
\left[ \frac{v}{2}\ c_j \left( \delta_{kl} - T_{1k} T_{1l} \right)
\right. \no & & \left.
+ \frac{1}{v} \left( m_+^2 - \frac{m_j^2}{2} \right) T_{1j} T_{1k} T_{1l}
+ \frac{m_k^2 - m_+^2}{v}\, T_{1j} \delta_{kl} \right].
\label{sbuir}
\ea
\es
where~\cite{Lavoura:1994fv}
\be
c_j \equiv T_{1j} \bar \lambda_3 + T_{2j}\, \mathrm{Re}\, \bar \lambda_7
- T_{3j}\, \mathrm{Im}\, \bar \lambda_7.
\label{mbkpq}
\ee
%
A thorough analysis of the trilinear Higgs couplings for the most general
2HDM potential, considering one-loop corrections, was undertaken
in~\cite{arXiv:0802.0060}.

\section{The oblique parameters}
\label{2_sec:ob}

\subsection{Definition}

Let the vacuum polarization tensors be written
\be
\Pi^{\mu \nu}_{V V^\prime} \left( q \right) =
g^{\mu \nu} A_{V V^\prime} \left( q^2 \right)
+ q^\mu q^\nu B_{V V^\prime} \left( q^2 \right),
\ee
where $V V^\prime$ may be either $\gamma \gamma$,
$\gamma Z^0$,
$Z^0 Z^0$,
or $W^+ W^-$,
and $q = \left( q^\alpha \right)$ is the four-momentum of the gauge bosons.
Let us moreover define
\be
\bar A_{V V^\prime} \left( q^2 \right) =
\left. A_{V V^\prime} \left( q^2 \right) \right|_\mathrm{2HDM}
- \left. A_{V V^\prime} \left( q^2 \right) \right|_\mathrm{SM},
\ee
where SM denotes the Standard Model with a Higgs particle of mass $m_H$.
Then,
the oblique parameters of the 2HDM are defined~\cite{Maksymyk:1993zm}
\bs
\ba
S &=& \frac{16 \pi c_W^2}{g^2} \left[
\frac{\bar A_{Z^0 Z^0} \left( m_Z^2 \right)
- \bar A_{Z^0 Z^0} \left( 0 \right)}{m_Z^2}
\right. \no & & \left.
- \left. \frac{\partial \bar A_{\gamma \gamma} \left( q^2 \right)}
{\partial q^2} \right|_{q^2 = 0}
+ \frac{c_W^2 - s_W^2}{c_W s_W}\,
\left. \frac{\partial \bar A_{\gamma Z^0} \left( q^2 \right)}
{\partial q^2} \right|_{q^2 = 0}
\right],
\\
T &=& \frac{4 \pi}{g^2 s_W^2} \left[
\frac{\bar A_{W^+ W^-} \left( 0 \right)}{m_W^2}
- \frac{\bar A_{Z^0 Z^0} \left( 0 \right)}{m_Z^2}
\right],
\\
U &=& \frac{16 \pi}{g^2} \left[
\frac{\bar A_{W^+ W^-} \left( m_W^2 \right)
- \bar A_{W^+ W^-} \left( 0 \right)}{m_W^2}
- c_W^2\, \frac{\bar A_{Z^0 Z^0} \left( m_Z^2 \right)
- \bar A_{Z^0 Z^0} \left( 0 \right)}{m_Z^2}
\right. \no & & \left.
- s_W^2 \left. \frac{\partial \bar A_{\gamma \gamma} \left( q^2 \right)}
{\partial q^2} \right|_{q^2 = 0}
+ 2 c_W s_W \left. \frac{\partial \bar A_{\gamma Z^0} \left( q^2 \right)}
{\partial q^2} \right|_{q^2 = 0}
\right],
\\
V &=& \frac{4 \pi}{g^2 s_W^2} \left[
\left. \frac{\partial \bar A_{Z^0 Z^0} \left( q^2 \right)}
{\partial q^2} \right|_{q^2 = m_Z^2}
- \frac{\bar A_{Z^0 Z^0} \left( m_Z^2 \right)
- \bar A_{Z^0 Z^0} \left( 0 \right)}{m_Z^2}
\right],
\\
W &=& \frac{4 \pi}{g^2 s_W^2} \left[
\left. \frac{\partial \bar A_{W^+ W^-} \left( q^2 \right)}
{\partial q^2} \right|_{q^2 = m_W^2}
- \frac{\bar A_{W^+ W^-} \left( m_W^2 \right)
- \bar A_{W^+ W^-} \left( 0 \right)}{m_W^2}
\right],
\\
X &=& \frac{4 \pi c_W}{g^2 s_W} \left[
\left. \frac{\partial \bar A_{\gamma Z^0} \left( q^2 \right)}
{\partial q^2} \right|_{q^2 = 0}
- \frac{\bar A_{\gamma Z^0} \left( m_Z^2 \right)}{m_Z^2}
\right].
\ea
\es
These parameters are finite and,
in principle,
observable.
In practice,
they will be measurable in practical electroweak experiments
on the 2HDM provided
\begin{description}
\item the measurements are performed
at one of the energy scales $q^2 \approx 0$,
$q^2 = m_W^2$,
or $q^2 = m_Z^2$,
and
\item the measurements are performed with light fermions
which couple mainly to the gauge bosons $\gamma$,
$Z^0$,
and $W^\pm$,
but couple only very weakly to the scalar particles.
\end{description}

\subsection{Formulae}

The expressions for the oblique parameters in multi-Higgs-doublet models
have been derived in~\cite{Grimus:2007if,Grimus:2008nb}.
We give here those expressions in the particular case of the 2HDM.
For $T$ one has~\cite{Grimus:2007if}
\bs
\ba
T &=& \frac{1}{16 \pi s_W^2 m_W^2} \Bigg\{
\sum_{j=1}^3 \left( 1 - T_{1j}^2 \right) F \left( m_+^2, m_j^2 \right)
\label{firstT} \\ & &
- T_{11}^2 F \left( m_2^2, m_3^2 \right)
- T_{12}^2 F \left( m_3^2, m_1^2 \right)
- T_{13}^2 F \left( m_1^2, m_2^2 \right)
\label{secondT} \\ & &
+ 3 \sum_{j=1}^3 T_{1j}^2 \left[ F \left( m_Z^2, m_j^2 \right)
- F \left( m_W^2, m_j^2 \right) \right]
\label{thirdT} \\ & &
- 3 \left[ F \left( m_Z^2, m_H^2 \right) - F \left( m_W^2, m_H^2 \right)
\right]
\Bigg\},
\label{fourthT}
\ea
\label{T}
\es
where
\be
F \left( x, y \right) = \left\{
\begin{array}{l}
\displaystyle{\frac{x + y}{2} - \frac{x y}{x - y}\, \ln{\frac{x}{y}}}
\ \Leftarrow x \neq y,
\\
0 \ \Leftarrow x = y.
\end{array}
\right.
\ee
Line~(\ref{firstT}) is a positive definite contribution to $T$
coming from the neutral scalars not having the same mass
as the charged scalar,
while line~(\ref{secondT}) is a negative definite contribution to $T$
resulting from the neutral scalars not all having the same mass.
Line~(\ref{fourthT}) is the subtraction of the SM equivalent
of line~(\ref{thirdT}).

The expressions for $S$,
$U$,
and $X$ involve the following two functions:
\ba
G \left( x, y \right) &=& - \frac{16}{3}
+ 5 \left( x + y \right) - 2 \left( x - y \right)^2
\no & &
+ 3 \left[ \frac{x^2 + y^2}{x - y}
- x^2 + y^2 + \frac{\left( x - y \right)^3}{3} \right] \ln{\frac{x}{y}}
\no & &
+ \left[ 1 - 2 \left( x + y \right) + \left( x - y \right)^2 \right]
f \left( x + y - 1,
1 - 2 \left( x + y \right) + \left( x - y \right)^2 \right),
\label{rtycn} \\
\hat G \left( x \right) &=&
- \frac{79}{3} + 9 x - 2 x^2
+ \left( - 10 + 18 x - 6 x^2 + x^3 - 9\, \frac{x + 1}{x - 1} \right) \ln{x}
\no & &
+ \left( 12 - 4 x + x^2 \right) f \left( x, x^2 - 4 x \right),
\label{nsdep}
\ea
where
\be
f \left( z, w \right) = \left\{
\begin{array}{l}
\displaystyle{\sqrt{w}\,
\ln{\left| \frac{z - \sqrt{w}}{z + \sqrt{w}} \right|}}
\ \Leftarrow w > 0,
\\*[3mm]
0 \ \Leftarrow w = 0,
\\
\displaystyle{2 \sqrt{-w}\, \arctan{\frac{\sqrt{-w}}{z}}}
\
\Leftarrow w < 0.
\end{array} \right.
\ee
One has
\ba
S &=& \frac{1}{24 \pi} \left\{
\left( s_W^2 - c_W^2 \right)^2 G \left( z_+, z_+ \right)
\right. \no & &
+ T_{11}^2\, G \left( z_2, z_3 \right)
+ T_{12}^2\, G \left( z_3, z_1 \right)
+ T_{13}^2\, G \left( z_1, z_2 \right)
\no & & \left.
+ \sum_{j=1}^3 \left[ T_{1j}^2\, \hat G \left( z_j \right)
+ \ln{\frac{m_j^2}{m_+^2}} \right]
- \hat G \left( z_H \right) - \ln{\frac{m_H^2}{m_+^2}}
\right\},
\\
U &=& \frac{1}{24 \pi} \left\{
\sum_{j=1}^3 \left( 1 - T_{1j}^2 \right) G \left( w_+, w_j \right)
- \left( s_W^2 - c_W^2 \right)^2 G \left( z_+, z_+ \right)
\right. \no & &
- T_{11}^2 G \left( z_2, z_3 \right)
- T_{12}^2 G \left( z_3, z_1 \right)
- T_{13}^2 G \left( z_1, z_2 \right)
\no & & \left.
+ \sum_{j=1}^3 T_{1j}^2 \left[
\hat G \left( w_j \right) - \hat G \left( z_j \right)
\right]
- \hat G \left( w_H \right) + \hat G \left( z_H \right)
\right\},
\\
X &=& \frac{c_W^2 - s_W^2}{48 \pi}\,
G \left( z_+, z_+ \right),
\ea
where
\be
z_a \equiv \frac{m_a^2}{m_Z^2}
\quad {\rm and} \quad
w_a \equiv \frac{m_a^2}{m_W^2}
\ee
for $a = +, 1, 2, 3, H$.

The expressions for the oblique parameters $V$ and $W$
involve the following two functions:
\ba
H \left( x, y \right) &=&
2 - 9 \left( x + y \right) + 6 \left( x - y \right)^2
\no & &
+ 3 \left[ - \frac{x^2 + y^2}{x - y} + 2 \left( x^2 - y^2 \right)
- \left( x - y \right)^3 \right] \ln{\frac{x}{y}}
\no & &
+ 3 \left[ x + y - \left( x - y \right)^2 \right]
f \left( x + y - 1,
1 - 2 \left( x + y \right) + \left( x - y \right)^2 \right),
\\
\hat H \left( x \right) &=&
47 - 21 x + 6 x^2
+ 3 \left( 7 - 12 x + 5 x^2 - x^3 + 3\, \frac{x + 1}{x - 1} \right) \ln{x}
\no & &
+ 3 \left( 28 - 20 x + 7 x^2 - x^3 \right)
\frac{f \left( x, x^2 - 4 x \right)}{x - 4}.
\ea
One has
\ba
V &=& \frac{1}{96 \pi c_W^2 s_W^2} \left[
\left( s_W^2 - c_W^2 \right)^2 H \left( z_+, z_+ \right)
\right. \no & &
+ T_{11}^2 H \left( z_2, z_3 \right)
+ T_{12}^2 H \left( z_3, z_1 \right)
+ T_{13}^2 H \left( z_1, z_2 \right)
\no & & \left.
+ \sum_{j=1}^3 T_{1j}^2 \hat H \left( z_j \right)
- \hat H \left( z_H \right)
\right],
\\
W &=& \frac{1}{96 \pi s_W^2} \left[
\sum_{j=1}^3 \left( 1 - T_{1j}^2 \right) H \left( w_+, w_j \right)
\right. \no & & \left.
+ \sum_{j=1}^3 T_{1j}^2 \hat H \left( w_j \right)
- \hat H \left( w_H \right)
\right].
\ea

\section{Renormalization-group equations}
\label{2_sec:rge}

The one-loop renormalization-group (RG) equations
for a general gauge theory
were presented in~\cite{Cheng:1973nv,Machacek:1981ic}.
For the specific case of the multi-Higgs-doublet
$SU(2) \times U(1)$ gauge theory,
they were given in~\cite{Haber:1993an,Grimus:2004yh,Ferreira:2010xe}.

Let $\mu$ be the mass parameter used in the regularization
of ultraviolet divergences in loop integrals.
Let $\mathcal{D}$ denote the dimensionless differential operator
$16 \pi^2 \left( \mathrm{d} / \mathrm{d} \ln{\mu} \right) =
16 \pi^2 \mu \left( \mathrm{d} / \mathrm{d} \mu \right)$.
Let $g_s$,
$g$,
and $g^\prime$ denote the gauge coupling constants
of $SU(3)_\mathrm{colour}$,
$SU(2)$,
and $U(1)$,
respectively.
The normalization of $g^\prime$ is such that
the $U(1)$ charge of the Higgs doublets is $+1/2$.
The one-loop RG equations for the gauge coupling constants
do not depend of the Yukawa and scalar couplings and are
\ba
\mathcal{D} g_s &=&
\left( - 11 + \frac{4}{3}\, n_F \right) g_s^3,
\\
\mathcal{D} g &=&
\left( - \frac{22}{3} + \frac{4}{3}\, n_F
+ \frac{1}{6}\, n_H \right) g^3,
\\
\mathcal{D} g^\prime &=&
\left( \frac{20}{9}\, n_F + \frac{1}{6}\, n_H \right) {g^\prime}^3,
\ea
where $n_F$ is the number of fermion generations
and $n_H$ is the number of Higgs doublets.
In the 2HDM with three families of fermions this is therefore
\ba
\mathcal{D} g_s &=& - 7 g_s^3,
\\
\mathcal{D} g &=& - 3 g^3,
\\
\mathcal{D} g^\prime &=& 7 {g^\prime}^3.
\ea

The one-loop RG equations for the Yukawa-coupling matrices
defined in eq.~\eqref{uvhfw}
do not depend on the scalar couplings and are
\ba
\mathcal{D} Y^d_j &=&
a_d Y^d_j + \sum_{k=1}^{n_H} T_{jk} Y^d_k
\nonumber
\label{uvhfw2}
\\ & &
+ \sum_{k=1}^{n_H} \left(
- 2\, Y^u_k {Y^u_j}^\dagger Y^d_k
+ \frac{1}{2}\, Y^u_k {Y^u_k}^\dagger Y^d_j
+ Y^d_j {Y^d_k}^\dagger Y^d_k
+ \frac{1}{2}\, Y^d_k {Y^d_k}^\dagger Y^d_j
\right), \\
\mathcal{D} Y^u_j &=&
a_u Y^u_j + \sum_{k=1}^{n_H} T_{jk}^\ast Y^u_k
\nonumber
\\ & &
+ \sum_{k=1}^{n_H} \left(
- 2\, Y^d_k {Y^d_j}^\dagger Y^u_k
+ \frac{1}{2}\, Y^d_k {Y^d_k}^\dagger Y^u_j
+ Y^u_j {Y^u_k}^\dagger Y^u_k
+ \frac{1}{2}\, Y^u_k {Y^u_k}^\dagger Y^u_j
\right), \\
\mathcal{D} Y^e_j &=&
a_e Y^e_j + \sum_{k=1}^{n_H} T_{jk} Y^e_k
+ \sum_{k=1}^{n_H} \left(
Y^e_j {Y^e_k}^\dagger Y^e_k
+ \frac{1}{2}\, Y^e_k {Y^e_k}^\dagger Y^e_j
\right),
\label{mkghq}
\ea
where
\ba
a_d &=& - 8 g_s^2 - \frac{9}{4}\, g^2 - \frac{5}{12}\, {g^\prime}^2,
\\
a_u &=& - 8 g_s^2 - \frac{9}{4}\, g^2 - \frac{17}{12}\, {g^\prime}^2,
\\
a_e &=& - \frac{9}{4}\, g^2 - \frac{15}{4}\, {g^\prime}^2,
\ea
and
\be
T_{jk} = 3\, \mathrm{tr} \left( Y^d_j {Y^d_k}^\dagger
+ {Y^u_j}^\dagger Y^u_k \right) + \mathrm{tr} \left(
Y^e_j {Y^e_k}^\dagger \right).
\ee
Of course,
in the context of the 2HDM we should set $n_H = 2$
in equations~(\ref{uvhfw2})--(\ref{mkghq}).

Let the scalar potential be
\be
V = \sum_{j,k}\, \mu^j_k \Phi_j^\dagger \Phi_k
+ \frac{1}{2} \sum_{j,k,l,m} \Lambda^{jl}_{km}\,
\Phi_j^\dagger \Phi_k\, \Phi_l^\dagger \Phi_m,
\ee
%
with $\Lambda^{jl}_{km} = \Lambda^{lj}_{mk}$.\footnote{
The correspondence with the notation of eq.~\eqref{2_VH2}
is $\Lambda^{jl}_{km} = \lambda_{jk,lm}$; using the
tensor $\Lambda$ allows for a more compact writing of
the formulae below.}
The one-loop RG equations for the quartic couplings are
\ba
\mathcal{D} \Lambda^{jl}_{km} &=& 2 \sum_{p,q=1}^{n_H} \left(
2 \Lambda^{jp}_{kq} \Lambda^{ql}_{pm}
+ \Lambda^{jp}_{kq} \Lambda^{lq}_{pm}
+ \Lambda^{jp}_{qk} \Lambda^{ql}_{pm}
+ \Lambda^{jl}_{pq} \Lambda^{pq}_{km}
+ \Lambda^{jq}_{pm} \Lambda^{pl}_{kq}
\right) \hspace*{8mm}
\nonumber
\\ & &
- \left( 9 g^2 + 3 {g^\prime}^2 \right) \Lambda^{jl}_{km}
\nonumber
\\ & &
+ \frac{9 g^4 - 6 g^2 {g^\prime}^2 + 3 {g^\prime}^4}{4}\,
\delta^j_k \delta^l_m
+ 3 g^2 {g^\prime}^2 \delta^j_m \delta^l_k
\nonumber
\\ & &
+ \sum_{p=1}^{n_H} \left(
T_{kp} \Lambda^{jl}_{pm} + T_{mp} \Lambda^{jl}_{kp}
+ T_{jp}^\ast \Lambda^{pl}_{km} + T_{lp}^\ast \Lambda^{jp}_{km}
\right)
\nonumber
\\ & &
- 4\, \mathrm{tr} \left( {Y^e_j}^\dagger Y^e_k {Y^e_l}^\dagger Y^e_m \right)
\no & &
- 12\, \mathrm{tr} \left(
{Y^d_j}^\dagger Y^d_k {Y^d_l}^\dagger Y^d_m
+ {Y^u_k}^\dagger Y^u_j {Y^u_m}^\dagger Y^u_l
+ {Y^d_j}^\dagger Y^u_l {Y^u_m}^\dagger Y^d_k
\right. \no & & \left.
+ {Y^u_k}^\dagger Y^d_m {Y^d_l}^\dagger Y^u_j
- {Y^u_m}^\dagger Y^d_k {Y^d_l}^\dagger Y^u_j
- {Y^d_j}^\dagger Y^u_l {Y^u_k}^\dagger Y^d_m
\right).
\ea
The one-loop RG equations for the quadratic couplings are
\be
\mathcal{D} \mu^j_k = 2 \sum_{p,q=1}^{n_H} \mu^p_q \left(
2 \Lambda^{jq}_{kp} + \Lambda^{jq}_{pk} \right).
\label{nvgfr}
\ee

In the specific case of the 2HDM,
with the notation
of eq.~\eqref{2_VH1},
one has
\be
\mu^1_1 = m_{11}^2, \quad
\mu^2_2 = m_{22}^2, \quad
\mu^1_2 = - m_{12}^2, \quad
\mu^2_1 = - {m_{12}^2}^\ast,
\ee
\be
\Lambda^{11}_{11} = \lambda_1, \quad
\Lambda^{22}_{22} = \lambda_2, \quad
\Lambda^{12}_{12} = \Lambda^{21}_{21} = \lambda_3, \quad
\Lambda^{12}_{21} = \Lambda^{21}_{12} = \lambda_4,
\ee
\be
\Lambda^{11}_{22} = \lambda_5, \quad
\Lambda^{22}_{11} = \lambda_5^\ast, \quad
\ee
\be
\Lambda^{11}_{21} = \Lambda^{11}_{12} = \lambda_6, \quad
\Lambda^{21}_{11} = \Lambda^{12}_{11} = \lambda_6^\ast, \quad
\Lambda^{12}_{22} = \Lambda^{21}_{22} = \lambda_7, \quad
\Lambda^{22}_{12} = \Lambda^{22}_{21} = \lambda_7^\ast.
\ee
Therefore,
from equation~(\ref{nvgfr}),
\ba
\mathcal{D} m_{11}^2 &=&
6 \lambda_1 m_{11}^2
+ \left( 4 \lambda_3 + 2 \lambda_4 \right) m_{22}^2
- 12\, \mathrm{Re} \left( m_{12}^2 \lambda_6^\ast \right),
\\
\mathcal{D} m_{22}^2 &=&
\left( 4 \lambda_3 + 2 \lambda_4 \right) m_{11}^2
+ 6 \lambda_2 m_{22}^2
- 12\, \mathrm{Re} \left( m_{12}^2 \lambda_7^\ast \right),
\\
\mathcal{D} m_{12}^2 &=&
- 6 \left( \lambda_6 m_{11}^2 + \lambda_7 m_{22}^2 \right)
+ \left( 2 \lambda_3 + 4 \lambda_4 \right) m_{12}^2
+ 6 \lambda_5 {m_{12}^2}^\ast.
\ea
%
For the quartic couplings,
in a 2HDM where the up-type quarks couple only to the doublet $\Phi_1$,
and if we only consider the contribution from
the top-quark Yukawa coupling $\lambda_t$
(with the normalization $m_t = \lambda_t v/\sqrt{2}$,
with $v = 246~\mathrm{GeV}$),
the one-loop RG equations are
\bs
\ba
\mathcal{D} \lambda_1 &=&
12 \lambda_1^2 + 4 \lambda_3^2 + 4 \lambda_3 \lambda_4 + 2 \lambda_4^2
+ 2 \left| \lambda_5 \right|^2 + 24 \left| \lambda_6 \right|^2 \nonumber \\
 & &
 +\ \frac{3}{4}(3g^4 + g^{\prime 4} +2 g^2 g^{\prime 2}) -
 3\lambda_1 (3 g^2 + g^{\prime
2} - 4 \lambda_t^2) - 12 \lambda_t^4, \\
\mathcal{D} \lambda_2 &=&
12 \lambda_2^2 + 4 \lambda_3^2 + 4 \lambda_3 \lambda_4 + 2 \lambda_4^2
+ 2 \left| \lambda_5 \right|^2 + 24 \left| \lambda_7 \right|^2 \nonumber \\
 & &
+\
\frac{3}{4}(3g^4 + g^{\prime 4} +2g^2 g^{\prime 2}) -3\lambda_2
(3g^2 +g^{\prime 2}), \\
\mathcal{D} \lambda_3  &=&
\left( \lambda_1 + \lambda_2 \right) \left( 6 \lambda_3 + 2 \lambda_4 \right)
+ 4 \lambda_3^2 + 2 \lambda_4^2
+ 2 \left| \lambda_5 \right|^2
+ 4 \left( \left| \lambda_6 \right|^2 + \left| \lambda_7 \right|^2 \right)
+ 16\, \mathrm{Re} \left( \lambda_6 \lambda_7^\ast \right) \nonumber \\
 & &
+\ \frac{3}{4}(3g^4 + g^{\prime 4} -2g^2 g^{\prime 2}) - 3\lambda_3
(3g^2 +g^{\prime 2}-2 \lambda_t^2), \\
\mathcal{D} \lambda_4  &=&
2 \left( \lambda_1 + \lambda_2 \right) \lambda_4
+ 8 \lambda_3 \lambda_4 + 4 \lambda_4^2
+ 8 \left| \lambda_5 \right|^2
+ 10 \left( \left| \lambda_6 \right|^2 + \left| \lambda_7 \right|^2 \right)
+ 4\, \mathrm{Re} \left( \lambda_6 \lambda_7^\ast \right) \nonumber \\
& &
+\ 3g^2 g^{\prime 2} - 3\lambda_4 (3g^2 +g^{\prime
2}-2\lambda_t^2),
\label{eq:dl4} \\
\mathcal{D} \lambda_5  &=&
\left( 2 \lambda_1 + 2 \lambda_2 + 8 \lambda_3 + 12 \lambda_4 \right) \lambda_5
+ 10 \left( \lambda_6^2 + \lambda_7^2 \right) + 4 \lambda_6 \lambda_7
\nonumber \\
 & &
- \ 3\lambda_5 (3g^2 +g^{\prime 2}-2 \lambda_t^2), \label{eq:dl5} \\
\mathcal{D} \lambda_6  &=&
\left( 12 \lambda_1 + 6 \lambda_3 + 8 \lambda_4 \right) \lambda_6
+ \left( 6 \lambda_3 + 4 \lambda_4 \right) \lambda_7
+ 10 \lambda_5 \lambda_6^\ast + 2 \lambda_5 \lambda_7^\ast \nonumber \\
 & &
-\ 3\lambda_6 (3g^2 +g^{\prime 2}-3\lambda_t^2), \\
\mathcal{D} \lambda_7  &=&
\left( 12 \lambda_2 + 6 \lambda_3 + 8 \lambda_4 \right) \lambda_7
+ \left( 6 \lambda_3 + 4 \lambda_4 \right) \lambda_6
+ 10 \lambda_5 \lambda_7^\ast + 2 \lambda_5 \lambda_6^\ast \nonumber \\
 & &
-\ 3\lambda_7 (3g^2 +g^{\prime 2}-\lambda_t^2).
\ea
\es

\section{Custodial symmetry}
\label{ap:cust}

The experimentally measured value for the observable
$\rho = m_W^2 \left/ \left(m_Z^2 \cos^2\theta_W \right) \right.$
is extremely close to one~\cite{Nakamura:2010zzi}.
In the SM,
the tree-level prediction for that observable is exactly one.
The fundamental reason for that prediction
is an approximate symmetry that the SM Lagrangian possesses.
Indeed,
if one writes the SM Higgs doublet as
$\Phi = \left( \begin{array}{c}
\varphi_1 + i \varphi_2 \\ \varphi_3 + i \varphi_4
\end{array} \right)$,
then the SM scalar potential only depends on
$\Phi^\dagger \Phi = \varphi_1^2 + \varphi_2^2 +
\varphi_3^2 + \varphi_4^2$.
Therefore,
the potential automatically has $SO(4)$ symmetry.
The group $SO(4)$ is isomorphic to $SU(2) \times SU(2)$,
which is larger than the SM gauge group $SU(2)_L \times U(1)_Y$.
On the other hand,
this symmetry $SO(4)$ is respected neither
by the scalar gauge-kinetic terms---specifically,
those involving the weak-hypercharge coupling $g^\prime$---nor
by the Yukawa terms,
linear in $\Phi$,
since the up-type and down-type quarks have different masses.
Thus,
$SO(4)$ is not a symmetry of the full SM Lagrangian,
rather a symmetry only of the scalar potential;
it is usually regarded as an \textit{approximate}\/ symmetry,
since in
the scalar sector it is only broken by (small) $g^\prime$ terms in
the kinetic energy,
and dubbed `custodial symmetry'~\cite{Sikivie:1980hm,Chanowitz:1985ug}.

In the 2HDM,
there is no $SO(4)$ symmetry in the most general scalar potential,
thus the possibility of large contributions to $\rho$
from that sector alone arises.
If one wants to avoid them,
one may impose custodial symmetry on the 2HDM potential.
Following the work of Pomarol and Vega~\cite{Pomarol:1993mu},
we first define the $2 \times 2$ matrices
\be
M_{ij} = \left( \tilde{\Phi}_i \ |\ \Phi_j  \right) = \left(
\begin{array}{cc}
{\varphi^0_i}^* & \varphi^+_j \\
- \varphi^-_i & \varphi^0_j \\
\end{array}
\right).
\ee
We further define an $SU(2)_L \times SU(2)_R$ group
under which these matrices transform as
\be
M_{ij} \rightarrow L M_{ij} R^\dagger,
\label{eq:lr}
\ee
with $L, R \in SU(2)$.
The quantities $\mbox{tr}(M_{ij}^\dagger M_{kl})$
are invariant under this $SO(4) = SU(2)_L \times SU(2)_R$.
This corresponds to the same $SO(4)$ custodial symmetry
of the SM scalar potential.

Pomarol and Vega have considered two separate situations.
In case 1 they have used only $M_{11}$ and $M_{22}$.
The most general scalar potential invariant under $SO(4)$ is then
\ba
V_1 &=& \tfrac{1}{2} m_{11}^2 \mbox{tr}(M_{11}^\dagger M_{11}) +
\tfrac{1}{2} m_{22}^2 \mbox{tr}(M_{22}^\dagger M_{22})
- m_{12}^2 \mbox{tr}(M_{11}^\dagger M_{22}) \nonumber \\
& &
+ \tfrac{1}{8} \lambda_1 \left[\mbox{tr}(M_{11}^\dagger M_{11})\right]^2 +
\tfrac{1}{8} \lambda_2 \left[\mbox{tr}(M_{22}^\dagger M_{22})\right]^2 +
\tfrac{1}{4} \lambda_3 \mbox{tr}(M_{11}^\dagger M_{11}) \mbox{tr}(M_{22}^\dagger M_{22})
\nonumber \\
 &  & + \tfrac{1}{2} \lambda_4 \left[\mbox{tr}(M_{11}^\dagger M_{22})\right]^2 +
\tfrac{1}{2} \left[\lambda_6 \mbox{tr}(M_{11}^\dagger M_{11}) +
\lambda_7 \mbox{tr}(M_{22}^\dagger M_{22})\right] \mbox{tr}(M_{11}^\dagger M_{22}).
\label{eq:case1}
\ea
Notice that $\mbox{tr}( M_{ii}^\dagger M_{jj})
= \Phi_i^\dagger \Phi_j + \Phi_j^\dagger \Phi_i$.
This allows us to identify,
in eq.~\eqref{eq:case1},
the relations that one obtains
for the usual parameters of the 2HDM potential in eq.~\eqref{2_VH1}.
We find
\begin{itemize}
\item Case 1: all the parameters are real,
and $\lambda_4 = \lambda_5$.
\end{itemize}
At a neutral vacuum $\langle \varphi_i^0 \rangle = v_i$,
one has
\be
\langle M_{ij} \rangle =
\left( \begin{array}{cc} v_i^\ast & 0 \\ 0 & v_j \end{array} \right).
\ee
This vacuum is {\em not} invariant
under the full group $SU(2)_L \times SU(2)_R$.
However,
if $v_i^\ast = v_j$,
then $\langle M_{ij} \rangle$ is proportional
to the $2 \times 2$ identity matrix
and the vacuum preserves a group $SU(2)_V$
(the ``V'' stands for ``vectorial''),
corresponding to identical matrices,
\textit{i.e.}\/ $L = R$,
in eq.~\eqref{eq:lr}.
This remaining group preserved by the vacuum
is the custodial-symmetry group.
However,
most authors refer to the potential invariant under $SU(2)_L \times SU(2)_R$
as displaying a custodial symmetry,
and we shall also employ that terminology.

In case 2 of Pomarol and Vega
the potential is built only with $M_{12}$.
It reads
\ba
V_2 &=& m_{11}^2 \mbox{tr}(M_{12}^\dagger M_{12})
- m_{12}^2 \left(\mbox{det} M_{12} + \mbox{h.c.} \right)  \nonumber \\
& &
+ \tfrac{1}{2} \lambda_1 \left[\mbox{tr}(M_{12}^\dagger M_{12})\right]^2 +
\lambda_4 \ \mbox{det} (M_{12}^\dagger M_{12})  +
\tfrac{1}{2} \left[\lambda_5 \ \mbox{det}(M_{12})^2 + \mbox{h.c.} \right]
\nonumber \\
 &  & + \left[\lambda_6 \,\mbox{det} M_{12} \,\mbox{tr}(M_{12}^\dagger M_{12})
 + \mbox{h.c.}\right] .
\label{eq:case2}
\ea
The following constraints on the parameters of the 2HDM scalar potential ensue:
\begin{itemize}
\item Case 2: $m^2_{11} = m^2_{22}$,
$\lambda_1 = \lambda_2 = \lambda_3$,
and $\lambda_6 = \lambda_7$.
\end{itemize}
Notice that $m^2_{12}$,
$\lambda_5$,
$\lambda_6$,
and $\lambda_7$ remain complex in this case.
The vacuum preserves $SU(2)_V$ if and only if $v_1 = v_2^\ast$.

In both cases,
there is a dramatic prediction for the scalar masses:
the charged Higgs $H^\pm$ is degenerate with the pseudoscalar $A$.
This is easy to see from eq.~\eqref{eq:ma} in the Case 1 model.
In both cases of Pomarol and Vega, the potential conserves CP,
even with the complex couplings of case 2.
There is thus a well-defined pseudoscalar particle.

G\'erard and Herquet~\cite{Gerard:2007kn}
have proposed a {\em twisted custodial symmetry}
which generalizes the formalism presented above.
They observed that the transformation matrix $R$
need not be the same for $M_{11}$ and $M_{22}$,
namely
\be
M_{11} \rightarrow L M_{11} R^\dagger, \quad
M_{22} \rightarrow L M_{22} {R^\prime}^\dagger .
\ee
This extra freedom has a limitation,
though:
since the hypercharge is proportional to the diagonal generator of $SU(2)_R$,
the matrices $R$ and $R^\prime$
must be related through $R^\prime = X^\dagger R X$,
with
$X = \mbox{diag} \left( e^{i \gamma/2}, e^{-i\gamma/2} \right)$.
A specific choice for the phase $\gamma$
yields the custodial mass relation $m^2_\pm = m^2_A$;
but a different choice imposes degeneracy
between the charged Higgs and one of the CP-even scalars,
$m^2_\pm = m^2_h$.

As we know,
in the general 2HDM the masses of the $H^\pm$,
$H$,
$A$,
and $h$ are arbitrary,
since they depend on arbitrary quartic terms in the potential.
In the MSSM,
the quartic terms are constrained to be gauge couplings,
and one finds the relationship $m_{H^\pm}^2 = m_A^2 + m_W^2$.
This would rule out decays $H^\pm \rightarrow W^\pm A$.
The twisted symmetry allows for a scenario where the pseudoscalar is light
(leading to possible decays $h\rightarrow AA$,
as discussed in Chapter~\ref{sec:nfc}).
A mass spectrum with $m_A < m_{H^\pm}, m_H < m_h$
leads to the possibility of the Standard Model-like Higgs
decaying into charged Higgs.
Any of the four flavour conserving models
can be accommodated in this scenario.
A detailed discussion of the phenomenology of the model
can be found in Ref.~\cite{deVisscher:2009zb}.

One might wonder about the apparent ambiguity
that one has in constructing the 2HDM custodial-symmetric potential
in the manner displayed above.
For instance, \\
$\mbox{tr} ( M_{11}^\dagger M_{12})$
is also invariant under $SU(2)_L\times SU(2)_R$,
and we might insert one such term in the potential,
together with many others.
So there would seem to be many more models to be considered.
Grzadkowski \etal~\cite{arXiv:1011.5228},
Haber and O'Neil~\cite{Haber:2010bw},
and Nishi~\cite{Nishi:2011gc} have shown that the models Case~1 and Case~2
are indeed related by a basis change---they are not different models,
rather they have the same physical predictions.
Furthermore,
basis-invariant methods were developed to identify
whether a given potential has custodial symmetry or not.
Grzadkowski \etal and Nishi have used the bilinear formalism
to find such conditions for the scalar sector alone.
Haber and O'Neil used a different formulation
and analysed the Yukawa sector as well.
The basis-invariant conditions of Grzadkowski \etal are particularly simple:
a necessary and sufficient condition for a potential
to possess custodial symmetry
is the existence of a three-vector $\tvec{v}$ such that
\be
E  \tvec{v} = 0, \quad
\tvec{\xi} \cdot \tvec{v} = 0, \quad
\tvec{\eta} \cdot \tvec{v} = 0,
\ee
with $E$,
$\tvec{\xi}$,
and $\tvec{\eta}$ defined in section~\ref{sec:cpvbil}.
Still regarding the twisted symmetry,
Haber and O'Neil~\cite{Haber:2010bw} argued that
the choice of the phase $\gamma$ discussed above
is tantamount to a basis choice.
They also showed that
in the region of parameter space where the twisted scenario arises
there is a fundamental indefinition on what constitutes CP,
and thus the distinction between scalar and pseudoscalar particles is not clear.

Extending the custodial symmetry
to the Yukawa sector~\cite{Haber:1992py,Haber:2010bw}
requires the equality of the up-type and down-type-quark mass matrices.
Thus,
the custodial symmetry is broken by the hypercharge interactions
and also by the mass differences amongst quarks.

Notice that,
even though we refer to the custodial symmetry as being a symmetry,
it does not correspond to any of the six proper symmetries
of the 2HDM potential discussed in section~\ref{sec:symmetries}.
For instance,
the renormalization-group (RG) running
preserves the relations among parameters of the potential
following from any of the six symmetries of table~\ref{2_master1},
whereas it does \emph{not} preserve the relations
following from custodial symmetry.
It is easy to see,
by using the $\beta$-functions of eqs.~\eqref{eq:dl4}--\eqref{eq:dl5},
that the Case 1 relation $\lambda_4 = \lambda_5$
is unstable under the RG,
\textit{i.e.}\/ $\beta_{\lambda_4} \neq \beta_{\lambda_5}$.\footnote{This occurs
even when one sets the Yukawa couplings to zero,
due to the $g^2 {g^\prime}^2$ term in $\beta_{\lambda_4}$.}
The fundamental reason why custodial symmetry
cannot appear in table~\ref{2_master1} is that
those symmetries were obtained by requiring invariance
of the scalar gauge-kinetic terms,
while the custodial transformations do not leave them invariant.
In fact,
the largest symmetry group of the 2HDM scalar gauge-kinetic terms
was identified in~\cite{arXiv:1007.1424} as $U(2) \times SU(2)$,
promoted to $Sp(2) \times SO(4)$
in the limit $g^\prime \rightarrow 0$.
If one relaxes the requirement of invariance of the gauge-kinetic terms,
then the list of possible symmetries of the potential increases,
as shown in~\cite{arXiv:1106.3482,arXiv:1109.3787},
and includes custodial symmetry.


\newpage

\begin{thebibliography}{999}
%
%
\bibitem{11883}
  P.~W.~Higgs,
  Phys.\ Rev.\ Lett.\ {\bf 13} (1964) 508.

\bibitem{50073}
  P.~W.~Higgs,
  Phys.\ Rev.\ {\bf 145} (1966) 1156.

\bibitem{12291}
  F.~Englert and R.~Brout,
  Phys.\ Rev.\ Lett.\ {\bf 13} (1964) 321.

\bibitem{12292}
  G.~S.~Guralnik, C.~R.~Hagen, and T.~W.~B.~Kibble,
  Phys.\ Rev.\ Lett.\ {\bf 13} (1964) 585.

\bibitem{51165}
  T.~W.~B.~Kibble,
  Phys.\ Rev.\ {\bf 155} (1967) 1554.

\bibitem{Langacker:1980js}
 P.~Langacker,
 Phys.\ Rep.\  {\bf 72} (1981) 185.

\bibitem{Nakamura:2010zzi}
 K.~Nakamura {\it et al.}  [Particle Data Group],
 J.\ Phys.\ G {\bf 37} (2010) 075021.

\bibitem{Chanowitz:1985ug}
 M.~S.~Chanowitz and M.~Golden,
 Phys.\ Lett.\  B {\bf 165} (1985) 105.

\bibitem{Lee:1973iz}
 T.~D.~Lee,
 Phys.\ Rev.\  D {\bf 8} (1973) 1226.

\bibitem{Haber:1984rc}
 H.~E.~Haber and G.~L.~Kane,
 Phys.\ Rep.\  {\bf 117} (1985) 75.

\bibitem{Kim:1986ax}
 J.~E.~Kim,
 Phys.\ Rep.\  {\bf 150} (1987) 1.

\bibitem{Peccei:1977hh}
 R.~D.~Peccei and H.~R.~Quinn,
 Phys.\ Rev.\ Lett.\  {\bf 38} (1977) 1440.

\bibitem{Trodden:1998qg}
 M.~Trodden,
 hep-ph/9805252.

 \bibitem{Turok:1990zg}
  N.~Turok and J.~Zadrozny,
  Nucl.\ Phys.\ B {\bf 358} (1991) 471.

\bibitem{Joyce:1994zt}
  M.~Joyce, T.~Prokopec, and N.~Turok,
  Phys.\ Rev.\ D {\bf 53} (1996) 2958
  [hep-ph/9410282].

\bibitem{Funakubo:1993jg}
  K.~Funakubo, A.~Kakuto, and K.~Takenaga,
  Prog.\ Theor.\ Phys.\ {\bf 91} (1994) 341
  [hep-ph/9310267].

\bibitem{Davies:1994id}
  A.~T.~Davies, C.~D.~Froggatt, G.~Jenkins, and R.~G.~Moorhouse,
  Phys.\ Lett.\ B {\bf 336} (1994) 464.

\bibitem{Cline:1995dg}
  J.~M.~Cline, K.~Kainulainen, and A.~P.~Vischer,
  Phys.\ Rev.\ D {\bf 54} (1996) 2451
  [hep-ph/9506284].

\bibitem{Cline:1996mga}
  J.~M.~Cline and P.-A.~Lemieux,
  Phys.\ Rev.\ D {\bf 55} (1997) 3873
  [hep-ph/9609240].

\bibitem{Laine:2000rm}
  M.~Laine and K.~Rummukainen,
  Nucl.\ Phys.\ B {\bf 597} (2001) 23
  [hep-lat/0009025].

 \bibitem{Fromme:2006cm}
  L.~Fromme, S.~J.~Huber, and M.~Seniuch,
  JHEP {\bf 0611} (2006) 038
  [hep-ph/0605242].

\bibitem{arXiv:1106.0790}
  A.~Kozhushko and V.~Skalozub,
  Ukr.\ J.\ Phys.\ {\bf 56} (2011) 431
  [arXiv:1106.0790 [hep-ph]].

\bibitem{arXiv:1107.3559}
  J.~M.~Cline, K.~Kainulainen, and M.~Trott,
  JHEP {\bf 1111} (2011) 089
  [arXiv:1107.3559 [hep-ph]].


\bibitem{Djouadi:2005gj}
 A.~Djouadi,
 Phys.\ Rep.\  {\bf 459} (2008) 1
 [hep-ph/0503173].

\bibitem{McWilliams:1980kj}
 B.~McWilliams and L.~F.~Li,
 Nucl.\ Phys.\  B {\bf 179} (1981) 62.

\bibitem{Shanker:1981mj}
 O.~U.~Shanker,
 Nucl.\ Phys.\  B {\bf 206} (1982) 253.

\bibitem{Glashow:1976nt}
 S.~L.~Glashow and S.~Weinberg,
 Phys.\ Rev.\  D {\bf 15} (1977) 1958.

\bibitem{Paschos:1976ay}
 E.~A.~Paschos,
 Phys.\ Rev.\  D {\bf 15} (1977) 1966.

\bibitem{Carena:2002es}
 M.~S.~Carena and H.~E.~Haber,
 Prog.\ Part.\ Nucl.\ Phys.\  {\bf 50} (2003) 63
 [hep-ph/0208209].

 \bibitem{Barger:1989fj}
 V.~D.~Barger, J.~L.~Hewett, and R.~J.~N.~Phillips,
 Phys.\ Rev.\  D {\bf 41} (1990) 3421.

\bibitem{Grossman:1994jb}
 Y.~Grossman,
 Nucl.\ Phys.\  B {\bf 426} (1994) 355
 [hep-ph/9401311].

\bibitem{Akeroyd:1994ga}
 A.~G.~Akeroyd and W.~J.~Stirling,
 Nucl.\ Phys.\  B {\bf 447} (1995) 3.

\bibitem{Akeroyd:1996di}
 A.~G.~Akeroyd,
 Phys.\ Lett.\  B {\bf 377} (1996) 95
 [hep-ph/9603445].

\bibitem{Barnett:1983mm}
 R.~M.~Barnett, G.~Senjanovi\'c, L.~Wolfenstein, and D.~Wyler,
 Phys.\ Lett.\  B {\bf 136} (1984) 191.

\bibitem{Barnett:1984zy}
 R.~M.~Barnett, G.~Senjanovi\'c, and D.~Wyler,
 Phys.\ Rev.\  D {\bf 30} (1984) 1529.

\bibitem{Aoki:2009ha}
 M.~Aoki, S.~Kanemura, K.~Tsumura, and K.~Yagyu,
 Phys.\ Rev.\  D {\bf 80} (2009) 015017
 [arXiv:0902.4665 [hep-ph]].

\bibitem{Pich:2009sp}
 A.~Pich and P.~Tuz\'on,
 Phys.\ Rev.\  D {\bf 80} (2009) 091702
 [arXiv:0908.1554 [hep-ph]].

\bibitem{Tuzon:2010vt}
 P.~Tuz\'on and A.~Pich,
 Acta Phys.\ Polon.\ Supp.\  {\bf 3} (2010) 215
 [arXiv:1001.0293 [hep-ph]].

\bibitem{Ferreira:2010xe}
 P.~M.~Ferreira, L.~Lavoura, and J.~P.~Silva,
 Phys.\ Lett.\  B {\bf 688} (2010) 341
 [arXiv:1001.2561 [hep-ph]].

\bibitem{arXiv:1111.5760}
  J.~Bijnens, J.~Lu, and J.~Rathsman,
  arXiv:1111.5760 [hep-ph].

\bibitem{Serodio:2011hg}
  H.~Ser\^odio,
  Phys.\ Lett.\ B {\bf 700} (2011) 133
  [arXiv:1104.2545 [hep-ph]].

\bibitem{arXiv:1104.2601}
  I.~de~M.~Varzielas,
  Phys.\ Lett.\ B {\bf 701} (2011) 597
  [arXiv:1104.2601 [hep-ph]].

\bibitem{DiazCruz:2010yq}
 J.~L.~D\'\i az-Cruz, A.~D\'\i az-Furlong, and J.~H.~Montes de Oca,
 arXiv:1010.0950 [hep-ph].

\bibitem{arXiv:1106.4039}
  G.~Cree and H.~E.~Logan,
  Phys.\ Rev.\ D {\bf 84} (2011) 055021
  [arXiv:1106.4039 [hep-ph]].

\bibitem{Braeuninger:2010td}
 C.~B.~Braeuninger, A.~Ibarra, and C.~Simonetto,
 Phys.\ Lett.\  B {\bf 692} (2010) 189
 [arXiv:1005.5706 [hep-ph]].

\bibitem{Mahmoudi:2009zx}
 F.~Mahmoudi and O.~Stal,
 Phys.\ Rev.\  D {\bf 81} (2010) 035016
 [arXiv:0907.1791 [hep-ph]].


\bibitem{Gunion:1989we}
 J.~F.~Gunion, H.~E.~Haber, G.~L.~Kane, and S.~Dawson,
 Front.\ Phys.\  {\bf 80} (2000) 1.

\bibitem{Akeroyd:1998ui}
 A.~G.~Akeroyd,
 J.\ Phys.\ G {\bf 24} (1998) 1983
 [hep-ph/9803324].

\bibitem{Akeroyd:1998sv}
 A.~G.~Akeroyd, M.~A.~D\'\i az, and J.~W.~F.~Valle,
 Phys.\ Lett.\  B {\bf 441} (1998) 224
 [hep-ph/9806382].

\bibitem{Barger:2009me}
 V.~Barger, H.~E.~Logan, and G.~Shaughnessy,
 Phys.\ Rev.\  D {\bf 79} (2009) 115018
 [arXiv:0902.0170 [hep-ph]].

\bibitem{Arhrib:2009hc}
 A.~Arhrib, R.~Benbrik, C.~H.~Chen, R.~Guedes, and R.~Santos,
 JHEP {\bf 0908} (2009) 035
 [arXiv:0906.0387 [hep-ph]].

\bibitem{Eriksson:2009ws}
 D.~Eriksson, J.~Rathsman, and O.~Stal,
 Comput.\ Phys.\ Commun.\  {\bf 181} (2010) 189
 [arXiv:0902.0851 [hep-ph]].

\bibitem{Logan:2010ag}
 H.~E.~Logan and D.~MacLennan,
 Phys.\ Rev.\  D {\bf 81} (2010) 075016
 [arXiv:1002.4916 [hep-ph]].


\bibitem{Su:2009fz}
 S.~Su and B.~Thomas,
 Phys.\ Rev.\  D {\bf 79} (2009) 095014
 [arXiv:0903.0667 [hep-ph]].

\bibitem{Logan:2009uf}
 H.~E.~Logan and D.~MacLennan,
 Phys.\ Rev.\  D {\bf 79} (2009) 115022
 [arXiv:0903.2246 [hep-ph]].

\bibitem{Goh:2009wg}
 H.~S.~Goh, L.~J.~Hall, and P.~Kumar,
 JHEP {\bf 0905} (2009) 097
 [arXiv:0902.0814 [hep-ph]].

\bibitem{arXiv:1106.3368}
  M.~S.~Boucenna and S.~Profumo,
  Phys.\ Rev.\ D {\bf 84} (2011) 055011
  [arXiv:1106.3368 [hep-ph]].

\bibitem{Cao:2009as}
 J.~Cao, P.~Wan, L.~Wu, and J.~M.~Yang,
 Phys.\ Rev.\  D {\bf 80} (2009) 071701
 [arXiv:0909.5148 [hep-ph]].

\bibitem{Jegerlehner:2009ry}
 F.~Jegerlehner and A.~Nyffeler,
 Phys.\ Rep.\  {\bf 477} (2009) 1
 [arXiv:0902.3360 [hep-ph]].

\bibitem{Aoki:2008av}
 M.~Aoki, S.~Kanemura, and O.~Seto,
 Phys.\ Rev.\ Lett.\  {\bf 102} (2009) 051805
 [arXiv:0807.0361 [hep-ph]].

\bibitem{Haber:1978jt}
 H.~E.~Haber, G.~L.~Kane, and T.~Sterling,
 Nucl.\ Phys.\  B {\bf 161} (1979) 493.

\bibitem{Ambroso:2008kb}
 M.~Ambroso, V.~Braun, and B.~A.~Ovrut,
 JHEP {\bf 0810} (2008) 046
 [arXiv:0807.3319 [hep-th]].

\bibitem{Stange:1993ya}
 A.~Stange, W.~J.~Marciano, and S.~Willenbrock,
 Phys.\ Rev.\  D {\bf 49} (1994) 1354
 [hep-ph/9309294].

\bibitem{Diaz:1994pk}
 M.~A.~D\'\i az and T.~J.~Weiler,
 hep-ph/9401259.

 \bibitem{hep-ph/9211234}
  V.~D.~Barger, N.~G.~Deshpande, J.~L.~Hewett, and T.~G.~Rizzo,
  In {\it Argonne 1993, Physics at current accelerators and supercolliders}
  [hep-ph/9211234].

\bibitem{Akeroyd:1995hg}
 A.~G.~Akeroyd,
 Phys.\ Lett.\  B {\bf 368} (1996) 89
 [hep-ph/9511347].

\bibitem{Brucher:1999tx}
 L.~Br\"ucher and R.~Santos,
 Eur.\ Phys.\ J.\  C {\bf 12} (2000) 87
 [hep-ph/9907434].

\bibitem{Brucher:2000qc}
 L.~Br\"ucher and R.~Santos,
 hep-ph/0002027.

\bibitem{Barroso:1999bf}
 A.~Barroso, L.~Br\"ucher, and R.~Santos,
 Phys.\ Rev.\  D {\bf 60} (1999) 035005
 [hep-ph/9901293].

\bibitem{Pois:1993ay}
 H.~Pois, T.~J.~Weiler, and T.~C.~Yuan,
 Phys.\ Rev.\  D {\bf 47} (1993) 3886
 [hep-ph/9303277].

\bibitem{Akeroyd:1998dt}
 A.~G.~Akeroyd,
 Nucl.\ Phys.\  B {\bf 544} (1999) 557
 [hep-ph/9806337].

\bibitem{Akeroyd:1998uw}
 A.~G.~Akeroyd, A.~Arhrib, and E.~M.~Naimi,
 Eur.\ Phys.\ J.\  C {\bf 12} (2000) 451
 [Erratum {\it ibid.}\ {\bf 14} (2000) 371]
 [hep-ph/9811431].

\bibitem{Akeroyd:2003xi}
 A.~G.~Akeroyd, M.~A.~D\'\i az, and F.~J.~Pacheco,
 Phys.\ Rev.\  D {\bf 70} (2004) 075002
 [hep-ph/0312231].

\bibitem{Akeroyd:2003bt}
 A.~G.~Akeroyd and M.~A.~D\'\i az,
 Phys.\ Rev.\  D {\bf 67} (2003) 095007
 [hep-ph/0301203].

\bibitem{Dedes:2001nx}
 A.~Dedes and H.~E.~Haber,
 JHEP {\bf 0105} (2001) 006
 [hep-ph/0102297].

\bibitem{Krawczyk:1996sm}
 M.~Krawczyk and J.~Zochowski,
 Phys.\ Rev.\  D {\bf 55} (1997) 6968
 [hep-ph/9608321].

\bibitem{Akeroyd:2001in}
 A.~G.~Akeroyd and S.~Baek,
 Phys.\ Lett.\  B {\bf 525} (2002) 315
 [hep-ph/0105228].

\bibitem{Litsey:2009rp}
 S.~Litsey and M.~Sher,
 Phys.\ Rev.\  D {\bf 80} (2009) 057701
 [arXiv:0908.0502 [hep-ph]].

\bibitem{Ciuchini:1997xe}
 M.~Ciuchini, G.~Degrassi, P.~Gambino, and G.~F.~Giudice,
 Nucl.\ Phys.\  B {\bf 527} (1998) 21
 [hep-ph/9710335].

\bibitem{Borzumati:1998tg}
 F.~Borzumati and C.~Greub,
 Phys.\ Rev.\  D {\bf 58} (1998) 074004
 [hep-ph/9802391].

\bibitem{Misiak:2006zs}
 M.~Misiak {\it et al.},
 Phys.\ Rev.\ Lett.\  {\bf 98} (2007) 022002
 [hep-ph/0609232].

\bibitem{Krawczyk:2007ne}
 M.~Krawczyk and D.~Sokolowska,
in {\it Proceedings of 2007 International Linear Collider Workshop (LCWS07
and ILC07), Hamburg, Germany, 30 May--3 Jun 2007}
 [arXiv:0711.4900 [hep-ph]].

\bibitem{Berger:2003sm}
 E.~L.~Berger, T.~Han, J.~Jiang, and T.~Plehn,
 Phys.\ Rev.\  D {\bf 71} (2005) 115012
 [hep-ph/0312286].

\bibitem{Alwall:2004xw}
 J.~Alwall and J.~Rathsman,
 JHEP {\bf 0412} (2004) 050
 [hep-ph/0409094].

\bibitem{Plehn:2002vy}
 T.~Plehn,
 Phys.\ Rev.\  D {\bf 67} (2003) 014018
 [hep-ph/0206121].

\bibitem{Asakawa:2005nx}
 E.~Asakawa, O.~Brein, and S.~Kanemura,
 Phys.\ Rev.\  D {\bf 72} (2005) 055017
 [hep-ph/0506249].

\bibitem{Krawczyk:2004na}
 M.~Krawczyk and D.~Temes,
 Eur.\ Phys.\ J.\  C {\bf 44} (2005) 435
 [hep-ph/0410248].

\bibitem{WahabElKaffas:2007xd}
 A.~W.~El Kaffas, P.~Osland, and O.~M.~Ogreid,
 Phys.\ Rev.\  D {\bf 76} (2007) 095001
 [arXiv:0706.2997 [hep-ph]].

\bibitem{Ma:2000cc}
 E.~Ma,
 Phys.\ Rev.\ Lett.\  {\bf 86} (2001) 2502
 [hep-ph/0011121].

\bibitem{Gabriel:2006ns}
 S.~Gabriel and S.~Nandi,
 Phys.\ Lett.\  B {\bf 655} (2007) 141
 [hep-ph/0610253].

\bibitem{Davidson:2009ha}
 S.~M.~Davidson and H.~E.~Logan,
 Phys.\ Rev.\  D {\bf 80} (2009) 095008
 [arXiv:0906.3335 [hep-ph]].

\bibitem{arXiv:1107.1026}
  T.~Morozumi, H.~Takata, and K.~Tamai,
  arXiv:1107.1026 [hep-ph].

\bibitem{Djouadi:2005gi}
 A.~Djouadi,
 Phys.\ Rep.\  {\bf 457} (2008) 1
 [hep-ph/0503172].

\bibitem{Barger:1992ac}
 V.~D.~Barger, M.~S.~Berger, and P.~Ohmann,
 Phys.\ Rev.\  D {\bf 47} (1993) 1093
 [hep-ph/9209232].

\bibitem{Carena:1994bv}
 M.~S.~Carena, M.~Olechowski, S.~Pokorski, and C.~E.~M.~Wagner,
 Nucl.\ Phys.\  B {\bf 426} (1994) 269
 [hep-ph/9402253].

\bibitem{Kanemura:1999tg}
 S.~Kanemura,
 Eur.\ Phys.\ J.\  C {\bf 17} (2000) 473
 [hep-ph/9911541].

\bibitem{Kanemura:1999xf}
 S.~Kanemura, T.~Kasai, and Y.~Okada,
 Phys.\ Lett.\  B {\bf 471} (1999) 182
 [hep-ph/9903289].

\bibitem{Akeroyd:2000wc}
 A.~G.~Akeroyd, A.~Arhrib, and E.~M.~Naimi,
 Phys.\ Lett.\  B {\bf 490} (2000) 119
 [hep-ph/0006035].

\bibitem{Arhrib:2000is}
 A.~Arhrib,
 hep-ph/0012353.

\bibitem{Gunion:2002zf}
 J.~F.~Gunion and H.~E.~Haber,
 Phys.\ Rev.\  D {\bf 67} (2003) 075019
 [hep-ph/0207010].

\bibitem{Mantry:2007ar}
 S.~Mantry, M.~Trott, and M.~B.~Wise,
 Phys.\ Rev.\  D {\bf 77} (2008) 013006
 [arXiv:0709.1505 [hep-ph]].

\bibitem{Randall:2007as}
 L.~Randall,
 JHEP {\bf 0802} (2008) 084
 [arXiv:0711.4360 [hep-ph]].

\bibitem{Amsler:2008zzb}
 C.~Amsler {\it et al.}  [Particle Data Group],
 Phys.\ Lett.\  B {\bf 667} (2008) 1.

 \bibitem{arXiv:1107.0975}
  M.~Baak, M.~Goebel, J.~Haller, A.~Hoecker, D.~Ludwig, K.~Moenig, M.~Schott, and J.~Stelzer,
  arXiv:1107.0975 [hep-ph].

\bibitem{Schael:2006cr}
 S.~Schael {\it et al.}  [ALEPH, DELPHI, L3, and OPAL Collaborations],
 Eur.\ Phys.\ J.\  C {\bf 47} (2006) 547
 [hep-ex/0602042].

 \bibitem{hep-ph/9506291}
  J.~Kalinowski and M.~Krawczyk,
  Phys.\ Lett.\ B {\bf 361} (1995) 66
  [hep-ph/9506291].

\bibitem{Gupta:2009wn}
 R.~S.~Gupta and J.~D.~Wells,
 Phys.\ Rev.\  D {\bf 81} (2010) 055012
 [arXiv:0912.0267 [hep-ph]].

\bibitem{Posch:2010hx}
 P.~Posch,
 Phys.\ Lett.\  B {\bf 696} (2011) 447
 [arXiv:1001.1759 [hep-ph]].

\bibitem{Barate:2003sz}
 R.~Barate {\it et al.}  [LEP Working Group for Higgs Boson Searches],
 Phys.\ Lett.\  B {\bf 565} (2003) 61
 [hep-ex/0306033].

\bibitem{Aaltonen:2009ga}
  T.~Aaltonen {\it et al.} [CDF Collaboration],
  Phys.\ Rev.\ Lett.\  {\bf 103} (2009)  061803
  [arXiv:0905.0413 [hep-ex]].

\bibitem{:2008it}
  V.~M.~Abazov {\it et al.} [D0 Collaboration],
  Phys.\ Rev.\ Lett.\  {\bf 101} (2008)  051801
  [arXiv:0803.1514 [hep-ex]].


\bibitem{Landsberg:2007mc}
  G.~L.~Landsberg [CDF and D0 Collaborations],
  arXiv:0705.2855 [hep-ex].

\bibitem{Akeroyd:2007yh}
 A.~G.~Akeroyd, M.~A.~D\'\i az, and M.~A.~Rivera,
 Phys.\ Rev.\  D {\bf 76} (2007) 115012
 [arXiv:0708.1939 [hep-ph]].

\bibitem{Abdallah:2004wy}
 J.~Abdallah {\it et al.}  [DELPHI Collaboration],
 Eur.\ Phys.\ J.\  C {\bf 38} (2004) 1
 [hep-ex/0410017].

\bibitem{ahmes}
Ahmes, the Rhind Papyrus, 1900 B.C.

\bibitem{conway}
J. Conway, private communication.

\bibitem{Dermisek:2005ar}
 R.~Dermisek and J.~F.~Gunion,
 Phys.\ Rev.\ Lett.\  {\bf 95} (2005) 041801
 [hep-ph/0502105].

\bibitem{Dermisek:2005gg}
 R.~Dermisek and J.~F.~Gunion,
 Phys.\ Rev.\  D {\bf 73} (2006) 111701
 [hep-ph/0510322].

\bibitem{Dermisek:2006wr}
 R.~Dermisek and J.~F.~Gunion,
 Phys.\ Rev.\  D {\bf 75} (2007) 075019
 [hep-ph/0611142].

\bibitem{Dermisek:2006py}
 R.~Dermisek, J.~F.~Gunion, and B.~McElrath,
 Phys.\ Rev.\  D {\bf 76} (2007) 051105
 [hep-ph/0612031].

\bibitem{Chang:2008cw}
 S.~Chang, R.~Dermisek, J.~F.~Gunion, and N.~Weiner,
 Ann.\ Rev.\ Nucl.\ Part.\ Sci.\  {\bf 58} (2008) 75
 [arXiv:0801.4554 [hep-ph]].

\bibitem{Abbiendi:2002qp}
  G.~Abbiendi {\it et al.} [OPAL Collaboration],
  Eur.\ Phys.\ J.\ C {\bf 27} (2003) 311
  [hep-ex/0206022].

\bibitem{Schael:2010aw}
 S.~Schael {\it et al.}  [ALEPH Collaboration],
 JHEP {\bf 1005} (2010) 049
 [arXiv:1003.0705 [hep-ex]].

\bibitem{Dermisek:2008uu}
 R.~Dermisek and J.~F.~Gunion,
 Phys.\ Rev.\  D {\bf 79} (2009) 055014
 [arXiv:0811.3537 [hep-ph]].

\bibitem{Kao:2003jw}
 C.~Kao, G.~Lovelace, and L.~H.~Orr,
 Phys.\ Lett.\  B {\bf 567} (2003) 259
 [hep-ph/0305028].

\bibitem{Kominis:1994fa}
 D.~Kominis,
 Nucl.\ Phys.\  B {\bf 427} (1994) 575
 [hep-ph/9402339].

\bibitem{Baer:1992uu}
 H.~Baer, C.~Kao, and X.~Tata,
 Phys.\ Lett.\  B {\bf 303} (1993) 284.

\bibitem{Abdullin:1996as}
 S.~Abdullin, H.~Baer, C.~Kao, N.~Stepanov, and X.~Tata,
 Phys.\ Rev.\  D {\bf 54} (1996) 6728
 [hep-ph/9603433].

 \bibitem{arXiv:1105.6095}
  S.~Bar-Shalom, S.~Nandi, and A.~Soni,
  Phys.\ Rev.\ D {\bf 84} (2011) 053009
  [arXiv:1105.6095 [hep-ph]].

\bibitem{ElKaffas:2007rq}
 A.~W.~El Kaffas, P.~Osland, and O.~M.~Ogreid,
 Nonlin.\ Phenom.\ Complex Syst.\  {\bf 10} (2007) 347
 [hep-ph/0702097].

 \bibitem{EFI-89-66}
  M.~A.~Luty,
  Phys.\ Rev.\ D {\bf 41} (1990) 2893.

\bibitem{arXiv:1109.3914}
  G.~Burdman and C.~E.~F.~Haluch,
  arXiv:1109.3914 [hep-ph].

\bibitem{Belyaev:2009zd}
 A.~Belyaev, R.~Guedes, S.~Moretti, and R.~Santos,
 JHEP {\bf 1007} (2010) 051
 [arXiv:0912.2620 [hep-ph]].

\bibitem{Belyaev:2010bc}
 A.~Belyaev, R.~Guedes, S.~Moretti, and R.~Santos,
 PoS D {\bf IS2010} (2010) 207
 [arXiv:1006.0144 [hep-ph]].

\bibitem{arXiv:1111.6089}
  S.~Kanemura, K.~Tsumura, and H.~Yokoya,
  arXiv:1111.6089 [hep-ph].

\bibitem{Plehn:2001qg}
 T.~Plehn and D.~L.~Rainwater,
 Phys.\ Lett.\  B {\bf 520} (2001) 108
 [hep-ph/0107180].

\bibitem{Han:2002gp}
 T.~Han and B.~McElrath,
 Phys.\ Lett.\  B {\bf 528} (2002) 81
 [hep-ph/0201023].

\bibitem{Su:2008bj}
 S.~Su and B.~Thomas,
 Phys.\ Lett.\  B {\bf 677} (2009) 296
 [arXiv:0812.1798 [hep-ph]].

\bibitem{Wang:2006jy}
 F.~Wang, W.~Wang, and J.~M.~Yang,
 Europhys.\ Lett.\  {\bf 76} (2006) 388
 [hep-ph/0601018].

\bibitem{Sher:2011mx}
 M.~Sher and C.~Triola,
 Phys.\ Rev.\ D {\bf 83} (2011) 117702
 [arXiv:1105.4844 [hep-ph]].

\bibitem{Zhou:2011rc}
 S.~Zhou,
 Phys. \ Rev.\ D {\bf 84} (2011) 038701
 [arXiv:1106.3880 [hep-ph]].

 \bibitem{Ma:2001mr}
 E.~Ma and M.~Raidal,
 Phys.\ Rev.\ Lett.\  {\bf 87} (2001) 011802
 [Erratum {\it ibid.}\  {\bf 87} (2001) 159901]
 [hep-ph/0102255].

\bibitem{Haba:2010zi}
 N.~Haba and M.~Hirotsu,
 Eur.\ Phys.\ J.\  C {\bf 69} (2010) 481
 [arXiv:1005.1372 [hep-ph]].

 \bibitem{Haba:2011fn}
 N.~Haba and T.~Horita,
 Phys.\ Lett.\ B {\bf 705} (2011) 98
 [arXiv:1107.3203 [hep-ph]].

\bibitem{Haba:2011nb}
 N.~Haba and K.~Tsumura,
 JHEP {\bf 1106} (2011) 068
 [arXiv:1105.1409 [hep-ph]].

\bibitem{Haba:2011ra}
 N.~Haba and O.~Seto,
 Prog.\  Theor.\  Phys.\ {\bf 125} (2011) 1155
 [arXiv:1102.2889 [hep-ph]].

\bibitem{Haba:2011yc}
 N.~Haba and O.~Seto,
 arXiv:1106.5354 [hep-ph].

\bibitem{Schaarschmidt:2007zz}
 J.~Schaarschmidt,
 PhD thesis (University of Dresden, 2007).

\bibitem{Aad:2008zzm}
 G.~Aad {\it et al.}  [ATLAS Collaboration],
 JINST {\bf 3} (2008) S08003.

\bibitem{Schaarschmidt:2008zz}
 J.~Schaarschmidt  [ATLAS Collaboration],
 AIP Conf.\ Proc.\  {\bf 1078} (2009) 220.

\bibitem{Plehn:1996wb}
 T.~Plehn, M.~Spira, and P.~M.~Zerwas,
 Nucl.\ Phys.\  B {\bf 479} (1996) 46
 [Erratum {\it ibid.}\ {\bf 531} (1998) 655]
 [hep-ph/9603205].

\bibitem{Baur:2002qd}
 U.~Baur, T.~Plehn, and D.~L.~Rainwater,
 Phys.\ Rev.\  D {\bf 67} (2003) 033003
 [hep-ph/0211224].

\bibitem{Baur:2003gp}
 U.~Baur, T.~Plehn, and D.~L.~Rainwater,
 Phys.\ Rev.\  D {\bf 69} (2004) 053004
 [hep-ph/0310056].

\bibitem{Moretti:2004dg}
 M.~Moretti, S.~Moretti, F.~Piccinini, R.~Pittau, and A.~D.~Polosa,
 hep-ph/0411039.

\bibitem{Djouadi:1999rca}
 A.~Djouadi, W.~Kilian, M.~M\"uhlleitner, and P.~M.~Zerwas,
 Eur.\ Phys.\ J.\  C {\bf 10} (1999) 45
 [hep-ph/9904287].

\bibitem{Dawson:1998py}
 S.~Dawson, S.~Dittmaier, and M.~Spira,
 Phys.\ Rev.\  D {\bf 58} (1998) 115012
 [hep-ph/9805244].

 \bibitem{Moretti:2004wa}
 M.~Moretti, S.~Moretti, F.~Piccinini, R.~Pittau, and A.~D.~Polosa,
 JHEP {\bf 0502} (2005) 024
 [hep-ph/0410334].

\bibitem{Moretti:2007ca}
 M.~Moretti, S.~Moretti, F.~Piccinini, R.~Pittau, and J.~Rathsman,
 JHEP {\bf 0712} (2007) 075
 [arXiv:0706.4117 [hep-ph]].

\bibitem{Moretti:2010kc}
 M.~Moretti, S.~Moretti, F.~Piccinini, R.~Pittau, and J.~Rathsman,
 JHEP {\bf 1011} (2010) 097
 [arXiv:1008.0820 [hep-ph]].

\bibitem{Deshpande:1977rw}
 N.~G.~Deshpande and E.~Ma,
 Phys.\ Rev.\  D {\bf 18} (1978) 2574.

\bibitem{Ma:2006km}
 E.~Ma,
 Phys.\ Rev.\  D {\bf 73} (2006) 077301
 [hep-ph/0601225].

\bibitem{Barbieri:2006dq}
 R.~Barbieri, L.~J.~Hall, and V.~S.~Rychkov,
 Phys.\ Rev.\  D {\bf 74} (2006) 015007
 [hep-ph/0603188].

\bibitem{Barbieri:2000gf}
 R.~Barbieri and A.~Strumia,
 hep-ph/0007265.

 \bibitem{Martinez:2011ua}
  H.~Mart\'\i nez, A.~Melfo, F.~Nesti, and G.~Senjanovi\'c,
  Phys.\ Rev.\ Lett.\ {\bf 106} (2011) 191802
  [arXiv:1101.3796 [hep-ph]].

\bibitem{Ma:2006fn}
 E.~Ma,
 Mod.\ Phys.\ Lett.\  A {\bf 21} (2006) 1777
 [hep-ph/0605180].

\bibitem{Majumdar:2006nt}
 D.~Majumdar and A.~Ghosal,
 Mod.\ Phys.\ Lett.\  A {\bf 23} (2008) 2011
 [hep-ph/0607067].

\bibitem{LopezHonorez:2006gr}
  L.~L.~Honorez, E.~Nezri, J.~F.~Oliver, and M.~H.~G.~Tytgat,
  JCAP {\bf 0702} (2007) 028
  [hep-ph/0612275].

\bibitem{Sahu:2007uh}
 N.~Sahu and U.~Sarkar,
 Phys.\ Rev.\  D {\bf 76} (2007) 045014
 [hep-ph/0701062].

\bibitem{Gustafsson:2007pc}
 M.~Gustafsson, E.~Lundstrom, L.~Bergstrom, and J.~Edsj\"o,
 Phys.\ Rev.\ Lett.\  {\bf 99} (2007) 041301
 [arXiv:astro-ph/0703512].

\bibitem{Lisanti:2007ec}
 M.~Lisanti and J.~G.~Wacker,
 arXiv:0704.2816 [hep-ph].

\bibitem{Hambye:2009pw}
 T.~Hambye, F.~S.~Ling, L.~L.~Honorez, and J.~Rocher,
 JHEP {\bf 0907} (2009) 090
 [Erratum {\it ibid.}\  {\bf 1005} (2010) 066]
 [arXiv:0903.4010 [hep-ph]].

 \bibitem{Melfo:2011ie}
  A.~Melfo, M.~Nemevsek, F.~Nesti, G.~Senjanovi\'c, and Y.~Zhang,
  arXiv:1105.4611 [hep-ph].

\bibitem{arXiv:1009.4593}
  I.~F.~Ginzburg, K.~A.~Kanishev, M.~Krawczyk, and D.~Sokolowska,
  Phys.\ Rev.\ D {\bf 82} (2010) 123533
  [arXiv:1009.4593 [hep-ph]].

\bibitem{Akerib:2005kh}
 D.~S.~Akerib {\it et al.}  [CDMS Collaboration],
 Phys.\ Rev.\ Lett.\  {\bf 96} (2006) 011302
 [arXiv:astro-ph/0509259].

\bibitem{Cao:2007rm}
 Q.~H.~Cao, E.~Ma, and G.~Rajasekaran,
 Phys.\ Rev.\  D {\bf 76} (2007) 095011
 [arXiv:0708.2939 [hep-ph]].

\bibitem{Lundstrom:2008ai}
 E.~Lundstrom, M.~Gustafsson, and J.~Edsj\"o,
 Phys.\ Rev.\  D {\bf 79} (2009) 035013
 [arXiv:0810.3924 [hep-ph]].

\bibitem{Eboli:2000ze}
 O.~J.~P.~Eboli and D.~Zeppenfeld,
 Phys.\ Lett.\  B {\bf 495} (2000) 147
 [hep-ph/0009158].

\bibitem{Grzadkowski:2009bt}
 B.~Grzadkowski, O.~M.~Ogreid, and P.~Osland,
 Phys.\ Rev.\  D {\bf 80} (2009) 055013
 [arXiv:0904.2173 [hep-ph]].

\bibitem{Grzadkowski:2010au}
 B.~Grzadkowski, O.~M.~Ogreid, P.~Osland, A.~Pukhov, and M.~Purmohammadi,
 arXiv:1012.4680 [hep-ph].

\bibitem{arXiv:1111.0963}
  S.~M.~Barr and T.~W.~Kephart,
  arXiv:1111.0963 [hep-ph].

\bibitem{Lee:1969fy}
 T.~D.~Lee and G.~C.~Wick,
 Nucl.\ Phys.\  B {\bf 9} (1969) 209.

\bibitem{Lee:1970iw}
 T.~D.~Lee and G.~C.~Wick,
 Phys.\ Rev.\  D {\bf 2} (1970) 1033.

\bibitem{Grinstein:2007mp}
 B.~Grinstein, D.~O'Connell, and M.~B.~Wise,
 Phys.\ Rev.\  D {\bf 77} (2008) 025012
 [arXiv:0704.1845 [hep-ph]].

\bibitem{Grinstein:2008bg}
 B.~Grinstein, D.~O'Connell, and M.~B.~Wise,
 Phys.\ Rev.\  D {\bf 79} (2009) 105019
 [arXiv:0805.2156 [hep-th]].

\bibitem{Rizzo:2007ae}
 T.~G.~Rizzo,
 JHEP {\bf 0706} (2007) 070
 [arXiv:0704.3458 [hep-ph]].

\bibitem{Dulaney:2007dx}
 T.~R.~Dulaney and M.~B.~Wise,
 Phys.\ Lett.\  B {\bf 658} (2008) 230
 [arXiv:0708.0567 [hep-ph]].

\bibitem{Alvarez:2008za}
 E.~Alvarez, L.~Da Rold, C.~Schat, and A.~Szynkman,
 JHEP {\bf 0804} (2008) 026
 [arXiv:0802.1061 [hep-ph]].

\bibitem{Underwood:2008cr}
 T.~E.~J.~Underwood and R.~Zwicky,
 Phys.\ Rev.\  D {\bf 79} (2009) 035016
 [arXiv:0805.3296 [hep-ph]].

\bibitem{Carone:2008bs}
 C.~D.~Carone and R.~F.~Lebed,
 Phys.\ Lett.\  B {\bf 668} (2008) 221
 [arXiv:0806.4555 [hep-ph]].

\bibitem{Carone:2008iw}
 C.~D.~Carone and R.~F.~Lebed,
 JHEP {\bf 0901} (2009) 043
 [arXiv:0811.4150 [hep-ph]].

\bibitem{Grinstein:2008qq}
 B.~Grinstein and D.~O'Connell,
 Phys.\ Rev.\  D {\bf 78} (2008) 105005
 [arXiv:0801.4034 [hep-ph]].

\bibitem{Carone:2009it}
 C.~D.~Carone,
 Phys.\ Lett.\  B {\bf 677} (2009) 306
 [arXiv:0904.2359 [hep-ph]].

\bibitem{Fornal:2009xc}
 B.~Fornal, B.~Grinstein, and M.~B.~Wise,
 Phys.\ Lett.\  B {\bf 674} (2009) 330
 [arXiv:0902.1585 [hep-th]].

\bibitem{Shalaby:2009re}
 A.~M.~Shalaby,
 Phys.\ Rev.\  D {\bf 80} (2009) 025006
 [arXiv:0812.3419 [hep-th]].

\bibitem{Wu:2008rr}
 F.~Wu and M.~Zhong,
 Phys.\ Rev.\  D {\bf 78} (2008) 085010
 [arXiv:0807.0132 [hep-ph]].

\bibitem{Espinosa:2007ny}
 J.~R.~Espinosa, B.~Grinstein, D.~O'Connell, and M.~B.~Wise,
 Phys.\ Rev.\  D {\bf 77} (2008) 085002
 [arXiv:0705.1188 [hep-ph]].

\bibitem{Krauss:2007bz}
 F.~Krauss, T.~E.~J.~Underwood, and R.~Zwicky,
 Phys.\ Rev.\  D {\bf 77} (2008) 015012
 [Erratum {\it ibid.}\  {\bf 83} (2011) 019902]
 [arXiv:0709.4054 [hep-ph]].

\bibitem{Carone:2009nu}
 C.~D.~Carone and R.~Primulando,
 Phys.\ Rev.\  D {\bf 80} (2009) 055020
 [arXiv:0908.0342 [hep-ph]].

\bibitem{Chivukula:2010nw}
 R.~S.~Chivukula, A.~Farzinnia, R.~Foadi, and E.~H.~Simmons,
 Phys.\ Rev.\  D {\bf 81} (2010) 095015
 [arXiv:1002.0343 [hep-ph]].

\bibitem{Alvarez:2011ah}
 E.~Alvarez, E.~C.~Leskow, and J.~Zurita,
 Phys.\ Rev.\ D {\bf 83} (2011) 115024
 [arXiv:1104.3496 [hep-ph]].

\bibitem{Cao:2011yt}
 Q.~H.~Cao, M.~Carena, S.~Gori, A.~Menon, P.~Schwaller,
 C.~E.~M.~Wagner, and L.~T.~M.~Wang,
 JHEP {\bf 1108} (2011) 002
 [arXiv:1104.4776 [hep-ph]].

\bibitem{Aaltonen:2011mk}
 T.~Aaltonen {\it et al.}  [CDF Collaboration],
 Phys.\ Rev.\ Lett.\  {\bf 106} (2011) 171801
 [arXiv:1104.0699 [hep-ex]].


\bibitem{Chen:2011wp}
 C.~H.~Chen, C.~W.~Chiang, T.~Nomura, and Y.~Fusheng,
 arXiv:1105.2870 [hep-ph].

\bibitem{Dutta:2011kg}
 B.~Dutta, S.~Khalil, Y.~Mimura, and Q.~Shafi,
 arXiv:1104.5209 [hep-ph].

\bibitem{Bjorken:1977vt}
 J.~D.~Bjorken and S.~Weinberg,
 Phys.\ Rev.\ Lett.\  {\bf 38} (1977) 622.

\bibitem{Cheng:1987rs}
 T.~P.~Cheng and M.~Sher,
 Phys.\ Rev.\  D {\bf 35} (1987) 3484.

 \bibitem{DiazCruz:2004tr}
  J.~L.~D\'\i az-Cruz, R.~Noriega-Papaqui, and A.~Rosado,
  Phys.\ Rev.\ D {\bf 69} (2004) 095002
  [hep-ph/0401194].

\bibitem{Antaramian:1992ya}
 A.~Antaramian, L.~J.~Hall, and A.~Ra\v{s}in,
 Phys.\ Rev.\ Lett.\  {\bf 69} (1992) 1871
 [hep-ph/9206205].

\bibitem{Hou:1991un}
 W.~S.~Hou,
 Phys.\ Lett.\  B {\bf 296} (1992) 179.

\bibitem{Gronau:1988qt}
 M.~Gronau, R.~Johnson, and J.~Schechter,
 Phys.\ Rev.\  D {\bf 39} (1989) 1913.

\bibitem{Sher:1991km}
 M.~Sher and Y.~Yuan,
 Phys.\ Rev.\  D {\bf 44} (1991) 1461.

\bibitem{Chang:1993kw}
 D.~Chang, W.~S.~Hou, and W.~Y.~Keung,
 Phys.\ Rev.\  D {\bf 48} (1993) 217
 [hep-ph/9302267].

\bibitem{Luke:1993cy}
 M.~E.~Luke and M.~J.~Savage,
 Phys.\ Lett.\  B {\bf 307} (1993) 387
 [hep-ph/9303249].

\bibitem{Kosmas:1993ch}
 T.~S.~Kosmas, G.~K.~Leontaris, and J.~D.~Vergados,
 Prog.\ Part.\ Nucl.\ Phys.\  {\bf 33} (1994) 397
 [hep-ph/9312217].

\bibitem{Wolfenstein:1994jw}
 L.~Wolfenstein and Y.~L.~Wu,
 Phys.\ Rev.\ Lett.\  {\bf 73} (1994) 2809
 [hep-ph/9410253].

\bibitem{Atwood:1996vj}
 D.~Atwood, L.~Reina, and A.~Soni,
 Phys.\ Rev.\  D {\bf 55} (1997) 3156
 [hep-ph/9609279].

\bibitem{Golowich:2007ka}
 E.~Golowich, J.~Hewett, S.~Pakvasa, and A.~A.~Petrov,
 Phys.\ Rev.\  D {\bf 76} (2007) 095009
 [arXiv:0705.3650 [hep-ph]].

\bibitem{Davidson:2010xv}
 S.~Davidson and G.~J.~Grenier,
 Phys.\ Rev.\  D {\bf 81} (2010) 095016
 [arXiv:1001.0434 [hep-ph]].

\bibitem{CorderoCid:2004vi}
 A.~Cordero-Cid, M.~A.~P\'erez, G.~Tavares-Velasco, and J.~J.~Toscano,
 Phys.\ Rev.\  D {\bf 70} (2004) 074003
 [hep-ph/0407127].

\bibitem{Larios:2006pb}
 F.~Larios, R.~Mart\'\i nez, and M.~A.~P\'erez,
 Int.\ J.\ Mod.\ Phys.\  A {\bf 21} (2006) 3473
 [hep-ph/0605003].

\bibitem{Aranda:2009cd}
 J.~I.~Aranda, A.~Cordero-Cid, F.~Ram\'\i rez-Zavaleta,
 J.~J.~Toscano, and E.~S.~Tututi,
 Phys.\ Rev.\  D {\bf 81} (2010) 077701
 [arXiv:0911.2304 [hep-ph]].

\bibitem{Joshipura:2010tz}
 A.~S.~Joshipura and B.~P.~Kodrani,
 Phys.\ Rev.\  D {\bf 82} (2010) 115013
 [arXiv:1004.3637 [hep-ph]].

\bibitem{Aubert:2007rn}
 B.~Aubert {\it et al.}  [BaBar Collaboration],
 Phys.\ Rev.\ Lett.\  {\bf 99} (2007) 201801
 [arXiv:0708.1303 [hep-ex]].

\bibitem{Nie:1998dg}
 S.~Nie and M.~Sher,
 Phys.\ Rev.\  D {\bf 58} (1998) 097701
 [hep-ph/9805376].

\bibitem{deRafael:2009zz}
 E.~de Rafael,
 PoS EFT09 (2009) 050.

\bibitem{Prades:2009qp}
 J.~Prades,
 Acta Phys.\ Polon.\ Supp.\  {\bf 3} (2010) 75
 [arXiv:0909.2546 [hep-ph]].

\bibitem{Diaz:2002uk}
 R.~A.~D\'\i az, R.~Mart\'\i nez, and J.~A.~Rodr\'\i guez,
 Phys.\ Rev.\  D {\bf 67} (2003) 075011
 [hep-ph/0208117].

\bibitem{Diaz:2004mk}
 R.~A.~D\'\i az, R.~Mart\'\i nez, and C.~E.~Sandoval,
 Eur.\ Phys.\ J.\  C {\bf 41} (2005) 305
 [hep-ph/0406265].

\bibitem{Hou:2008yb}
 W.~S.~Hou, F.~F.~Lee, and C.~Y.~Ma,
 Phys.\ Rev.\  D {\bf 79} (2009) 073002
 [arXiv:0812.0064 [hep-ph]].

\bibitem{Li:2008xx}
 W.~Li, Y.~Ma, G.~Liu, and W.~Guo,
 arXiv:0812.0727 [hep-ph].

\bibitem{Li:2010vf}
 W.~J.~Li, Y.~Y.~Fan, G.~W.~Liu, and L.~X.~Lu,
 Int.\ J.\ Mod.\ Phys.\  A {\bf 25} (2010) 4827
 [arXiv:1007.2894 [hep-ph]].

\bibitem{BowserChao:1998yp}
 D.~Bowser-Chao, K.~M.~Cheung, and W.~Y.~Keung,
 Phys.\ Rev.\  D {\bf 59} (1999) 115006
 [hep-ph/9811235].

\bibitem{Xiao:2003ya}
 Z.~J.~Xiao and L.~Guo,
 Phys.\ Rev.\  D {\bf 69} (2004) 014002
 [hep-ph/0309103].

\bibitem{Idarraga:2005ia}
 J.~P.~Idarraga, R.~Mart\'\i nez, J.~A.~Rodr\'\i guez, and N.~Poveda,
 hep-ph/0509072.

\bibitem{Huang:2004pk}
 C.~S.~Huang and J.~T.~Li,
 Int.\ J.\ Mod.\ Phys.\  A {\bf 20} (2005) 161
 [hep-ph/0405294].

 \bibitem{Blechman:2010cs}
  A.~E.~Blechman, A.~A.~Petrov, and G.~Yeghiyan,
  JHEP {\bf 1011} (2010) 075
  [arXiv:1009.1612 [hep-ph]].

\bibitem{Branco:1996bq}
 G.~C.~Branco, W.~Grimus, and L.~Lavoura,
 Phys.\ Lett.\  B {\bf 380} (1996) 119
 [hep-ph/9601383].

\bibitem{D'Ambrosio:2002ex}
  G.~D'Ambrosio, G.~F.~Giudice, G.~Isidori, and A.~Strumia,
  Nucl.\ Phys.\  B {\bf 645} (2002) 155
  [hep-ph/0207036].

\bibitem{Chivukula:1987py}
 R.~S.~Chivukula and H.~Georgi,
 Phys.\ Lett.\  B {\bf 188} (1987) 99.

\bibitem{Buras:2000dm}
 A.~J.~Buras, P.~Gambino, M.~Gorbahn, S.~J\"ager, and L.~Silvestrini,
 Phys.\ Lett.\  B {\bf 500} (2001) 161
 [hep-ph/0007085].

\bibitem{Blanke:2006ig}
 M.~Blanke, A.~J.~Buras, D.~Guadagnoli, and C.~Tarantino,
 JHEP {\bf 0610} (2006) 003
 [hep-ph/0604057].

\bibitem{arXiv:1103.2915}
  R.~Alonso, M.~B.~Gavela, L.~Merlo, and S.~Rigolin,
  JHEP {\bf 1107} (2011) 012
  [arXiv:1103.2915 [hep-ph]].

\bibitem{Hall:1993ca}
 L.~J.~Hall and S.~Weinberg,
 Phys.\ Rev.\  D {\bf 48} (1993) 979
 [hep-ph/9303241].

\bibitem{Jo1991}
A.~S.~Joshipura,
Mod.\ Phys.\ Lett.\ A {\bf 6} (1991) 1693.

\bibitem{Joshipura:1990pi}
 A.~S.~Joshipura and S.~D.~Rindani,
 Phys.\ Lett.\  B {\bf 260} (1991) 149.

\bibitem{Jung:2010ik}
 M.~Jung, A.~Pich, and P.~Tuz\'on,
 JHEP {\bf 1011} (2010) 003
 [arXiv:1006.0470 [hep-ph]].

\bibitem{Jung:2010ab}
 M.~Jung, A.~Pich, and P.~Tuz\'on,
 Phys.\ Rev.\  D {\bf 83} (2011) 074011
 [arXiv:1011.5154 [hep-ph]].

\bibitem{Lavoura:1994ty}
 L.~Lavoura,
 Int.\ J.\ Mod.\ Phys.\  A {\bf 9} (1994) 1873.

\bibitem{Botella:2009pq}
 F.~J.~Botella, G.~C.~Branco, and M.~N.~Rebelo,
 Phys.\ Lett.\  B {\bf 687} (2010) 194
 [arXiv:0911.1753 [hep-ph]].

\bibitem{Botella:2011ne}
 F.~J.~Botella, G.~C.~Branco, M.~Nebot, and M.~N.~Rebelo,
 arXiv:1102.0520 [hep-ph].

\bibitem{Joshipura:2007sf}
  A.~S.~Joshipura and B.~P.~Kodrani,
  Phys.\ Lett.\ B {\bf 670} (2009) 369
  [arXiv:0706.0953 [hep-ph]].

\bibitem{Botella:2004ks}
 F.~J.~Botella, M.~Nebot, and O.~Vives,
 JHEP {\bf 0601} (2006) 106
 [hep-ph/0407349].

\bibitem{Buras:2010mh}
 A.~J.~Buras, M.~V.~Carlucci, S.~Gori, and G.~Isidori,
 JHEP {\bf 1010} (2010) 009
 [arXiv:1005.5310 [hep-ph]].

\bibitem{Cirigliano:2005ck}
 V.~Cirigliano, B.~Grinstein, G.~Isidori, and M.~B.~Wise,
 Nucl.\ Phys.\  B {\bf 728} (2005) 121
 [hep-ph/0507001].

\bibitem{Davidson:2006bd}
 S.~Davidson and F.~Palorini,
 Phys.\ Lett.\  B {\bf 642} (2006) 72
 [hep-ph/0607329].

\bibitem{Branco:2006hz}
 G.~C.~Branco, A.~J.~Buras, S.~J\"ager, S.~Uhlig, and A.~Weiler,
 JHEP {\bf 0709} (2007) 004
 [hep-ph/0609067].

\bibitem{Georgi:1994qn}
 H.~Georgi,
 Ann.\ Rev.\ Nucl.\ Part.\ Sci.\  {\bf 43} (1993) 209.

\bibitem{Tomozawa:1977ea}
 Y.~Tomozawa,
 Phys.\ Rev.\  D {\bf 18} (1978) 2556.

\bibitem{Donoghue:1978cj}
 J.~F.~Donoghue and L.~F.~Li,
 Phys.\ Rev.\  D {\bf 19} (1979) 945.

\bibitem{charged06}
http://www.grid.tsl.uu.se/chargedhiggs2006/

\bibitem{charged08}
http://www.grid.tsl.uu.se/chargedhiggs2008/

\bibitem{charged10}
http://www.grid.tsl.uu.se/chargedhiggs2010/

\bibitem{Guedes:2011ki}
 R.~Guedes, S.~Kanemura, S.~Moretti, R.~Santos, and K.~Yagyu,
 PoS CHarged 2010 (2010) 037
 [arXiv:1102.3791 [hep-ph]].

\bibitem{hep-ph/9808278}
  J.~A.~Coarasa P\'erez, J.~Guasch, J.~Sol\`a, and W.~Hollik,
  Phys.\ Lett.\ B {\bf 442} (1998) 326
  [hep-ph/9808278].

\bibitem{Bertolini:1986th}
 S.~Bertolini, F.~Borzumati, and A.~Masiero,
 Phys.\ Rev.\ Lett.\  {\bf 59} (1987) 180.

\bibitem{Deshpande:1987nr}
 N.~G.~Deshpande, P.~Lo, J.~Trampetic, G.~Eilam, and P.~Singer,
 Phys.\ Rev.\ Lett.\  {\bf 59} (1987) 183.

\bibitem{Grinstein:1987pu}
 B.~Grinstein and M.~B.~Wise,
 Phys.\ Lett.\  B {\bf 201} (1988) 274.

\bibitem{Grinstein:1987vj}
 B.~Grinstein, R.~P.~Springer, and M.~B.~Wise,
 Phys.\ Lett.\  B {\bf 202} (1988) 138.

\bibitem{Grigjanis:1988iq}
 R.~Grigjanis, P.~J.~O'Donnell, M.~Sutherland, and H.~Navelet,
 Phys.\ Lett.\  B {\bf 213} (1988) 355
 [Erratum {\it ibid.}\  {\bf 286} (1992) 413].

\bibitem{Cella:1990sh}
 G.~Cella, G.~Curci, G.~Ricciardi, and A.~Vicere,
 Phys.\ Lett.\  B {\bf 248} (1990) 181.

\bibitem{Cella:1994px}
 G.~Cella, G.~Curci, G.~Ricciardi, and A.~Vicere,
 Phys.\ Lett.\  B {\bf 325} (1994) 227
 [hep-ph/9401254].

\bibitem{Cella:1994np}
 G.~Cella, G.~Curci, G.~Ricciardi, and A.~Vicere,
 Nucl.\ Phys.\  B {\bf 431} (1994) 417
 [hep-ph/9406203].

\bibitem{Ciuchini:1993ks}
 M.~Ciuchini, E.~Franco, G.~Martinelli, L.~Reina, and L.~Silvestrini,
 Phys.\ Lett.\  B {\bf 316} (1993) 127
 [hep-ph/9307364].

\bibitem{Ciuchini:1993fk}
 M.~Ciuchini, E.~Franco, L.~Reina, and L.~Silvestrini,
 Nucl.\ Phys.\  B {\bf 421} (1994) 41
 [hep-ph/9311357].

\bibitem{Buras:1993xp}
 A.~J.~Buras, M.~Misiak, M.~M\"unz, and S.~Pokorski,
 Nucl.\ Phys.\  B {\bf 424} (1994) 374
 [hep-ph/9311345].

\bibitem{Ali:1993ct}
 A.~Ali and C.~Greub,
 Z.\ Phys.\  C {\bf 60} (1993) 433.

\bibitem{Misiak:1992bc}
 M.~Misiak,
 Nucl.\ Phys.\  B {\bf 393} (1993) 23
 [Erratum {\it ibid.}\ {\bf 439} (1995) 461].

\bibitem{Adel:1993ah}
 K.~Adel and Y.~P.~Yao,
 Phys.\ Rev.\  D {\bf 49} (1994) 4945
 [hep-ph/9308349].

\bibitem{Greub:1996tg}
 C.~Greub, T.~Hurth, and D.~Wyler,
 Phys.\ Rev.\  D {\bf 54} (1996) 3350
 [hep-ph/9603404].

\bibitem{Buras:1997bk}
 A.~J.~Buras, A.~Kwiatkowski, and N.~Pott,
 Phys.\ Lett.\  B {\bf 414} (1997) 157
 [Erratum {\it ibid.}\ {\bf 434} (1998) 459]
 [hep-ph/9707482].

\bibitem{Haisch:2008ar}
 U.~Haisch,
 arXiv:0805.2141 [hep-ph].

\bibitem{Hurth:2010tk}
 T.~Hurth and M.~Nakao,
 Ann.\ Rev.\ Nucl.\ Part.\ Sci.\  {\bf 60} (2010) 645
 [arXiv:1005.1224 [hep-ph]].

\bibitem{Hurth:2011jc}
 T.~Hurth,
 PoS CHarged 2010 (2010) 020
 [arXiv:1104.5123 [hep-ph]].

\bibitem{Ciafaloni:1997un}
 P.~Ciafaloni, A.~Romanino, and A.~Strumia,
 Nucl.\ Phys.\  B {\bf 524} (1998) 361
 [hep-ph/9710312].

\bibitem{Borzumati:2003rr}
 F.~Borzumati, C.~Greub, and Y.~Yamada,
 Phys.\ Rev.\  D {\bf 69} (2004) 055005
 [hep-ph/0311151].

\bibitem{Bobeth:1999mk}
 C.~Bobeth, M.~Misiak, and J.~Urban,
 Nucl.\ Phys.\  B {\bf 574} (2000) 291
 [hep-ph/9910220].

\bibitem{:2009qg}
 A.~Limosani {\it et al.}  [Belle Collaboration],
 Phys.\ Rev.\ Lett.\  {\bf 103} (2009) 241801
 [arXiv:0907.1384 [hep-ex]].

\bibitem{Hou:1992sy}
 W.~S.~Hou,
 Phys.\ Rev.\  D {\bf 48} (1993) 2342.

\bibitem{Grossman:1994ax}
 Y.~Grossman and Z.~Ligeti,
 Phys.\ Lett.\  B {\bf 332} (1994) 373
 [hep-ph/9403376].

\bibitem{Hara:2010dk}
 K.~Hara {\it et al.}  [Belle Collaboration],
 Phys.\ Rev.\  D {\bf 82} (2010) 071101
 [arXiv:1006.4201 [hep-ex]].

 \bibitem{hep-ex/0604018}
  K.~Ikado {\it et al.} [Belle Collaboration],
  Phys.\ Rev.\ Lett.\ {\bf 97} (2006) 251802
  [hep-ex/0604018].

\bibitem{Aubert:2009wt}
 B.~Aubert {\it et al.}  [BaBar Collaboration],
 arXiv:0912.2453 [hep-ex].

\bibitem{hfag}
Heavy Flavor Averaging Group, http://www.slac.stanford.edu/xorg/hfag/

\bibitem{Czarnecki:1998tn}
 A.~Czarnecki and W.~J.~Marciano,
 Phys.\ Rev.\ Lett.\  {\bf 81} (1998) 277
 [hep-ph/9804252].

\bibitem{Tanaka:2010se}
 M.~Tanaka and R.~Watanabe,
 Phys.\ Rev.\  D {\bf 82} (2010) 034027
 [arXiv:1005.4306 [hep-ph]].

\bibitem{Grossman:1995yp}
 Y.~Grossman, H.~E.~Haber, and Y.~Nir,
 Phys.\ Lett.\  B {\bf 357} (1995) 630
 [hep-ph/9507213].

\bibitem{Haber:1999zh}
 H.~E.~Haber and H.~E.~Logan,
 Phys.\ Rev.\  D {\bf 62} (2000) 015011
 [hep-ph/9909335].

\bibitem{Degrassi:2010ne}
 G.~Degrassi and P.~Slavich,
 Phys.\ Rev.\  D {\bf 81} (2010) 075001
 [arXiv:1002.1071 [hep-ph]].

\bibitem{Geng:1988bq}
 C.~Q.~Geng and J.~N.~Ng,
 Phys.\ Rev.\  D {\bf 38} (1988) 2857
 [Erratum {\it ibid.}\ {\bf 41} (1990) 1715].

\bibitem{Deschamps:2009rh}
 O.~Deschamps, S.~Descotes-Genon, S.~Monteil, V.~Niess, S.~T'Jampens, and V.~Tisserand,
 Phys.\ Rev.\  D {\bf 82} (2010) 073012
 [arXiv:0907.5135 [hep-ph]].

\bibitem{Djouadi:1997yw}
 A.~Djouadi, J.~Kalinowski, and M.~Spira,
 Comput.\ Phys.\ Commun.\  {\bf 108} (1998) 56
 [hep-ph/9704448].

\bibitem{Kanemura:2009mk}
 S.~Kanemura, S.~Moretti, Y.~Mukai, R.~Santos, and K.~Yagyu,
 Phys.\ Rev.\  D {\bf 79} (2009) 055017
 [arXiv:0901.0204 [hep-ph]].

\bibitem{:2001xy}
   LEP Working Group for Higgs Boson Searches,
 hep-ex/0107031.



\bibitem{Aaltonen:2009ke}
 T.~Aaltonen {\it et al.}  [CDF Collaboration],
 Phys.\ Rev.\ Lett.\  {\bf 103} (2009) 101803
 [arXiv:0907.1269 [hep-ex]].

\bibitem{:2009zh}
 V.~M.~Abazov {\it et al.}  [D0 Collaboration],
 Phys.\ Lett.\  B {\bf 682} (2009) 278
 [arXiv:0908.1811 [hep-ex]].

 \bibitem{arXiv:1107.1268}
  V.~M.~Abazov {\it et al.} [D0 Collaboration],
  Phys.\ Rev.\ D {\bf 84} (2011) 092002
  [arXiv:1107.1268 [hep-ex]].

\bibitem{arXiv:1110.5349}
  A.~Buzatu,
  arXiv:1110.5349 [hep-ex].

\bibitem{Aad:2009wy}
 G.~Aad {\it et al.}  [ATLAS Collaboration],
 arXiv:0901.0512 [hep-ex].

\bibitem{Baarmand:2006dm}
 M.~Baarmand, M.~Hashemi, and A.~Nikitenko,
 J.\ Phys.\ G {\bf 32} (2006) N21.

 \bibitem{arXiv:1103.1827}
  A.~Ali, F.~Barreiro, and J.~Llorente,
  Eur.\ Phys.\ J.\ C {\bf 71} (2011) 1737
  [arXiv:1103.1827 [hep-ph]].

\bibitem{arXiv:1109.5356}
  M.~Hashemi,
  arXiv:1109.5356 [hep-ph].

\bibitem{arXiv:1111.4530}
  S.~Yang and Q.~-S.~Yan,
  arXiv:1111.4530 [hep-ph].

\bibitem{DiazCruz:1992gg}
 J.~L.~D\'\i az-Cruz and O.~A.~Sampayo,
 Phys.\ Rev.\  D {\bf 50} (1994) 6820.

\bibitem{Gunion:1986pe}
 J.~F.~Gunion, H.~E.~Haber, F.~E.~Paige, W.~K.~Tung, and S.~S.~D.~Willenbrock,
 Nucl.\ Phys.\  B {\bf 294} (1987) 621.

\bibitem{Moretti:1996ra}
 S.~Moretti and K.~Odagiri,
 Phys.\ Rev.\  D {\bf 55} (1997) 5627
 [hep-ph/9611374].

\bibitem{He:1998ie}
 H.~J.~He and C.~P.~Yuan,
 Phys.\ Rev.\ Lett.\  {\bf 83} (1999) 28
 [hep-ph/9810367].

\bibitem{Dittmaier:2007uw}
 S.~Dittmaier, G.~Hiller, T.~Plehn, and M.~Spannowsky,
 Phys.\ Rev.\  D {\bf 77} (2008) 115001
 [arXiv:0708.0940 [hep-ph]].

\bibitem{BarrientosBendezu:1998gd}
  A.~A.~B.~Bendezu and B.~A.~Kniehl,
  Phys.\ Rev.\  D {\bf 59} (1999) 015009
  [hep-ph/9807480].

\bibitem{Moretti:1998xq}
 S.~Moretti and K.~Odagiri,
 Phys.\ Rev.\  D {\bf 59} (1999) 055008
 [hep-ph/9809244].

\bibitem{Brein:2000cv}
 O.~Brein, W.~Hollik, and S.~Kanemura,
 Phys.\ Rev.\  D {\bf 63} (2001) 095001
 [hep-ph/0008308].

\bibitem{Eriksson:2006yt}
 D.~Eriksson, S.~Hesselbach, and J.~Rathsman,
 Eur.\ Phys.\ J.\  C {\bf 53} (2008) 267
 [hep-ph/0612198].

\bibitem{Hashemi:2010ce}
 M.~Hashemi,
 Phys.\ Rev.\  D {\bf 83} (2011) 055004
 [arXiv:1008.3785 [hep-ph]].

 \bibitem{arXiv:1104.0889}
  R.~Enberg and R.~Pasechnik,
  Phys.\ Rev.\ D {\bf 83} (2011) 095020
  [arXiv:1104.0889 [hep-ph]].

\bibitem{arXiv:1112.0086}
  S.-S.~Bao, X.~Gong, H.-L.~Li, S.-Y.~Li, and Z.-G.~Si,
  arXiv:1112.0086 [hep-ph].


\bibitem{Kanemura:2001hz}
 S.~Kanemura and C.~P.~Yuan,
 Phys.\ Lett.\  B {\bf 530} (2002) 188
 [hep-ph/0112165].

\bibitem{Cao:2003tr}
 Q.~H.~Cao, S.~Kanemura, and C.~P.~Yuan,
 Phys.\ Rev.\  D {\bf 69} (2004) 075008
 [hep-ph/0311083].

\bibitem{Belyaev:2006rf}
 A.~Belyaev, Q.~H.~Cao, D.~Nomura, K.~Tobe, and C.~P.~Yuan,
 Phys.\ Rev.\ Lett.\  {\bf 100} (2008) 061801
 [hep-ph/0609079].

\bibitem{arXiv:hep-ph/0306045}
  A.~G.~Akeroyd,
  Phys.\ Rev.\ D {\bf 68} (2003) 077701
  [hep-ph/0306045].

\bibitem{hep-ph/0412365}
  D.~K.~Ghosh and S.~Moretti,
  Eur.\ Phys.\ J.\ C {\bf 42} (2005) 341
  [hep-ph/0412365].

\bibitem{Alves:2005kr}
 A.~Alves and T.~Plehn,
 Phys.\ Rev.\  D {\bf 71} (2005) 115014
 [hep-ph/0503135].

\bibitem{Moretti:2001pp}
 S.~Moretti,
 J.\ Phys.\ G {\bf 28} (2002) 2567
 [hep-ph/0102116].

\bibitem{arXiv:1012.0527}
  K.~Huitu, S.~K.~Rai, K.~Rao, S.~D.~Rindani, and P.~Sharma,
  JHEP {\bf 1104} (2011) 026
  [arXiv:1012.0527 [hep-ph]].

\bibitem{arXiv:1109.2420}
  J.~Baglio, M.~Beccaria, A.~Djouadi, G.~Macorini, E.~Mirabella,
  N.~Orlando, F.~M.~Renard, and C.~Verzegnassi,
  Phys.\ Lett.\ B {\bf 705} (2011) 212
  [arXiv:1109.2420 [hep-ph]].

\bibitem{arXiv:1011.1409}
  S.-S.~Bao, Y.~Tang, and Y.-L.~Wu,
  Phys.\ Rev.\ D {\bf 83} (2011) 075006
  [arXiv:1011.1409 [hep-ph]].

\bibitem{Bayatian:2006zz}
 G.~L.~Bayatian {\it et al.}  [CMS Collaboration],
 ``CMS physics: Technical design report.''

\bibitem{Borzumati:1998xr}
 F.~Borzumati and A.~Djouadi,
 Phys.\ Lett.\  B {\bf 549} (2002) 170
 [hep-ph/9806301].

\bibitem{Akeroyd:2000xa}
 A.~G.~Akeroyd, A.~Arhrib, and E.~Naimi,
 Eur.\ Phys.\ J.\  C {\bf 20} (2001) 51
 [hep-ph/0002288].

\bibitem{Carena:2000yx}
 M.~S.~Carena {\it et al.}  [Higgs Working Group Collaboration],
 hep-ph/0010338.

\bibitem{Drees:1999sb}
 M.~Drees, M.~Guchait, and D.~P.~Roy,
 Phys.\ Lett.\  B {\bf 471} (1999) 39
 [hep-ph/9909266].

\bibitem{Assamagan:2002ne}
 K.~A.~Assamagan, Y.~Coadou, and A.~Deandrea,
 Eur.\ Phys.\ J.\ direct C {\bf 4} (2002) 9
 [hep-ph/0203121].

  \bibitem{aker}
  A.~G.~Akeroyd, in~\cite{charged10}.

\bibitem{Field:1997gz}
 J.~H.~Field,
 Mod.\ Phys.\ Lett.\  A {\bf 13} (1998) 1937
 [hep-ph/9801355].

\bibitem{Urban:1997gw}
 J.~Urban, F.~Krauss, U.~Jentschura, and G.~Soff,
 Nucl.\ Phys.\  B {\bf 523} (1998) 40
 [hep-ph/9710245].

\bibitem{Buras:2008nn}
 A.~J.~Buras and D.~Guadagnoli,
 Phys.\ Rev.\  D {\bf 78} (2008) 033005
 [arXiv:0805.3887 [hep-ph]].

\bibitem{Buras:2010pza}
 A.~J.~Buras, D.~Guadagnoli, and G.~Isidori,
 Phys.\ Lett.\  B {\bf 688} (2010) 309
 [arXiv:1002.3612 [hep-ph]].

\bibitem{Abdallah:2003wd}
 J.~Abdallah {\it et al.}  [DELPHI Collaboration],
 Eur.\ Phys.\ J.\  C {\bf 34} (2004) 399
 [hep-ex/0404012].

 \bibitem{arXiv:0812.0267}
  G.~Abbiendi {\it et al.} [OPAL Collaboration],
  arXiv:0812.0267 [hep-ex].

\bibitem{arXiv:1104.3178}
  M.~Aoki, R.~Guedes, S.~Kanemura, S.~Moretti, R.~Santos, and K.~Yagyu,
  Phys.\ Rev.\ D {\bf 84} (2011) 055028
  [arXiv:1104.3178 [hep-ph]].

\bibitem{Park:2006gk}
 J.~H.~Park,
 JHEP {\bf 0610} (2006) 077
 [hep-ph/0607280].

\bibitem{Abbiendi:2003ji}
 G.~Abbiendi {\it et al.}  [OPAL Collaboration],
 Eur.\ Phys.\ J.\  C {\bf 32} (2004) 453
 [hep-ex/0309014].

\bibitem{Davidson:2010sf}
 S.~M.~Davidson and H.~E.~Logan,
 Phys.\ Rev.\  D {\bf 82} (2010) 115031
 [arXiv:1009.4413 [hep-ph]].

 \bibitem{Heister:2002ev}
  A.~Heister \etal  [ALEPH Collaboration],
  Phys.\ Lett.\  B {\bf 543} (2002) 1
  [hep-ex/0207054].

\bibitem{ATLASstudy}
ATLAS note ATLAS-PHYS-PUB-2010-009.

\bibitem{901632}
  A.~Ferrari [ATLAS Collaboration],
  PoSCHARGED\ {\bf 2010} (2010) 010.

\bibitem{hep-ph/9509203}
  A.~G.~Akeroyd,
  hep-ph/9509203.

\bibitem{Miao:2010rg}
 X.~Miao, S.~Su, and B.~Thomas,
 Phys.\ Rev.\  D {\bf 82} (2010) 035009
 [arXiv:1005.0090 [hep-ph]].

\bibitem{Dolle:2009ft}
 E.~Dolle, X.~Miao, S.~Su, and B.~Thomas,
 Phys.\ Rev.\  D {\bf 81} (2010) 035003
 [arXiv:0909.3094 [hep-ph]].

\bibitem{Huitu:2010uc}
 K.~Huitu, K.~Kannike, A.~Racioppi, and M.~Raidal,
 JHEP {\bf 1101} (2011) 010
 [arXiv:1005.4409 [hep-ph]].

\bibitem{Davidson:2005cw}
 S.~Davidson and H.~E.~Haber,
 Phys.\ Rev.\  D {\bf 72} (2005) 035004
 [Erratum {\it ibid.}\ {\bf 72} (2005) 099902]
 [hep-ph/0504050].

\bibitem{Haber:2006ue}
 H.~E.~Haber and D.~O'Neil,
 Phys.\ Rev.\  D {\bf 74} (2006) 015018
 [hep-ph/0602242].

 \bibitem{arXiv:1105.4951}
  A.~Cordero-Cid, O.~Felix-Beltr\'an, J.~Hern\'andez-S\'anchez,
  and R.~Noriega-Papaqui,
  PoSCHARGED {\bf 2010} (2010) 042
  [arXiv:1105.4951 [hep-ph]].

\bibitem{arXiv:1106.5035}
  J.~Hern\'andez-S\'anchez, L.~L\'opez-Lozano, R.~Noriega-Papaqui,
  and A.~Rosado,
  arXiv:1106.5035 [hep-ph].

\bibitem{Abazov:2008rn}
 V.~M.~Abazov {\it et al.}  [D0 Collaboration],
 Phys.\ Rev.\ Lett.\  {\bf 102} (2009) 191802
 [arXiv:0807.0859 [hep-ex]].

\bibitem{arXiv:0810.3046}
  H.~Cardenas and J.~A.~Rodr\'\i guez,
  Mod.\ Phys.\ Lett.\ A\ {\bf 26} (2011) 1869
  [arXiv:0810.3046 [hep-ph]].


\bibitem{Martinez:2002tn}
 R.~Mart\'\i nez, J.~A.~Rodr\'\i guez, and M.~Rozo,
 Phys.\ Rev.\  D {\bf 68} (2003) 035001
 [hep-ph/0212236].

\bibitem{Xiao:2003vq}
 Z.~J.~Xiao and C.~Zhuang,
 Eur.\ Phys.\ J.\  C {\bf 33} (2004) 349
 [hep-ph/0310097].

\bibitem{Diaz:2005rv}
 R.~A.~D\'\i az, R.~Mart\'\i nez, and C.~E.~Sandoval,
 Eur.\ Phys.\ J.\  C {\bf 46} (2006) 403
 [hep-ph/0509194].

\bibitem{Idarraga:2008zz}
 J.~P.~Idarraga, R.~Mart\'\i nez, J.~A.~Rodr\'\i guez, and N.~Poveda,
 Braz.\ J.\ Phys.\  {\bf 38} (2008) 531.

\bibitem{hep-ph/9409421}
  Y.~L.~Wu and L.~Wolfenstein,
  Phys.\ Rev.\ Lett.\ {\bf 73} (1994) 1762
  [hep-ph/9409421].

\bibitem{Botella:1994cs}
 F.~J.~Botella and J.~P.~Silva,
 Phys.\ Rev.\  D {\bf 51} (1995) 3870
 [hep-ph/9411288].

\bibitem{Velhinho:1994vh}
 J.~Velhinho, R.~Santos, and A.~Barroso,
 Phys.\ Lett.\  B {\bf 322} (1994) 213.

\bibitem{Nagel:2004sw}
 F.~Nagel,
PhD thesis (University of Heidelberg, 2004).

\bibitem{Maniatis:2006fs}
 M.~Maniatis, A.~von Manteuffel, O.~Nachtmann, and F.~Nagel,
 Eur.\ Phys.\ J.\  C {\bf 48} (2006) 805
 [hep-ph/0605184].

\bibitem{Maniatis:2006jd}
 M.~Maniatis, A.~von Manteuffel, and O.~Nachtmann,
 Eur.\ Phys.\ J.\  C {\bf 49} (2007) 1067
 [hep-ph/0608314].

\bibitem{Maniatis:2007vn}
 M.~Maniatis, A.~von Manteuffel, and O.~Nachtmann,
 Eur.\ Phys.\ J.\  C {\bf 57} (2008) 719
 [arXiv:0707.3344 [hep-ph]].

\bibitem{Nishi:2006tg}
 C.~C.~Nishi,
 Phys.\ Rev.\  D {\bf 74} (2006) 036003
 [Erratum {\it ibid.}\ {\bf 76} (2007) 119901]
 [hep-ph/0605153].

\bibitem{Nishi:2007nh}
 C.~C.~Nishi,
 Phys.\ Rev.\  D {\bf 76} (2007) 055013
 [arXiv:0706.2685 [hep-ph]].

\bibitem{Nishi:2007dv}
 C.~C.~Nishi,
 Phys.\ Rev.\  D {\bf 77} (2008) 055009
 [arXiv:0712.4260 [hep-ph]].

\bibitem{Ivanov:2005hg}
 I.~P.~Ivanov,
 Phys.\ Lett.\  B {\bf 632} (2006) 360
 [hep-ph/0507132].

\bibitem{Ivanov:2006yq}
 I.~P.~Ivanov,
 Phys.\ Rev.\  D {\bf 75} (2007) 035001
 [Erratum {\it ibid.}\ {\bf 76} (2007) 039902]
 [hep-ph/0609018].

\bibitem{arXiv:0910.4492}
  A.~Degee and I.~P.~Ivanov,
  Phys.\ Rev.\ D {\bf 81} (2010) 015012
  [arXiv:0910.4492 [hep-ph]].

\bibitem{Neufeld:1987wa}
 H.~Neufeld, W.~Grimus, and G.~Ecker,
 Int.\ J.\ Mod.\ Phys.\  A {\bf 3} (1988) 603.


\bibitem{Ecker:1987qp}
 G.~Ecker, W.~Grimus, and H.~Neufeld,
 J.\ Phys.\ A  {\bf 20} (1987) L807.

 \bibitem{Ecker:1989ay}
  G.~Ecker, W.~Grimus, and H.~Neufeld,
  Phys.\ Lett.\ B {\bf 228} (1989) 401.

  \bibitem{Grimus:1995zi}
  W.~Grimus and M.~N.~Rebelo,
  Phys.\ Rep.\ {\bf 281} (1997) 239
  [hep-ph/9506272].


\bibitem{Ferreira:2009wh}
 P.~M.~Ferreira, H.~E.~Haber, and J.~P.~Silva,
 Phys.\ Rev.\  D {\bf 79} (2009) 116004
 [arXiv:0902.1537 [hep-ph]].


\bibitem{Maniatis:2009vp}
 M.~Maniatis and O.~Nachtmann,
 JHEP {\bf 0905} (2009) 028
 [arXiv:0901.4341 [hep-ph]].


\bibitem{Maniatis:2009by}
 M.~Maniatis and O.~Nachtmann,
 JHEP {\bf 1004} (2010) 027
 [arXiv:0912.2727 [hep-ph]].

 \bibitem{arXiv:1009.1869}
  M.~Maniatis, O.~Nachtmann, and A.~von Manteuffel,
  arXiv:1009.1869 [hep-ph].

\bibitem{Haber:2010bw}
  H.~E.~Haber and D.~O'Neil,
  Phys.\ Rev.\ D {\bf 83} (2011) 055017
  [arXiv:1011.6188 [hep-ph]].

\bibitem{Ferreira:2010bm}
 P.~M.~Ferreira and J.~P.~Silva,
 Eur.\ Phys.\ J.\  C {\bf 69} (2010) 45
 [arXiv:1001.0574 [hep-ph]].

\bibitem{Ferreira:2010jy}
 P.~M.~Ferreira, H.~E.~Haber, and J.~P.~Silva,
 Phys.\ Rev.\  D {\bf 82} (2010) 016001
 [arXiv:1004.3292 [hep-ph]].

\bibitem{arXiv:0909.2855}
  E.~Ma and M.~Maniatis,
  Phys.\ Lett.\ B {\bf 683} (2010) 33
  [arXiv:0909.2855 [hep-ph]].

\bibitem{arXiv:1106.1436}
  M.~Maniatis and O.~Nachtmann,
  arXiv:1106.1436 [hep-ph].

\bibitem{arXiv:1106.3482}
  R.~A.~Battye, G.~D.~Brawn, and A.~Pilaftsis,
  JHEP {\bf 1108} (2011) 020
  [arXiv:1106.3482 [hep-ph]].

\bibitem{arXiv:1109.3787}
  A.~Pilaftsis,
  arXiv:1109.3787 [hep-ph].

\bibitem{Ferreira:2010hy}
 P.~M.~Ferreira, M.~Maniatis, O.~Nachtmann, and J.~P.~Silva,
 JHEP {\bf 1008} (2010) 125
 [arXiv:1004.3207 [hep-ph]].


\bibitem{Ferreira:2010yh}
 P.~M.~Ferreira, H.~E.~Haber, M.~Maniatis, O.~Nachtmann, and J.~P.~Silva,
 Int.\ J.\ Mod.\ Phys.\  A {\bf 26} (2011) 769
 [arXiv:1010.0935 [hep-ph]].



\bibitem{Lee:1966ik}
 T.~D.~Lee and G.~C.~Wick,
 Phys.\ Rev.\  {\bf 148} (1966) 1385.

\bibitem{Bernabeu:1986fc}
 J.~Bernab\'eu, G.~C.~Branco, and M.~Gronau,
 Phys.\ Lett.\  B {\bf 169} (1986) 243.


\bibitem{Ecker:1981wv}
 G.~Ecker, W.~Grimus, and W.~Konetschny,
 Nucl.\ Phys.\  B {\bf 191} (1981) 465.


\bibitem{Ecker:1983hz}
 G.~Ecker, W.~Grimus, and H.~Neufeld,
 Nucl.\ Phys.\  B {\bf 247} (1984) 70.

\bibitem{Haber:1993an}
 H.~E.~Haber and R.~Hempfling,
 Phys.\ Rev.\  D {\bf 48} (1993) 4280
 [hep-ph/9307201].


\bibitem{Ferreira:2008zy}
 P.~M.~Ferreira and J.~P.~Silva,
 Phys.\ Rev.\  D {\bf 78} (2008) 116007
 [arXiv:0809.2788 [hep-ph]].




\bibitem{Ferreira:2004yd}
 P.~M.~Ferreira, R.~Santos, and A.~Barroso,
 Phys.\ Lett.\  B {\bf 603} (2004) 219
 [Erratum {\it ibid.}\ {\bf 629} (2005) 114]
 [hep-ph/0406231].

 \bibitem{Klimenko:1984qx}
  K.~G.~Klimenko,
  Theor.\ Math.\ Phys.\ {\bf 62} (1985) 58.

\bibitem{Ferreira:2009jb}
 P.~M.~Ferreira and D.~R.~T.~Jones,
 JHEP {\bf 0908} (2009) 069
 [arXiv:0903.2856 [hep-ph]].

\bibitem{Lindner:1985uk}
 M.~Lindner,
 Z.\ Phys.\  C {\bf 31} (1986) 295.

\bibitem{Sher:1988mj}
 M.~Sher,
 Phys.\ Rep.\  {\bf 179} (1989) 273.

\bibitem{Sher:1993mf}
 M.~Sher,
 Phys.\ Lett.\  B {\bf 317} (1993) 159
 [Addendum {\it ibid.}\ {\bf 331} (1994) 448]
 [hep-ph/9307342].

\bibitem{Casas:1994qy}
 J.~A.~Casas, J.~R.~Espinosa, and M.~Quir\'os,
 Phys.\ Lett.\  B {\bf 342} (1995) 171
 [hep-ph/9409458].

\bibitem{Casas:1996aq}
 J.~A.~Casas, J.~R.~Espinosa, and M.~Quir\'os,
 Phys.\ Lett.\  B {\bf 382} (1996) 374
 [hep-ph/9603227].

\bibitem{Espinosa:1995se}
 J.~R.~Espinosa and M.~Quir\'os,
 Phys.\ Lett.\  B {\bf 353} (1995) 257
 [hep-ph/9504241].

\bibitem{Kreyerhoff:1989fa}
G.~Kreyerhoff and R.~Rodenberg,
Phys.\ Lett.\ B {\bf 226} (1989) 323.

\bibitem{Freund:1992yd}
J.~Freund, G.~Kreyerhoff, and R.~Rodenberg,
Phys.\ Lett.\ B {\bf 280} (1992) 267.

\bibitem{Kastening:1992by}
B.~M.~Kastening,
hep-ph/9307224.

\bibitem{Nie:1998yn}
S.~Nie and M.~Sher,
Phys.\ Lett.\ B {\bf 449} (1999) 89
[hep-ph/9811234].

\bibitem{Barroso:2007rr}
 A.~Barroso, P.~M.~Ferreira, and R.~Santos,
 Phys.\ Lett.\  B {\bf 652} (2007) 181
 [hep-ph/0702098].

\bibitem{Barroso:2006pa}
 A.~Barroso, P.~M.~Ferreira, R.~Santos, and J.~P.~Silva,
 Phys.\ Rev.\  D {\bf 74} (2006) 085016
 [hep-ph/0608282].

\bibitem{DiazCruz:1992uw}
  J.~L.~D\'\i az-Cruz and A.~M\'endez,
  Nucl.\ Phys.\ B {\bf 380} (1992) 39.

\bibitem{Ivanov:2007de}
 I.~P.~Ivanov,
 Phys.\ Rev.\  D {\bf 77} (2008) 015017
 [arXiv:0710.3490 [hep-ph]].

\bibitem{Frere:1983ag}
 J.-M.~Fr\`ere, D.~R.~T.~Jones, and S.~Raby,
 Nucl.\ Phys.\  B {\bf 222} (1983) 11.

\bibitem{Barroso:2005sm}
 A.~Barroso, P.~M.~Ferreira, and R.~Santos,
 Phys.\ Lett.\  B {\bf 632} (2006) 684
 [hep-ph/0507224].

\bibitem{hep-ph/9902371}
  A.~Pilaftsis and C.~E.~M.~Wagner,
  Nucl.\ Phys.\ B {\bf 553} (1999) 3
  [hep-ph/9902371].

\bibitem{Lavoura:1994fv}
 L.~Lavoura and J.~P.~Silva,
 Phys.\ Rev.\  D {\bf 50} (1994) 4619
 [hep-ph/9404276].

\bibitem{Lavoura:1994yu}
 L.~Lavoura,
 Phys.\ Rev.\  D {\bf 50} (1994) 7089
 [hep-ph/9405307].

\bibitem{Ferreira:2011xc}
 P.~M.~Ferreira, L.~Lavoura, and J.~P.~Silva,
 Phys.\ Lett.\  B {\bf 704} (2011) 179
 [arXiv:1102.0784 [hep-ph]].

\bibitem{Ferreira:2010ir}
 P.~M.~Ferreira and J.~P.~Silva,
 Phys.\ Rev.\  D {\bf 83} (2011) 065026
 [arXiv:1012.2874 [hep-ph]].

\bibitem{Maniatis:2007de}
 M.~Maniatis, A.~von Manteuffel, and O.~Nachtmann,
 Eur.\ Phys.\ J.\  C {\bf 57} (2008) 739
 [arXiv:0711.3760 [hep-ph]].

\bibitem{Branco:1999fs}
 G.~C.~Branco, L.~Lavoura, and J.~P.~Silva,
 {\it CP Violation},
 Oxford University Press (1999).

\bibitem{Branco:2005em}
 G.~C.~Branco, M.~N.~Rebelo, and J.~I.~Silva-Marcos,
 Phys.\ Lett.\  B {\bf 614} (2005) 187
 [hep-ph/0502118].

\bibitem{Gunion:2005ja}
 J.~F.~Gunion and H.~E.~Haber,
 Phys.\ Rev.\  D {\bf 72} (2005) 095002
 [hep-ph/0506227].

\bibitem{Branco:1985aq}
 G.~C.~Branco and M.~N.~Rebelo,
 Phys.\ Lett.\  B {\bf 160} (1985) 117.

\bibitem{Weinberg:1990me}
 S.~Weinberg,
 Phys.\ Rev.\  D {\bf 42} (1990) 860.
 
\bibitem{Ginzburg:2004vp}
 I.~F.~Ginzburg and M.~Krawczyk,
 Phys.\ Rev.\ D {\bf 72} (2005) 115013 [hep-ph/0408011].  

\bibitem{Weinberg:1976hu}
 S.~Weinberg,
 Phys.\ Rev.\ Lett.\  {\bf 37} (1976) 657.

\bibitem{Branco:1979pv}
 G.~C.~Branco,
 Phys.\ Rev.\ Lett.\  {\bf 44} (1980) 504.

\bibitem{Branco:1980sz}
 G.~C.~Branco,
 Phys.\ Rev.\  D {\bf 22} (1980) 2901.

\bibitem{Branco:1985pf}
 G.~C.~Branco, A.~J.~Buras, and J.-M.~G\'erard,
 Nucl.\ Phys.\  B {\bf 259} (1985) 306.

\bibitem{Luders:1954zz}
 G.~Luders,
 Kong.\ Dan.\ Vid.\ Sel.\ Mat.\ Fys.\ Med.\  {\bf 28N5} (1954) 1.

\bibitem{Pauli}
W. Pauli,
in {\it Niels Bohr and the Development of Physics},
Pergamon Press, London (1955).

\bibitem{Jost:1957zz}
 R.~Jost,
 Helv.\ Phys.\ Acta {\bf 30} (1957) 409.

\bibitem{Streater:1989vi}
 R.~F.~Streater and A.~S.~Wightman,
{\it PCT, spin and statistics, and all that},
Addison-Wesley Publishing Company, Redwood City (1989).

\bibitem{Branco:1983tn}
 G.~C.~Branco, J.-M.~G\'erard, and W.~Grimus,
 Phys.\ Lett.\  B {\bf 136} (1984) 383.

\bibitem{Branco:2000dq}
 G.~C.~Branco, F.~Kr\"uger, J.~C.~Rom\~ao, and A.~M.~Teixeira,
 JHEP {\bf 0107} (2001) 027
 [hep-ph/0012318].

\bibitem{Branco:2006pj}
 G.~C.~Branco, D.~Emmanuel-Costa, and J.~C.~Rom\~ao,
 Phys.\ Lett.\  B {\bf 639} (2006) 661
 [hep-ph/0604110].

\bibitem{Derman:1978rx}
 E.~Derman,
 Phys.\ Rev.\  D {\bf 19} (1979) 317.

\bibitem{Botella:2005fc}
 F.~J.~Botella, G.~C.~Branco, M.~Nebot, and M.~N.~Rebelo,
 Nucl.\ Phys.\  B {\bf 725} (2005) 155
 [hep-ph/0502133].

\bibitem{Bento:1991ez}
 L.~Bento, G.~C.~Branco, and P.~A.~Parada,
 Phys.\ Lett.\  B {\bf 267} (1991) 95.

 \bibitem{'tHooft:1976up}
  G.~'t Hooft,
  Phys.\ Rev.\ Lett.\ {\bf 37} (1976) 8.

\bibitem{'tHooft:1976fv}
  G.~'t Hooft,
  Phys.\ Rev.\ D {\bf 14} (1976) 3432.

\bibitem{Nelson:1983zb}
  A.~E.~Nelson,
  Phys.\ Lett.\ B {\bf 136} (1984)  387.

\bibitem{Nelson:1984hg}
  A.~E.~Nelson,
  Phys.\ Lett.\ B {\bf 143} (1984)  165.


\bibitem{Barr:1984qx}
S.~M.~Barr,
Phys.\ Rev.\ Lett.\  {\bf 53} (1984) 329.

\bibitem{Peccei:2006as}
For a review,
see  R.~D.~Peccei,
  Lect.\ Notes Phys.\ {\bf 741} (2008) 3
  [hep-ph/0607268].


\bibitem{Bento:1990wv}
 L.~Bento and G.~C.~Branco,
 Phys.\ Lett.\  B {\bf 245} (1990) 599.

\bibitem{delAguila:1985mk}
  F.~del Aguila and J.~Cort\'es,
  Phys.\ Lett.\  B {\bf 156} (1985) 243.

\bibitem{Branco:1986my}
  G.~C.~Branco and L.~Lavoura,
  Nucl.\ Phys.\  B {\bf 278} (1986) 738.

\bibitem{delAguila:1985ne}
  F.~del Aguila, M.~K.~Chase, and J.~Cort\'es,
  Nucl.\ Phys.\  B {\bf 271} (1986) 61.

\bibitem{Nir:1990yq}
  Y.~Nir and D.~J.~Silverman,
  Phys.\ Rev.\  D {\bf 42} (1990) 1477.

\bibitem{Silverman:1991fi}
  D.~Silverman,
  Phys.\ Rev.\  D {\bf 45} (1992) 1800.

\bibitem{Choong:1993gq}
  W.~S.~Choong and D.~Silverman,
  Phys.\ Rev.\  D {\bf 49} (1994) 2322.

\bibitem{Barger:1995dd}
  V.~D.~Barger, M.~S.~Berger, and R.~J.~N.~Phillips,
  Phys.\ Rev.\  D {\bf 52} (1995) 1663
  [hep-ph/9503204].

\bibitem{Gronau:1996rv}
  M.~Gronau and D.~London,
  Phys.\ Rev.\  D {\bf 55} (1997) 2845
  [hep-ph/9608430].

\bibitem{delAguila:1997vn}
  F.~del Aguila, J.~A.~Aguilar-Saavedra, and G.~C.~Branco,
  Nucl.\ Phys.\  B {\bf 510} (1998) 39
  [hep-ph/9703410].

\bibitem{Branco:1992wr}
  G.~C.~Branco, T.~Morozumi, P.~A.~Parada, and M.~N.~Rebelo,
  Phys.\ Rev.\  D {\bf 48} (1993) 1167.

  \bibitem{Barenboim:1997qx}
  G.~Barenboim, F.~J.~Botella, G.~C.~Branco, and O.~Vives,
  Phys.\ Lett.\  B {\bf 422} (1998) 277
  [hep-ph/9709369].

\bibitem{Barenboim:2000zz}
  G.~Barenboim, F.~J.~Botella, and O.~Vives,
  Phys.\ Rev.\  D {\bf 64} (2001) 015007
  [hep-ph/0012197].

\bibitem{Barenboim:2001fd}
  G.~Barenboim, F.~J.~Botella, and O.~Vives,
  Nucl.\ Phys.\  B {\bf 613} (2001) 285
  [hep-ph/0105306].

\bibitem{AguilarSaavedra:2002kr}
  J.~A.~Aguilar-Saavedra,
  Phys.\ Rev.\  D {\bf 67} (2003) 035003
  [Erratum {\it ibid.}\ {\bf 69} (2004) 099901]
  [hep-ph/0210112].

\bibitem{AguilarSaavedra:2004mt}
  J.~A.~Aguilar-Saavedra, F.~J.~Botella, G.~C.~Branco, and M.~Nebot,
  Nucl.\ Phys.\  B {\bf 706} (2005) 204
  [hep-ph/0406151].

\bibitem{Botella:2006va}
  F.~J.~Botella, G.~C.~Branco, and M.~Nebot,
  Nucl.\ Phys.\  B {\bf 768} (2007) 1
  [hep-ph/0608100].

\bibitem{Botella:2008qm}
  F.~J.~Botella, G.~C.~Branco, and M.~Nebot,
  Phys.\ Rev.\  D {\bf 79} (2009) 096009
  [arXiv:0805.3995 [hep-ph]].

\bibitem{Mendez:1991gp}
 A.~M\'endez and A.~Pomarol,
 Phys.\ Lett.\  B {\bf 272} (1991) 313.

\bibitem{Gun2004}
J.\ F.\ Gunion,
talk given  at the CPNSH, CERN, Switzerland,
December 2004.

\bibitem{Sokolowska:2008bt}
D.~Sokolowska, K.~A.~Kanishev, and M.~Krawczyk,
PoS {\bf CHARGED2008} (2008) 016 [arXiv:0812.0296 [hep-ph]]. 

\bibitem{Davidson:1996cc}
 S.~Davidson and J.~R.~Ellis,
 Phys.\ Lett.\  B {\bf 390} (1997) 210
 [hep-ph/9609451].

\bibitem{roldan}
J.\ Rold\'an,
PhD thesis (University of Val\`encia, 1991).

\bibitem{Pilaftsis:1999qt}
  A.~Pilaftsis and C.~E.~M.~Wagner,
  Nucl.\ Phys.\ B {\bf 553} (1999) 3
  [hep-ph/9902371].

\bibitem{brazil}
CERN notes ATLAS-CONF-2011-157 and CMS PAS HIG-11-023,
Hadron Collider Physics Symposium, November 2011.

\bibitem{newatlas}
ATLAS collaboration, ATLAS note ATLAS-CONF-2011-161.

\bibitem{newcms}
CMS collaboration, talk given by G. Tonelli at CERN on Dec 13 2011.

\bibitem{fs^3} P.M. Ferreira, R. Santos, M. Sher and J.P Silva,
arXiv:1112.3277 [hep-ph].

\bibitem{Lee:1977yc}
 B.~W.~Lee, C.~Quigg, and H.~B.~Thacker,
 Phys.\ Rev.\ Lett.\  {\bf 38} (1977) 883.

\bibitem{Lee:1977eg}
 B.~W.~Lee, C.~Quigg, and H.~B.~Thacker,
 Phys.\ Rev.\  D {\bf 16} (1977) 1519.

\bibitem{Casalbuoni:1986hy}
 R.~Casalbuoni, D.~Dominici, R.~Gatto, and C.~Giunti,
 Phys.\ Lett.\  B {\bf 178} (1986) 235.

\bibitem{Casalbuoni:1987eg}
 R.~Casalbuoni, D.~Dominici, F.~Feruglio, and R.~Gatto,
 Phys.\ Lett.\  B {\bf 200} (1988) 495.

\bibitem{Maalampi:1991fb}
 J.~Maalampi, J.~Sirkka, and I.~Vilja,
 Phys.\ Lett.\  B {\bf 265} (1991) 371.

\bibitem{Kanemura:1993hm}
 S.~Kanemura, T.~Kubota, and E.~Takasugi,
 Phys.\ Lett.\  B {\bf 313} (1993) 155
 [hep-ph/9303263].

\bibitem{Horejsi:2005da}
 J.~Ho\v{r}ej\v{s}i and M.~Kladiva,
 Eur.\ Phys.\ J.\  C {\bf 46} (2006) 81
 [hep-ph/0510154].

\bibitem{Ginzburg:2005dt}
 I.~F.~Ginzburg and I.~P.~Ivanov,
 Phys.\ Rev.\  D {\bf 72} (2005) 115010
 [hep-ph/0508020].

\bibitem{Grimus:2007if}
 W.~Grimus, L.~Lavoura, O.~M.~Ogreid, and P.~Osland,
 J.\ Phys.\ G {\bf 35} (2008) 075001
 [arXiv:0711.4022 [hep-ph]].

 \bibitem{arXiv:0802.0060}
  P.~Osland, P.~N.~Pandita, and L.~Selbuz,
  Phys.\ Rev.\ D {\bf 78} (2008) 015003
  [arXiv:0802.0060 [hep-ph]].

\bibitem{Maksymyk:1993zm}
 I.~Maksymyk, C.~P.~Burgess, and D.~London,
 Phys.\ Rev.\  D {\bf 50} (1994) 529
 [hep-ph/9306267].

\bibitem{Grimus:2008nb}
 W.~Grimus, L.~Lavoura, O.~M.~Ogreid, and P.~Osland,
 Nucl.\ Phys.\  B {\bf 801} (2008) 81
 [arXiv:0802.4353 [hep-ph]].

\bibitem{Cheng:1973nv}
 T.~P.~Cheng, E.~Eichten, and L.~F.~Li,
 Phys.\ Rev.\  D {\bf 9} (1974) 2259.

\bibitem{Machacek:1981ic}
 M.~E.~Machacek and M.~T.~Vaughn,
 Phys.\ Lett.\  B {\bf 103} (1981) 427.

\bibitem{Grimus:2004yh}
 W.~Grimus and L.~Lavoura,
 Eur.\ Phys.\ J.\  C {\bf 39} (2005) 219
 [hep-ph/0409231].


\bibitem{Sikivie:1980hm}
  P.~Sikivie, L.~Susskind, M.~B.~Voloshin, and V.~I.~Zakharov,
  Nucl.\ Phys.\ B {\bf 173} (1980) 189.

\bibitem{Pomarol:1993mu}
  A.~Pomarol and R.~Vega,
  Nucl.\ Phys.\ B {\bf 413} (1994) 3
  [hep-ph/9305272].

  \bibitem{Gerard:2007kn}
  J.-M.~G\'erard and M.~Herquet,
  Phys.\ Rev.\ Lett.\  {\bf 98} (2007) 251802
  [hep-ph/0703051].

  \bibitem{deVisscher:2009zb}
  S.~de Visscher, J.-M.~G\'erard, M.~Herquet, V.~Lemaitre, and F.~Maltoni,
  JHEP {\bf 0908} (2009) 042
  [arXiv:0904.0705 [hep-ph]].

\bibitem{arXiv:1011.5228}
  B.~Grzadkowski, M.~Maniatis, and J.~Wudka,
  JHEP {\bf 1111} (2011) 030
  [arXiv:1011.5228 [hep-ph]].

\bibitem{Nishi:2011gc}
  C.~C.~Nishi,
  Phys.\ Rev.\ D {\bf 83} (2011) 095005
  [arXiv:1103.0252 [hep-ph]].

  \bibitem{Haber:1992py}
  H.~E.~Haber and A.~Pomarol,
  Phys.\ Lett.\ B {\bf 302} (1993) 435
  [hep-ph/9207267].

  \bibitem{arXiv:1007.1424}
  K.~Olaussen, P.~Osland, and M.~A.~.Solberg,
  JHEP {\bf 1107} (2011) 020
  [arXiv:1007.1424 [hep-ph]].

\end{thebibliography}
\end{document}